\documentclass[12pt,a4paper]{article}

\usepackage[utf8]{inputenc}
\usepackage{amssymb}
\usepackage{multicol}\usepackage[a4paper,left=3cm,right=3cm,top=3cm,bottom=3cm]{geometry}
\usepackage{setspace}
\usepackage{graphicx}
\usepackage{color}
\usepackage[hidelinks]{hyperref}
\graphicspath{ {images/} }
\usepackage{tikz}
\usepackage{amsmath}
\usepackage{physics}
\usepackage{ulem}
\usepackage{xcolor}
\usepackage{centernot}
\usepackage{amsthm}

\usepackage{fancyhdr}

\usepackage{centernot}

\begin{document}


\title{\Large \textbf{Rules and Meaning  in Quantum Mechanics} \bigskip \bigskip
}

\author{\Large  Iulian D. Toader}

\date{\Large  University of Vienna}

\maketitle

\newpage

\begin{center}
   \Large  \textbf{Preface}
\end{center}

\bigskip \bigskip \bigskip

\noindent This book concerns the metasemantics of quantum mechanics (QM). Roughly, it
pursues an investigation at the intersection of philosophy of physics and philosophy of language, and it offers a critical analysis of rival explanations of the semantic facts of standard QM. Such facts, about the meaning of its formalism, are typically
taken to be grounded in language-world relations or in the rule-based linguistic
practices of scientific communities. These relations and practices raise a series of important questions, two of which are taken up and discussed at length. The problem of completeness, understood as a concern with the categoricity of rules, is one focus
of the book, and it is discussed not only as a problem for standard QM but also for quantum logic (QL). The problem of permanence, understood as a concern with the preservation of rules, as for instance in the transition from classical physics to QM,
is another focus.

Both of these problems are analyzed in their historical contexts, but it should be clear that their philosophical significance transcends these contexts. The new results of this analysis include a rigorous reconstruction of Einstein’s incompleteness argument, which concludes that a local, separable, and categorical QM cannot exist; a reinterpretation of Bohr’s principle of correspondence, grounded in a conception of permanence as a metatheoretical principle; also, an improved meaning-variance argument about QL, which follows a line of critical reflections that was initiated by
Weyl; and an argument concerning semantic indeterminacy, raised against a recent version of inferentialism about QM, and inspired by Carnap’s categoricity problem for inferentialism about classical logic. Some of these arguments are drawn from
my previous papers, as indicated in the right places in the book, but most of that material has been rewritten here. Any remaining errors are, of course, entirely my responsibility.

\newpage

\tableofcontents

\newpage

\section{The metasemantics of quantum mechanics}

One cannot fully understand a theory -- \textit{any} theory -- without knowing how its meaning is determined, without figuring out the purported relationship between the language of the theory and its semantic attributes. Explaining this relationship is the main task of metasemantics. This chapter introduces this task, in general terms, and then presents representationalism and non-representationalism as rival metasemantic views about quantum mechanics (QM). The analysis of the overall hypothesis of the book, that metasemantics contributes to an adequate philosophical understanding of QM, will focus on two problems for these views: categoricity and permanence.

\subsection{Hilbert's question}

The metasemantics of QM, as understood in this book, is concerned with clarifying the relationship between the formalism of the physical theory and its semantics, quite apart from the nature or existence of the reality that the theory might be supposed to describe. In this sense, it is an attempt to address a particular case of a general question formulated by David Hilbert in his ``Axiomatic Thinking'', a lecture delivered in 1917 at the ETH in Zurich. He described there two development strategies that he took to characterize the axiomatic method: a progressive strategy, which leads from a system of axioms to what can be logically derived from them, and a regressive strategy, which leads to deeper axioms, uncovered by a critical examination of the system. Hilbert further mentioned a series of ``difficult epistemological questions'', such as the questions about the internal and external consistency, the independence, and the decidability of an axiomatic system, then he also specified ``the question of the relationship between \textit{content} and \textit{formalism} in mathematics and logic.'' (Hilbert 1918, 1113) Although this last question expresses a concern with classical mathematics and logic, it can be extended to physics as well. Indeed, Hilbert himself would later become very much concerned with the relationship between content and formalism in QM (Hilbert \textit{et al.} 1928). 

In the present book, I take up Hilbert's question by pursuing an investigation of the metasemantics of standard QM and quantum logic (QL). This investigation should be expanded, I believe, to the numerous interpretations and extensions, the information-theoretic reconstructions, and the various reformulations of QM, but these projects will not be considered here.\footnote{The only place where I will tentatively go beyond this scope is in section 5.3.3, in which I put forward some exploratory ideas about a metasemantics for QBism. I will also touch upon the relation between interpretations and reconstructions of QM, in section 5.2.1, where I consider several arguments that reconstructionists have proposed against the rational acceptability of realist interpretations, like the many-worlds, hidden variables, or spontaneous collapse interpretations. For an overview of these interpretations, with a special metaphysical knack, see Lewis 2016. For all things interpretational, see Freire Jr. \textit{et al.} 2022.} I will begin with a few introductory remarks on metasemantics (in section 1.2), as understood in philosophy of language, and then present two rival views on the metasemantics of QM in philosophy of physics: representationalism and non-representationalism. Afterwards, I will describe the structure of the book (in section 1.3) and the way I plan to address the two problems for these views that I will be concerned with.

\subsection{Representationalism vs. non-representationalism}

What is metasemantics? Roughly, just as metaethics is the philosophy of ethics, and metaphilosophy is the philosophy of philosophy, and perhaps just as metaphysics is (or ought to be, at least in part) philosophy of physics, so metasemantics is the philosophy of semantics. Here is how David Kaplan characterized it:

\begin{quote}

The fact that a word or phrase has a certain meaning clearly belongs to semantics. On the other hand, a claim about the basis for ascribing a certain meaning to a word or phrase does not belong to semantics. ``Ohsnay" means snow in Pig-Latin. That's a semantic fact about Pig-Latin. The reason why ``ohsnay" means snow is not a semantic fact; ... because it is a fact about semantics, as part of the Metasemantics of Pig-Latin (or perhaps, for those who prefer working from below to working from above, as part of the Foundations of semantics of Pig-Latin). (Kaplan 1989, 573sq) 

\end{quote}

Working from below, Robert Stalnaker considered the explanation of semantic facts as an essentially foundational investigation: 

\begin{quote}
    
First, there are questions of what I will call ``descriptive semantics''. A descriptive‐semantic theory is a theory that says what the semantics for the language is, without saying what it is about the practice of using that language that explains why that semantics is the right one. ... Second, there are questions ... of ``foundational semantics'', about what the facts are that give expressions their semantic values, or more generally, about what makes it the case that the language spoken by a particular individual or community has a particular descriptive semantics.\footnote{Cf. Stalnaker 1997, 903. Both quotations, from Kaplan and Stalnaker, are given in García-Carpintero 2012a (see also García-Carpintero 2012b) and in Burgess and Sherman 2014, which also describe other possible approaches to metasemantics.} 

\end{quote}

Quite generally, semantics assigns semantic attributes, i.e., content, meaning, and truth values, to expressions in a language (singular terms, sentences, etc.). To put it differently, semantics states or describes semantic facts, such as the fact that ``oshnay'' means snow, or the fact that ``snow'' means snow, for that matter. Metasemantics is the part of philosophy concerned with what explains such assignments, with what determines semantic facts. It is concerned, more exactly, with understanding the nature of the relationship between linguistic expressions and their semantic attributes, with uncovering the grounds of semantic facts. 

Typical questions in metasemantics ask for reasons why expressions have the meaning they have: what exactly makes it the case that they have that meaning, or as I prefer to put it, in virtue of what do they have that meaning? Two general answers to such questions are formulated in terms of a language-world relation of representation or, alternatively, in terms of rule-based linguistic practices. 

According to one answer, expressions have meaning in virtue of their representational properties. Why does ``Findus is on the mat'' mean that Findus is on the mat? It is because ``Findus is on the mat'' represents a state of affairs in which Findus is on the mat. Semantic facts are determined by representation relations. This answer characterizes what I will call a representationalist metasemantics. 

According to the alternative answer, expressions have meaning independently of their representational properties (if they have any), only in virtue of their use properties. Why does ``Findus is on the mat'' mean that Findus is on the mat? It is because ``Findus is on the mat'' is used in accordance with the communication protocols and practices of a community that shares the relevant language. Semantic facts are determined by communicational uses. This answer characterizes what I will call a non-representationalist metasemantics. 

The difference between representationalist and non-representationalist metasemantics can also be explicated in terms of rules, if one takes representation relations to be specified by semantic rules that postulate correspondences between linguistic expressions and extra-linguistic states of affairs, while communicational uses are specified by non-semantic rules which state conditions for the proper employment of expressions within a community of speakers. 

Note, however, that a non-representationalist metasemantics, according to which non-semantic rules do all the metasemantic work that is to be done, need not require that there be no semantic rules specifying the representational properties of expressions. For a semantic fact, e.g., ``Findus is on the mat'' means that Findus is on the mat, might be determined by use in accordance with communication protocols and practices, even though ``Findus is on the mat'' does actually represent a state of affairs in which Findus is on the mat. But the further exploration of this apparent gap between semantics and metasemantics falls beyond the scope of this book.

The two metasemantic views just presented can be easily identified in the case of QM. In a Schrödinger's cat scenario, the semantic fact that ``Findus is both dead and alive'' means that Findus is both dead and alive may be understood in virtue of that expression's representing a superposition of my cat's states.\footnote{With thanks and apologies to Sven Nordqvist.} A representationalist metasemantics of QM, just like a representationalist metasemantics of natural language, is intended to explain the relationship between expressions and their semantic attributes via representational properties. Semantic rules that specify representational properties for quantum expressions have been sometimes insistently demanded especially by philosophers who defend realist metaphysical views. 

For example, Tim Maudlin has maintained that ``The mathematical object [i.e., the $\psi$-function] must be supplemented with a physical ontology and semantic rules specifying how the physical ontology is represented by the wavefunction.'' (Maudlin 2016, 8) Similar statements can be found in his recent book on quantum theory: ``a wavefunction is a mathematical item -- as `function' testifies -- and the quantum state is whatever real physical feature of an individual system (if any) obtains iff the system is represented by a given wavefunction.'' (Maudlin 2019, 81) On this view, semantic rules postulate a correspondence between physical states and $\psi$-functions, between physical magnitudes and self-adjoint operators, etc. 

But what explains the semantic attributes of these mathematical expressions? In virtue of what do they have the meaning they are supposed to have? One way in which the representationalist can attempt to answer this question is by pointing to the representational properties specified by semantic rules. These rules are then taken to do double work: they \textit{both} assign semantic attributes to expressions \textit{and} explain in virtue of what these expressions have such attributes.

A non-representationalist metasemantics of QM, like a non-representationalist metasemantics of natural language, understands the relationship between expressions and their semantic attributes via use properties. In the Schrödinger's cat scenario above, the fact that ``Findus is both dead and alive'' means that Findus is both dead and alive is determined in virtue of the communicational use of this expression, rather than in virtue of its representing a superposition of cat states. A version of non-representationalism, dispensing with any semantic rules for quantum concepts, was arguably endorsed by Niels Bohr and then adopted by neo-Bohrians, implicitly or explicitly, ever since. 

For example, Richard Healey has declared that ``All our concepts, including those of physics, derive their content from use in human communication.'' (Healey 2017, 203) On this view, the semantic attributes of mathematical expressions involving $\psi$-functions and self-adjoint operators are determined in virtue of their inferential properties, rather than any purported representational properties. Semantic rules, if they are formulated at all, are simply idle: they \textit{neither} assign semantic attributes to quantum expressions, \textit{nor} explain in virtue of what these expressions have such attributes.

As in the case of natural language, a third view about the formalism of QM seems possible. Although semantic rules do not do any metasemantic work, they are not entirely idle, either, for on this view they specify the representational properties of quantum expressions even if non-semantic rules are needed to explain in virtue of what these expressions have such properties. Some mixture of Healey's and Maudlin's views might illustrate this third view, which allows a significant gap between the semantics and metasemantics of QM. But I will not be particularly concerned with such special possible mixtures (e.g., Bohmian inferentialism, Everretian QBism, etc.). 

In the philosophy of language, as explained above, metasemantics concerns the factors that determine the meaning of linguistic expressions in the natural language. These often include expressions with evaluative communicational uses, such as deontic modals (``should'', ``ought'', etc.) As already noted, I will focus on the meaning of expressions in the standard formalism of QM, as well as the meaning of the QL connectives. But if QM is reformulated so as to include expressions that have evaluative communicational uses, then a metasemantics of QM must further consider the factors that would explain the semantics of this reformulation. Again, I will not attempt to do this here in detail, although I will say something about it towards the end of the book when I envisage how a non-representationalist metasemantics for QBism might be developed, and explain why it should be developed in that way. 

The two explanations of the semantic facts of standard QM described above, the representationalist and the non-representationalist metasemantics, raise some important philosophical questions. A good grasp of their features, their problems, and their implications is beneficial, I would say even indispensable, to a philosophical understanding of at least some aspects of QM. Two particular problems that I will be concerned with in this book are the problem of \textit{categoricity} and the problem of \textit{permanence}, the descriptions of which I include in the presentation of the structure of the book, in the next section. A detailed consideration of these problems will throw some new light on, e.g., Einstein completeness as a notion of descriptive completeness, and on Bohr's correspondence principle as a metatheoretical principle. It will also help reject some deficient reasons against Putnam's quantum logical revisionism, and formulate new challenges for more recent views about QM.  

\subsection{Two problems: categoricity and permanence}

Categoricity has been extensively discussed in philosophy of mathematics and philosophy of logic, though not so much in philosophy of science, and almost not at all in philosophy of physics. As we will see, Hermann Weyl understood categoricity as the expression of a certain limitation of the descriptive completeness of a theory and took it to have important epistemological and metaphysical consequences. But
a categorical theory, i.e., a theory the models of which belong to one and the same isomorphism class, also has metasemantic consequences: the semantic attributes of
all expressions in its language can be precisely determined. Thus, I will say that a theory has a categoricity problem when its language is semantically indeterminate, i.e., the semantic attributes of its vocabulary cannot be precisely determined \textit{because} the theory allows non-isomorphic models. The problem of categoricity, in the case of standard QM, is the main focus of chapter 2.

Adopting a distinction between global and local models that a scientific theory can admit of, I will address the question whether, for any physical system, standard QM can admit of local models, i.e., models that are assigned to that individual system relative to a given physical location, such that all of them belong to the same isomorphism class. This makes the question about the categoricity of QM significantly different than the question about the categoricity of mathematical theories like, say, arithmetic. To address the former, I will first discuss some worries about whether the question is well posed, or whether its well posedness requires an interpretation of QM, where the notion of interpretation is understood as in the foundations of physics. But I will also discuss worries about whether well posedness demands a rational reconstruction of QM, i.e., a reformulation (or as some like to put it, less sympathetically, a regimentation) of the theory within a formal (ideally, first-order) language. Then I will turn to a critical discussion of the intuitive claim that the Stone-von Neumann theorem can be understood as a categoricity result. I will argue that this claim remains unjustified, if some of the assumptions behind the relation established by the theorem are not properly discharged. Another point that I will emphasize is that only that relation, but not its relata, can be formalized in a
first-order language.

Following up on an earlier suggestion by Don Howard that one could read Einstein completeness as categoricity, I will subsequently give a rigorous reconstruction of Einstein's incompleteness argument that clarifies the assumptions behind that suggestion and shows that his no-go result may be interpreted as claiming that a local, separable, and categorical QM cannot exist. The idea that the Stone-von Neumann theorem expresses a necessary, but not a sufficient, condition for categoricity, and the algebraic reasoning which supports the claim that, for the EPR system originally considered by Einstein, this condition fails, are central to my reconstruction. I take this to imply, as against a representationalist metasemantics, that the semantic attributes of a local and separable QM cannot be precisely determined by semantic rules because this theory allows non-isomorphic models. This suggests, at the very least, a new perspective on the Bohr-Einstein controversy, but I think it offers a better understanding of Einstein completeness than others have provided. The implications of my reconstruction for other debates in the foundations of QM, e.g., concerning the significance of Bell's theorem, are to be drawn elsewhere.

The main focus of chapter 3 is the problem of permanence. This is concerned with the application of the principle of permanence in the development of QM. This principle, which stipulates the preservation of central elements of theories (such as their rules or laws) when the theories and their domains are to be extended, has been almost exclusively a topic for the history of philosophy of mathematics. It expresses a type of methodological conservatism that still needs to be properly understood. However, in philosophy of physics, the principle of permanence has been largely ignored. Conceived of as a problem in metasemantics, its application raises concerns about the explanation of the semantic attributes of a theory developed in accordance with this principle. I will say that a theory has a permanence problem if preserving its predecessor's rules obscures the explanation of semantic facts.

There is good reason to believe that Bohr provided a solution to this problem. For, as we will see, he appears to have understood his own correspondence principle as grounded in a version of the principle of permanence that had been crucial to the development of 19th century mathematics, e.g., in the works of Hermann Hankel and many others. The analysis of the latter, and the emphasis on its metasemantic implications, and in particular the claim that preserving the rules of a theory plays a role in determining the meaning of its successor theory, will help explain why Bohr believed that QM should be regarded as a ``rational generalization'' of classical physics. Even more importantly, it will clarify, to the extent that this can be done, why he said that the meaning of QM is to be determined by its rules, and what he may have meant by that. Tersely expressed in his reply to the EPR paper, the non-representationalist metasemantics that I attribute to Bohr, as a solution to the permanence problem of QM, exposed his related doctrine of the necessity of classical concepts to a type of criticism that, I will argue, can be understood if one pays attention to the relationship between the principle of permanence and Bohr's correspondence principle. 

The principle of permanence was also applied in the transition from finite- to infinite-dimensional QM, in von Neumann's work with Garrett Birkhoff leading to their introduction of QL. Although partly justified by this principle, the preservation of QL rules was, however, not taken to imply a non-representationalist metasemantics. Thus, von Neumann's solution to the problem of permanence is different than Bohr's. This points to a fundamental difference between their approaches to QM, beyond the usual observation that they were almost opposite figures: Bohr, well known for his physical insight and his (alleged) lack of mathematical rigor; von Neumann, highly regarded for his mathematical genius, despite his (alleged) lack of physical insight. It turns out that a significant difference between them concerns the way in which they appear to have understood the relationship between rules and meaning in QM.  

Like the problem of permanence, the problem of categoricity arises not only for mathematical and physical theories; it arises also for logical theories. Here, instead of asking whether the models of a theory are all mutually isomorphic, the issue is whether a logical calculus admits of (classes of) truth valuations that are all mutually isomorphic (or, more precisely, whether all valuations are homomorphisms from the set of sentences in the language to the two-element set $\{0,1\}$ of possible truth values). The fact that this is not the case for the standard systems of classical logic (CL), i.e., finitary and single conclusions calculi, was proved by Carnap, who solved this categoricity problem by introducing a rejection rule and allowing valid arguments with multiple conclusions for propositional logic, and infinitary rules for first-order logic. But the fact that it is not the case for the QL introduced by Birkhoff and von Neumann had been already shown, as we will see, by Weyl.

The line of critical reflection initiated by Weyl, concerning the truth-valuational semantics of QL, will be further pursued in chapter 4, where I focus on arguments developed after Hilary Putnam enrolled QL in the service of realist metaphysics, which was intended to oust what he took to be some unpalatable metaphysical hypotheses, e.g., the existence of hidden variables. One of these arguments, leveled against Putnam's logical revisionism by Geoffrey Hellman, is based on an assumption about the meaning of logical connectives and their truth-functionality. I will explain how, on the same assumption, one can significantly improve the argument and explain why the semantic attributes of QL connectives are indeed different than the semantic attributes of their CL counterparts. Putnam's revisionism, i.e., the view that QL is an alternative to CL, in the sense of asserting different truths about the same connectives, can finally be buried, metasemantically.

When properly taken into account, the improved argument also takes the steam out of Ian Rumfitt's more recent claim that Putnam's revisionism is not only unnecessary on empirical grounds, but impossible on rational grounds. The claim is based on the observation that the proof against distributivity in QM is rule-circular, which arguably shows that it cannot be used to rationally adjudicate against CL. But this conclusion can be resisted, and I will demonstrate that it remains possible to rationally adjudicate against CL in QM even if QL connectives were considered semantically equivalent to their CL counterparts. I will then explain how to solidly withstand Timothy Williamson's imperious attack on non-classical logics, in the particular case of QL, by recalling the simple, though seemingly forgotten fact that QL is not inconsistent with the application of classical mathematics in QM. Rather, QL was actually introduced as a result of this application.

In chapter 5, I will return to a non-representationalist metasemantics for QM. For the case of natural language, non-representationalism was defended by Wilfried Sellars, who proposed a global inferentialist metasemantics: ``There is nothing to a conceptual apparatus that isn't determined by its rules.'' (Sellars 1953, 337) Drawing on Robert Brandom's further development of Sellars' view, Healey has embraced inferentialism, as already noted above, as a metasemantics for standard QM. After describing his version of quantum inferentialism in some detail, and then comparing it with the more encompassing expressivist view defended by Huw Price for the whole of science, I will present two preliminary objections to Healey's view, one from recent information-theoretic reconstructions of QM, and the other from relativistic extensions of Wigner's friend scenario. 

The first objection points out that the metasemantics of reconstructed QM cannot be inferentialist, since without correspondence rules that assign information-theoretic properties to quantum concepts the semantic attributes of the latter cannot be explained. Thus, further work would be required to show that inferentialism and reconstructionism are compatible. The second objection shows that Healey's attempt to make the relativistic extensions of Wigner's friend scenario compatible with the objectivity of QM weakens his inferentialism, since it adds a semantic constraint on truth valuations, which is not strictly based on inferential rules. Likewise, the quantum inferentialist has further work to do to save objectivity inferentially.

Inspired by the categoricity problem for logical inferentialism, I will then formulate a potentially more devastating problem for Healey's inferentialism. This new categoricity problem, I will argue, is raised by the failure of the rules of material inference in standard QM to determine, relative to a system and a physical location, a quantum state assignment -- a local model -- that is unique up to isomorphism. Against inferentialist metasemantics, I take this to imply that the semantic attributes of the standard formalism of QM cannot be precisely determined by its inferential rules because the theory allows non-isomorphic models. The problem shows that the objection according to which inferentialism is ``unable to nail down a claim's content sufficiently to explain how this may be unambiguously communicated in public discourse'' (Healey 2017, 211) has not been successfully addressed. To overcome this semantic indeterminacy, the inferentialist should address the categoricity problem in a way that is fully compatible with inferentialism. 

The semantic indeterminacy of standard QM may, of course, be inconsequential for the scientific practice, just like it is negligible for our daily argumentative practice that logical connectives are semantically indeterminate. But for inferentialism as a metasemantics of QM, it is a serious problem, since it points out that the meaning of quantum expressions fails to be articulated in just the way the inferentialist claims it is in fact articulated. However, I will further make a suggestion, on behalf of the inferentialist, as to how the problem might be solved. The suggestion is to reformulate both Schrödinger's equation and decoherence explicitly as inferential rules. 

There is another possible answer to the categoricity problem, which is to simply ignore it precisely because it seems inconsequential for the scientific practice. But I will argue that, even if one is willing to bite the bullet of semantic indeterminacy, as a QBist might be inclined to do on practical grounds, one would still need to develop a metasemantics, on philosophical grounds. Further, I will propose and motivate the development of a non-representationalist, non-inferentialist metasemantics for the QBist reformulation of standard QM, which purports to explain the meaning of quantum expressions on the basis of their decision-theoretic properties, rather than their representational or inferential properties.

In the concluding chapter of the book, I will describe two projects of my own, which I regard as very much worth pursuing. The first project starts from the observation that categoricity can, and perhaps should, be conceptualized as the expression of an ideal of logical perfection. Historically, this indicates a line that still needs to be carefully drawn from Einstein's characterization of principle theories to a recent development due to Boris Zilber. The latter illustrates and defends the view that first-order languages should not be discounted in the metasemantic analysis of QM, despite their L\"owenheim-Skolem properties. This will demand a reconceptualization of the very notion of categoricity, but this looks like a small price if the gain might be, among other things, the semantic determinacy of quantum expressions. 

The second project reconsiders the seemingly unworkable project of a Carnapian rational reconstruction of QM, to show that it can be rescued from the deadly confines of 20th century criticism, and that it could be taken as a basis for the metasemantic analysis of QM after all. One of the major obstacles to be overcome in this project, beside the (in)famous division between observational and theoretical terms to which any rational reconstruction is usually taken to be committed, is the implications of G\"odel's second incompleteness theorem. For the theorem implies that the consistency of any viable rational reconstruction cannot be proved in an informative way, which seems to further imply that a viable rational reconstruction of QM cannot even be given. Nonetheless, progress in this direction can be made, I believe, if one weakens the epistemic goals associated with, and the epistemic claims derived from, this Carnapian project. 

The book ends by emphasizing that a metasemantic analysis of scientific theories, which was here motivated by Hilbert's question about the relationship between the content and formalism of classical logic and mathematics, should also take into account his own views on the matter. The claim that metasemantics, conceived of as a foundational investigation, contributes to an adequate philosophical understanding of theories could, thus, ultimately be credited to Hilbert.

\newpage 

\section{The problem of categoricity}

This chapter introduces the categoricity problem for a representationalist metasemantics of quantum mechanics (QM). Some intended limitations of the descriptive completeness of scientific theories are first presented, motivated by the view advocated by Hertz and Weyl, that some things are beyond what a complete theory ought to tell us about its domains of application. After specifying a proper notion of categoricity for standard QM, and responding to some possible objections, an intuitive reading of the Stone-von Neumann theorem as a categoricity result is critically discussed. Drawing on this, a new reconstruction of Einstein's incompleteness argument is proposed.\footnote{This chapter is partly based on Toader 2018, 2021a, and 2023.}

\subsection{Intended limitations of descriptive completeness}

Various types of completeness were formulated, before and after the early 1930s, when von Neumann presented his system of axioms for QM in his \textit{Mathematical Foundations of Quantum Mechanics}. For instance, Russell and Whitehead had stated that a system of axioms for pure mathematics, like the one presented in their \textit{Principia Mathematica}, is complete just in case it is able to capture ``as much as may seem necessary'' of the domain that it aims to describe, or more precisely, just in case it logically derives the whole class of theorems of ordinary mathematics. Similarly, von Neumann maintained that a system of axioms for QM is complete just in case it is able to derive all statistical formulas of QM. 

There are, of course, limitations to the kind of descriptive completeness that Russell and Whitehead, and von Neumann had in mind, limitations that are clearly unintended by the proponent of an axiomatic system. G\"{o}del proved, on some arguably reasonable assumptions, that there are statements in the class of theorems which cannot be derived in the system of \textit{Principia Mathematica}, and he took this to show that system necessarily incomplete.\footnote{For discussion of Russell and Whitehead's notion of descriptive completeness, and an argument that G\"{o}del's own understanding of completeness, as negation-completeness, is in fact relevantly different than their descriptive variety, see Detlefsen 2014.} Likewise, von Neumann noted that the statistical nature of QM formulas could be thought to betray ``an ambiguity (i.e., incompleteness) in our description of nature'', and this could motivate the elimination of this incompleteness by the addition of statements about hidden dynamical variables to the theory.\footnote{Nevertheless, von Neumann rejected such additions as inconsistent with his system of axioms. See Acuña 2021 and Mitsch 2022 for recent re-evaluations of (the debate on) von Neumann's proof of the descriptive completeness of QM.} 

There are also intended limitations of descriptive completeness. This is indicated not by true statements that are provably impossible to derive in a system of axioms, but by those that are epistemically defective or illegitimate, and thus ought not to be derivable in a system of axioms. One argument to this effect (presented in section 2.1.1) was given by Hermann Hertz, who justified an intended limitation of descriptive completeness on the basis of a notion of mental comfort. Another argument to a similar effect (discussed in section 2.1.2) was offered by Hermann Weyl, who thought that descriptive completeness should be constrained by an invariantist notion of objectivity. Both arguments concern the nature of the representation relation between the language of a scientific theory and the phenomena within its domain of applicability. Even a cursory look at these arguments will help us better appreciate, when we later (in sections 2.2.1 -- 2.3.2) turn to QM, the overall significance of the question about its categoricity.

\subsubsection{Hertz on illegitimate questions}

Hertz famously diagnosed a certain tendency to ask what he called ``illegimate questions'' in science, e.g., questions asked in mechanics about the essence (\textit{Wesen}) of force or electricity, expressed his skepticism that they can be properly answered, and offered a solution as to how to avoid asking such questions in the first place. Here is the relevant passage in full: 

\begin{quote}
    
Weighty evidence seems to be furnished by the statements which one hears with wearisome frequency, that the essence of force is still a mystery, that one of the chief problems of physics is the investigation of the essence of force, and so on. In the same way electricians are continually attacked as to the essence of electricity. Now, why is it that people never in this way ask what is the essence of gold, or what is the essence of velocity? Is the essence of gold better known to us than that of electricity, of the essence of velocity better than that of force? Can we by our conceptions, by our words, completely represent the essence of any thing? Certainly not. I fancy the difference must lie in this. With the terms ``velocity'' and ``gold'' we connect a number of relations to other terms; and between all these relations we find no contradictions which offend us. We are therefore satisfied and ask no further questions. But we have accumulated around the terms ``force'' and ``electricity'' more relations than can be reconciled amongst themselves. We have an obscure feeling of this and want to have things cleared up. Our confused wish finds expression in the confused question as to the essence of force and electricity. But the answer which we want is
not really an answer to this question. It is not by finding out more and fresh relations and connections that it can be answered; but by removing the contradictions existing between those already known, and thus perhaps by reducing their number. When these painful contradictions are removed, the question as to the essence of force will not have been answered; but our minds, no longer vexed, will cease to ask illegitimate questions.\footnote{Cf. Hertz 1899, 7sq. I have replaced the term ``nature'' in the translation of ``\textit{Wesen}'' by ``essence''.} 
\end{quote}

Contradictions are identified when the relations between the statements of science, including statements about force, are carefully attended to. Once identified, they are perceived as painful because, as Hertz assumed, they create a certain kind of mental vexation. But since, as seems natural to believe, one desires mental comfort, and since introducing more relations by adding statements about the essence of force cannot remove those contradictions, one should instead focus primarily on removing the contradictions. And one should continue to remove contradictions until mental comfort is achieved, to the extent that the very tendency to ask illegitimate questions is eliminated.\footnote{See Eisenthal 2021, for an interesting discussion of a particular ambiguity in Newton's laws of motion, which appears to be due to the use of a slightly different notion of force in the third law than the notion used in the first and second laws, and which Hertz thought created mental vexation with respect to the meaning of ``force''.} 

This argument, it seems to me, can be best understood on the background of a 19th century view according to which the removal of contradictions is a necessary and sufficient goal of scientific inquiry. Peirce expressed this view when he characterized consciousness as perceived consistency: ``Again, consciousness is sometimes used to signify the \textit{I think}, or unity in thought; but the unity is nothing but consistency, or the recognition of it.'' (Peirce 1868, CP 5.313) He argued that scientific inquiry lacks proper motivation when there is no doubt that is caused by inconsistencies: ``When doubt ceases, mental action on the subject comes to an end; and, if it did go on, it would be without a purpose.'' (Peirce 1877, CP 5.376) Once contradictions are removed, mental vexation ceases, belief settles, and purpose-driven inquiry stops. 

Unlike Peirce, however, Hertz thought that mental comfort was not the end of scientific inquiry, but just a precondition for asking only legitimate questions. It is worth emphasizing how Hertz believed one may achieve such mental comfort in science. His proposal, as I see it, is not to remove contradictions by reducing the number of relations between the statements of a theory, but rather the other way around, to reduce the number of such relations by eliminating contradictions.\footnote{On this point, one commentator writes: ``The solution is to remove the contradictions by reducing the number of relations in which the term `force' stands -- in effect by determining a single meaning for the term `force'.'' (Kremer 2012, 204sq) But it seems to me that this gets Hertz's ``solution'' backwards. Indeed, how could a precise meaning determination, even assuming that this can be achieved by a reduction of the number of relations between statements, be enough for consistency? In any case, the removal of contradictions requires a delicate consideration of essential and inessential relations (cf. Hüttemann 2009).} For by eliminating all identified contradictions, the number of relations between the statements about force and other statements in the axiomatic system of mechanics would be reduced. And it could in principle be reduced to such an extent that a precise meaning for the term ``force'' would be determined. Having determined, in this way, a precise meaning for all scientific terms, illegitimate questions about essences would not arise any more. The real question remained, of course, whether consistency would be enough for a precise determination of meaning. 

\subsubsection{Weyl's universal boundary}

Like Hertz, but on rather different grounds, Weyl later justified a similar limitation of descriptive completeness. He gave an argument to the effect that, in science, statements about the essence (\textit{Wesen}) of any objects are epistemically insignificant to such an extent that science should not really care about such statements at all. Also like Hertz, Weyl further expressed skepticism that questions about essence could actually be answered in scientific terms, and then offered a solution as to how science could stop or avoid asking such questions:    

\begin{quote}
A science can determine its domain only up to an isomorphic mapping. In particular it remains entirely indifferent as to the `essence' of its objects. That which distinguishes the real points in space from number triples or other interpretation of geometry one can only \textit{know} by immediate, living intuition. ... The idea of
isomorphism designates the natural insurmountable boundary of scientific cognition. This thought has clarificatory value for the metaphysical speculations about a world of things in themselves behind the phenomena. ... Thus, even if we do not know the things in themselves, still we have just as much cognition about them as we do about the phenomena. (Weyl 1949, 26)
\end{quote}

Unlike Hertz, however, Weyl did not build his argument on a notion of mental comfort. Rather, he started from an invariance-based conception of scientific objectivity, which he imposed as a constraint on a theory's descriptive completeness. According to him, a scientific theory is objective only if it does not ask questions about the essence of objects, about their intrinsic or non-relational properties. All possible answers to such questions would require immediate intuition, but this violates the objectivity constraint. This is because such answers can be true in one interpretation of a theory, but false in another interpretation of the same theory. If objectivity is taken as the primary epistemic ideal of science, and so if objectivity gives the measure of epistemic significance, all statements ascribing intrinsic properties to objects must be rejected as epistemically insignificant. 

But they are not only epistemically insignificant, for in science all questions concerning the essence of objects are actually illegitimate. Weyl thought the notion of isomorphism indicated a way of demarcating science from pseudo-science. In a lecture titled ``Scientific Method'', he noted:

\begin{quote}

Before any closer discussion I should like to point to that universal boundary of all theories as revealed in the idea of isomorphism. ... With science thus proclaiming its own barriers, we must always be afraid that some high-sounding soothsayers may come along and attempt to settle with their pseudo-science beyond the border. They believe they have the arcanum disclosing to them the essence of things, which science is admittedly unable to grasp. However, the border which we have pointed out here is not a border towards a different brand of knowledge; it is rather a border beyond which there is nothing. (Weyl 1953) 

\end{quote}

Since relations can remain invariant across the interpretations of a theory, science can offer objective knowledge only about the structure of relations, i.e., about the structure of relational properties of all objects in a domain. It is exactly this intended limitation of descriptive completeness that led Weyl to identify categoricity as a more appropriate notion of completeness for scientific theories:

\begin{quote}
    
One might have thought of calling a system of axioms complete if the meaning of the basic concepts present in them were univocally fixed through the requirement that the axioms be valid. But this ideal cannot be realized, for the isomorphic mapping of a contentual interpretation is surely just another contentual interpretation. The final formulation is therefore this: a system of axioms is complete, or \textit{categorical}, if any two contentual interpretations of it are necessarily isomorphic. (Weyl 1949, 25) 
\end{quote} 

In light of the unintended limitations of descriptive completeness, already mentioned above, one must ask whether Weyl's universal boundary can be reached at all, and if so, under what conditions exactly. Can any scientific theory describe the relations between objects in its domain as Weyl thought should be described, i.e., uniquely up to isomorphism? Is it possible to characterize QM, in particular, as a  categorical theory? How should the very notion of categoricity be understood to make it applicable to a theory like QM? And what might QM, thus characterized, tell us about the relationship between its formalism and its semantics?

Note that, on Weyl's view, non-categoricity, that is the existence of mutually non-isomorphic interpretations of a theory, has primarily epistemological implications. If QM, for example, failed to be categorical, then it would fail to be objective, since there would be statements in its language that are true in one interpretation of the theory, and false in another interpretation.\footnote{Note that here I am following Weyl's use of  ``interpretation" to designate what we would call a model of a theory.} While it is not clear what his view on the categoricity of QM was, Weyl seems to have thought that at least some physical theories are non-categorical and, as a consequence, lack objectivity. He lamented theis as a failure:

\begin{quote}
    
Objective Being, reality, becomes elusive; and science no longer claims to erect a sublime, truly objective world above the Slough of Despond in which our daily life moves. ... the objective Being that we hoped to construct as one big piece of cloth each time tears off; what is left in our hands are -- rags.\footnote{Cf. Weyl 1954, 627. For my analysis of what might have motivated this view, see Toader 2018. Part of that analysis assumes that it is the failure of the Stone-von Neumann theorem (discussed further below, in section 2.2.3) in quantum field theory that led Weyl to think that science remains unable to achieve objectivity. It is unclear what his view was about the relationship between this theorem and the categoricity of QM.}

\end{quote}

The remainder of this chapter deals with the question about the categoricity of standard QM, raised as a problem for a representationalist metasemantics, that is one in which semantic rules do all metasemantic work. I will be primarily interested in how to formulate an appropriate notion of categoricity for QM, and I will then respond to several possible objections, before assessing whether QM might be considered categorical or not. Unlike Weyl, however, I do not propose to support a structuralist view of science.\footnote{For such views, more or less following Weyl, see, e.g., van Fraassen 2008, on the anti-realist side, and French 2016, on the realist side.} My focus will remain mostly on metasemantics, rather than the epistemology or metaphysics of science.

\subsection{The categoricity of quantum mechanics}

A theory is categorical with respect to an isomorphism class of models if and only if all models of the theory are in that class. A model is a relational structure on a particular domain, the elements of which can be assigned by an interpretation function to elements of the language of the theory. The language, together with the rules for manipulating its symbols, constitute the syntax of the theory, while models constitute its semantics. Thus, I will say that a categorical theory has a unique semantics up to a suitable isomorphism. In this sense, for example, first-order Peano arithmetic is not categorical and, thus, it does not have a unique semantics because it allows both standard and non-standard models. Likewise, classical logic is not categorical and, thus, it does not have a unique semantics, either, since it allows both normal and non-normal models, or more exactly, non-normal valuations, on which some of the logical connectives behave differently than on the normal valuations.\footnote{See Carnap 1943 for the construction of non-normal valuations of classical logic (to which I will return in section 5.3.1), and Skolem 1955 for the construction of a non-standard model of arithmetic. The close relationship between model theory and the study of semantics goes back to Tarski, and was then widely adopted: e.g., ``The modern form of semantics is the theory of models.'' (Mostowksi 1965, 115) and ``The name `model theory' obviously refers literally to any discussion of the relationship: (structure $\mathcal{A}$) is a model of (sentence $\sigma$). Thus it is more or less synonymous with `scientific semantics'.'' (Vaught 1974, 153) More recently, however, Wilfrid Hodges professed to challenge this relationship, arguing that ``the three areas of research -- model theory, the definition of logical consequence and model-theoretic semantics -- are quite different and they have hardly anything in common beyond a connection with models in the sense associated with Alfred Tarski.'' (Hodges 2016, 174) For a book-length discussion of the philosophical significance of model theory, including the metasemantic significance of categoricity, see Button and Walsh 2018.}

But is QM categorical? Can the notion of categoricity be applied to its standard axioms? In the historical development of quantum theory, categoricity concerns appear to have been first raised in two particular contexts, or so I will argue: one is Einstein's 1935 argument for the incompleteness of QM, and the other is Weyl's 1940 argument for the incompleteness of quantum logic (QL). A rigorous reconstruction of the former, which identifies the assumptions behind an understanding of Einstein completeness as categoricity, will be given in section 2.3.2. An analysis of the latter will be offered in section 3.3.2.  

I shall start with a few preliminary considerations (in section 2.2.1) about the kind of categoricity that can be attributed to QM, and then specify a proper notion of categoricity for its standard axioms (in section 2.2.2). My main interest, for the moment, is to point out that the well posedness of the categoricity problem for QM depends on what one takes the theory to be, as well as on what one takes its models to be. Related to this, I also respond to two immediate objections: that QM cannot be categorical because it incorporates first-order Peano arithmetic, which as just noted is not categorical; and that if QM were categorical, it could not have the wide applicability that the scientific practice tells us it actually does have. Then I will take up (in section 2.2.3) the question whether the Stone-von Neumann theorem can be read as a categoricity result for QM, and if so, in what sense.

\bigskip

\subsubsection{Preliminary considerations}

The general problem of categoricity, formulated independently of any particular theory, has been indicated by Tarski already in the 1930s:

\begin{quote}
A non-categorical set of sentences (especially if it is used as an axiom system of a deductive theory) does not give the impression of a closed and organic unity and does not seem to determine precisely the meaning of the concepts contained in it. (Tarski 1934, 311)     
\end{quote}

Thus, a theory has a categoricity problem when the meaning of expressions in its language is not determined precisely by its axioms because the latter admit of non-isomorphic models. As I see it, the problem is, thus, a metasemantic one. Whatever Tarski might have meant by ``closed and organic unity'', such an impression would be given, I take it, by a precisely determined semantics. But the language of a non-categorical theory is, or at least seems to be, semantically indeterminate, i.e., indeterminate with respect to the semantic attributes of its linguistic expressions. It is semantically indeterminate with respect to the truth values of its sentences: some sentences are true in some models, but false in others. In contrast, a categorical theory determines (or seems to determine) the meaning of its concepts precisely, and its models agree on all theorems in its language. 

For example, the meaning of arithmetical expressions is precisely determined if arithmetic is categorical with respect to the isomorphism class of an omega sequence, that is if it allows only models in which arithmetical vocabulary is used in a natural way, i.e., in agreement with the natural numbers. Likewise, the meaning of classical logical connectives (such as disjunction and negation) is precisely determined if classical logic is categorical with respect to the isomorphism class of a two-element Boolean algebra, that is if it allows only valuations on which the connectives are used in the normal way, i.e., in agreement with the normal truth tables. Categoricity appears, therefore, to be a strongly desirable property of theories. At least one reason for this is that it blocks semantic indeterminacy by allowing us to pin down a unique semantics for the language of a theory, like the natural numbers for arithmetic and the normal truth tables for classical logic. 

However, in the case of first-order theories with infinite models, metatheoretical properties like compactness or L\"owenheim-Skolem, as well as G\"odel incompleteness, frustrate the attempt to establish categoricity. Therefore, any success in this respect requires either augmenting the language of a theory or adjusting its logic. Full second-order logic, for example, proves the categoricity of arithmetic, although there are strong and well-known arguments against such proofs, e.g., based on the fact that second-order logic is not a deductively complete logic. Furthermore, since second-order quantification is non-categorical, in this case the categoricity problem just gets pushed up a level: in order to pin down the natural numbers up to isomorphism, one must first be able to pin down a unique semantics for the second-order quantifiers up to isomorphism. Furthermore, it also turns out that any attempt to prove the categoricity of arithmetic by means of a logic weaker than second-order, but stronger than first-order, similarly presupposes the categoricity of that underlying logic (Read 1997). 

Such considerations suggest that, despite its desirability, categoricity must remain beyond reach for most theories: one can never actually pin down the unique semantics of a theory without first having to pin down the unique semantics of a background theory. Linguistic expressions must, therefore, remain semantically indeterminate. But is categoricity really indispensable? Some argued that the existence of non-standard models for first-order Peano arithmetic is no real concern, because any two of its non-isomorphic models are arithmetically indiscernible, in the sense that the application of induction within the language of the theory yields arithmetical results that hold in any model. Thus, all models of arithmetic agree on all theorems in its language (Resnik 1996). Arguably, this would be enough for a precise determination of the semantic attributes of arithmetical expressions as well. But the fact that provability obliterates some differences between non-isomorphic models does not guarantee that one pins down a unique semantics for arithmetic. For there necessarily exist unprovable sentences in the language of the theory on which non-isomorphic models would in fact disagree. Semantic indeterminacy might remain a problem.   

More importantly, it has also been noted that some other mathematical theories, like the theories of fields, rings, groups, algebras, etc., do not actually have any categoricity problem. For their non-categoricity is a result of design and mathematical acumen, and as such, it should not be considered a failure, but a theoretical success. No unique semantics is typically supposed to exist for such theories, and no unique semantics can be intended in such cases. Normally, these theories have a variety of distinct non-isomorphic models, and they may all be thought of as intended. Of course, this does not mean that categoricity, if it can be obtained, would necessarily be a disadvantage. Some categorical algebraic theories have some rather nice features like quantifier elimination. But the semantic indeterminacy of their linguistic expressions does not seem to be considered a problem, although the reasons for this are not always clear. This view, which has been extended also to arithmetic and set theory, stems from considerations put forth by logicians and mathematicians in the early part of the 20th century, who believed that non-categoricity is a theoretical virtue, rather than a liability, and that one should formulate axioms that are weak enough to admit a multitude of non-isomorphic models (see, e.g., Zermelo 1930). 

Now, I think it's fair to say that, by comparison to philosophy of mathematics and philosophy of logic, the problem of categoricity does not loom large in philosophy of science, and is unfortunately rather absent in philosophy of physics. Is categoricity a desirable feature of scientific theories, as Weyl took it to be, even if perhaps for different reasons than he emphasized? Should philosophers of science be concerned with what Tarski called the ``closed and organic unity'' of a theory? Are metatheoretical properties like the ones mentioned above (i.e., compactness, L\"owenheim-Skolem, and G\"odel incompleteness) relevant for purported attempts to establish categoricity in this context? Might categoricity be dispensable? Is the physical indiscernibility of the non-isomorphic models of a physical theory, if it has such models, all one should care about? Should non-categoricity be regarded as a theoretical success, rather than a problem? Or would the existence of non-isomorphic models of a physical theory be enough reason to be concerned about the semantic indeterminacy of its language? Such questions, as far as I can tell, have remained largely unexamined.

To the extent that they have been noted, however, they have been treated rather inadequately. On the one hand, categoricity has been rejected by philosophers of science as a kind of ``rigidity or inflexibility'' that should have no place in physics (see, e.g., Bunge 1973). I will return to this general idea below, when I discuss the objection from the wide applicability of QM. But on the other hand, non-categorical scientific theories have been considered too abstract, or more abstract than categorical ones. This is because the class of models of the former is too large and diverse to allow a simple mental image of a typical model (like the natural numbers, say) despite dropping some of its complex properties. Such level of abstractness should arguably be avoided in science (Suppes 1993, 74). As I am especially interested in the particular case of QM, I can only begin to address the questions above with respect to this theory. And the first thing that needs to be clarified is what we might mean to say when we say that QM is a categorical theory.

What does it mean to say that QM has a unique semantics up to isomorphism? Does it mean to say that it has a unique \textit{interpretation}, where this notion is understood as in the foundations of physics? Since there are so many mutually non-isomorphic interpretations of QM, such as $\psi$-ontic and $\psi$-epistemic interpretations, the categoricity problem would turn out to be really trivial: QM \textit{is} a non-categorical theory, admitting models as distinct as, e.g., the pilot wave model, dynamical reduction models, etc. My answer to this question is that it is actually misleading to consider any of the $\psi$-ontic and $\psi$-epistemic interpretations of QM as models, since such interpretations are typically extensions of QM, extensions that add, for example, extra dynamical laws to the Schrödinger equation. But the extensions of a theory are surely not its models, although these extensions -- as theories in their own right -- may admit as models some suitable expansions of the models of QM. The pilot wave model, dynamical reduction models, etc., are precisely such expansions. They are models of QM's dynamical extensions.

Furthermore, is the categoricity problem well posed in the case of standard QM? Is it possible to even ask whether its formalism can give the impression of a ``closed and organic unity''? Shouldn't \textit{this} require, at the very least, interpretation? My answer is that surely categoricity would be a well posed problem even if QM were only ``a very effective and accurate \textit{recipe} for making certain sorts of predictions'' (Maudlin 2019, 2), not a full-fledged theory, but rather ``a general method, a framework in which many theories can be developed'' (Lalo\"e 2019, 13). This view of QM is typically motivated by the belief that a full-fledged physical theory must provide a physical ontology, which implies that QM can be considered a theory in its own right only if QM has been given an appropriate interpretation. However, this view is highly controversial (see, e.g., Fuchs and Perez 2000). It is, therefore, contentious to say that only after an ontology has been given can the categoricity problem be well posed for QM. Actually, in chapter 5, I will formulate a categoricity problem for a neo-Bohrian approach to QM that rejects any quantum ontology. In any case, and perhaps more importantly, shouldn't the well posedness of the categoricity problem require, if not an interpretation, then a rational reconstruction of QM, i.e., its reformulation in a formal (ideally, first-order) language? I will come back to this question in section 2.2.3, where I discuss the Stone-von Neumann theorem.

\subsubsection{The standard axioms and their local models}

The categoricity problem, as understood here, concerns the standard axioms or rules of QM, which can be stated as follows: 

1. The state space of a physical system corresponds to an infinite-dimensional complex Hilbert space $\mathcal{H}$, such that the quantum state of a physical system is a mathematical function $\psi : T \rightarrow \mathcal{H}$ (or an equivalence class of unit-norm vectors of $\mathcal{H}$), defined at each time instant $t \in T$, where $T \subseteq \mathbb{R}$ denotes a time interval. 

2. The set $\mathcal{A}$ of dynamical quantities of a physical system (e.g., position, momentum, spin, polarization, etc.) corresponds to the set of self-adjoint operators acting on $\mathcal{H}$, such that the possible values of any variable are contained in the spectrum of its corresponding operator. 

3. The unitary time evolution of a physical system is described by the Schrödinger equation $i \partial_t \ket{\psi}_t=H\ket{\psi}_t$, for some operator (the Hamiltonian) $H \in \mathcal{A}$. 

4. For each unit-norm vector $\ket{\phi} \in \mathcal{H}$, $\mu_{\phi}$ is a measure on the set of subspaces of $\mathcal{H}$ such that for an arbitrary subspace $S \subseteq \mathcal{H}$: $\mu_{\phi}(S)=\bra{\phi}P_S\ket{\phi}$, where $P_S$ is the projector on $S$.

The language of these axioms includes terms for physical concepts like ``state
space”, ``quantum state”, ``dynamical quantities”, etc., as well as symbols for their associated mathematical objects like ``Hilbert space”, ``functions”, ``self-adjoint
operators”, etc. The axioms are semantic rules that postulate correspondences
such as between state spaces and Hilbert spaces, between quantum states and
$\psi$-functions, . between dynamical quantities and self-adjoint operators, etc. On a representationalist metasemantics, these rules themselves are taken to do metasemantic work, i.e., to explain in virtue of what the formalism has semantic attributes.
Supplementing these rules, as a $\psi$-ontic or a $\psi$-epistemic interpretation might do,
is thus not metasemantically necessary.  

Alternatively, as Howard Stein once suggested, one could take the mathematical expressions in the formalism to have ``the status of theoretical terms, whose empirical application has to be given by some sort of `correspondence rules’.” (Stein 1970, 96) On this view, correspondence rules will do metasemantic work by explaining the semantic attributes of the formalism via the representational properties of suitable observational terms, rather than via the representational properties of theoretical terms. This view assumes, however, a distinction between
theoretical and observational vocabulary that has been long considered problematic
by philosophers (for more on this, see section 6.2 below). In any case, there are more recent accounts of the representational capacities of QM, which do not require a solution to the measurement problem (such as, e.g., the account given in van Fraassen 2008). But whatever the account, I think that representationalism cannot
avoid the categoricity problem.

In order to be able to address this problem, and thereby assess whether the meaning of quantum expressions can be precisely determined, one needs to further clarify exactly what the models of the standard axioms are. An influential view in the philosophy of quantum physics, which I will adopt here, is the following: ``Quantum mechanics, we may say, uses the \textit{models} supplied by Hilbert spaces.'' (Hughes 1989, 79) Similarly, and more recently: ``The models of NRQM [i.e., non-relativistic quantum mechanics] are Hilbert spaces, along with a suitable subalgebra of the bounded operators on that Hilbert space.'' (Weatherall 2019, 7) This view suggests that one should take a model of QM to be a complex structure $\langle \mathcal{H}, \mathcal{A}, T, H, \psi, \mu_{\phi} \rangle$ over a physical domain. Postponing rigorous definitions until the next section, let me note that since such structures are taken to describe individual physical systems, they are to be understood as \textit{local}, rather than \textit{global}, models (Hughes 2010).

Two immediate objections can now be raised against the claim that QM can be a categorical theory. The first objection goes as follows: 

\begin{quote}
    
[A] corollary of G\"odel's first incompleteness theorem asserts that any theory as powerful as or more powerful than Peano arithmetic -- in first-order formulation -- will be not only deductively incomplete but also non-categorical, which is to say that it will have models that are not isomorphic to one another. And since any moderately sophisticated theory in physics will incorporate a mathematical apparatus as powerful as or more powerful than arithmetic, the same will be true of our physics. (Howard 2012)

\end{quote}

According to this objection, even if the question of categoricity is well posed for standard QM, one can only settle it in the negative: QM cannot be categorical because, like most physical theories, it implicitly incorporates first-order Peano arithmetic, which allows non-isomorphic models and is, thus, non-categorical. More generally, the categoricity of QM would seem to require that all mathematical theories, including its background logic, incorporated as components in the physical theory, must be categorical. While a lot more needs to be said about this extremely important issue, I think that this objection assumes a questionable transfer of non-categoricity from the mathematical (and the logical) components to the physical theory. Conversely, it assumes a questionable transfer of categoricity from the physical theory to all of its mathematical  (and logical) components. It is doubtful that a categorical physical theory could entail the categoricity of its mathematics (or its logic): in the case of QM, this would require that the isomorphism that holds between its local models, relative to a physical system, could define an isomorphism that holds between the models of its mathematics (as well as one that holds between the models of its logic). But this requirement obviously fails even within mathematics itself: there are categorical theories, like second-order arithmetic, that incorporate first-order, and so non-categorical, arithmetic. In other words, the requirement fails because the isomorphism that holds between second-order models cannot define one that holds between the first-order models of arithmetic.

The second objection states that if QM were categorical, then it would be incapable of application to the wide variety of physical systems to which in fact it can be, and has been, successfully applied. The wide applicability of QM seems to require that its models be non-isomorphic to one another. Indeed, even if they were constructed on infinite-dimensional Hilbert spaces, which are indeed mutually isomorphic, the algebras of operators corresponding to the dynamical magnitudes of different physical systems must still have different structures. 

This objection is in line with the general idea noted above that a categorical theory would be too rigid or inflexible to be acceptable as a physical theory. But I think that the objection simply calls for a more exact specification of the notion of categoricity that can be properly attributed to standard QM. The specification is as follows: QM is categorical with respect to an isomorphism class of models if and only if all models of the theory, \textit{relative to an individual physical system}, are in that class. The relativity clause is justified by the fact, already emphasized above, that an algebra of operators describes the dynamical magnitudes of an individual physical system. This is why in QM, as already noted, the relevant notion of a model is that of a local model, which always specifies the algebra of operators and the Hamiltonian for an individual system. We might dispense with the relativity clause, and instead introduce a new notion, \textit{absolute categoricity}, defined as follows. Let $\mathcal{QM}$ be an abstract quantum theory, which leaves the algebra of operators and the Hamiltonian unspecified. Let QM($\mathcal{A}, H$) be a concrete quantum theory, which specifies the algebra of operators and the Hamiltonian for any individual system. Then we can say that $\mathcal{QM}$ is absolutely categorical if and only if, for any algebra of operators and any $H \in \mathcal{A}$, QM($\mathcal{A}, H$) is categorical. Obviously, if categoricity fails, then absolute categoricity fails as well. But for my purposes, the specification that includes the relativity clause will do.

I turn now to the main question: for any physical system, is there an isomorphism class of quantum models such that all local models of the system are in that class? For simplicity, I will address here the following version of the question: is there an isomorphism class of Hilbert space representations such that all representations of the Weyl algebra associated to that system are in that class? The existence of such an isomorphism class is usually taken to be established by the Stone-von Neumann theorem, first conjectured by Marshall Stone in 1930 and then proved by von Neumann in the following year. Whether one is justified to interpret this theorem as a categoricity result is discussed in the next section. The outcome will be used in my subsequent analysis and reconstruction of Einstein's argument for the incompleteness of QM.

\subsubsection{The Stone-von Neumann theorem}

The Stone-von Neumann theorem states that any irreducible, faithful, and regular Hilbert space representation of the Weyl algebra, which describes a quantum mechanical system (or, more generally, any system with a finite number of degrees of freedom), is uniquely determined up to a unitary transformation.\footnote{Cf. Stone 1930 and von Neumann 1931. For more details, see Ruetsche 2011, esp. section 2.3.} A Weyl algebra $\mathfrak{A}$ is a C$^{*}$-algebra generated by the Weyl form of the canonical commutation relations. A Hilbert space representation $(\mathcal{H}, \pi)$ is a $^{*}$-homomorphism $\pi : \mathfrak{A} \rightarrow B(\mathcal{H})$, where $B(\mathcal{H})$ is the set of bounded linear self-adjoint operators on an infinite-dimensional complex Hilbert space $\mathcal{H}$. A representation is irreducible if no (nontrivial) subspace of $\mathcal{H}$ is invariant under the operators in $\pi(\mathfrak{A})$. If $\pi$ is a $^{*}$-isomorphism, then the representation is also faithful. A faithful and irreducible representation is also regular if the operators in $\pi(\mathfrak{A})$ are weakly continuous. In the representation theory of C$^{*}$-algebras, the Stone-von Neumann theorem entails that any two irreducible, faithful, and regular representations $(\mathcal{H}_{1}, \pi_{1})$ and $(\mathcal{H}_{2}, \pi_{2})$ of $\mathfrak{A}$ are unitarily equivalent if and only if there is an element $U\in\mathfrak{A}$ which acts as an operator $U:\mathcal{H}_{1}\rightarrow\mathcal{H}_{2}$ such that $UU^{*}=U^{*}U=1$ and $\pi_{1}(A)$ = $U\pi_{2}(A)U^{*}$ for all elements $A\in\mathfrak{A}$. If $\mathcal{H}_{1}$ and $\mathcal{H}_{2}$ are separable, infinite-dimensional spaces, so their orthonormal bases are both countably infinite, $U$ intertwines all bounded linear self-adjoint operators on the two spaces. Thus, there is an isometric isomorphism between $\mathcal{H}_{1}$ and $\mathcal{H}_{2}$ that underlies the unitary equivalence of $(\mathcal{H}_{1}, \pi_{1})$ and $(\mathcal{H}_{2}, \pi_{2})$. 

Unitary equivalence is typically taken to entail physical equivalence, in the sense that the quantum states described as density matrices in unitarily equivalent representations assign the same expectation values to corresponding physical observables (Weyl 1930, 407; Ruetsche 2011, 24sq). This is, for example, the sense in which one has come to speak of the physical equivalence of the Schr\"odinger and the Heisenberg representations of a quantum mechanical system: time evolution on states, in the Schr\"odinger representation, is the same as time evolution on observables, in the Heisenberg representation.\footnote{See Perovic 2008 for discussion of the sense and scope of the equivalence before von Neumann's 1931 proof of Stone's 1930 conjecture.} For this reason, the Stone-von Neumann theorem has been sometimes intuitively read as proving the categoricity of standard QM.\footnote{Cf. St\"oltzner 2002, 45. More recently, I also maintained that the Stone-von Neumann theorem ``can be naturally read as a categoricity result.'' (Toader 2018, 21; see also Toader 2021a) The present section offers a more nuanced view on the matter.} 

But in what sense can this theorem be read in this way, if at all? More precisely, is the relation of unitary equivalence sufficient for constraining all local models of QM, relative to an individual physical system into one isomorphism class? The central question is if the kind of isomorphism that underlies unitary equivalence might be enough for categoricity, understood as in the previous section. This question is raised here for standard QM, and will be answered negatively, but I will presently consider the point that the very consideration of the question would actually require a rational reconstruction of QM, i.e., a reformulation of the theory in a formal language. More specifically, I will argue that, on the one hand, the Stone-von Neumann theorem could be intuitively read as establishing categoricity only if all its assumptions, including regularity, could be unproblematically justified. On the other hand, if one insisted on rational reconstruction, the theorem could be rigorously read as establishing categoricity only if all components of standard QM could be formalized, ideally within a first-order language. However, this is at least \textit{prima facie} questionable.

The intuitive reading of the theorem, just mentioned, fails to take into account a crucial assumption concerning the nature of Hilbert space representations: the regularity assumption. Thus, it ignores the existence of representations of the Weyl algebra, e.g. the position representation (with position eigenstates, but no operator for momentum) and the momentum representation (with momentum eigenstates, but no operator for position), for which there is no intertwiner, and thus no isometric isomorphism exists between them. These are the so-called non-regular representations, which are unitarily inequivalent to the Schrödinger representation, as well as to one another. The expectation values of corresponding physical observables are different in such representations than in the Schrödinger representation. The existence of non-regular representations strongly suggests that the Stone-von Neumann theorem cannot be understood as a categoricity result for standard QM: if non-regular representations are allowed by the standard formalism, and they are considered as non-isomorphic local models of the theory, this would be enough to make QM non-categorical.

Can non-regular representations be eliminated in standard QM? Arguably, these representations have physical significance, and despite their non-regularity, position and momentum representations can be taken to describe the behavior of a physical system with a finite number of degrees of freedom. Actually, they have been used to rigorously articulate Bohr's notion of complementarity (Halvorson 2004). Against this view, the objection has been recently raised that, if non-regular representations describe a physical system, then its dynamics cannot be the unitary dynamics governed by Schrödinger's equation. This is because non-regular states, i.e., the states on the Hilbert spaces of these representations in which positions or momenta have determinate values, are unitarily inaccessible from regular states (as well as mutually inaccessible). The claim that non-regularity is incompatible with unitarity builds on an unpublished result by David Malament, which states that if a free dynamics is assumed in the position representation, then exact localizability is violated (Feintzeig \textit{et al.} 2019, 127). Thus, one either changes the standard formalism of QM by replacing the Weyl algebra with a different mathematical structure that does not allow non-regular representations,\footnote{Cf. Feintzeig and Weatherall 2019. The introduction of a different structure than the Weyl algebra may nevertheless allow the recovery of non-regular states as approximations or idealizations of regular ones: ``non-regular quantum states should be considered unphysical for essentially the same reasons that classical states at infinity are considered unphysical.'' (Feintzeig 2022, 472)} or one eliminates non-regular representations in standard QM as physically insignificant on account of their incompatibility with the unitarity of quantum dynamics. If the latter option turned out to be unproblematic, then we would have a sense in which the Stone-von Neumann theorem could be intuitively read as a categoricity result for standard QM.

The intuitive reading is, however, further questionable if one insists that the well posedness of the categoricity problem actually demands a formal axiomatization of QM, or what Carnap called a rational reconstruction. For it might seem doubtful that categoricity, as a notion of mathematical logic, can be applied to QM, since the language of its standard axioms (as formulated in the previous section) is not strictly speaking a formal language. But if one demands that, before the problem can be even properly discussed, one should provide a rational reconstruction of standard QM as a formal system in which a consequence relation is defined and logical rules and the quantum-mechanical rules are formally expressed, then the trouble is of course that there exists no such reconstruction of QM yet, and even worse, that no such reconstruction seems possible. 

I will return to the question whether a rational reconstruction of QM is possible, in chapter 6, where I will argue against the two most usual philosophical objections to the Carnapian project of rational reconstruction, pointing out that if one assumes a global non-representationalist metasemantics, then a viable rational reconstruction of QM seems entirely possible. But here I want to make two related points. 

The first point is that I see no reason against specifying a notion of categoricity for standard QM -- the notion that I have actually specified in the previous section -- even in the absence of a rational reconstruction of the theory. After all, the concept of categoricity, and even the term ``categoricity'', historically predate the emergence of formalized languages and theories. So in this book I will continue to say that QM is categorical with respect to an isomorphism class of models if and only if all local models of the theory, relative to an individual physical system, are in that class. And I will take standard QM as a basis for metasemantic analysis, rather than waiting until a rational reconstruction becomes available.

One might, of course, consider the notion I have specified as a mere analogue of the ``legitimate'' notion of categoricity, and allow the latter to be strictly applied to formalized theories only. In this case, I suggest that the relationship between the two notions, in the case of interest, should be understood as follows: a formalized QM is not categorical, in the strict sense, if standard QM is not categorical, in the sense I have specified. This is because any two mutually non-isomorphic models of the standard axioms (or any two unitarily inequivalent Hilbert space representations) must be formalized as non-isomorphic models of a rational reconstruction of QM. For otherwise, one would be committed to the unacceptable claim that formalization can make unitarily inequivalent representations, including non-regular position and momentum representations, indistinguishable (from one another, as well as from the Schrödinger representation). 

However, one might also be unfazed by the possibility that a formalization of QM makes unitarily inequivalent representations indistinguishable as models of a formalized QM. Then one might perhaps argue that formalized QM can be categorical, in the strict sense, even if standard QM is non-categorical, in the sense I have specified. In this case, I would say that standard QM is \textit{weakly non-categorical} (adopting a notion from Bernays 1966). But this would surely not change what I have to say about the metasemantics of standard QM, although (as Bernays maintained, for the case of mathematical theories) it might provide some insight into formalization.

The second point is this. Even if one accepts my suggestion that a formalized QM is categorical, in the strict sense, only if standard QM is categorical, in the sense I have specified, there might be additional reasons to suspect that one cannot show a formalized QM categorical, beside the existence of non-regular Hilbert space representations in standard QM. Put differently, even if standard QM would allow just regular representations, and would thus be categorical, in the sense I specified, that is not enough to establish the categoricity of formalized QM. This is because it is impossible to formalize QM \textit{in first-order logic}: while the relation of unitary equivalence established by the Stone-von Neumann theorem can be formalized as a first-order relation, representations cannot be formalized as first-order models.

It is worth developing this argument in a bit more detail. Consider the claim that the relation of unitary equivalence makes the Heisenberg representation of the Weyl algebra of a quantum system, and its Schrödinger representation, intertranslatable. More specifically, let's assume that intertranslatability is meant to account for the fact that these representations are physically equivalent. How should one understand intertranslatability more precisely? And what could be its semantic counterpart?

What intertranslatability may be taken to mean is that ``[a]ny data which elements of the [one representation] accommodate, counterpart elements of the [other representation] accommodate as well – and \textit{mutatis mutandis} for falsifying data''.\footnote{Cf. Ruetsche 2011, 45. Note that Ruetsche speaks of Heisenberg and Schrödinger \textit{theories}, rather than representations. Their intertranslatability is understood in the more general sense that had been articulated by Clark Glymour: ``[The intertwiner of representations] provides the translation manual [Glymour] is after.'' (\textit{loc. cit.}) Glymour's manual helps translating between theories and it ``guarantees that all and only theorems of [one theory] are translated as theorems of [another theory], and conversely.'' (Glymour 1970, 279)} The data accommodated by unitarily equivalent representations, as also noted above, are the expectation values assigned to corresponding physical observables in states described as density matrices in those representations. Thus, taking the statements about expectation values as theorems (Clifton and Halvorson 2001, 430), we can say that representations are intertranslatable in the sense that they agree on all theorems. More exactly, they are intertranslatable in the sense that they have a ``common definitional extension'' (Glymour 1970, 279), i.e., they have definitional extensions that derive the same theorems. In other words, unitarily equivalent representations are intertranslatable in the sense that they are definitionally equivalent.

Note that if we consider Hilbert space representations \textit{qua} local models of QM, the semantic counterpart of intertranslatability, whatever relation that might turn out to be, can obtain only if the representations have expansions that satisfy the same theorems. This implies that the semantic counterpart cannot be understood as categoricity, since non-isomorphic structures can satisfy the same theorems. Furthermore, if understood as definitional equivalence, intertranslatability is a first-order relation, which implies that its semantic counterpart could hold between Hilbert space representations \textit{qua} local models only in the sense that they satisfy all and only first-order theorems. But representations cannot be formalized as first-order models, on account of the metric completeness of the Hilbert space, a property that cannot be expressed in a first-order language. If some higher-order statements have physical significance, and there is no immediate reason to think that they cannot have such significance, then unitary equivalence, when reconstructed as definitional equivalence, is insufficient to account for physical equivalence. 

Alternatively, one might want to reconstruct unitary equivalence as Morita equivalence, rather than definitional equivalence, as suggested by Jonathan Rosenberg: ``The `modern' approach to the Stone-von Neumann Theorem, which is somewhat more algebraic, is due to Rieffel ... The key observation of Rieffel is that the theorem is really about an equivalence of categories of representations, or in the language of ring theory, a Morita equivalence.'' (Rosenberg 2004, 342) The Stone-von Neumann theorem, as formulated by Marc Rieffel, states the following: ``Every irreducible Heisenberg G-module is unitarily equivalent to the Schrödinger G-module.'' (Rieffel 1972) Unitary equivalence is reconstructed as the Morita equivalence of G-modules, which are Hilbert space representations together with the collection of all intertwining operators between them. But this does not fare better than the previous one, if one looks for a reconstruction of unitary equivalence that can account for physical equivalence within the confines of a first-order language. 

Note that the semantic counterpart of Morita equivalence could obtain between G-modules considered \textit{qua} local models of QM only if they have Morita expansions that satisfy the same theorems. However, for the same reason as above, this implies that the semantic counterpart cannot be categoricity. Moreover, Morita equivalence is actually a generalized definitional equivalence relation (Barrett and Halvorson 2016), and thus it is also first-order, but Rieffel's G-modules, just like the Hilbert space representations, and for the same reason, cannot be formalized as first-order models. This suggests, once more, that while the Stone-von Neumann theorem might be intuitively read as a categoricity result for standard QM, in the sense and under conditions that I specified above, this reading could hardly be maintained if one demanded a first-order formalization of QM.\footnote{This raises further questions about the possibility and, of course, the acceptability of higher-order formalizations of QM, but these are questions that I will not address here. Nevertheless, in section 6.1, I will point to one contemporary project, which attempts to formalize QM in \textit{continuous first-order logic} (which can express the completeness of the Hilbert space).}  

Nevertheless, as in the case of mathematical theories (discussed in section 2.2.1), one might be inclined to respond by insisting that the categoricity of QM, formalized or not, is not really indispensable. Indeed, one might be inclined to adopt the attitude that logicians and mathematicians have adopted in the case of set theory and arithmetic, learn to appreciate non-categoricity as a theoretical virtue of QM, and simply ignore or otherwise minimize its metasemantic consequences. Although he considered the categoricity of set theory a desirable property (von Neumann 1925, 412), von Neumann himself never considered the Stone-von Neumann theorem (as far as I have been able to determine) as a result that established the categoricity of his QM. Moreover, when he later developed his axiomatic theory of games, for example, he expressly noted that this is intended as non-categorical (von Neumann and Morgenstern 1944, section 10.2). 

Along quite similar lines, Miklós Rédei has recently introduced the apt notion of ``intended non-categoricity'' to characterize the objective to construct physical theories that allow for multiple non-isomorphic models (Rédei 2014, 80). This objective is typically exemplified by theories like quantum field theory (QFT), which clearly invalidates the Stone-von Neumann theorem, and so allows for unitarily inequivalent representations (e.g.,  representations of the C$^{*}$-algebra on a free field, but also representations on an interacting field). This failure of the Stone-von Neumann theorem for quantum systems with an infinite number of degrees of freedom has been correctly, I think, considered as an indication that theories describing such systems, like quantum statistical mechanics (in the thermodynamic limit) and QFT, are not categorical: 

\begin{quote}
    
Our best current theory is QFT. It is a relativistic theory (in the sense of special, not general relativity), and it is a theory of systems with an infinite number of degrees of freedom. As such, in its most natural algebraic form, it can be shown to possess representations that are, of necessity, unitarily inequivalent. This is the algebraist's way of saying that the theory is not categorical, that it does not constrain the class of its models up to the point of isomorphism. (Howard 2011, 231)

\end{quote}

As I have argued in this section, it appears to be the case that QM is not categorical, either. More exactly, due to the existence of non-regular representations of the Weyl algebra in standard QM, which are also necessarily unitarily inequivalent, the Stone-von Neumann theorem cannot be understood as a categoricity result for standard QM, unless such representations are unproblematically eliminated. If this cannot be done, on account of their physical significance, then the theorem makes no metasemantic difference between QM and QFT. In particular, one cannot claim that the meaning of standard, non-relativistic QM is more precisely determined than that of QFT.

In the next section, drawing on my analysis so far, I will provide a new, rigorous reconstruction of Einstein's argument for the incompleteness of QM, which is one of the few historical contexts in which categoricity concerns appear to have been raised about QM. As I will explicate it, the argument establishes that a local, separable, and categorical QM cannot exist. If this is correct, then one of the most famous no-go results in the history of the theory turns out, rather intriguingly, to have metasemantic significance.

\subsection{Einstein completeness as categoricity}

Einstein's argument for the incompleteness of QM, which did not make it into the EPR paper (Einstein, Podolski, and Rosen 1935) in the way Einstein thought it should have, was clearly formulated in letters to Schrödinger and Popper, as well as in several publications (e.g. Einstein 1936). After Arthur Fine brought it to philosophical attention (Fine 1981), Don Howard suggested that the argument might be understood as deploying a notion of completeness known as categoricity (Howard 1990). This suggestion was motivated by Einstein's claim that QM fails to assign a unique wavefunction to the real state of one subsystem of an EPR system, since the assignment depends on the measurement that could be performed on the other subsystem. If multiple wavefunctions can be assigned to the same subsystem, and if one is justified in considering them as (parts of) non-isomorphic models, then this would be enough to show QM non-categorical. If Howard's suggestion is taken seriously, then Einstein completeness turns out to be a rather different type of completeness than the one articulated in the EPR paper. 

In this section, I will provide a reconstruction of Einstein's argument, which I think can clarify the assumptions underlying an understanding of Einstein completeness as categoricity. The key idea to be rigorously articulated is the following: ``if one understands a theoretical state as, in effect, a model for a set of equations plus boundary conditions ..., then Einstein's conception of a completeness requirement should really be understood as a categoricity requirement.'' (Howard 1992, 208) To stay as close as possible to Einstein's own argument, I will focus on the original EPR state, with observables having a continuous spectrum, suitably defined within an algebraic framework (Arens and Varadarajan 2000, Werner 1999), and I will explain under what conditions one would be justified to read Einstein completeness as categoricity. On my reconstruction, the argument assumes that representations on (tensor products of) Hilbert spaces are the models of QM of (composite) systems. It assumes as well that the unitary equivalence of such representations is a necessary, though not sufficient, condition for categoricity.\footnote{For a discussion of these assumptions, see sections 2.2.2 and 2.2.3 above.} The argument then points out that there are representations of the (tensor product of) algebras describing a subsystem of an EPR system that are not unitarily equivalent. As such, Einstein's argument concludes that categoricity is inconsistent with separability and locality.

It is, of course, difficult to say that the reconstruction I propose is actually entirely faithful to Einstein's own thought. One worry one might raise is that while my reconstruction works for the original EPR state, it does not work for entangled spins, since on finite-dimensional Hilbert spaces there are no unitarily inequivalent representations. But Einstein, as is well known, never cared much about Bohm's version of the EPR argument.\footnote{The only place where Einstein formulated a spin version of his argument appears to be in a late manuscript from around 1955. For discussion, see Sauer 2007.} Focusing on the infinite-dimensional case is thus historically reasonable. Moreover, for Einstein's objection to stand, it is of course sufficient that his argument goes through in one case; it is not required that it should do so in all cases. In any event, I will argue that my reconstruction is preferable to others, according to which Einstein's argument should be taken to establish  ``overcompleteness'' (Lehner 2014) or unsoundness (Gömöri and Hofer-Szabó 2021), rather than non-categoricity.  

Furthermore, I will suggest that my reconstruction sheds some new light on the Bohr-Einstein controversy. As mentioned in the previous section, Bohr's doctrine of complementarity has been rigorously interpreted in terms of the unitary inequivalence of non-regular Hilbert space representations, which vindicates the view that Bohr's notion of completeness was significantly distinct from the descriptive completeness articulated in the EPR paper.\footnote{For an expression of this view, see, e.g., Norsen 2017, 148.} On my reconstruction, Einstein completeness fails precisely due to this unitary inequivalence. Thus, from an algebraic point of view, it appears that the sense in which Bohr thought QM was complete is exactly the sense in which Einstein argued it wasn't. From a metasemantic perspective, their views are precisely antithetical: whereas Einstein deplored QM's non-categoricity and the ensuing semantic indeterminacy, Bohr embraced non-categoricity as a theoretical asset and arguably ignored semantic indeterminacy as insignificant.\footnote{As we will see in section 5.3.3, QBism seems to be closest to Bohr on this matter.}

\subsubsection{Misconstruals of Einstein completeness}

Recall the completeness condition that the EPR paper purported to argue it is not satisfied by QM: ``Whatever the meaning assigned to the term \textit{complete}, the following requirement for a complete theory seems to be a necessary one: \textit{every element of the physical reality must have a counterpart in the physical theory}.'' (Einstein, Podolski, and Rosen 1935, 777) This condition has typically been understood to formulate a type of descriptive completeness, since it applies to a physical theory just in case that theory is able to describe all of the physical reality that it aims to describe. The EPR paper argued that QM does not satisfy the completeness condition because in the case of a system in an EPR state there are elements of physical reality, i.e., properties of a subsystem of that system, that the theory aims to describe, but fails to do so. 

Crucial to the EPR argument is the following criterion of reality: ``If, without in any way disturbing a system, we can predict with certainty (i.e., with probability equal to unity) the value of a physical quantity, then there exists an element of physical reality corresponding to this physical quantity.'' (\textit{loc. cit.}; italics removed) The role and character of this criterion, as well as the formal structure of the argument, have been often addressed.\footnote{Perhaps the most detailed formal reconstruction of the EPR argument has been given in McGrath 1978, where the following is also noted: ``Regrettably EPR equate two notions of completeness: `complete representation by a wave function' and `complete theory' are used interchangeably.'' (560) See also Gömöri and Hofer-Szabó 2021, for a nice discussion of the EPR criterion of reality, its indispensable role within the EPR argument for incompleteness, and its absence from Einstein’s own argument. For a more recent and comprehensive discussion of the EPR argument in its historical context, see Bacciagaluppi and Crull 2024.} But what has been often ignored is the fact that EPR completeness and Einstein completeness are not the same type of \textit{descriptive} completeness.\footnote{The fact that EPR completeness and Einstein completeness are not the same type of completeness has been, to my knowledge, first explicitly noted by Arthur Fine:  ``Einstein does not give [the latter] a catchy name, but ... [we can] call this more technical conception \textit{bijective completeness}.'' (Fine 1981, 72)} 

In one widely quoted passage from his letter to Schrödinger, Einstein explained his notion of completeness in the following way: 

\begin{quote}
    
In the quantum theory, one describes a real state of a system through a normalized function, $\psi$, of the coordinates (of the configuration-space). ... Now one would like to say the following: $\psi$ is correlated one-to-one with the real state of the real system. ... If this works, then I speak of a complete description of reality by the theory. But if such an interpretation is not feasible, I call the theoretical description `incomplete'. (Letter to Schrödinger, 19 June 1935; translated in Howard 1985, 179)

\end{quote}

 Einstein completeness can thus be attributed to QM if and only if there exists a one-to-one correlation between a $\psi$-function and the real state of a system the theory aims to describe. Einstein's point that a one-to-one correlation does not exist in the case of an EPR system is meant to be supported by his separability and locality assumptions, so his argument can be formulated in the following way:

\smallskip

1. Spacelike separated physical systems have real states, which cannot causally influence one another.

2. Consider a system with two subsystems, A and B, in an EPR state.

3. Thus, each subsystem has a real state, no matter what measurements can be carried out on its other subsystem.

4. QM assigns different $\psi$-functions to A, depending on which observable one can choose to measure on B.

5. But if QM is complete, these $\psi$-functions should be identical.

6. Thus, QM is incomplete.

\smallskip

 Most of my discussion below will be focused on premises 4 and 5. It should be clear that the notion of completeness in premise 5 is directly implied by that defined in the letter to Schrödinger. In a letter to Pauli, Schrödinger put it in the following metaphorical terms: ``He [i.e., Einstein] has a model of that which is real consisting of a map with little flags. To every real thing there must correspond on the map a little flag, and vice versa.'' (Schrödinger to Pauli, July 1935, translated in Howard 1990, 106) It immediately follows that a map with little flags is an Einstein complete description of physical reality only if the map is essentially unique, in the sense that there is a bijection between any two maps describing that reality. In the case of QM, by analogy, Einstein completeness requires that the $\psi$-function representing the real state of a system be essentially unique.

Howard's suggestion, that Einstein completeness is rather similar to categoricity, was first offered in a footnote to his presentation of Einstein's argument, in the following passage: 

\begin{quote}
    
A complete theory assigns one and only one theoretical state to each real state of a physical system. [Footnote: This is a curious conception of completeness, more akin to what is called in formal semantics ``categoricity.''] But in EPR-type experiments involving spatio-temporally separated, but previously interacting systems, A and B, quantum mechanics assigns different theoretical states, different `psi-functions,' to one and the same real state of A, say, depending upon the kind of measurement we choose to carry out on B. Hence quantum mechanics is incomplete. (Howard 1990, 64)

\end{quote}

 The same suggestion, a bit more provocatively formulated (hence, I take it, the ``outrageous'' qualification), was then uplifted to the main text of a later paper: 

\begin{quote}
    
Let me conclude with one really outrageous suggestion. ... the operative criterion of completeness in Einstein's thinking was this: a theory is complete if and only if it assigns a unique theoretical state, such as a psi-function, to every unique real physical state. But if one understands a theoretical state as, in effect, a model for a set of equations plus boundary conditions ..., then \textit{Einstein's conception of a completeness requirement should really be understood as a categoricity requirement}. In other words, Einstein is saying that a `complete' (read `categorical') theory is one that determines a unique (\textit{eindeutige}) model for the reality it aims to represent. (Howard 1992, 208; my emphasis) 

\end{quote}

The conception of completeness that Howard came to think should be credited to Einstein is that according to which a theory is complete if and only if all its models belong to one and the same isomorphism class. On this conception, what Einstein argued for is that QM is incomplete in the sense that, roughly, even though it may be able to describe all of the physical reality that it aims to describe, it ends up describing a lot more besides that as well.

My reconstruction of Einstein's argument, in an algebraic framework, will be given further below. That will require a preliminary consideration of premise 2, in particular, a properly algebraic definition of the original EPR state, for continuous observables. It will also require a discussion of premise 4, that is an algebraic account of the difference between the $\psi$-functions assigned to susbsystem A. The outcome of all that will, I hope, be three-fold: (i) a good understanding of the assumptions behind Howard's suggestion to read Einstein completeness as categoricity, (ii) an explanation of what makes Einstein completeness different from EPR completeness as a type of descriptive completeness, and (iii) an unequivocal sense of where some of the misreadings of Einstein completeness have gone wrong. Although my real emphasis will be on (i), let me briefly address (iii) right away, in order to candidly raise the reader's interest in what is to come later. 

Einstein's argument for incompleteness is sometimes considered too confused to take seriously. Klaas Landsman, for example, maintained that it is a ``muddled'' argument (Landsman 2006, 234) for the following reason: 

\begin{quote}
    
Unfortunately, Einstein (and EPR) insisted on a further elaboration of this disjunction [i.e., completeness or separability], namely the idea that there exists some version of quantum mechanics that is separable ... at the cost of assigning more than one state to a system (two in the simplest case). It is this unholy version of quantum mechanics that Einstein (and EPR) called `incomplete'. Now, within the formalism of quantum mechanics such a multiple assignment of states (except in the trivial sense of wave functions differing by a phase factor) makes no sense at all, for the entanglement property lying at the root of the non-separability of quantum mechanics is so deeply entrenched in its formalism that it simply cannot be separated from it. (\textit{op. cit.}, 227)

\end{quote}

However, if it implies that Einstein's argument requires that entanglement be relinquished, this is simply missing the point. One would have thought it obvious that, quite the contrary, the argument is essentially based on entanglement. Clearly, Landsman's criticism fails insofar as it neglects the fact that Einstein separability (expressed in premise 1 above) and entanglement (implicit in premise 2) are compatible.\footnote{See Murgueitio Ram\'irez 2020, for the point that Einstein's argument would be trivially unsound, if separability and entanglement were not compatible. Landsman argued that, as a consequence of Raggio's theorem, Einstein separability is mathematically equivalent to Bohr's doctrine of the necessity of classical concepts. But his argument takes Einstein separability as state decomposability, which according to Murgueitio Ram\'irez was not what Einstein actually had in mind.} The alleged reason why multiple assignment of states is nonsensical doesn't stand: there is no ``unholy version'' of QM, i.e., a version of the theory without entanglement, and Einstein did not think otherwise. 

Furthermore, and this is a more significant problem, Landsman seems to misconstrue Einstein's view on the difference between the $\psi$-functions assigned to subsystem A as a trivial phase difference. As we will see further below, the multiple assignment of $\psi$-functions that Einstein actually considered can be aptly interpreted algebraically in terms of the unitary inequivalence of non-regular Hilbert space representations. Were this nonsensical, Bohr's complementarity doctrine would appear to be nonsensical as well, but this is a doctrine that Landsman professes to defend.

Einstein completeness, and Einstein's argument that this cannot be attributed to QM, have been characterized by Christoph Lehner in these terms: ``Einstein ... concludes that the quantum mechanical description is not biunique because it is incomplete. ... This conclusion is not warranted logically.'' (Lehner 2014, 334) A description that is not biunique is one that assigns to subsystem A a $\psi$-function which is not correlated one-to-one with its real state. On this reconstruction, Einstein's second half of the argument would be as follows:

\smallskip

4$_{L}$. QM assigns different $\psi$-functions to A, depending on which observable one can choose to measure on B.

5$_{L}$. Thus, QM is incomplete.

6$_{L}$. Thus, QM is not biunique.

\smallskip

Lehner maintained that the inference from 5$_{L}$ to 6$_{L}$ is not justified. But this clearly puts Einstein's cart before his horse: the one-to-one correlation does not fail because of incompleteness; rather, its failure is the actual reason for incompleteness, just as the reason why the one-to-one correlation fails is provided by the possible assignment of different $\psi$-functions to A (expressed by 4$_{L}$). Moreover, since a $\psi$-function is just a theoretical description and such ``a description that is not invariant is not necessarily incomplete ... it is `overcomplete' or nonabsolute'' (\textit{loc. cit.}), i.e., it assigns to A a theoretical state that is correlated many-to-one with its real state, then what Einstein should have allegedly argued is the following (statements 1-3 in the argument are, again, unchanged):

\smallskip

4$^{L}$. QM assigns different theoretical states to A, depending on which observable one can choose to measure on B.

5$^{L}$. Thus, QM is overcomplete.

6$^{L}$. Thus, QM is not biunique.

\smallskip

The inference from 5$^{L}$ to 6$^{L}$ is immediate, but trivial. However, Lehner appears to reduce Einstein's incompleteness to a quite uninteresting case of empirical underdetermination. Jos Uffink has recently pointed out that reading Einstein incompleteness as overcompleteness is ``surprising'', since ``cases of overcompleteness are ubiquitous in physics... Indeed one might wonder whether overcompleteness
is a worrisome issue at all in theoretical physics.'' (Uffink 2020, 557) In QM, as Uffink emphasized, such cases of overcompleteness as illustrated by phase differences between $\psi$-functions are not worrisome at all. 

But, of course, Einstein did not think otherwise. As already noted, we will see presently that the essential differences between $\psi$-functions that Einstein did think worrisome can be aptly interpreted algebraically in terms of the unitary inequivalence of non-regular Hilbert space representations. If this is enough to conclude that QM is not categorical, then Lehner's reconstruction of Einstein's argument turns out to conflate the problems of categoricity and empirical underdetermination.

Along different lines, Márton Gömöri and Gábor Hofer-Szabó argued for a similar conclusion, that Einstein's argument should not really be taken to establish incompleteness: ``According to Einstein's later [than the EPR] argument, the Copenhagen interpretation is committed to the existence of elements of reality that cannot be out there in the world –- under the assumptions of locality and no-conspiracy. Hence, given these assumptions, the Copenhagen interpretation is \textit{unsound} –- as opposed to being incomplete.'' (Gömöri and Hofer-Szabó 2021, 13453) Supposing it right that Einstein intended his argument against the Copenhagen interpretation of QM, the second half of Einstein's argument would change as follows:

\smallskip

4$_{G}$. QM assigns different $\psi$-functions to A, depending on which observable one can choose to measure on B.

5$_{G}$. Thus, at least some $\psi$-functions represent states of A that cannot exist.

6$_{G}$. Thus, QM is unsound.

\smallskip

Gömöri and Hofer-Szabó readily acknowledge that their reconstruction of Einstein's argument is in conflict with his own understanding of completeness. Pursuing ``logical reconstruction'' in spite of what ``Einstein actually thought to argue'', they find his own notion ``not quite apt'', in part because it is not identical to the EPR notion of completeness (\textit{loc. cit.}). But one would have thought that the difference between EPR completeness and Einstein completeness cannot be sufficient to deny the latter as an apt notion, since there is nothing to indicate that EPR completeness is the only possible type of descriptive completeness attributable to physical theories. Still, their main reason for having Einstein's argument conclude that QM is unsound, rather than incomplete, is the incompatibility of the different $\psi$-functions assigned to subsystem A of an EPR system. This is correct as an understanding of the essential difference between $\psi$-functions, expressed by premise 4$_{G}$. In the next section, this incompatibility will be precisely interpreted algebraically as the unitary inequivalence of non-regular Hilbert space representations. If my interpretation is adequate, I do not see why that incompatibility would make QM unsound, rather than sound but non-categorical. To argue that my interpretation is adequate, I turn now to reconstructing Einstein's argument in the framework of C$^{*}$-algebras. 

\subsubsection{Reconstruction of Einstein's argument}

Taking Howard's suggestion seriously requires that one consider the theoretical states assigned to a system by QM as (parts of the) models of the physical theory. But it's not immediately clear how a $\psi$-function, and in particular one that represents the EPR state of a system, can be considered in this way. For one thing, such a $\psi$-function cannot be construed as a unit-norm vector in an infinite-dimensional Hilbert space $\mathcal{H}$. This is for multiple related reasons, such as that the $\psi$-function has infinite norm, and the (operators associated to) observables (like position and momentum) of the subsystems of an EPR system have continuous spectra, so their probability distribution will have a probability density that cannot be concentrated on a single point in $\mathcal{H}$. This is why the definition of the EPR state for a system composed of subsystems A and B as a unit vector in $\mathcal{H}_{A} \otimes \mathcal{H}_{B}$ for observables in subalgebras $\textbf{B} (\mathcal{H}_{A}) \otimes \mathcal{I}$ and $\mathcal{I} \otimes \textbf{B}(\mathcal{H}_{B})$ is not adequate (Arens and Varadarajan 2000, 638). But that definition has been generalized, and the EPR state has been identified with a normalized positive linear functional on $\mathfrak{A}_{A} \otimes \mathfrak{A}_{B}$ for observables in the von Neumann algebras $\pi (\mathfrak{A}_{A} \otimes \mathcal{I})''$ and $\pi (\mathcal{I} \otimes \mathfrak{A}_{B})''$ (Werner 1999, Halvorson 2000). The generalized definition will be used below to reformulate premises 4 and 5 in Einstein's argument. The essential difference between the theoretical states assigned to subsystem A of the EPR system will be interpreted in terms of the unitary inequivalence of non-regular representations of $\mathfrak{A}_{A} \otimes \mathcal{I}$ on $\mathcal{H}_{A} \otimes \mathcal{I}$. 

Let's start by considering observables $O_{A}\otimes I \in \textbf{B} (\mathcal{H}_{A}) \otimes \mathcal{I}$ and $ I \otimes O_{B} \in \mathcal{I} \otimes \textbf{B} (\mathcal{H}_{B})$. As defined by Richard Arens and Veeravalli S. Varadarajan, $\omega$ is an EPR state if and only if the joint distribution of $O_{A}\otimes I$ and $ I \otimes O_{B}$ is a measure on $\mathbb{R}^{2}$ concentrated on the diagonal, $\mu_{\omega}^{O_{A}\otimes I, I \otimes O_{B}}(\left\{ (x,x) | x\in \mathbb{R}\right\}) = 1$. Such pairs of observables are typically called EPR-doubles: the outcome of measuring one predicts with certainty the outcome of measuring the other. These EPR-doubles form type I factors (isomorphic to von Neumann algebras $\textbf{B} (\mathcal{H}_{A}) \otimes \mathcal{I}$ and $\mathcal{I} \otimes \textbf{B} (\mathcal{H}_{B})$ respectively), so they have a discrete spectrum (their distributions relative to $\omega$ take discrete values only; see Arens and Varadarajan 2000, 647). Thus, this definition is not enough to characterize the original EPR state -- the state that Einstein was concerned with. 

In order to overcome this limitation, Reinhard F. Werner considered $\mathfrak{A}_{A} \otimes \mathcal{I}$ and $\mathcal{I} \otimes \mathfrak{A}_{B}$ as the mutually commuting subalgebras of $\mathfrak{A}_{A} \otimes \mathfrak{A}_{B}$, and took $\pi (\mathfrak{A}_{A} \otimes \mathcal{I})''$ as a self-adjoint unital subalgebra of $\textbf{B} (\mathcal{H}_{A}) \otimes \mathcal{I}$, and $\pi (\mathcal{I} \otimes \mathfrak{A}_{B})''$ as a self-adjoint unital subalgebra of $\mathcal{I} \otimes \textbf{B} (\mathcal{H}_{B})$, both closed in the weak operator topology. Then $\pi(O_{A}\otimes I) \in \pi (\mathfrak{A}_{A} \otimes \mathcal{I})''$ and $\pi( I \otimes O_{B}) \in \pi (\mathcal{I} \otimes \mathfrak{A}_{B})''$ are EPR-doubles and have continuous spectra, since $\pi (\mathfrak{A}_{A} \otimes \mathcal{I})''$ and $\pi (\mathcal{I} \otimes \mathfrak{A}_{B})''$ are type II$_{1}$ factors (cf. Werner 1999). 

Now, consider the representations $(\pi, \mathcal{H}_{A} \otimes I)$ and $(\pi, I \otimes \mathcal{H}_{B})$ of $\mathfrak{A}_{A} \otimes \mathfrak{A}_{B}$. If $\omega$ is an original EPR state and $\pi$ is faithful, then there is a state $\tau$ in each of these representations such that for any element $O_{A} \otimes O_{B} \in \mathfrak{A}_{A} \otimes \mathfrak{A}_{B}$, we will have $\omega (O_{A} \otimes O_{B}) = \tau(\pi(O_{A} \otimes O_{B}))$ (cf. Halvorson 2000, 327). This entails, on the same conditions, that for any two representations $(\pi_{1}, \mathcal{H}_{A} \otimes I)$ and $(\pi_{2}, \mathcal{H}_{A} \otimes I)$ of $\mathfrak{A}_{A} \otimes \mathcal{I}$, there are different states $\tau_{1}$ and $\tau_{2}$ in these representations, respectively, such that for two different elements $O^{1}_{A} \otimes I, O^{2}_{A} \otimes I \in \mathfrak{A}_{A} \otimes \mathcal{I}$, we have the corresponding restrictions of $\omega$, that is $\omega(O^{1}_{A} \otimes I) = \tau_{1}(\pi_{1} (O^{1}_{A} \otimes I))$ and $\omega(O^{2}_{A} \otimes I) =\tau_{2}(\pi_{2} (O^{2}_{A} \otimes I))$. This suggests the following reformulation of the second half of Einstein's argument:

\smallskip

4$_{T}$. QM assigns different states, $\tau_{1}$ or $\tau_{2}$, to subsystem A, depending on which EPR-double, $\pi_{1} ( I \otimes O^{1}_{B})$ or $\pi_{2} ( I \otimes O^{2}_{B})$, one can choose to measure on B.

5$_{T}$. But if QM is complete, $\tau_{1}$ and $\tau_{2}$ should be identical.

6$_{T}$. Thus, QM is incomplete.

\smallskip

How should we understand premise 4$_{T}$? Within this framework, what makes $\tau_{1}$ and $\tau_{2}$ different states? As we have seen above, some commentators took Einstein's argument to be confused on this very point, for the reason that the only differences between such states allowed by standard QM are ``trivial'' phase differences. But this is not what Einstein actually had in mind, as Howard already pointed out: 

\begin{quote}
    
Might there not be situations in which the differences between two $\psi$-functions (phase differences, for example) are inessential from the point of view of the system whose real state they aim to describe? Einstein's completeness condition would, indeed, be too strong if it required that literally \textit{every} difference between $\psi$-functions mirror a difference in the real state of the system in question; but such was not Einstein’s intention. (Howard 1985, 181)

\end{quote}

 But what are then the non-trivial differences between theoretical states that Einstein did have in mind? Here is Howard, again: 

\begin{quote}
    
The kind of difference with which Einstein was concerned is clear from his argument: [$\tau_{1}$] and [$\tau_{2}$] differ in the predictions they yield for the results of certain objective, local measurements on A. ... (For example, if [$\tau_{1}$] attributed a definite position to [A], but not a definite momentum, it would be incomplete in its description of [A]’s momentum; but, of course, Einstein's argument does not require any such reference to specific parameters or `elements of reality'.) (\textit{loc. cit.}, modified for uniform notation)

\end{quote}

 The essential differences between  $\tau_{1}$ and $\tau_{2}$ concern their predictions of the measurement outcomes for A's observables. For instance, $\tau_{1}$ and $\tau_{2}$ are essentially different if one of them ``lives'' in $(\pi_{1}, \mathcal{H}_{A} \otimes I)$, say a position representation, and the other in $(\pi_{2}, \mathcal{H}_{A} \otimes I)$, say a momentum representation. More generally then, $\tau_{1}$ and $\tau_{2}$ are essentially different if $(\pi_{1}, \mathcal{H}_{A} \otimes I)$ and $(\pi_{2}, \mathcal{H}_{A} \otimes I)$ are unitarily inequivalent representations of $\mathfrak{A}_{A} \otimes \mathcal{I}$. Taking this into account, one obtains the following reconstruction of the second half of Einstein's argument:

\smallskip

4$^{T}$. QM allows unitarily inequivalent representations of $\mathfrak{A}_{A} \otimes \mathcal{I}$, depending on which EPR-double, $\pi_{1} ( I \otimes O^{1}_{B})$ or $\pi_{2} ( I \otimes O^{2}_{B})$, one can choose to measure on subsystem B.

5$^{T}$. But if QM is complete, the representations $(\pi_{1}, \mathcal{H}_{A} \otimes I)$ and $(\pi_{2}, \mathcal{H}_{A} \otimes I)$ should be unitarily equivalent.

6$^{T}$. Thus, QM is incomplete.

\smallskip

I believe that there is a good chance that this reconstruction captures exactly what Einstein thought on the matter. But one immediate criticism against this reconstruction concerns the apparent conflict between premise 4$^{T}$ and the Stone-von Neumann theorem (discussed in section 2.2.3), a theorem that Einstein was no doubt fully aware of by 1935. This asserts, recall, that any irreducible, faithful and regular Hilbert space representation of the Weyl algebra describing a quantum-mechanical system is uniquely determined up to a unitary transformation, and in fact unitarily equivalent to the Schrödinger representation. The theorem applies, of course, to EPR systems. But the conflict with my reconstruction is merely apparent, since the representations mentioned in premise 4$^{T}$ are non-regular and, therefore, not in the range of the Stone-von Neumann theorem.

This reply, however, just seems to have opened the door to further criticism. For if Einstein's argument required that regularity be dropped, then one might seem justified to consider it muddled after all, for it is not at all clear what physical meaning can be given to non-regular representations.\footnote{As pointed out in section 2.2.3, section, if position and momentum representations are taken to be physically significant despite their non-regularity, then unitary dynamics must be given up (cf. Feintzeig \textit{et al.} 2019).} But in fact, I think that this adequately captures precisely the sense of incompatibility that Gömöri and Hofer-Szabó, as we have seen above, associate with the multiple assignment of $\psi$-functions to subsystem A. The \textit{dynamical} incompatibility between (regular and) non-regular representations not only does not muddle Einstein's argument, but rather helps clarify his justification for rejecting the completeness of QM.

Now, even if my algebraic reconstruction is correct, one might insist that in order to fully support Howard's suggestion, a formalization in a suitable language should be given. Until we consider more closely the challenges for a rational reconstruction of QM, the following version of the argument should be enough here:

\smallskip

4$_{H}$. QM allows non-isomorphic models of its theoretical description of A, depending on which EPR-double one can choose to measure on subsystem B.

5$_{H}$. But if QM is categorical, these models should be isomorphic.

6$_{H}$. Thus, QM is not categorical.

\smallskip

As noted already, this version of the argument assumes that unitarily inequivalent representations (featuring in premise 4$^{T}$) can be construed as non-isomorphic models of the quantum mechanical description of subsystem A (mentioned in premise 4$_{H}$). More exactly, it assumes that unitary equivalence is a necessary, but not a sufficient, condition for categoricity. On this assumption, which I consider unproblematic, if the validity of the argument is to be preserved, premise 5$^{T}$ must be reformulated as premise 5$_{H}$. Thus, Einstein completeness can be understood as categoricity. I think that this clarifies and supports, at least in part, Howard's suggestion that it should be so understood. Einstein's no-go result may indeed be taken to say that a local, separable, and categorical QM cannot exist.

Reconstructing Einstein's argument in this way points to a series of general consequences often discussed, in other theoretical contexts, throughout the rich history of philosophical concerns with the categoricity of logic and mathematics. For instance, as noted in the preliminary considerations above (section 2.2.1), the meaning of arithmetical terms is precisely determined if Peano arithmetic is categorical with respect to the isomorphism class of an omega sequence. And the meaning of classical logical connectives is precisely determined if classical logic is categorical with respect to the isomorphism class of a two-element Boolean algebra. Quite similarly, the meaning of quantum expressions is precisely determined if QM is categorical with respect to the relevant isomorphism class of models. Thus, Einstein's argument, reconstructed as I proposed above, attributes to QM a kind of semantic indeterminacy that may help explain the difference between EPR completeness and Einstein completeness as types of descriptive completeness. The question is what kind of descriptive failure is the incompleteness that Einstein attributed to QM? As a non-categorical theory, QM is descriptively incomplete, just not in the sense that it fails to describe \textit{all} of the physical reality that it aims to describe, but rather in the sense that it describes \textit{more} than the physical reality that it aims to describe. This does not necessarily imply that the theory describes states that cannot exist; rather, it describes dynamically incompatible states, construed as elements of non-isomorphic models.

Before I conclude this chapter, let me briefly note some implications I take my reconstruction of Einstein's argument to have for our understanding of his debate with Bohr. Recall that one important point in Bohr's reply to the EPR paper, as was well understood by Einstein, is to deny separability (Howard 2007). Thus, it also applies to Einstein's own argument (by rejecting the first half of premise 1). Einstein famously commented on this point as follows: 

\begin{quote}
    
By this way of looking at the matter it becomes evident that the paradox forces us to relinquish one of the following two assertions:
(1) the description by means of the $\psi$-function is \textit{complete}.
(2) the real states of spatially separated objects are independent of each other. (Einstein 1949, 682) 

\end{quote}

Einstein thought that relinquishing assertion (2), i.e., denying that spatially separated objects have real states, would make physics impossible. But a second important point made by Bohr is that EPR-doubles are complementary. What does this point amount to, when considered in the algebraic framework described above? 

As noted above, Hans Halvorson has rigorously interpreted Bohr's complementarity in terms of the unitary inequivalence of non-regular Hilbert space representations. Since Einstein's argument, on my algebraic reconstruction, points out that there are representations describing subsystem A of an EPR system that are unitarily inequivalent (premise 4$^{T}$), Bohr's reply to the EPR paper can be understood as a rejection of premise 5$^{T}$: a complete QM does not require unitary equivalence. Furthermore, if one is justified to reformulate premise 5$^{T}$ as premise 5$_{H}$, as I have done above, then the whole controversy can be seen in a different, arguably clearer light: while Einstein deplored the non-categoricity of QM, Bohr embraced it as a theoretical asset. As a consequence, the metasemantic contrast between them could not have been more stark: the semantic indeterminacy of the standard formalism was considered a serious problem by Einstein, but was ignored as insignificant by Bohr.

In any case, my discussion in this section has clarified, I hope, at least some of the assumptions behind Howard's suggestion that Einstein completeness should be read as categoricity. But I also pointed to some of the misconstruals of Einstein completeness in the literature, and provided an explanation of the sense in which Einstein completeness, as a type of descriptive completeness, is different than EPR completeness. Finally, I suggested that the Bohr-Einstein controversy may be considered as a debate in the metasemantics of QM. 

At this point in my narrative, I want to go back to an even earlier stage in the history of mathematical science -- more precisely, to the development of symbolic algebra in the 19th Century -- in order to then return to QM and explain the nature and metasemantic implications of another central piece in its conceptual development: Bohr's correspondence principle. 

\newpage 

\section{The problem of permanence}

This chapter presents Bohr’s solution to the problem of permanence, based on a reinterpretation of his correspondence principle. The reinterpretation is grounded in Hankel's principle of permanence, it elucidates the sense in which quantum mechanics (QM) can be considered a rational generalization of classical physics, and justifies the attribution to Bohr of a non-representationalist metasemantics for QM. After reviewing von Neumann’s use of Hankel's principle in quantum logic (QL), the chapter revisits Weyl’s criticism that QL is not categorical because its connectives are not truth-functional.\footnote{This chapter is partly based on Toader 2021b and 2024.}

\subsection{Historical roots and metasemantic implications}

The problem of permanence concerns the metasemantic implications of the principle of permanence. This principle stipulates that central elements of a scientific theory, e.g., its rules or laws, should be preserved when extending or generalizing the theory, or when expanding its domain. It expresses a methodological conservatism that has long been presupposed and often tacitly applied by scientists. In the history of mathematics, the principle played a crucial role in the development of algebra, e.g., in the works of De Morgan and Boole. It was instrumental in articulating \textit{formalism} -- a cluster of philosophical views on the foundations of mathematics, including views expressed in the 19th century by George Peacock and Hermann Hankel, but also that defended by Hilbert and Bernays in the 20th century. Furthermore, as I will show in this chapter, the principle of permanence is centrally connected with the development of modern physics as well, especially in the work of Bohr, where it was taken to guide the transition from classical physics to QM, and von Neumann, where it was used in further generalizations of the latter. Before I can do so (in sections 3.2.1 -- 3.2.3), I give an overview (in sections 3.1.1 -- 3.1.2) of the emergence of the principle of permanence in the works of Peacock and Hankel, emphasizing what I take to be their views on its metasemantic implications.

\subsubsection{Peacock's principle of permanence}

While the actual roots of the principle of permanence may lie deep into the philosophy and science of the 18th century, or even earlier ones, its first explicit formulation  -- as the principle of permanence of equivalent forms (PPF) -- is due to Peacock, mathematician at Trinity College, Cambridge:

\begin{quote}

Direct proposition: Whatever form is algebraically equivalent to another when expressed in general symbols, must continue to be equivalent whatever those symbols denote.

Converse proposition: Whatever equivalent form is discoverable in arithmetical algebra considered as the science of suggestion, when their symbols are general in their form, though specific in their value, will continue to be an equivalent form when the symbols are general in their nature as well as in their form. (Peacock 1833, 198sq, emphasis removed)

\end{quote}

Peacock's arithmetical algebra, or what had also been called ``specious or universal arithmetic'', is a theory the language of which includes not only constants and signs for operations, like elementary or common arithmetic, but also variables ranging over the domain of positive integers. He contrasted this theory with what he called ``symbolical algebra'', the language of which further includes variables that are allowed to range over any domain of objects whatsoever, but most importantly over negative, rational, and imaginary numbers such as the ``impossible'' roots of equations of second or higher degree. 

The direct proposition of the PPF was taken to be rather self-evident: ``The direct proposition must be true, since the laws of combination of symbols by which such equivalent forms are deduced, have no reference to the specific values of the symbols.'' (\textit{op. cit.}, 199) But the converse proposition stipulates that some equivalent forms, i.e., at least the ones that can be discovered and expressed in the language of arithmetical algebra, will be preserved as equivalent forms when expressed in the language of symbolical algebra. A prime example of such forms are the ``laws of combination'', which Peacock indiscriminately also considered as the principles or the rules of a theory. For instance, the rule of distributivity in arithmetical algebra, which he wrote as $ma+na=(m+n)a$, is to be preserved in symbolical algebra as such. But he intended the notion of equivalent forms to be more general, so what is to be preserved in passing from one theory to another are not only such rules, but theorems as well. In any case, as I read him, he clearly rejected the universal validity of the PPF, for he saw that those equivalent forms of arithmetical algebra that are ``essentially arithmetical'' cannot be preserved when passing to symbolical algebra.\footnote{For this point, see Toader 2021b, 80. Similarly, the rule of commutativity will be later considered an essentially classical rule, since it cannot be preserved as a relation between operators corresponding to physical variables when passing from classical physics to QM. For discussions of Peacock's algebra and his PPF, see e.g. Pycior 1981, Lambert 2013, and Bellomo 2025.} Whenever the PPF could be applied, however, it was so useful for solving problems and proving theorems that other mathematicians, such as De Morgan, came to regard it not merely as a heuristic principle, but as a necessary mathematical truth.\footnote{For De Morgan's 1837 formulation of the principle of permanence, and for his discussion of this with Ada Lovelace, who questioned the validity of the principle and his insistence on ``the necessity of its truth'', see Hollings \textit{et al.} 2017.} 

The sense in which arithmetical algebra can be considered as a ``science of suggestion'' seems, however, rather obscure, so a brief explanation is in order. Peacock's main concern in his ``Report'' was the clarification of the nature of the principles of symbolical algebra, which he thought had been misunderstood as being founded on, i.e., deduced from, arithmetical principles -- a misunderstanding that he took to be responsible for the view of symbolical algebra as a mere generalization of arithmetical algebra. The PPF, and in particular the notion of a science of suggestion, was meant to help with Peacock's refutation of this view. On his own view, the principles of symbolical algebra are suggested by, rather than deduced from, the principles of arithmetical algebra and, as a consequence, the former is something other than a mere generalization of the latter. To say otherwise was, to him, an abuse of terminology, for the following reason: 

\begin{quote}
    
The operations in arithmetical algebra can be previously defined, whilst those in symbolical algebra, though bearing the same name, cannot: their meaning, however, when the nature of the symbols is known, can be generally, but by no means necessarily, interpreted. The process, therefore, by which we pass from one science to the other is not an ascent from particulars to generals, which is properly called \textit{generalization}, but one which is essentially arbitrary, though restricted with a specific view to its operations and their results admitting of such interpretations as may make its applications most generally useful. (\textit{op. cit.}, 194) 
\end{quote}

Peacock dismissed the view according to which one arrives at symbolical algebra from arithmetical algebra via \textit{generalization}, properly so called, for he thought this would entail that the operations in symbolical algebra must admit of an interpretation over the domain of arithmetical algebra. On his view, one reaches the former from the latter via what he called \textit{suggestion}, which would not entail that the operations in symbolical algebra must admit of an interpretation over the domain of arithmetical algebra. Indeed, they need not admit of any interpretation at all. Intriguingly, he then added that arithmetical algebra ``necessarily \textit{suggests}'' the principles of symbolic algebra. This might be interpreted to say that arithmetical algebra is indispensable as a science of suggestion for symbolical algebra. But actually Peacock considered arithmetical algebra as merely the ``most convenient'' science of suggestion for symbolical algebra, rather than an indispensable one. 

As I understand Peacock's view here, his intriguing addition should be interpreted to say that it is the intertheoretical \textit{relation} of suggestion that is necessary. In general, on his view, any ``arbitrary assumptions'' that are logically consistent, or as he put it, ``as far as they can exist in common'', could be stipulated as principles of symbolical algebra. But since any useful application of symbolic algebraic operations and results requires their interpretation over the domain of the application, the arbitrary assumptions must be restricted such that symbolic operations and results can admit of such an interpretation. This restriction is satisfied by necessarily assuming some subordinate
science of suggestion, like arithmetical algebra. Thus, for example, if the principles of symbolic algebra are not entirely arbitrary, but they are both consistent with and actually suggested by the principles of arithmetical algebra, then the operations of symbolic algebra and their results can be interpreted over the positive integers. More generally, without satisfying this interpretability condition, no useful application of symbolical algebra would be possible. 

Peacock's view is, therefore, that when arithmetical algebra is actually taken as a science of suggestion, although it is not the case that the operations in symbolical algebra must admit of an arithmetical interpretation, it must be the case that they can admit of an arithmetical interpretation. \textit{The arithmetical interpretation is not necessary; it is only necessarily possible.} This condition of necessary interpretability is dictated by the requirement of usefulness in applications. But it allows that other sciences of suggestion might be considered instead, and consequently that other interpretations of the symbolic operations are admissible. Peacock's point is that some such particular interpretation must be possible and, as is clear from the direct proposition of the PPF, any such interpretation is admissible. Moreover, if one does take the PPF as a proper guide for the development of symbolical algebra, then its principles, albeit not deducible from the principles of arithmetical algebra, are deducible from the conjunction of the principles of arithmetical algebra and the PPF. This is why Peacock said the PPF is the ``real foundation'' of symbolical algebra. But this also explains why the latter's principles, despite their character as arbitrary assumptions, are also ``necessary consequences'' of the PPF.

What are the metasemantic implications of Peacock's principle? Could one recover, from his reflections on the PPF, an answer to Hilbert's question about metasemantics (see above, section 1.1), the question about the relationship between the language of a theory and its semantic attributes? If the operations in symbolical algebra cannot be ``previously defined'', if their meaning cannot be determined before its principles or rules are determined, then what is it that gives meaning to the operations afterwards? Peacock's answer is as follows:

\begin{quote} 
    
    In arithmetical algebra, the definitions of the operations determine the rules; in symbolical algebra, the rules determine the meaning of the operations, or more properly speaking, they furnish the means of interpreting them (\textit{op. cit.}, 200)
    
\end{quote}

I take this answer as an expression of the metasemantic revolution brought about by the PPF, a revolution in the metasemantics of 19th century mathematics. This is a revolution insofar as it represents a radical change from a traditional view, according to which the meaning of mathematical operations is determined by their definitions relative to a particular domain, as is the case in arithmetical algebra, to a novel view, according to which the meaning of operations is determined by their very rules, as in the case of Peacock's symbolical algebra. One way to understand his claim that the rules of symbolical algebra determine the meaning of its operations, in the more proper sense that these rules ``furnish the means of interpreting'' the operations, is to say that the rules determine the interpretation of the operations. But if arithmetical algebra is taken as a science of suggestion, and if one understands this notion as I proposed above, then an interpretation over the positive integers of the operations of symbolical algebra is necessarily possible, without being necessary. Thus, to say that the rules of symbolical algebra determine the meaning of its operations is to say that the rules determine their interpretability, or more precisely, the interpretability of the results obtained by those operations.\footnote{Peacock's view was endorsed, among others, by William Whewell, who emphasized that the PPF should not be considered only heuristically, as a useful guide for developing new mathematical theories, but as a principle that characterizes the interpretability relation between theories (cf. Whewell 1840, 143). See also Gregory 1840, 323. For discussion of the PPF in the context of formalism, see Detlefsen 2005. More needs to be said about this metasemantic revolution, and in particular about the kind of conservatism expressed by the PPF. I offer a comprehensive discussion in my current book-in-progress, \textit{Mathematical Conservatism}.}

\subsubsection{Hankel's  principle of permanence}

A view that is similar to Peacock's would be later defended by Hankel, and will be discussed in this section, before I turn to Bohr's and von Neumann's applications of the PPF in QM. Hankel adopted Peacock's view of the PPF when he set up to develop the formal theory of complex numbers. He formulated his own version of the principle, which he initially characterized as a methodological or heuristic principle -- ``das hodegetische Princip der Permanenz der formalen Gesetze'' (Hankel 1867, vii) -- as follows: 

\begin{quote}

If two forms expressed in general signs of the universal arithmetic are equal to one another, they should remain equal if the signs cease to denote simple quantities and the operations thereby receive some different content as well. (\textit{op. cit.}, 11)

\end{quote}

This corresponds, roughly, to Peacock's converse proposition of the PPF. For example, the rule of (left and right) distributivity in universal arithmetic, $a(b+c)=ab+ac$ and $(b+c)a=ba+ca$, where the quantities denoted by its variables are the positive integers, should be preserved when they denote any other objects, like the negative, rational, or imaginary numbers. Just what rules (forms or formal laws) are exactly to be preserved when extending the number domain is a question that Hankel carefully considered. Like Peacock, Hankel also warned against the universal application of the PPF, emphasizing that certain rules that hold for the real numbers are essentially real, and as such cannot be extended to complex and hypercomplex numbers.\footnote{Cf. Hankel 1867, 195. Furthermore, he proved that there can exist no extension beyond the complexes that preserves the commutativity of basic operations (Detlefsen 2005, 286). For recent discussions of Hankel's philosophy of mathematics, especially in connection to Frege's criticism of it, see Tappenden 2019 and Lawrence 2021.} This justified his initial characterization of PPF as a methodological or heuristic (\textit{hodegetisches}) principle. By stipulating that forms should be preserved as much as possible when generalizing an arithmetical theory, Hankel also took the PPF to characterize the relations between theories. Thus, this was not merely a heuristic principle, but a metatheoretical one as well. Furthermore, as we will presently see, he came to think of it as a ``metaphysical'' principle, a view that was suggested to him by his envisaged application of the PPF to the physics of mechanical quantities (forces, momenta, etc.).

Before we can identify in Hankel's writings the same revolutionary metasemantics noted above in Peacock's, an account of his view of numbers should be in order. As Hankel remarked, the question about whether certain numbers, like the imaginary ones, are possible should first clarify the assumed notion of possibility. He understood this notion in terms of logical consistency: numbers exist only if their concept is clearly and distinctly defined without any contradiction. The question about the existence of numbers, he then suggested, reduces to the question about the existence of the thinking subject or the objects of thought, since numbers represent the relations between such objects. He then distinguished, in Kantian terms, between two main types of numbers: on the one hand, what he called ``transcendent, purely mental, purely intellectual or purely formal'' numbers are those representing relations between objects of thought that cannot be constructed
in intuition. On the other hand, actual numbers, or what Hankel called
``actuelle Zahlen'', are those representing relations between objects of thought that are constructed in intuition. However, he considered this distinction to be ``not a rigid, but a blurred distinction'' (``kein starrer,
sondern ein fliessender'',  \textit{op. cit.}, 8). Indeed, he further characterized as ``potentielle Zahlen'' those numbers that, although initially taken as transcendent, incapable of any construction in intuition, eventually become actual numbers, just as the complexes did after receiving a geometrical representation. Potential
numbers formally represent relations between objects of thought, which
are such that an intuitive construction of them turns out to be nevertheless possible. 

With his classification of types of numbers in place, Hankel characterized formal mathematics as a pure doctrine of forms (\textit{reine Formenlehre}):

\begin{quote}
 
The condition for the establishment of a general arithmetic is therefore a purely intellectual mathematics, detached from all intuition, a pure doctrine of form, in which it is not quantities or their representations [\textit{Bilder}], the numbers, that are combined, but intellectual objects, objects of thought, to
which actual [\textit{actuelle}] objects or relations thereof can, but do not have to, correspond. (\textit{op. cit.}, 10)
\end{quote}

Thus, Hankel's formal mathematics stipulates rules of combination for potential numbers, i.e., for numbers representing relations an intuitive construction of which is possible, though not necessary. These rules are arbitrary to an extent limited solely by logical consistency, which one could ascertain, Hankel thought, by establishing their mutual independence. But nothing like Peacock's intertheoretical relation of suggestion seems to be explicitly required to impose further restrictions on the rules of formal mathematics. 

Nevertheless, Hankel believed that a system of operations obeying formal rules remains ``empty'', if no applications of results are possible. An empty system allows no relations between objects of thought to be constructed in intuition. Thus, Hankel understood formal mathematics as a theory of potential numbers, conceived of as purely formal representations of relations, for which it must be the case that a construction in intuition -- an interpretation -- of its formal results can be given. While he also thought that no such particular interpretation was necessary, he took the interpretability (or, more exactly, constructibility) of formal results to be a crucial requirement. He believed that it must be the case, for practical reasons, that formal mathematics can be interpreted over the domain of actual numbers. In this, Hankel closely followed Peacock's view. 

Indeed, more traces of the latter can be identified in Hankel's writings, including statements that endorse Peacock’s rejection of generalization:

\begin{quote}

The purely formal mathematics, whose principles we have stated here, does not consist in a generalization of the usual arithmetic; it is a completely new science, the rules of which are not proved, but only exemplified, insofar as the formal operations, applied to actual numbers, give the same results as the intuitive operations of common arithmetic. In the latter the definitions of the operations determine their rules, in the former the rules [determine] the meaning of the operations, or to put it another way, they give the instruction for their interpretation and their use. (\textit{op. cit.}, 12)
 
\end{quote}

Hankel clearly embraced Peacock's revolutionary metasemantics. One way to understand the claim that the rules of formal mathematics determine
the meaning of its operations, in the sense that they give instructions
for their interpretation and application, is by taking the formal rules to provide instructions for the interpretation of formal operations over the domain of actual numbers, where these instructions include a condition of numerical identity. Thus, formal operations are meaningful only if the actual results obtained by interpreting the results of formal mathematics over the domain of common arithmetic are numerically identical with the actual results derivable by the actual operations
of arithmetic. This view is similar to that expressed by Peacock. The
interpretability of formal results is necessary for applications, and sufficient to render the formal operations meaningful, provided that the numerical identity condition is generally satisfied.

But Hankel went, in fact, further than Peacock. He considered PPF not only as a guide or a heuristic principle, and not only as a metatheoretical principle characterizing interpretability relations, either. Rather, Hankel came to think that it was a ``metaphysical'' principle (\textit{op. cit.}, 12), a claim that he attempted to justify by pointing to non-arithmetical interpretations of formal mathematics, e.g., geometrical and physical interpretations. Indeed, Hankel took classical mechanics to be a theory of actual relations between physical quantities that is, just like universal arithmetic, merely subordinate to the pure doctrine of forms. And he took the PPF to stipulate that the relations between physical quantities -- the physical laws or forms, as it were -- should be preserved as much as possible in passing from classical mechanics to the pure doctrine of forms. The operations of the latter are meaningful only if the actual results derivable by means of the operations of mechanics are the same as the actual results obtained by interpreting formal results over the domain of physical quantities. On this ground, he criticized Peacock's conception of the PPF as ``too narrow'' (\textit{op. cit.}, 15). Investigations in the natural science, Peacock had maintained, proceed in two directions: from principles to results, but also towards deeper principles, in a series that terminates only in the ``mystery of the first cause''.\footnote{Cf. Peacock 1833, 186. I will return to Peacock's view on principles in the natural sciences in section 6.1, in my brief discussion of Russell's regressive method.} Since he thought that the first cause could not be understood as a set of ultimate natural facts, he believed that the deepest principles of the natural sciences could not be conceived of in relation to any formal principles. According to Hankel's view, by contrast, mechanics, just like arithmetic, is another necessarily possible interpretation of the pure theory of forms. Without this possibility, formal operations could have no physical meaning.

As we will see presently, this view had a deep influence on Bohr's conception of the relation between classical physics and QM: the latter, he believed, is meaningful only if any description of its experimental results is ``essentially equivalent'' with their classical description, where essential equivalence is just Hankel's condition of numerical identity. 

\subsection{Bohr's correspondence principle}

The relationship between the PPF and Bohr's correspondence principle (CP) has been so far rather neglected in the literature. Having looked at the historical and theoretical context of the former, I now want to revisit the latter. In particular, without attempting to provide an exhaustive analysis, I will recall (in section 3.2.1) the emergence of CP in the old quantum theory and some relevant changes that occurred in Bohr's thinking until the application of CP in his approach to quantum
mechanics.\footnote{For more comprehensive accounts of CP, see e.g. Darrigol 1997, Tanona 2004, Bokulich and Bokulich 2005, Bokulich 2008, Jähnert 2019, and Perovic 2021. For a succinct account, see Bokulich and Bokulich 2020.} But I should note that it is not clear when Bohr actually became aware of PPF: it is possible that he knew about it from the very beginning of his articulation of CP, but it is also possible that he found out about PPF only later, when as we will see he shared it with his students.\footnote{I should note here that one very plausible source on PPF could have been Bohr’s brother, Harald, who would have been aware of this principle and its importance in the history of mathematics.} It is quite remarkable, nevertheless, that the development of CP in Bohr’s thinking matches rather accurately the views about PPF that we have seen Peacock and Hankel to have, respectively, developed. In any case, let me stress here already the expected benefits of my point, in order to properly motivate this inquiry. If CP is understood as grounded in PPF, then one can explain several aspects of Bohr's thinking, including his claim that QM is a ``rational generalization'' of classical physics. As we will see (in section 3.2.2), Bohr's notion of rational generalization is based on Hankel's notion of generalization, which was essentially based on Peacock’s concept of suggestion. Furthermore, if CP is understood as grounded in PPF, then one can also explain a crucial element of Bohr's approach to QM, i.e., his view that the meaning of QM is determined by its rules, a view rather tersely expressed in his reply to the EPR paper. As a bonus, we will also be able to make sense of a seemingly strange criticism directed by Feyerabend and Bohm against this approach, and to show (in section 3.2.3) that Howard's reconstruction of Bohr's doctrine of the necessity of classical concepts can avoid a conflict with CP.

\subsubsection{The nature of the correspondence principle}

Bohr's early use of CP was related to his analysis of radiation into harmonic components. The analysis was concerned with the description of the so-called quantum jumps, i.e., the kind of transitions an electron undergoes between stationary states, which unlike its motion in a particular stationary state, could not be accounted for by classical electrodynamics. The radiation emitted during such transitions allows, according to Bohr's analysis, values of the frequencies in the harmonic components different from the classical values. But he noted an approximate agreement between quantum and classical frequency values and stipulated such an agreement between transition probabilities and the amplitudes of the harmonic components of the classical motion, in the limit of large quantum numbers. He also stipulated an agreement between transition probabilities and the amplitudes of the harmonic components, in the case of small quantum numbers. Bohr appears to have included all these possible relations in his early notion of correspondence:

\begin{quote} 

This correspondence between frequencies determined by the two methods must have a deeper significance and we are led to anticipate that it will also apply to the intensities. This is equivalent to the statement that, when the quantum numbers are large, the relative probability of a particular transition is connected in a simple manner with the amplitude of the corresponding harmonic component in the motion. 

This peculiar relation suggests a \textit{general law for the occurrence of transitions between stationary states}. Thus we shall assume that even when the quantum numbers are small the possibility of transition between two stationary states is connected with the presence of a certain harmonic component in the motion of the system. (Bohr 1920, 249sq.)
    
\end{quote}

As Bohr emphasized here clearly enough, however, the correspondence between transition probabilities and the amplitudes of harmonic components, when quantum numbers are large, is anticipated on the basis of the correspondence between frequencies, and should be preserved even when the quantum numbers are small. Note that, as a \textit{general law}, one that is valid for \textit{all} quantum numbers, this correspondence is not deduced from, but is said to be \textit{suggested} by the correspondence that holds for large quantum numbers. Thus, even though CP may have emerged in Bohr's thinking about radiation and his atomic model, as ``a result of gradual bottom-up hypothesis-building from the experimental context within the confines of the model'' (Perovic 2021, 89), and even though it was concerned primarily with the relation between physical quantities, it seems fair to say that here it was also assumed as a guide in the development of Bohr's radiation theory, as a heuristic principle instrumental for his generalization to the case of all quantum numbers. 

Furthermore, the stipulated preservation of the relation between transition probabilities and the amplitudes of harmonic components underscores a relation between the classical theory of radiation and Bohr's own radiation theory, a metatheoretical relation that he denoted as a ``formal analogy'', despite his suspicion that the idea of correspondence as formal analogy ``might cause misunderstanding''. Indeed, it's difficult to see how a formal analogy between the classical and the quantum theory could be justified, if one took CP to concern exclusively a relation between physical quantities. But what he meant, I think, is that it is the stipulated preservation of this relation that justifies the formal analogy. Of course, it is a further question what else and precisely how much of the classical theory can be preserved in the transition to quantum theory, and I will return to this question below.

Now, if CP is also understood as stipulating the preservation of certain relations between physical quantities, and if it justifies Bohr's claim of formal analogy, then it further justifies a conception of the quantum theory as a certain kind of generalization of the classical theory. Bohr, himself, implied as much at the third Solvay Congress in 1921, when he noted

\begin{quote} 

on the one hand, the radical departure of the quantum theory from our ordinary ideas of mechanics and electrodynamics as well as, on the other hand, the formal analogy with these ideas. ... [T]he analogy is of such a type that in a certain respect we are entitled in the quantum theory to see an attempt of a natural generalisation of the classical theory of electromagnetism. (quoted in Bokulich and Bokulich 2005, 348)
\end{quote}

 What did Bohr mean here by a ``natural generalization'' of a classical theory? And did he mean the same thing when he later referred to QM as a  ``rational generalization'' of classical physics? After QM received a coherent mathematical formulation, did Bohr consider this as a generalization justified by a formal analogy between classical physics and QM, in turn justified by CP, understood to require the preservation of correspondences between physical quantities? An answer to such questions would require a detailed analysis of Bohr's writings after 1925, and especially of the development of his conception of CP, which I cannot give here. But I should, nevertheless, note that the view that there is a correspondence in the sense of formal analogy between classical mechanics and QM was expressed by Dirac:

\begin{quote} 
The correspondence between the quantum and classical theories lies not so much in the limiting agreement when $h \rightarrow $ 0 as in the fact that the mathematical operations on the two theories obey in many cases the same laws. (Dirac 1925, 649)\footnote{Later, Dirac attempted to rigorously justify this view of the correspondence between classical mechanics and Heisenberg's QM by setting up a general theory of functions of non-commuting variables (see Dirac 1945).}

\end{quote}

 Dirac appears to have justified the formal analogy via the preservation of laws (many, though not all of them) rather than, like Bohr, via the preservation of correspondences between physical quantities. Despite differences in their views, such as they were, Bohr also came to characterize the metatheoretical relation between classical mechanics and QM in terms of preservation of rules (many, though not all of them):

\begin{quote}

In this formalism, the canonical equations of classical mechanics ... are maintained unaltered, and the quantum of action is only introduced in the so-called commutation rules... for any pair of canonically conjugate variables. While in this way the whole scheme reduces to classical mechanics in the case h = 0, all the exigencies of the correspondence argument are fulfilled also in the general case... (Bohr 1939, 14)

\end{quote}

 Bohr clearly indicates that this characterization is in accordance with his CP, which shows that, as recent commentators aptly put it, ``Bohr is not simply saying that the quantum theory should `go over' to the classical theory in the appropriate limit. Rather, he is maintaining that that quantum mechanics should be a theory that \textit{departs as little as possible from} classical mechanics.'' (Bokulich and Bokulich 2005, 349, emphasis added) The idea is that CP, and more precisely the inherent notion of generalization, should be understood as the requirement that QM should preserve \textit{as many classical rules as possible}. It is this very notion of generalization, then, that Bohr thought characterized the relation between classical physics and QM. Obviously, this does not imply that classical physics is just a particular case of QM, which \textit{would} be implied if QM were a mere or properly called generalization of classical physics. 

Having focused on Bohr's notion of rational generalization, I now want to argue that this is exactly the notion of generalization that Peacock and Hankel had thought characterized the relation between arithmetic and arithmetical algebra, on the one hand, and symbolical algebra and formal mathematics, on the other hand. If this is true, then it is only in part correct to say that ``Through his rational generalization thesis, Bohr is offering us a new way of viewing the relationship between classical and quantum mechanics.'' (Bokulich and Bokulich 2005, 354) Bohr's notion of a natural or rational generalization was new only in the sense that it had not been applied, before him, to the relation between classical and quantum physics. Before I present evidence for the connection between Bohr's view and that of Hankel, I want to defend the viability of my interpretation.

Recall that Peacock rejected the view that symbolical algebra is a generalization of arithmetical algebra, for he thought this would imply that the principles of symbolical algebra could then be deduced from, rather than just suggested by the principles of arithmetical algebra. He also thought that the relation of suggestion between these theories implied that an arithmetical interpretation of symbolical algebraic results is not necessary, but only necessarily possible. Later, Hankel held a similar view and maintained that the laws of formal mathematics cannot be proved by the rules of arithmetic, but must nevertheless be interpretable in arithmetical language, in the sense that formal results must, when arithmetically interpreted, be numerically identical with the results of arithmetic. This shows that, on both of these 19th century views, their notion of generalization implied a certain interpretability relation between the theories they were concerned with. In order to test the viability of my interpretation of Bohr's notion of rational generalization in the same vein, even before I adduce any evidence for the connection between his view and that of Hankel, one should ask whether Bohr's notion also implies a certain interpretability relation between QM and classical physics. 

But the fact that it actually does so is, of course, well known. A crucial element of Bohr's approach to QM, which will be discussed further below, is the requirement that any description of experimental results must be ``essentially equivalent'' to their classical description. This requirement was taken to have metasemantic implications: the necessarily classical description of experimental results settled, on Bohr's view, the question about the meaning of QM, which became puzzling in the context of the measurement problem. As he put it in his reply to the EPR paper, ``there can be no question of any unambiguous interpretation of the symbols of quantum mechanics other than that embodied in the well-known rules which allow to predict the results to be obtained by a given experimental arrangement described in a totally classical way.'' (Bohr 1935, 701) This sounds a lot like Peacock and Hankel: rules determine the meaning of symbols, rather than the other way around. However, for the rules of QM to determine, or ``embody'', its meaning, Bohr demanded that experimental results \textit{must be} classically interpreted. Here he deviated from Peacock and Hankel in a significant way, and this exposed him to criticism from Feyerabend and Bohm, as we will see. But before that, let me turn to the imminent question about textual evidence. 

\subsubsection{Correspondence as permanence of rules}

Is there any textual evidence that Bohr understood CP as an expression of PPF? If there is, then it has remained by and large unnoticed by Bohr scholars, so far as I have been able to determine. Max Jammer, who seems to have been the first commentator to read CP as a metatheoretical principle, though not also as a quantum law, mentioned PPF in one of his ``digressions'' from a streamlined presentation of the conceptual development of QM:

\begin{quote} 
    
[M]atrices, multidimensional vectors, and quaternions are extensions of the concept of real numbers. Beyond the domain of complex numbers, however, extensions are possible only at the expense of Hankel's principle of permanence, according to which \textit{generalized entities should satisfy the rules of calculation pertaining to the original mathematical entities from which they have been abstracted}. Thus, while associativity and distributivity could be preserved, commutativity had to be sacrificed. It was the price which had to be paid to obtain the appropriate mathematical apparatus for the description of atomic states. (Jammer 1966, 217, emphasis added)

\end{quote}

Jammer implied that the mathematical description of atomic states in QM, just like the extension of mathematics beyond the complexes, was possible only at the expense of PPF, which suggests that he thought PPF was invalidated by the development of QM. But this would assume that it is a universally valid principle, a view that we have seen both Peacock and Hankel had rejected. At the same time, Jammer thought that ``there was rarely in the history of physics a comprehensive
theory which owed so much to one principle as quantum mechanics owed to Bohr's correspondence principle.'' (\textit{op. cit.}, 118) A proper understanding of CP would show that there is a ``logical rupture'' between classical mechanics and QM, which he described in the following terms:

\begin{quote}
    
[T]he correspondence principle, while leading to numerical agreements between
quantum mechanical and classical deductions, affirmed no longer a conceptual convergence of the results but established merely a formal, symbolic analogy between conclusions derived within the context of two disparate and mutually irreducible theories. It only showed that under certain conditions (for instance, for high quantum numbers or, in classical terms, for great distances from the nucleus) the formal treatments in both theories converge to notationally identical expressions (and numerically equal results) even though the symbols, corresponding to each other, differ strikingly in their conceptual contents. (\textit{op. cit.}, 227)\footnote{Others appear to have followed Jammer in making a similar point, that CP could not and was not meant to close the conceptual gap between classical physics and QM. To give just two examples:  Olivier Darrigol claimed that ``permanent formal schemes allow transfers of knowledge between successive theories even if their basic concepts appear to be incommensurable.'' (Darrigol 1986, 198sq) Also, Hans Radder wrote: ``Generally speaking, intertheoretical correspondence is primarily of a formal-mathematical and empirical but not of a conceptual nature.'' (Radder 1991, 195)}

\end{quote}

A logical connection or harmony (or whatever the opposite of logical rupture might be) between QM and classical mechanics would establish, Jammer believed, a conceptual convergence of their experimental results, by which he meant an identity of conceptual contents, rather than the numerical identity of results. He took CP to imply that only the latter must obtain between the two theories, while presumably thinking that PPF would require the former. However, as we have seen above, Hankel had rejected this understanding of PPF, when (following Peacock's view on symbolic algebra) he denied that formal mathematics is a generalization of universal arithmetic. Recall that, according to the reading I proposed, Hankel emphasized that PPF implies only that formal mathematics must be interpretable in the language of arithmetic, and when thus interpreted, all formal results should be numerically identical with results derived in arithmetic. What Jammer wrote about CP is correct then, provided that one takes CP as a version of PPF properly understood. 

The fact that Bohr did take CP as a version of PPF has been reported by Paul Feyerabend, who reminisced that, some time between 1949--1952, in some of his seminars,

\begin{quote}

Bohr ... talked about the discovery that the square root of two cannot be an integer or a fraction. To him this seemed an important event, and he kept returning to it. As he saw it, the event led to \textit{an extension of the concept of a number that retained some properties of integers and fractions and changed others}. Hankel, whom Bohr mentioned, had called the idea behind such an extension the principle of the permanence of rules of calculation. \textit{The transition from classical mechanics to quantum mechanics, said Bohr, was carried out in accordance with precisely this principle.} That much I understood. The rest was beyond me. (Feyerabend 1995, 76-78, emphasis added)

\end{quote}

Note that, according to Feyerabend, Bohr's reading of PPF did not assume the universal validity of the principle. In accordance with Hankel's own formulation, Bohr knew well that when domains are extended in mathematics, the rules of calculation are always preserved as far as this is possible. He also explicitly emphasized the significance of PPF for the development of QM as he saw it: it was the very transition from classical mechanics that he saw carried out in accordance with PPF, a significance that he typically attributed, in print, to CP. It is in fact utterly remarkable that, in his published works, Bohr always maintained that this transition was carried out in accordance with CP, and as far as I have been able to determine, he never mentioned PPF or Hankel at all. In any case, as Feyerabend's reminiscence indicates, Bohr also knew too well that, as in mathematics, classical rules are preserved in QM as far as this is possible. The same point is clearly made in his 1938 Warsaw conference talk, already quoted above:

\begin{quote} 

In the search for the formulation of such a generalization [of the customary classical description of phenomena], \textit{our only guide} has just been the so-called correspondence argument, which gives expression for \textit{the exigency of upholding the use of classical concepts to the largest possible extent compatible with the quantum postulates}. (Bohr 1939, 13; emphasis added)
    
\end{quote}

 After one seminar meeting, Feyerabend confessed his lack of understanding to Bohr's assistant, Aage Petersen. In a decade or so, Feyerabend returned to that conversation:

\begin{quote}

As Aage Petersen has pointed out to me, Bohr’s ideas may be compared with Hankel’s principle of the permanence of rules of calculation in new domains...  According to Hankel’s principle \textit{the transition from a domain of mathematical entities to a more embracing domain should be carried out in such a manner that as many rules of calculation as possible are taken over from the old domain to the new one}. For example, the transition from natural numbers to rational numbers should be carried out in such a manner as to leave unchanged as many rules of calculation as possible.  In the case of mathematics, this principle has very fruitful applications. (Feyerabend 1962, 120)

\end{quote}

What Bohr had surely explained to his students, Feyerabend now finally understood correctly: PPF should not be taken to hold universally. It stipulates, just like Peacock and Hankel emphasized, that rules or laws are to be preserved to the largest extent possible. Thus, if applied to QM under the guise of CP, as I take it to have been the case, then PPF allows that those classical laws that are essentially classical, like the commutativity of operations, can be given up. 

As already noted in the previous section, this evidence, based on Feyerabend's recollections of Bohr's lectures and Petersen's explanations, does not tell us precisely when Bohr actually became aware of the connection between PPF and CP: it is possible that he knew about it from the very beginning of his articulation of CP, in his theory of radiation, but it is also possible that he found out about PPF only later, maybe after 1925, or even later than that in the late 1930s or early 1940s. But I find it quite remarkable that Bohr's characterizations of CP as a methodological principle or a guide -- the ``only guide'', as he specified in Warsaw in 1938 -- as well as as a metatheoretical principle concerning the ``rational generalization'' of classical physics in QM, match rather accurately the views about PPF that we have seen Peacock and Hankel to have,  respectively, developed in 19th century mathematics. The evidence supports at least the claim that Bohr's notion of rational generalization was grounded in Hankel's notion of generalization, which in turn was grounded, as we have seen, in Peacock's notion of suggestion. This clarifies, I hope, what so far has been a rather enigmatic detail in Bohr's works. 

In fact, more can be explained on the basis of my interpretation of the connection between PPF and CP. For Feyerabend had, of course, a lot more to say about Bohr's approach to QM. One particular weakness with this approach that he immediately identified was described as follows: 

\begin{quote}
    
A complete replacement of the classical formalism seems therefore to be unnecessary. All that is needed is a modification of that formalism which retains the laws that have found to be valid and makes room for those new laws which express the specific behavior of the quantum mechanical entities. ... [The new laws] must allow for the description of any conceivable experiment in classical terms -- for it is, in classical terms that results of measurement and experimentation are expressed; ... [this requirement] is needed \textit{if we want to retain the idea... that experience must be described in classical terms}. (Feyerabend 1962, 120; emphasis added)

\end{quote}

As Feyerabend correctly observed, Bohr insisted that the classical description of experimental results, i.e., presumably their interpretation in the language of classical physics, is necessary. But Feyerabend rejected this necessity claim, which might seem rather strange. His criticism, further elaborated in the same paper, emphasizes the point that in principle a different language could be developed at least as adequate for the description of experimental results as the language of classical physics. Feyerabend's point, that a classical description is not necessary, allows however that this may be necessarily possible. In light of the view articulated by Peacock and Hankel, Feyerabend's criticism appears to be justified. It clearly emphasizes that Bohr's view deviated from that of Peacock and Hankel, for whom interpretability, though no particular interpretation, was necessary. But it is this deviation precisely that exposed Bohr to Feyerabend's criticism.\footnote{To be sure, Feyerabend's opinions about Bohr's approach to QM have also evolved over time (see Kuby 2021).} 

This very criticism was later pressed by Bohm as well: ``What is called for, in my view, is therefore a movement in which physicists freely explore novel forms of language, which take into account Bohr’s very significant insights but which do not remain fixed statically to Bohr's adherence to the need for classical language forms.'' (Bohm 1985, 159; quoted in Bokulich and Bokulich 2005, 368). That the description of experimental results must be given in a classical language, Bohm might have added, just because they are in practice presented in this language is not merely a static fixation, but a downright fallacy.

Having offered some evidence that Bohr understood CP as a version of PPF, or at least that he (and some of his assistants and, eventually, Feyerabend) thought that a comparison of the former with the latter would be fruitful for understanding the transition from classical physics to QM, and having also admitted that this could only justify the necessary possibility of interpreting experimental results in classical terms, rather than Bohr's insistence on its necessity, I want to turn to the question of what exactly Bohr meant by ``classical''. Some commentators think that he took such concepts to be simply concepts of classical mechanics and electrodynamics (cf. Bokulich and Bokulich 2005, 351), but others maintain that, for Bohr, a classical description meant ``a description in terms of what physicists call `mixtures''' (Howard 1994, 203). I want to argue that, despite appearances, if the connection between CP and PPF is taken seriously, then one can rather nicely accommodate the latter view.

\subsubsection{Howard on Bohr's essential equivalence}

In his reconstruction of Bohr's philosophy of physics, Don Howard emphasized that the doctrine of the necessity of classical terms was upheld by Bohr in an attempt to overcome a problem for objectivity that arises in QM. The problem is that what is generally considered a necessary condition for objectivity -- the metaphysical independence of observer and observed reality, and more precisely their separability, which Einstein thought was indispensable to the very formulation and testing of physical laws -- cannot be preserved when passing from classical physics to QM. As Howard presented it, Bohr's doctrine was meant as a purported solution to this problem. Classical terms are necessary because they ``embody'' the separability condition, which despite being false in QM allows for an unambiguous communicability of experimental results.\footnote{Cf. Howard 1994, 207. Note that separability is understood as state decomposability. To say that classical concepts ``embody'' the separability condition is taken to mean that separability is mathematically equivalent to Bohr's doctrine of the necessity of classical concepts (see Landsman 2006 for a proof of this equivalence). This entails that separability and entanglement are incompatible. As Howard emphasized, this is precisely the reason Bohr's solution to the problem of objectivity is unacceptable. But note also that separability, as Einstein himself appears to have conceived of it, may be a weaker condition than state decomposability and, thus, compatible with entanglement (see Murgueitio Ram\'irez 2020 for an argument to this effect).}

Further, Howard distinguished two ways of understanding Bohr's doctrine: one of them ``leaves open the possibility that, as our language develops, we might outgrow this dependence'' on classical concepts; the other, which is considered preferable, takes ``the necessity of classical concepts to be an enduring one, not to be overcome at a later stage in the evolution of language.'' (Howard 1994, 209) However, barring the potential fallacy mentioned in the previous section, it is not clear why the latter view should be preferred. What exactly might explain Bohr's insistence on the enduring character of classical language? What reasons did he have, and what reasons might anyone have, for excluding the possibility that other languages, as both Feyerabend and Bohm suggested, could at least in principle be developed to communicate quantum-mechanical results unambiguously and at least as adequately as the language of classical physics? If CP is understood as grounded in PPF, as I have suggested, then it is the interpretability, rather than any particular interpretation, of experimental results that should be required by Bohr's doctrine. This would leave open the possibility envisaged by Feyerabend and Bohm.\footnote{It is of course also possible that Bohr had other unstated reasons, unrelated to CP's grounding in PPF, that he took to justify his doctrine (see Faye 2017).} 

More importantly, however, Howard argued that, in demanding a classical description of experimental results, Bohr's doctrine does not require that a measuring instrument must be described \textit{entirely} in classical terms. Rather, only some of its properties are to be described classically, i.e., those that are correlated with the properties of the quantum system undergoing measurement (\textit{op. cit.}, 216). Howard took this to imply that what Bohr meant by a classical description should be most plausibly reconstructed as a description in terms of mixtures, rather than pure states; mixtures that must always be appropriate to a given experimental context. The reason for this is that, unlike pure states, mixtures are considered to ``embody'' the separability condition, in the sense that they allow the separability of measuring instrument and measured object with regard to exactly those properties of the object one is looking to determine in a particular measurement.

What is the role of CP on this reconstruction of Bohr's doctrine? As Howard noted, this doctrine requires an ``essential equivalence'', i.e., an equivalence between, on the one hand, the QM description of the properties of the measuring instrument that are correlated with the measured properties of the system undergoing measurement and, on the other hand, the classical description of those properties of the measuring instrument. Indeed, the main goal of Bohr's 1938 Warsaw conference paper was to discuss ``certain novel epistemological aspects'' involved in what he called ``the observation problem'' and, more specifically, certain aspects regarding ``the analysis and synthesis of physical experience.'' (Bohr 1939, 19) What were these aspects? The main outcome of the analysis was an emphasis on the necessity of taking the whole experimental arrangement, i.e., measured object plus measuring instrument, into consideration. Without this, said Bohr, no unambiguous meaning could be given to the QM formalism (\textit{op. cit.}, 20). The outcome of the synthesis was presented as follows:

\begin{quote} 

In the system to which the quantum mechanical formalism is applied, it is of course possible to
include any intermediate auxiliary agency employed in the measuring process. Since, however, all those properties of such agencies which, according to the aim of the measurement, have to be compared with the corresponding properties of the object, must be described on classical lines, their quantum mechanical treatment will for this purpose be essentially equivalent with a classical description. (\textit{op. cit.}, 23sq)

\end{quote}

 Thus, Bohr's insight was that giving a classical description of experimental results can only mean establishing the essential equivalence of the two descriptions of the relevant subset of properties of the measuring instrument. But establishing such an equivalence, Howard maintained, is at odds with Bohr's CP:

\begin{quote}

[W]hat kind of ``classical'' description could be ... ``essentially equivalent'' to a quantum mechanical description. In the sense intended by the \textit{correspondence principle}, quantum mechanics might agree with Newtonian mechanics or with Maxwell's electrodynamics in the limit of large quantum numbers, but that is not an ``essential'' equivalence. Moreover, the kind of convergence between quantum and classical descriptions demanded by the correspondence principle is a wholesale convergence, not an equivalence between selected sets of properties. ...

How can a classical description be `essentially equivalent' to a quantum mechanical one? Bohr's correspondence principle is what first comes to mind, but it cannot provide the answer, for two reasons. First, the correspondence principle asserts that quantum and classical descriptions
agree in the limit of large quantum numbers, that, is, in phenomena where the quantum of action is negligible. ...

Second, what the correspondence principle says about the relationship between classical and quantum descriptions is that they give \textit{approximately} the same predictions in the limit of large quantum numbers. But approximate agreement is hardly essential equivalence. The appropriate mixtures model gives a quite different answer. A quantum mechanical description, in terms of a pure case, and a `classical' description, in terms of
an appropriate mixture, give \textit{exactly} the same predictions for those observables measurable in the context that determines the appropriate mixture.\footnote{Cf. Howard 1994, 217-225. See also Howard 2021, 166 for a more recent iteration of this view.}

\end{quote}

As we have seen above, there is evidence (and an apparent consensus today) that CP should be read as asserting not merely an approximate agreement that holds in the limit of large quantum numbers, but an agreement that also holds more generally, for small quantum numbers as well. But I think that this poses no problem for Howard's reconstruction of Bohr's doctrine of classical concepts. This is because my account of CP as grounded in PPF entails that there is no conflict at all between CP and Bohr's demand of an essential equivalence. Quite the opposite, this account can nicely accommodate the fact that a ``wholesale convergence'', i.e., an equivalence between the QM description of \textit{all} properties of the measuring instrument and their classical description, cannot be established, and that an essential equivalence, as reconstructed by Howard, is necessary for the classical \textit{interpretability} of experimental results. 

This point can be succinctly clarified by appeal to Landsman's Bohrification strategy.\footnote{Cf. Landsman 2017. Howard's own formal explication of what it means for a classical description to be essentially equivalent to a QM description is based on his 1979 theorem concerning context-dependent mixtures. For details, see Howard 2021, 162-170.} Quantum measurement results, considered as physically significant aspects of a noncommutative algebra of observables (NAO), are accessible only if they can be described classically, i.e., only if they can be considered as aspects of a commutative algebra (CA). But NAO should be considered as a rational generalization of CA, in Bohr's sense. This kind of generalization requires exactly the essential equivalence that Bohr demanded, which can be established if and only if the elements of CA are a proper subset of the elements of NAO -- the subset determined by the particular experimental context. A wholesale convergence, which would require that CA and NAO be coextensive, is mathematically impossible.

In any case, I hope that my analysis above goes at least some of the way towards placing ``Bohr's views on the role of classical concepts ... in their proper historical context, especially as regards the relevant philosophical context.'' (Howard 1994, 227) As always, this is merely part of the whole story, more details of which await to be uncovered. But as I will show in the next section, a careful consideration of the PPF can help place also von Neumann's views about QL in their proper historical and relevant philosophical context.

\subsection{The permanence of the rule of modularity}

QL was the outcome of Birkhoff and von Neumann's project ``to discover what logical structure one may hope to find in physical theories which, like quantum mechanics, do not conform to classical logic'' (Birkhoff and von Neumann 1936, 823). The project was motivated by what they saw as the ``novelty of the logical notions`` of QM, e.g., the fact that logical operations cannot be defined for all pairs of experimental propositions, independently of experimental context. More specifically, Birkhoff and von Neumann argued (as recalled below in section 3.3.1) that the rule of distributivity must be considered essentially classical, and that due to the non-distributive lattice structure of the closed linear subspaces on Hilbert space, the rule of modularity must instead be adopted in the logic of QM. 

But the preservation of modularity in von Neumann's generalization of QM to an infinite number of dimensions, i.e., in the transition from QM$_{n}$ to QM$_{\infty}$, was explicitly justified by an appeal to the PPF. This justification raises the question about its metasemantic implications: what does the use of PPF entail about the semantic attributes of the formalism of QM$_{\infty}$, and in particular about the meaning of its logical connectives? Was the meaning of connectives in the logic of QM$_{\infty}$ to be determined by the rules of the calculus or by appropriately formulated semantic rules? Was von Neumann committed, like Bohr seems to have been, to a non-representationalist metasemantics?

\subsubsection{Birkhoff and von Neumann's quantum logic}

In their 1936 joint paper, Birkhoff and von Neumann followed the latter's own, then recent axiomatization of QM, and assumed a relation of correspondence between the mathematical terms in the axioms and the physical variables of a system: ``Before a phase-space can become imbued with reality, its elements and subsets must be correlated in some way with `experimental propositions'.'' (\textit{op. cit.}, 825) More exactly, ``closed linear subspaces of Hilbert space correspond one-many to experimental propositions, but one-one to physical qualities.'' (\textit{op. cit.}, 828) The main goal of their paper was to show that the experimental propositions describing the possible values of the physical variables of the system form a non-distributive lattice, and thus cannot admit a classical logical calculus. 

Let me start by revisiting their semantic proof against classical logic (CL) for the case of QM$_n$, that is, finite-dimensional QM. The quantum logical calculus, $\mathcal{QL}$, that Birkhoff and von Neumann envisaged includes a sentential language with variables for atomic sentences, $p$, $q$, ..., and symbols for logical connectives, $\neg, \wedge, \vee $, such that the set $S_\mathcal{QL}$ of sentences is defined, as usual, by induction: for any sentences $p$ and $q$ in the set, $\neg p$, $p \wedge q$, and $p \vee q$ are also in the set. $\mathcal{QL}$ also includes some of the rules of the classical calculus $\mathcal{CL}$, such as double negation and De Morgan. But unlike the semantics of $\mathcal{CL}$, which they took to be a Boolean algebra, $\mathcal{BA}$ (\textit{op. cit.}, 826), the semantics of $\mathcal{QL}$ is a non-Boolean lattice, $\mathcal{LA}$, i.e., a partially ordered set $S_\mathcal{LA}$ of elements, together with operations $^{\bot}, \cap, \cup$, for orthocomplementation, meet, and join, respectively. Meet is set-theoretic intersection, join is the operation that yields the smallest element containing the joined elements, i.e., their span, and orthocomplementation is defined in terms of a map $h : S_\mathcal{QL} \longrightarrow S_\mathcal{LA}$ as set-theoretic complementation plus partial order reversal: 

\begin{center}
    
$h(p) = h(q)^{\bot}$ iff $\{x : x \subseteq h(p)\} = \{x : x {\bot} h(q) \}$. 

\end{center}

 To say that $\mathcal{LA}$ is the semantics of $\mathcal{QL}$ is to say that \textit{h} is a homomorphism, with the following statements as definitions of the connectives, i.e., their semantic rules: 

\begin{center}
$h(\neg p) = h(p)^{\bot}$ \\ $h(p \wedge q) = h(p) \cap h(q)$ \\ $h(p \vee q) = h(p) \cup h(q)$.
\end{center}

 The relation of logical consequence for $\mathcal{QL}$ is defined in terms of partial order: 

\begin{center}
    
$\Gamma \models_{\mathcal{QL}} q$ iff $\cap \{h(p):p \in \Gamma \} \subseteq h(q)$.

\end{center}

 As Birkhoff and von Neumann noted, the mathematical structure of $\mathcal{LA}$ is that of the set of closed linear subspaces of a finite-dimensional Hilbert space, and can thus be taken to represent the physical properties of a quantum system. This implies that sentences in $S_\mathcal{QL}$ can be interpreted as ``experimental propositions'', i.e., statements that specify the possible values of the physical variables of a quantum system when this is in an appropriate state. However, $\mathcal{LA}$ does not validate the rule of distributivity. Their semantic proof that distributivity breaks down in QM$_n$ is formulated as follows (\textit{op. cit.}, 831):

\begin{quote}

if \textit{a} denotes the experimental observation of a wave-packet $\psi$ on one side of a plane in ordinary space, $a^{\bot}$ correspondingly the observation of $\psi$ on the other side, and \textit{b} the observation of $\psi$ in a state symmetric about the plane, then (as one can readily check): $b \cap (a \cup a^{\bot}) = b \cap 1 = b > 0 = (b \cap a) = (b \cap a^{\bot}) = (b \cap a) \cup (b \cap a^{\bot})$.

\end{quote}

 This means that if the quantum system is observed in a state represented by a vector (in a finite-dimensional Hilbert space) contained in the subspace $b \cap (a \cup a^{\bot})$, that vector cannot be contained in the subspace $(b \cap a) \cup (b \cap a^{\bot})$. The fact that there are such elements $a, b \in S_\mathcal{LA}$ entails that there are sentences $p, q \in S_\mathcal{QL}$ such that

\begin{center}
    
$h(p) \cap (h(q) \cup h(q)^{\bot}) \nsubseteq (h(p) \cap h(q)) \cup (h(p) \cap h(q)^{\bot})$,

\end{center} 

 which, given the stipulated semantic rules and the relation of logical consequence, invalidates the rule of distributivity: $p \wedge (q \vee \neg q) \nvDash_{\mathcal{QL}} (p \wedge q) \vee (p \wedge \neg q)$. 

Birkhoff and von Neumann further mentioned as ``a salient fact'' that a ``generalized'' rule of distributivity is invalidated also by the lattice of the closed subspaces of an infinite dimensional Hilbert space, i.e., for the case of QM$_{\infty}$. This was then explained in more detail by von Neumann at a conference in Warsaw in 1938, in his remarks following Bohr's lecture mentioned already in the previous section. The generalized rule was formulated as follows (modified for uniform notation): 

\begin{center}
    
$(a \cup b \cup c \cup ...) \cap (a^{\bot} \cup b^{\bot} \cup c^{\bot} \cup ...)$ = \\ 

$(a \cap a^{\bot}) \cup (a \cap b^{\bot}) \cup (a \cap c^{\bot})$ $\cup ... \cup$ \\ 

$(b \cap a^{\bot}) \cup (b \cap b^{\bot}) \cup (b \cap c^{\bot}) \cup ... \cup$ \\

$(c \cap a^{\bot}) \cup (c \cap b^{\bot}) \cup (c \cap c^{\bot}) \cup ... \cup ... $ 

\end{center}

This rule breaks down, von Neumann explicitly pointed out, as a consequence of Heisenberg's Uncertainty Principle. Here is his argument. Let the elements $a, b, c, ... $ be such that their corresponding sentences $p, q, r, ... $ state ``$m \Delta \textbf{p} \le \textbf{p} < (m+1)$'', for \textbf{p} the momentum of a system and $m=0, \pm 1, \pm 2, ...$, and such that their negations $\neg p, \neg q, \neg r, ... $ state ``$n \Delta \textbf{q} \le \textbf{q} < (n+1)$'', for \textbf{q} the position of that system and $n=0, \pm 1, \pm 2, ...$. It follows then that 

\begin{center}
    
$a \cup b \cup c \cup ... = a^{\bot} \cup b^{\bot} \cup c^{\bot} \cup ... = 1$,

\end{center}

  but since $\Delta \textbf{p} \Delta \textbf{q} < < h$, then 

\begin{center}
    
$a \cap a^{\bot} = a \cap b^{\bot} = a \cap c^{\bot}$ = ... = \\ 

$b \cap a^{\bot} = b \cap b^{\bot} = b \cap c^{\bot}$ = ... = \\ 

$c \cap a^{\bot} = c \cap b^{\bot} = c \cap c^{\bot} = ... = 0$.

\end{center}

 The conclusion of von Neumann's argument is this:

\begin{quote}

The `principle of indeterminacy' means that the `distributive law' of logics fails. The current view of quantum mechanics therefore forbids us to form both $a \cup b \cup c \cup ... $ and $a^{\bot} \cup b^{\bot} \cup c^{\bot} \cup ... $ in the same consideration. We have freed ourselves of such restrictions, but had to sacrifice the `distributive law` of logics instead. (Bohr 1939, 37)
\end{quote}

Already in 1936, Birkhoff and von Neumann maintained that $\mathcal{QL}$ is a \textit{modular} calculus, since the lattice of ``the \textit{finite dimensional} subspaces of any topological linear space such as Cartesian \textit{n}-space or Hilbert space'' (\textit{op. cit., 832}) satisfies the rule of modularity (i.e., a version of Dedekind's rule of modular identity, first formulated in 1871). This is a weaker form of distributivity, since it must satisfy an additional condition, i.e., $b \subseteq a^{\bot}$ (for the case illustrated in their semantic proof). But they provided a counterexample to the generalized rule of modularity in QM$_{\infty}$, by arguing that the lattice of closed linear subspaces of an infinite-dimensional Hilbert space cannot be modular, and then suggesting the following two possibilities:

\begin{quote}

One can ... construct for every dimension-number n a model $P_{n}(F)$ [that is, a projective geometry isomorphic to the modular lattice $\mathcal{LA}$], having all of the properties of the propositional calculus suggested by quantum-mechanics. One can also construct infinite-dimensional models $P_{\infty}(F)$ whose elements consist of all closed linear subspaces of normed infinite-dimensional spaces. But philosophically, Hankel's principle of the `perseverance of formal laws' (which leads one to try to preserve [modularity]) and mathematically, technical analysis of spectral theory in Hilbert space, lead one to prefer a continuous-dimensional model $P_{c}(F)$.

\end{quote}

The generalization to infinite dimensions, that is the transition from QM$_{n}$ to QM$_{\infty}$, could be done in two different ways, but only one of them preserves the rule of modularity in the logical calculus. This is because while $P_{c}(F)$, as a semantics of $\mathcal{QL}$, is isomorphic to a modular lattice, the other semantics they mentioned, $P_{\infty}(F)$, preserves the Hilbert space but is not isomorphic to a modular lattice. One reason they specified in support of their choosing the model $P_{c}(F)$ rather than $P_{\infty}(F)$ explicitly invokes Hankel's principle of permanence. The more explicit reasoning behind this choice had been explained by von Neumann in a 1935 letter to Birkhoff: 

\begin{quote}

I would like to make a confession which may seem immoral: I do not believe absolutely in Hilbert space any more. ... (as far as quantum-mechanical things are concerned) ... footing on the principle of `conserving the validity of all formal rules', Hilbert space is the straightforward generalization of Euclidean space, if one considers the \textit{vectors} as the essential notions. Now we begin to believe, that it is not the \textit{vectors} which matter but the \textit{lattice of all linear (closed) subspaces}. Because: (1) The vectors ought to represent the physical \textit{states}, but they do it redundantly, up to a complex factor, only. (2) And besides the \textit{states} are merely a derived notion, the primitive (phenomenologically given) notion being the \textit{qualities}, which correspond to the \textit{linear closed subspaces}. But if we wish to generalize the lattice of all linear closed subspaces from a Euclidean space to infinitely many dimensions, then one does not obtain Hilbert space, but that configuration, which Murray and I called `case \textbf{II}$ _{1}$' [i.e., the modular lattice isomorphic to $P_{c}(F)$]. ... And this is chiefly due to the presence of the rule [of modularity]. This `formal rule' would be lost, by passing to Hilbert space!'' (quoted in Rédei 1998, 112)

\end{quote}

This confession has been extensively discussed in the literature. For example, Miklos Rédei wrote: ``[T]he moral of the presented story of von Neumann's intellectual move from the Hilbert space formalism towards the type \textbf{II}$ _{1}$ (and even more general) algebras is that what drove him was not the desire to have a mathematically unobjectionable theory -- there was nothing wrong with Hilbert space formalism as a mathematical theory. What von Neumann wanted was conceptual understanding.'' (Rédei 1998, 116) But how exactly does Hankel's PPF, which is invoked not only in the Hilbert space generalization of Euclidean space, but also as a philosophical justification for the preservation of the rule of modularity in the transition from QM$_{n}$ to QM$_{\infty}$, help in this pursuit of what Rédei called conceptual understanding? To be sure, modularity was taken by Birkhoff and von Neumann to allow a well-defined dimension function \textit{d(a)}, which they noted to be formally equivalent to a finite probability measure.\footnote{Cf. Birkhoff and von Neumann 1936, 832. See Rédei 1998 for extensive discussion of von Neumann's conception of probability. The assumption of the well-definedness of \textit{d(a)} was questioned by Weyl in 1940, as we will see in the next section.} As Rédei further pointed out, modularity excludes not only infinite, and thus ``pathological'', probability measures, but pathological operators as well, i.e., unbounded operators on Hilbert space. But was this enough for reaching the kind of understanding that von Neumann was looking for? How did his pursuit relate to the overall project of discovering not only the non-classical logical structure of QM$_n$, but that of QM$_\infty$ as well? 

As avowed from the very outset, I think that one cannot reach a full understanding of a scientific theory without knowing what determines the meaning of its linguistic expressions, including both its logical and non-logical terms. I also think, more particularly, that von Neumann's move against the Hilbert space formalism in QM$_{\infty}$ can be better understood if one considers the metasemantics of QM$_{\infty}$, and in particular, what he thought determined the semantic attributes of its non-distributive logical calculus. The relevant question here is whether the application of Hankel's PPF to what von Neumann took to be a ``proper'' generalization of QM$_n$ led him to believe that the meaning of QM$_\infty$ is determined by its rules, and in particular, that the rules of its logical calculus determine the meaning of its connectives. 

As we have seen above, in QM$_{n}$, the logical connectives are defined by semantic rules in terms of operations on the modular lattice $\mathcal{LA}$ of the closed linear subspaces of finite-dimensional Hilbert space, via the homomorphism $h : S_\mathcal{QL} \longrightarrow S_\mathcal{LA}$. But these semantic rules cannot remain in place in QM$_{\infty}$, since preserved modularity eliminates the possibility of defining logical connectives in terms of operations on the lattice of the closed linear subspaces of infinite-dimensional Hilbert space. This is because, as we will see in chapter 4, this lattice turns out to be orthomodular, rather than modular. In other words, no homomorphism $h_\infty : S_\mathcal{QL} \longrightarrow P_{\infty}(F)$ can be constructed that can help define the logical connectives and the relation of logical consequence, while at the same time having $P_{\infty}(F)$ validate the generalized rule of modularity. But if the semantic rules that define the connectives in QM$_{n}$ cannot remain in place in QM$_{\infty}$, then what determines the meaning of its connectives? 

Had von Neumann embraced what Hankel (following Peacock) thought were the metasemantic implications of the PPF, he would have argued, I think, that it is the rules of the logical calculus that determine the meaning of connectives in QM$_{\infty}$. However, despite his invoking the PPF to select $P_{c}(F)$ in the transition to infinite dimensions, von Neumann argued differently: the logical connectives in QM$_{\infty}$ must be defined by stipulating new semantic rules in terms of operations on the algebra that he called ``type \textbf{II}$ _{1}$''. In other words, a new homomorphism must be constructed, $h_{c} : S_\mathcal{QL} \longrightarrow P_{c}(F)$ such that the generalized rule of modularity is validated by $P_{c}(F)$ and the connectives are then defined by semantic rules:

\begin{center}
    
$h_{c}(\neg p) = h_{c}(p)^{\bot}$ \\ $h_{c}(p \wedge q) = h_{c}(p) \cap h_{c}(q)$ \\ $h_{c}(p \vee q) = h_{c}(p) \cup h_{c}(q)$, 

\end{center}

This raises important metasemantic questions. Do these new semantic rules determine the meaning of connectives precisely? Is the logical calculus of QM$_{\infty}$ categorical? More exactly, are its logical connectives truth-functional under all truth valuations $v : S_\mathcal{QL} \longrightarrow \{0,1\}$? In a paper that has unfortunately drawn less attention than it deserves, Weyl argued that this was not shown to be the case even for the logical calculus of QM$_{n}$ (Weyl 1940). As we will see presently (section 3.2.2), he argued that the connectives of the latter, although they may be truth-functional under some valuations, it is doubtful that they are truth-functional under all valuations. Although he could have made the same argument with respect to the logical calculus of QM$_{\infty}$, that argument had to wait 40 years to be made explicitly. This will be critically discussed further below (in section 4.2). Before I do so, I want to close this section with a brief historical note concerning what I take to be an important difference between Bohr's and von Neumann's views of QM. 

It is well known that, while many criticized Bohr's CP, von Neumann described it as a ``striking point'' of QM, ``which played a decisive role in the final clarification of the problem'' of a coherent formulation of the theory (von Neumann 1932, 6). He also regarded the CP as a not fully quantum-mechanical principle, ``belonging half to classical mechanics and electrodynamics'' (\textit{loc. cit.}). Returning the favor, Bohr praised von Neumann's ``elegant axiomatic exposition'' and even pointed out the ``evident'' character of his completeness proof (Bohr 1939, 16). However, although their views on QM were, for some time, considered parts of the same interpretation -- the Copenhagen interpretation -- Bohr never endorsed or even mentioned (in print) von Neumann’s collapse postulate (Bokulich and Bokulich 2005, 367sq). Bohr did speak against collapse at the 1938 Warsaw conference, where he deftly remarked, in the discussion following his talk, that it was ``a question of choosing the most adequate description of the experiment.'' (Bohr 1939, 45) The implication was that the collapse postulate failed to provide such a description. 

Beyond this, the difference between Bohr's and von Neumann's view on QM is usually characterized in terms of their personal abilities. A meeting between them was described by Léon Rosenfeld in a 1963 interview by Thomas Kuhn and John Heilbron as ``rather disastrous'', and Bohr's (private) comments on a talk by von Neumann as ``very disparaging''. In 1968, in his letters to Jeffrey Bub, Bohm portrayed them as almost opposite figures: Bohr, well known for his physical insight and his (alleged) lack of mathematical rigor; von Neumann, highly regarded for his mathematical genius, despite his (alleged) lack of physical insight. Such characterizations may express significant and complex aspects of their science. Nevertheless, I think that the difference in the conception of metasemantics that can be attributed to each of them is no less important. While Bohr followed Hankel in adopting a non-representationalist metasemantics, von Neumann did not do so. They had different solutions to what I called the permanence problem. If I am right to think that metasemantics has consequences for our philosophical understanding of QM, then this difference can help us grasp what might have well been the real rift between Bohr and von Neumann.     

In any case, in the next section, I will reconstruct Weyl's argument against Birkhoff and von Neumann's non-distributive QL, which clearly shows that metasemantics has consequences for our understanding of the logic of QM.

\subsubsection{Weyl's objections to quantum logic}

Weyl's paper, ominously titled ``The Ghost of Modality'', deals with the question whether a universal logic of modality exists. Although overlooked, the paper did not go completely unnoticed. Bas van Fraassen mentioned it as an antecedent to the semantic approach to scientific theories that, on his view, only began in earnest with Evert Willem Beth's 1948 \textit{Natuurphilosophie} (van Fraassen 1987, 105). Likewise, he later wrote: ``Weyl gave in rudimentary but prescient form the outline of the semantic analysis that would eventually unify modal, quantum, and intuitionistic logic.'' (van Fraassen 1991, 129) To answer the question whether a universal logic of modality exists, Weyl did indeed carefully examine several ``models'' in which modal operators combine ``unambiguously'' with logical connectives. Given the existence of different sets of axioms for different such models, among which was Birkhoff and von Neumann's QL, Weyl concluded, however, that (as one might already guess from the title of the paper) there is no universal logic of modality.

In his analysis of QL, Weyl took up one of the questions that Birkhoff and von Neumann had asked at the very end of their paper: ``What experimental meaning can one attach to the meet and join of two given experimental propositions?'' (Birkhoff and von Neumann 1936, 837) Weyl's answer to this question was negative: no experimental meaning can be given to the meets and joins of any two experimental propositions of QM. This is because, he maintained, $\mathcal{QL}$ was not shown to be a complete calculus. By this he may be taken to have meant two things: first, it was not shown to be a categorical calculus, i.e., it may allow for non-isomorphic valuations, including valuations that would make the logical connectives non-truth-functional; and secondly, but relatedly, despite what von Neumann believed to be the case, the modular lattice, $\mathcal{LA}$, was not shown to admit of a well-defined probability measure. 

In the first step of his argument for the incompleteness of $\mathcal{QL}$, Weyl considered CL:

\begin{quote}

In classical logic there is no doubt about the meaning of any combination of arbitrary propositions \textit{p}, \textit{q}, \textit{r}, ... by the operators $\neg, \wedge, \vee, $... however complicated the structure may be, and we have a perfectly clear combinatorial criterion by which to decide whether such a combined proposition is generally (analytically) true: if its value turns out to be 1 whatever combination of values 1, 0 one assigns to the arguments \textit{p}, \textit{q}, \textit{r}, ... .\footnote{Cf. Weyl 1940, 288. I have modified the text of this quotation for uniform notation. For simplicity, I have also ignored the conditional.}

\end{quote}

The combinatorial criterion for identifying classical tautologies would seem to imply that the CL connectives are truth-functional, i.e., they are such that for any truth valuation $v$: 

\begin{center}
    
$v(\lnot p)=1$ if and only if $v(p)=0$, \\ $v(p \land q)=1$ if and only if $v(p)=v(q)=1$, \\ $v(p \vee q)=0$ if and only if $v(p)=v(q)=0$.

\end{center}

The truth-functionality of CL connectives requires the existence of a homomorphism $f : S_\mathcal{CL} \longrightarrow S_\mathcal{BL}$ where, as Birkhoff and von Neumann noted, $\mathcal{BL}$ is a Boolean algebra, as well as that of a homomorphism $g : S_\mathcal{BL} \longrightarrow \{0,1\}$. If these two conditions were met, then the CL connectives would be truth-functional because the valuation $v : S_\mathcal{CL} \longrightarrow \{0,1\}$ is a homomorphism, too, since $v = g \circ f$. Weyl must have thought that no other valuations are possible, and in particular, no valuation that would make the CL connectives non-truth-functional while obeying the combinatorial criterion for identifying classical tautologies.\footnote{As it turned out, Weyl was not right about this, but the reason why he was not right would only become better known a few years after his paper was published, when Carnap constructed what he called ``non-normal interpretations'' of the classical logical calculus (Carnap 1943). I will have more to say about these interpretations, in section 5.3.1.} 

In the second step of the argument, when Weyl compared CL with QL, he claimed that
what he referred to as a ``parallelism'' holds in the former but not in the latter:

\begin{quote}

The classical logic of propositional functions with its \textit{variables} \textit{p}, \textit{q}, ... has a much greater flexibility, due to the parallelism between the operators $\sim, \cap, \cup$ for sets and for (truth or probability) values, a feature prevailing in classical logic which breaks down completely in quantum logic.\footnote{Cf. Weyl 1940, 299. I modified the text of this quotation for uniform notation, and I ignored the quantifiers for simplicity.}

\end{quote}

Weyl's view is that, unlike \textit{g}, the following two maps, $g' : S_\mathcal{LA} \longrightarrow \{0,1\}$ and $g'' : S_\mathcal{LA} \longrightarrow [0,1]$, are not homomorphisms. Therefore, even if $h : S_\mathcal{QL} \longrightarrow S_\mathcal{LA}$ is a homomorphism, there may be at least one truth valuation $v' : S_\mathcal{QL} \longrightarrow \{0,1\}$, with $v' = g' \circ h$, which would not be a homomorphism. In this case, the QL connectives would not be truth-functional under valuation $v'$. This may not exclude the possibility that there are other valuations under which QL connectives are truth-functional, but Weyl was right that if $g'$ is not a homomorphism, then this is enough to justify the claim that Birkhoff and von Neumann did not establish that $\mathcal{QL}$ is a complete, i.e., categorical calculus. 

To see why Weyl thought that $g'$ was not, or at least was not shown to be a homomorphism, consider any elements $a, b \in S_\mathcal{LA}$. What is the truth value that $g'$ assigns to their meet and join? Weyl thought that what Birkhoff and von Neumann said about the modular lattice $\mathcal{LA}$ provided no reason to believe that 

\begin{center}
    
$g'(a \cap b)=1$ if and only if $g'(a)=g'(b)=1$; \\ $g'(a \cup b)=0$ if and only if $g'(a)=g'(b)=0$,

\end{center}

 and so, no reason to believe that, for any $p,q \in S_\mathcal{QL}$,

\begin{center}
    
$g'(h(p \wedge q))=1$ if and only if $g'(h(p))=g'(h(q))=1$; \\ $g'(h(p \vee q))=0$ if and only if $g'(h(p))=g'(h(q))=0$.

\end{center}

 Therefore, Weyl had no reason to believe that QL conjunction and QL disjunction are truth-functional, i.e., no reason to believe that

\begin{center}
    
$v'(p \wedge q)=1$ if and only if $v'(p)=v'(q)=1$; \\ $v'(p \vee q)=0$ if and only if $v'(p)=v'(q)=0$.

\end{center}

 Birkhoff and von Neumann seem to have assumed precisely the parallelism that Weyl saw broken. For not only they believed that $\mathcal{LA}$ admits of a well-defined probability measure, which will be presently discussed, but they also maintained that, for any system whose physical variables are described by sentences \textit{A} and \textit{B}, it is natural to assume that when \textit{A} is true at precisely the states in \textit{h(A)}, and when \textit{B} is true at precisely the states in \textit{h(B)}, then $ A \wedge B $ is true at precisely the states in their meet $ h(A) \cap h(B) $.\footnote{Cf. Birkhoff and von Neumann 1936, 829n. They also probably believed, although never stated explicitly, that it would be as natural to assume that when \textit{A} is true at precisely the states in \textit{h(A)}, and when \textit{B} is true at precisely the states in \textit{h(B)}, then $ A \vee B $ is true at precisely the states in their span $h(A) \cup h(B)$.} More exactly, this assumption says that, for any valuation $v$, if $v(A)=1$ precisely when $A$ is mapped to $h(A)$, and $v(B)=1$ precisely when $B$ is mapped to $h(B)$, then $v(A \wedge B)=1$ precisely when $A \wedge B$ is mapped to $h(A \wedge B)$. This entails that there is a map $g'$ such that if $g'(h(A))=1$ and $g'(h(B))=1$, then $g'(h(A \wedge B))=1$. But as just discussed, Weyl pointed out that Birkhoff and von Neumann had given no reason that would justify this condition on $g'$.

Furthermore, Weyl also argued that the modular lattice, $\mathcal{LA}$, does not, or at least was not shown to admit of a well-defined probability measure. Let $v'' : S_\mathcal{QL} \longrightarrow [0,1]$ be a probability valuation. He noted that $v'' = g'' \circ h$ cannot be a homomorphism, even if $h : S_\mathcal{QL} \longrightarrow S_\mathcal{LA}$ is a homomorphism. This is because $g'' : S_\mathcal{LA} \longrightarrow [0,1]$ is not a homomorphism. To see that $g''$ is not a homomorphism, consider again any elements $a, b \in S_\mathcal{LA}$. What is the probability value that $g''$ assigns to their meet and join? Weyl pointed out that the values of $a \cap b$ and $a \cup b$ are not uniquely determined by the values of $a$ and $b$. Instead, 
\begin{center}
    
$g''(a \cup b) \in [0, max(g''(a), g''(b))]$, \\
$g''(a \cap b) \in [0, min(g''(a), g''(b))]$.

\end{center}

This immediately implies that a probability measure on $\mathcal{LA}$ is not well-defined. Recall that Birkhoff and von Neumann maintained that it is precisely its modularity that guarantees the existence of a probability measure, which is formally identical to their dimension function \textit{d(a)}. But Weyl's criticism clearly entails that one property of this function,

\begin{center}
    
$d(a) + d(b) = d(a \cap b) + d(a \cup b)$)

\end{center}

fails to hold. This is because, since $d(a \cap b) + d(a \cup b)$ lacks a unique value, $d(a) + d(b)$ lacks a unique value. He also thought that it would be simply arbitrary to fix a unique value by merely stipulating that 

\begin{center}
    
$g''(a \cup b) = max(g''(a), g''(b))$, \\ $g''(a \cap b) = min(g''(a), g''(b))$. 

\end{center}

As he put it, ``by enforcing the arbitrary rules ... we sold our birthright of reality for the pottage of a nice formal game.'' (Weyl 1940, 299) As a consequence of this analysis, Weyl concluded that $\mathcal{QL}$ has very little extrinsic significance, for quantum physics, despite its attractive intrinsic mathematical features.

Weyl's criticism is a clear answer to that important question from the very end of Birkhoff and von Neumann's paper: the meets and joins of any two experimental propositions of QM can be given no experimental meaning because it has not been shown that such a meaning can be precisely determined. More exactly, it has not been shown that the semantic attributes, i.e., truth values and probability values, of the conjunctions and disjunctions of any two sentences in $S_\mathcal{QL}$ are precisely determined by the stipulated semantic rules.    

This is important for the metasemantics of QL, for the following reason. While it would later be established that an orthomodular lattice admits a well-defined (essentially unique) probability measure,\footnote{Cf. Gleason 1957, but see also Beltrametti \& Cassinelli 1981, ch. 11 for a concise presentation.} this development, i.e., the replacement of the rule of modularity by the weaker rule of orthomodularity, fails to address Weyl's worries about the truth-functionality of QL connectives. In fact, it shows his extraordinary prescience, for after the rule of modularity was replaced by that of orthomodularity, QL turned out to be provably non-categorical. This was established by Geoffrey Hellman, who demonstrated that QL disjunction and conjunction are non-truth-functional connectives (Hellman 1980).

Wrapping up, I hope to have shown in this chapter that the significance of the PPF for the development of QM is undeniable. In one case, I have argued that Bohr's CP was grounded in the PPF as a metatheoretical principle. This explains why he thought that QM was a rational generalization of classical physics, and that this notion should be understood in the sense of generalization that had been articulated by Peacock and Hankel in the 19th century. I have also attributed a non-representationalist metasemantics to Bohr, who maintained that the physical meaning of symbolic expressions in QM is determined by its rules of calculation. In the other case, I have argued that Birkhoff and von Neumann's work on QL, and especially the latter's generalization of QM$_n$ to QM$_{\infty}$, despite being driven by an application of the PPF, does not challenge their representationalist metasemantics. This reveals what I take to be a significant difference between Bohr's and von Neumann's views on QM. Finally, I have presented Weyl's metasemantic reasons for rejecting modular QL as a ``formal pottage''.

In the second half of the book, my investigation will split in the following way. Chapter 4 will follow the line of critical reflections on QL initiated by Weyl and will analyze more recent arguments concerning orthomodular QL. Chapter 5 will then return to non-representationalism about QM, and consider the inferentialist metasemantics proposed by Richard Healey, as well as the project of a non-inferentialist metasemantics for QBism.

\newpage

\section{Logical revisionism}

This chapter focuses on the logical revisionism initiated by Putnam’s
rejection of CL in QM. One argument, due to Rumfitt, denies the very possibility
of revisionism, on rational grounds. Another, due to Williamson, claims that
revisionism is either inconsistent or explanatorily too costly. Both of these recent arguments are shown to be unsound, although neither QL, nor revisionism is thereby supported. An older argument, due to Hellman, opposes revisionism on semantic grounds, related to the non-truth-functionality of QL connectives. But only a significantly improved version of this argument is shown to provide a sufficient
reason to refute Putnam’s revisionism.\footnote{This chapter is partly based on Toader 2025a, Horvat and Toader 2023 and 2024.}

\subsection{Rational adjudication against classical logic}

The transition from classical physics to QM was taken by Birkhoff and von Neumann to require a replacement of classical logic (CL) by a non-distributive quantum logic (QL). The present chapter follows the line of critical reflection on QL initiated by Weyl, focusing on arguments developed after Putnam enrolled QL in the service of realist metaphysics (Putnam 1968). One important argument, right along that line, was developed by Geoffrey Hellman, who proved (as recalled in section 4.2.3) that QL disjunction is not truth-functional, and took this to justify the claim that QL connectives are semantically inequivalent to their CL counterparts (Hellman 1980). Rather than a revision of CL, as Putnam saw it, QL is therefore just an alternative to CL, as Birkhoff and von Neumann appear to have intended it. An objection recently raised by Ian Rumfitt against Putnam's revisionism contends that the proof that the rule of distributivity fails in QM (henceforth referred to as ``the Proof'') is rule-circular: it cannot justify the claim that distributivity fails because it actually assumes this very rule in the metalanguage (Rumfitt 2015). This objection thus purports to reject the claim that quantum logicians can rationally adjudicate against CL. I will describe in detail a way to resist this objection by minimally revising the so-called eigenstate-eigenvalue link of standard QM, although an immediate response could also be that the objection actually equivocates on the meaning of ``distributivity''.

This response is based on Hellman's argument: if one takes semantic inequivalence seriously, then distributivity in the metalanguage is a classical rule, while that which the Proof shows to fail in QM is not. Thus, there really is no rule-circularity. But then, of course, the real problem with the Proof is that, despite Putnam's intention, it fails to justify the claim that it is classical distributivity that fails in QM. I will start by presenting the Proof (section 4.1.1), discussing Rumfitt's objection (section 4.1.2), and explaining how this can be resisted (section 4.1.3). Turning to Hellman's argument, I will focus on his assumption that QL negation is truth-functional and, thus, classical. He justified this by noting that quantum logicians already accept CL in mathematics and in the language about the macro-physical world, rather than by appeal to any realist metaphysical grounds. However, we will see (in sections 4.2.1 -- 4.2.3) that there are strong reasons for taking QL conjunction to be truth-functional, which entails that QL disjunction and negation are non-truth-functional. This backs the claim, first made by Arthur Fine in 1972, that the Proof does not actually justify the failure of \textit{classical} distributivity. But one might claim, as Timothy Williamson has done, that if quantum logicians accept CL in mathematics, then the application of classical mathematics to QM leads to an inconsistency with the use of QL in the language about the micro-physical world (Williamson 2018). However, as we will see (in sections 4.3.1 -- 4.3.3), no such inconsistency exists.

\subsubsection{Putnam's logical revisionism}

Birkhoff and von Neumann's argument for replacing the package QM \& CL by the alternative package QM \& QL does not explicitly say anything about metaphysics. Taking this up in the late 1960s, Putnam argued for replacing QM \& CL \& IM by QM \& QL \& TM, where intolerable metaphysics is part of the former package and tolerable metaphysics part of the latter. He pointed out that QM \& QL \& TM is preferable to QM \& CL \& IM, since the absence of worldly indeterminacy is more tolerable than its presence, just as the absence of hidden variables is more tolerable than their presence, on any measure of tolerance for realist metaphysical hypotheses. In addition, QM \& CL \& IM makes stronger logical assumptions, since it demands that conjunction distribute over disjunction. On this basis, Putnam defended the view that CL must be revised by QL, holding on to this view for almost 25 years.\footnote{Putnam also allowed a bivalent semantics for QL, so his favorite package was actually QM \&  BQL \& TM. Putnam 1991, written in Italian and never translated, may be the first paper where he retracted his view on QL, following criticism by Michael Dummett. See also Putnam 1994, where his retraction is triggered by Michael Redhead's criticism. For an analysis of Putnam's views on QL, see Bacciagaluppi 2009. For more on QL, see e.g. Dalla Chiara \textit{et al.} 2004.} Later on, Putnam came to acknowledge packages that preserve CL and include realist metaphysical hypotheses that are not that intolerable: ``Surely, before we accept views that require us to revise our logic, we need to be sure that it is \textit{necessary} to go \textit{that} far to make sense of quantum phenomena. And we now know that it is not. ... I was wrong to think that a tenable realistic interpretation must give up classical logic.`` (Putnam 2012, 175-177) 

However, it remains unclear whether Putnam was ready to concede more than this, and in particular that he was perhaps also wrong to think that it is even \textit{possible} to go that far, that a tenable realist interpretation of quantum mechanics \textit{can} give up CL and adopt QL instead. Of course, one might think that the possibility of adopting QL in QM should not hang on the tenability of a realist interpretation, especially if -- as some philosophers claim -- a QL realist interpretation of QM has already been shown untenable and is now of merely historical interest (Maudlin 2010, 2022). Indeed, as announced above, my focus in this section will be on Rumfitt's objection, which can be analyzed independently of the issue of the tenability of a realist interpretation, that Putnam was wrong to believe that the rule of distributivity can be ditched. To facilitate the analysis of this objection, this section offers a formal proof that distributivity fails in QM, as well as its validation by the semantics assumed by Putnam.

Let $\mathcal{QL}$ be the quantum logical calculus, which includes a sentential language, $L$, with variables for atomic sentences, $p$, $q$, ..., and symbols for logical connectives, $\neg, \wedge, \vee $, such that the set $S_\mathcal{QL}$ of sentences is defined inductively. Let $\mathcal{QL}$ have most rules of a classical natural deduction system, such as double negation, De Morgan rules, as well as introduction and elimination rules for all connectives (with the exception of unrestricted $\vee$-elimination).\footnote{One could further let $\mathcal{QL}$ include a conditional defined, say, as follows: $ p \rightarrow q $ iff $ \neg p \vee ( p \wedge q ) $. This is a counterfactual conditional, and it is weaker than the classical one, as for example it does not admit strengthening the antecedent. But again, for simplicity, the conditional will be ignored.} For any $i \in \{1, ..., n\}$, let $A_{i}$ be sentences in $S_\mathcal{QL}$ stating the possible positions of a quantum system \textit{P}, and for any $j \in \{1, ..., m\}$, let $B_{j}$ be sentences in $S_\mathcal{QL}$ stating its possible momenta. Assume that the state space associated with \textit{P} is a finite-dimensional Hilbert space, and assume the standard eigenstate-eigenvalue link (EEL), i.e., that \textit{P}'s observables can have precise values only at their eigenstates. Suppose \textit{P} is in a position eigenstate, and $A_{z}$ states \textit{P}'s measured position. Then the Proof is given by the following formal derivation:

\bigskip

$ \begin{array}{llr}

1.\  A_{z} \wedge (B_{1} \vee ... \vee B_{m}) & \text{premise} \\

2.\  \neg (A_{i} \wedge B_{j}) & \text{premise} \\

3.\  A_{z} & 1, \wedge\text{-elimination} \\

4.\  \neg(A_{z} \wedge B_{1}) \wedge ... \wedge \neg (A_{z} \wedge B_{m}) & 2, 3, \text{substitution}, \wedge\text{-introduction} \\

5.\  \neg( (A_{z} \wedge B_{1}) \vee ... \vee (A_{z} \wedge B_{m})) & 4, \text{de Morgan}

\end{array} $

\bigskip

The Proof provides us with a counterexample to the rule of distributivity in \textit{L}, which can be stated as $ \ulcorner A_{z} \wedge (B_{1} \vee ... \vee B_{m}) \urcorner \nvdash \ulcorner (A_{z} \wedge B_{1}) \vee ... \vee (A_{z} \wedge B_{m})\urcorner $. To see why Putnam took this counterexample to be validated by \textit{L}'s semantics, let's have a look at the semantics he considered.\footnote{In his 1968 paper, Putnam also suggested an operational semantics, where the logical connectives of the calculus are defined in terms of tests that can in principle be performed in a lab, rather than abstract operations on the elements of a lattice. More details about this are given in section 4.3.2.} 

Let $\mathcal{OO}$ be an orthomodular ortholattice, i.e., a partially ordered set $S_\mathcal{OO}$ of elements, such as the closed subspaces of the Hilbert space associated with \textit{P}, together with operations $^{\bot}, \cap, \cup$, for orthocomplementation, meet, and join (or span), respectively. This lattice satisfies the rule of orthomodularity -- an even weaker form of distributivity than modularity -- which states that, for any $a, b \in S_\mathcal{OO}$,

	\begin{center}
		
		if $a \subseteq b$, then $b = a \cup (a^{\bot} \cap b)$. 
		
	\end{center}
	
	We say that $\mathcal{OO}$ is a model of $\mathcal{QL}$ if and only if there is a map $h : S_\mathcal{QL} \longrightarrow S_\mathcal{OO}$ such that \textit{h} is a homomorphism:
	
	\begin{center}
		
		$h(\neg p) = h(p)^{\bot}$ \\ 
		$h(p \wedge q) = h(p) \cap h(q)$ \\ 
		$h(p \vee q) = h(p) \cup h(q) $ 
		
	\end{center}
	
	On the basis of this homomorphism and the partial order of $\mathcal{OO}$, one can then define orthocomplementation as set-theoretic complementation plus partial order reversal, and logical consequence:
	
	\begin{center}
		
		$h(p) = h(q)^{\bot}$ iff $\{x : x \subseteq h(p)\} = \{x : x {\bot} h(q) \}$
		
		$\Gamma \models_\mathcal{QL} q$ iff $\cap \{h(p):p \in \Gamma \} \subseteq h(q)$.
	\end{center}
	
	The question is whether $\mathcal{OO}$ verifies the Proof. Does it validate the above counterexample to distributivity? Indeed, let $\varnothing$ be the empty space and \textbf{1} be the entire Hilbert space. Then the validation goes as follows:
	
	\begin{center}
		
		$ h(A_{z}) \cap (h(B_{1}) \cup ... \cup h(B_{m})) \nsubseteq (h(A_{z}) \cap h(B_{1})) \cup ... \cup (h(A_{z}) \cap h(B_{m})) $ 
		
		$ h(A_{z}) \cap \textbf{1} \nsubseteq \varnothing \cup ... \cup \varnothing $ 
		
		$ h(A_{z}) \nsubseteq  \varnothing $. 
		
	\end{center}
	
The failure of distributivity is a direct consequence of the fact that the sentences in $S_\mathcal{QL}$, interpreted as experimental propositions in QM, form the non-distributive lattice $\mathcal{OO}$, which can be arguably understood as a consequence of the non-commutativity of the algebra of quantum-mechanical observables. The most important thing to note, for present purposes, is that $\mathcal{OO}$ makes the QL connectives non-truth-functional.\footnote{For more discussion of the non-truth-functionality of QL connectives, see section 4.2 below. Non-truth-functionality arguably follows from the Kochen-Specker theorem as well, which guarantees that the partial Boolean algebra of the Hilbert space has no homomorphic Boolean extension, i.e., there is no homomorphism from the partial Boolean algebra to the two-valued Boolean algebra $\{0,1\}$, on the assumption that the space is of dimension $d>2$ (Kochen and Specker 1967). For some discussion, see Dickson 1998.} But Putnam does not seem to have been worried at all about non-truth-functionality, or in any case did not think that this had any implications with respect to the semantic attributes of QL connectives, for he famously, though controversially, claimed that ``adopting quantum logic is not changing the meaning of the logical connectives, but merely changing our minds about the [distributive] law''.\footnote{Cf. Putnam 1968, 190. This controversial claim will be also further discussed in section 4.2 below.} 	

Non-truth-functionality is, however, the very reason why Rumfitt rejects $\mathcal{OO}$ as a semantics without any logical significance:

\begin{quote}

The operations of intersection, span, and orthocomplement on the subspaces of a Hilbert space indeed form a non-distributive lattice, and that non-distributive lattice is useful in quantum-theoretic calculations. But we have as yet been given no reason to assign any \textit{logical} significance to the lattice. To do so, it would need to be argued that, when the statement \textit{A} is true at precisely the states in \textit{h(A)}, and when the statement \textit{B} is true at precisely the states in \textit{h(B)}, then the disjunction $\ulcorner A \vee B \urcorner $ is true at precisely the states in the span of \textit{h(A)} and \textit{h(B)}. Putnam gives no such argument, so his claim that the proposed rules have any logical significance is unsupported.'' (Rumfitt 2015, 176; modified for uniform notation)

\end{quote}

 But to argue, as Rumfitt demanded, that when \textit{A} is true at precisely the states in \textit{h(A)}, and when \textit{B} is true at precisely the states in \textit{h(B)}, then their disjunction $\ulcorner A \vee B \urcorner $ is true at precisely the states in the span of \textit{h(A)} and \textit{h(B)} is to argue exactly for the existence of the parallelism that already Weyl in 1940 saw was broken, in the case of modular QL, and Hellman in 1980 proved cannot exist, in the case of orthomodular QL. For Rumfitt would have liked Putnam to argue not only that $h : S_\mathcal{QL} \longrightarrow S_\mathcal{OO}$ is a homomorphism, but also that $g : S_\mathcal{OO} \longrightarrow \{0,1\}$ is a homomorphism as well. But this would entail that any truth valuation is a homomorphism $v : S_\mathcal{QL} \longrightarrow \{0,1\}$, since $v = g \circ h$. In the case of QL disjunction, this would entail that 

\begin{center}
    
$v(A \vee B) =1$ if and only if $v(A) = 1$ or $v(B) = 1$.

\end{center}

 Thus, what Rumfitt would have wanted Putnam to argue is that QL disjunction is truth-functional. Since Putnam did not so argue, for it is actually impossible to do so, Rumfitt rejected $\mathcal{OO}$ as a logically insignificant semantics. But this raises a difficult question in the philosophy of logic: what justifies the assignment of logical significance to a certain semantics? More specifically, why should logical significance be dictated, as Rumfitt appears to assume, by our pre-theoretic intuitions about truth-functionality, as opposed to our most successful scientific theories? Leaving this for later, I want now to discuss his objection against the Proof.

\subsubsection{Rumfitt's rule-circularity objection}

While simplicity and strength are significant, perhaps even predominant criteria for choosing between rival scientific theories, such criteria are not indispensable when it comes to choosing between rival logics. This view, recently defended by Rumfitt, adopts Dummett's stability condition for the mutual understanding between logicians of different denominations: ``How can the classical logician and the non-standard logician come to understand one another? Not, obviously, by defining the logical constants. They have to give a semantic theory; and they need one as stable as possible under changes in the underlying logic of the metalanguage.'' (Dummett 1987: 254) Rumfitt takes Dummett to have ``pointed the way'' towards identifying what is necessary for rational adjudication between logics (Rumfitt 2015: 9). Thus, in the case of interest here, if $\mathcal{QL}$ were given a classical semantics, then the Proof cannot adjudicate against CL, for a classical semantics would not be stable enough if one adopted QL (rather than CL) in the metalanguage. If one did adopt CL in the metalanguage, the Proof still cannot adjudicate against CL, because a non-classical semantics for $\mathcal{QL}$ would be also unstable (for reasons we are about to discuss). 

These violations of Dummett's stability condition justify the classical logician's rejection of the Proof on rational grounds, or so Rumfitt believes. But I don't think that his view is correct, and in this section I want to provide some reasons for doubting it's tenability. 
	
Let $\mathcal{TG}_{1}$ be a non-classical semantics, what Rumfitt calls a truth-ground semantics. The basic notion of this semantics is that of a truth-ground for a statement, that is a closed set of possibilities at which the statement is true. A set of possibilities, \textit{U}, is closed if and only if it is its own closure, \textit{Cl(U)}, where this is ``the smallest set of possibilities at which every statement that is true throughout \textit{U} is true.'' (Rumfitt 2015: 162) Let $S_{\mathcal{TG}_{1}}$ be a set of truth-grounds, together with operations $^{\bot}, \cap, \cup$, for orthocomplementation, intersection, and union, respectively. Then $\mathcal{TG}_{1}$ is a model of $\mathcal{QL}$ if and only if the truth-grounds can be taken as the closed subspaces of a finite-dimensional Hilbert space associated with our physical system \textit{P} and there is a homomorphism $r_{1} : S_\mathcal{QL} \longrightarrow S_{\mathcal{TG}_{1}}$ such that:
	
\begin{center}
		
		$r_{1}(\neg p) = r_{1}(p)^{\bot}$ \\ $r_{1}(p \wedge q) = r_{1}(p) \cap r_{1}(q)$ \\ $r_{1}(p \vee q) = Cl(r_{1}(p) \cup r_{1}(q)) $ 
		
\end{center}
	
The closure operation, \textit{Cl}, which is taken here as a primitive relation, has some nice lattice-theoretic properties such as
	
	\begin{center}
		
		$ r_{1}(p) \subseteq Cl(r_{1}(p))$, \\ $ClCl(r_{1}(p))=Cl(r_{1}(p))$, and  \\ if $r_{1}(p) \subseteq r_{1}(q)$, then $Cl(r_{1}(p)) \subseteq Cl(r_{1}(q))$     
	\end{center}
	
	Importantly, however, \textit{Cl} does not have the following topological property:
	
	\begin{center}
		
		$Cl(\varnothing \cup ... \cup \varnothing) = Cl(\varnothing)= \varnothing$,     
		
	\end{center}
	
	so as a consequence, it cannot verify that a disjunction of false sentences is false (Rumfitt 2015: 135-6, 162-3). It turns out that, precisely because it lacks this property, $\mathcal{TG}_{1}$ by itself cannot validate our prior counterexample to distributivity:
	
	\begin{center}
		
		$ r_{1}(A_{z}) \cap Cl((r_{1}(B_{1}) \cup ... \cup r_{1}(B_{m})) \nsubseteq Cl((r_{1}(A_{z}) \cap r_{1}(B_{1})) \cup ... \cup (r_{1}(A_{z}) \cap r_{1}(B_{m}))) $ 
		
		\bigskip
		
		$ r_{1}(A_{z}) \cap Cl(\varnothing \cup ... \cup \varnothing) \nsubseteq Cl(\varnothing \cup ... \cup \varnothing) $
		
	\end{center}
	
	In order to be able to validate the counterexample, $\mathcal{TG}_{1}$ must be supplemented by a metalogical proof that a disjunction of false sentences is false, i.e., a proof in the metalanguage to the effect that, on line 5 of the Proof, disjunction is truth-functional. As Rumfitt puts it, one needs a proof of ``the semantic principle that a true disjunction must contain at least one true disjunct'' (Rumfitt 2015: 174-5). This is supposed to make up for $\mathcal{TG}_{1}$'s missing the topological property above.

However, Rumfitt notes, the metalogical proof requires the unrestricted rule of $\vee$-elimination and, thus, assumes CL.\footnote{Putnam allowed the restricted form of $\vee$-elimination in QL in order to support the claim that its connectives have the same meaning as their classical counterparts since they have the same introduction and elimination rules (Putnam 1968, 189) So he may perhaps be taken to have implicitly acknowledged that the unrestricted form of the rule is valid only in classical logic.} The Proof is, therefore, rule-circular (since the unrestricted rule of $\vee$-elimination is logically equivalent to distributivity). Because of this rule-circularity, Rumfitt rejected $\mathcal{TG}_{1}$ as an unacceptable semantics for $\mathcal{QL}$ on rational grounds. For convenience, here is the metalogical proof (including a tacit correction of Rumfitt's justification for its last step): 

\bigskip

$\begin{array}{lll}

1.\ Tr(\ulcorner A \vee B\urcorner) & \text{premise} \\

2.\ Tr(\ulcorner A \vee B \urcorner) \longrightarrow A \vee B & \text{principle about truth} \\

3.\ A \vee B & \text{1, 2, modus ponens} \\

4.\ A & \text{assumption} \\

5.\ A \longrightarrow Tr(\ulcorner A \urcorner) & \text{principle about truth} & \text{(side premise)} \\

6.\ Tr(\ulcorner A \urcorner) & \text{4, 5, modus ponens} \\

7.\ Tr(\ulcorner A \urcorner) \vee Tr(\ulcorner B \urcorner) & \text{6,} \vee\text{-introduction} \\

8.\ B & \text{assumption} \\

9.\ B \longrightarrow Tr(\ulcorner B \urcorner) & \text{principle about truth} & \text{(side premise)} \\

10.\ Tr(\ulcorner B \urcorner) & \text{8, 9, modus ponens} \\

11.\ Tr(\ulcorner A \urcorner) \vee Tr(\ulcorner B \urcorner) & \text{10,} \vee\text{-introduction} \\

12.\ Tr(\ulcorner A \urcorner) \vee Tr(\ulcorner B \urcorner) & \text{3, 5, 9,} \vee\text{-elimination}

\end{array}$

\bigskip

Note, first, that since the calculus $\mathcal{QL}$ is not, itself, a quantum system, there can be no surprise that the metalanguage has a truth-functional and, thus, classical  disjunction. Assuming that \textit{L} is a countable language, one could also prove that the semantics of the metalanguage of \textit{L} is a distributive lattice, based on Dunn's classical recapture of distributivity as an arithmetical theorem in $PA_{QL}^{1}$, i.e., first-order Peano arithmetic with QL (Dunn 1980). Of course, one would expect the classical logician to insist that, despite such classical recapture results, one should still not use CL in an argument against CL. Secondly, recall that we only needed a classical metalogical proof because $\mathcal{TG}_{1}$ was unable, by itself, to validate Putnam's proof. But a quantum logician can, and I think should, reject this semantics. 

Rumffit assumes that $\mathcal{TG}_{1}$ is ``acceptable to adherents of many rival logical schools; these principles [i.e., the three semantic rules for logical connectives], then, have a good claim to articulate the commonly understood senses of the sentential connectives.'' (Rumfitt 2015: 167) But they do not have a good enough claim. In order to have a better claim to articulate the semantic attributes of QL connectives, and in particular those of QL disjunction, $\mathcal{TG}_{1}$ must be revised. In particular, instead of being taken as a primitive, the \textit{Cl} operation on the set of truth-grounds must be defined in terms of quantum incompatibility, i.e., the relation that obtains between quantum observables in virtue of the fact that the operators representing them on the Hilbert space associated with \textit{P} do not commute. I turn now to consider Rumfitt's second, revised truth-ground semantics for $\mathcal{QL}$.
 
Let $\mathcal{TG}_{2}$ be the revised truth-ground semantics, with a set $S_{\mathcal{TG}_2}$ of truth-grounds and operations on them, $^{\bot}, \cap, \cup$, for orthocomplementation, intersection, and union, respectively. Then $\mathcal{TG}_{2}$ is a model of $\mathcal{QL}$ if and only if the truth-grounds are the closed subspaces of a finite-dimensional Hilbert space associated with system \textit{P} and there is a map $r_{2} : S_\mathcal{QL} \longrightarrow S_{\mathcal{TG}_2}$ such that $r_{2}$ is a homomorphism:
	
	\begin{center}
		
		$r_{2}(\neg p) = r_{2}(p)^{\bot}$ \\ $r_{2}(p \wedge q) = r_{2}(p) \cap r_{2}(q)$ \\ $r_{2}(p \vee q) = (r_{2}(p) \cup r_{2}(q))^{\bot \bot} $ 
		
	\end{center}
	
	Orthocomplementation and logical consequence are, as before, defined lattice-theoretically in terms of partial order. But the \textit{Cl} operation is now defined as double orthocomplementation: $Cl(r_{2}(p)) = r_{2}(p)^{\bot \bot}$. Furthermore, orthocomplementation has the topological property that was lacking in the previous $\mathcal{TG}_{1}$ semantics:
	
	\begin{center}
		
		$(\varnothing \cup ... \cup \varnothing)^{\bot \bot} = \varnothing^{\bot \bot} = \varnothing$
		
	\end{center}
	
	It would seem that precisely because orthocomplementation has this property, the truth-functionality of disjunction on line 5 of the Proof can be justified, so then $\mathcal{TG}_{2}$ could validate our counterexample to distributivity:
	
	\begin{center}
		
		$ r_{2}(A_{z}) \cap (r_{2}(B_{1}) \cup ... \cup r_{2}(B_{m}))^{\bot \bot} \nsubseteq ((r_{2}(A_{z}) \cap r_{2}(B_{1})) \cup ... \cup (r_{2}(A_{z}) \cap r_{2}(B_{m})))^{\bot \bot} $ 
		
		$ r_{2}(A_{z}) \cap (\varnothing \cup ... \cup \varnothing)^{\bot \bot} \nsubseteq  (\varnothing \cup ... \cup \varnothing)^{\bot \bot} $
		
		$ r_{2}(A_{z}) \nsubseteq \varnothing $
		
	\end{center}
	
	Yet, Rumfitt argues that this is still not the case, despite the fact that no metalogical proof of the truth-functionality of disjunction is needed any longer, so that rule-circularity is now avoided. The reason given for $\mathcal{TG}_{2}$'s inability to validate the counterexample to distributivity is that, on the first line of the Proof, $A_{z}$ is false, since $r_{2}(A_{z}) = \varnothing$. 
	
	In order to see why Rumfitt maintains that $r_{2}(A_{z}) = \varnothing$, a justified revision of the Proof is needed. What he believes justifies the revision are some features of standard QM that Putnam professed to ignore (Putnam 1968: 178). More exactly, Rumfitt notes the following: 
	
	\begin{quote} 
		
		The truth principle [i.e., the eigenstate-eigenvalue link] tells us that a statement ... that attributes a precise value to an observable quantity, will be true only at eigenstates of that observable. In a finite-dimensional Hilbert space, any self-adjoint operator will possess eigenstates, so in such a space there will be a state at which [such a statement] is true. In an infinite-dimensional space, however, even self-adjoint operators need have no eigenstates at all. ... once an infinite-dimensional Hilbert space is endowed with observables for position and momentum, there are no eigenvalues for position. (Rumfitt 2015: 180)
		
	\end{quote}
	
	Indeed, on an infinite-dimensional Hilbert space, those operators corresponding to observables with continuous spectra, like position and momentum, have no eigenstates in a Schrödinger representation, and thus such observables can have no precise values. (As discussed in chapter 2, position and momentum do have eigenstates, but only in the so-called non-regular representations.) Consequently, the Proof has to be revised. Needless to say, the revision does not make the (unrevised) Proof superfluous. Since QM makes extensive use of finite-dimensional Hilbert spaces in modeling observables like spin, which have discrete spectra, the Proof is enough to provide a counterexample to distributivity. The revision is only needed if one looks for a counterexample that involves observables with continuous spectra.

\subsubsection{Revising the eigenstate-eigenvalue link}

Let $\mathcal{QL}$ be our calculus, as before. Assume, again, that the EEL is true, but let the state space associated with our quantum system \textit{P} be an infinite-dimensional Hilbert space. For any $i \in \{1, ..., n, ...\}$, let $A_{i} \in S_{\mathcal{QL}}$ be sentences that state the possible positions of \textit{P}, and for any $j \in \{1, ..., m, ...\}$, let $B_{j} \in S_{\mathcal{QL}}$ be sentences that state its possible momenta. Then the revised Proof is the following derivation:
	
\bigskip

$\begin{array}{ll}

1.\ A_{z} \wedge (B_{1} \vee ... \vee B_{m} \vee ...) & \text{premise} \\

2.\ \neg (A_{i} \wedge B_{j}) & \text{premise} \\

3.\ A_{z} & \text{1,} \wedge\text{-elimination} \\

4.\ \neg(A_{z} \wedge B_{1}) \wedge ... \wedge \neg (A_{z} \wedge B_{m}) \wedge ... & \text{2, 3, substitution,} \wedge\text{-introduction} \\

5.\ \neg( (A_{z} \wedge B_{1}) \vee ... \vee (A_{z} \wedge B_{m}) \vee ...) &  \text{4, de Morgan}

\end{array}$

\bigskip

This derivation provides us with a counterexample to distributivity in \textit{L}, which can be stated as $\ulcorner A_{z} \wedge (B_{1} \vee ... \vee B_{m} \vee ...) \urcorner \nvdash \ulcorner (A_{z} \wedge B_{1}) \vee ... \vee (A_{z} \wedge B_{m}) \vee ... \urcorner$. This counterexample is, however, not validated by $\mathcal{TG}_{2}$:
	
	\begin{center}
		
		$ r_{2}(A_{z}) \cap (r_{2}(B_{1}) \cup ... \cup r_{2}(B_{m}) \cup ...)^{\bot \bot} \nsubseteq ((r_{2}(A_{z}) \cap r_{2}(B_{1})) \cup ... \cup (r_{2}(A_{z}) \cap r_{2}(B_{m})) \cup ...)^{\bot \bot} $ 
		
		$ r_{2}(A_{z}) \cap \textbf{1} \nsubseteq (\varnothing \cup ... \cup \varnothing \cup ...)^{\bot \bot} $
		
		$ \varnothing \nsubseteq \varnothing $
		
	\end{center}
 
As already pointed out, $ r_{2}(A_{z}) = \varnothing $ because \textit{P}'s position has no eigenstates. This entails that $A_{z}$ is false and, thus, that the revised Proof is unsound with respect to $\mathcal{TG}_{2}$. 
	
However, the soundness of the revised Proof might easily be restored by simply denying the EEL. This is not unheard of, and some have contended that denying the EEL is fully justified, since it ``has nothing much to do with quantum theory. ... It’s an interpretive assumption. Its motivation (I think) comes from the idea that measurement must be discovering some preexisting measured value, in which case that value must be possessed by a system iff it is certain to give that value as a result of measurement.  But this isn’t realized in any realist interpretation of quantum mechanics ... And it is anyway incompatible with the actual physics of quantities with continuously many measurement outcome possibilities, like position and momentum.'' (Wallace 2013: 215) Of course, this view conflicts with the Kochen-Specker theorem, and the conflict resolves by dropping distributivity (Demopoulos 1976). But this solution is not available here, since whether distributivity can be dropped is the very problem that denying the EEL attempts to address. 
 
Less radically perhaps, one could also just revise the EEL, so that instead of allowing that observables have precise values at all states, one modifies the notion of a state such that observables like position and momentum have precise values even though the state space of the system is an infinite-dimensional Hilbert space. This could be implemented, for example, by coarse-graining the state space, an operation which allows that, for system \textit{P}, an observable $O$ with a continuous spectrum can have precise values when the state of \textit{P} assigns probability 1 to some cells or regions, rather than to points, in that space. For any such observable, these regions are determined, say, by the interval $[x - \epsilon, x + \epsilon]$, for some $\epsilon > 0$ and for any possible value \textit{x} in the spectrum of that observable. The revised EEL then stipulates that a physical system $P$ has a precise value for observable $O$ if and only if $P$ is in \textit{a coarse-grained state} of $O$.\footnote{For a defense of this kind of revision, see Halvorson 2001, which critically discusses Teller 1979. Such a revision has been first considered in Fine 1971.} Consequently, the statement $A_{z}$ on line 1 of the revised Proof will be true at some coarse-grained position state. Thus, the revised Proof is not unsound after all. 

Rumfitt considers a similar revision of the EEL, but rejects it as an ``unsatisfactory constraint on truth'' (Rumfitt 2015: 180). \textit{Mutatis mutandis}, the revision of the EEL that I just suggested would also be considered unsatisfactory, on the ground that having a precise position at a coarse-grained state really amounts to nothing else than having an imprecise position. For if the statement $A_{z}$ is taken to describe a precise position of \textit{P}, then it can be only approximately true. But since logic is concerned with the preservation of truth, not that of approximate truth, revising the EEL in the way coarse-graining allows would change the subject of logic. However, in response to this, one can surely revise the EEL even further, so that it stipulates that an observable $O$ of $P$ has \textit{a coarse-grained value} if and only if $P$ is in a coarse-grained state of $O$. In this case then, when $A_{z}$ is taken to describe a coarse-grained position of \textit{P}, the statement $A_{z}$ will be true, rather than approximately true, at some coarse-grained position states. Therefore, coarse-graining the infinite-dimensional Hilbert space and then revising the EEL accordingly can successfully block Rumfitt's argument against the revised Proof. This is done \textit{without changing the subject of logic}, just as he correctly demands. 

My response admittedly requires some slight changes to the standard formalism of QM, e.g., an application of mathematical notions from coarse geometry on metric spaces.\footnote{For an introduction to coarse geometry on metric spaces, see Roe 2003.} But one might object that the response is beside the point, since the (revised) Proof assumes standard QM. However, my goal is not to defend standard QM. Rather, it is to show that rational adjudication against CL in QM is possible. If this requires moving beyond standard QM, that's fine by me. After all, classical logicians, too, usually go beyond standard QM, invoking one of its dynamical extensions or interpretations, when they insist that distributivity need not be dropped, as indeed Putnam himself ultimately did.

Two more points, before I move on to discuss Hellmann's argument. The first point concerns Rumfitt's claim, noted in section 4.1.1, that the orthomodular ortholattice $\mathcal{OO}$ should be rejected as logically insignificant because it makes disjunction not-truth-functional. This raises a fundamental question: What is a proper justification of our attribution of logical significance to a given semantics? Should such a justification turn to our intuitions and dispositions or to our theories and theorems? Should our considered reasons for assigning logical significance then necessarily align with truth-functionality, or not? My take on this comes close to the view on logic suggested, in the following passage, by Michael Dickson: 

\begin{quote}
    
What counts as an example of good reasoning, and how do we come to know what those examples are? Do we `just know' one when we see it? No. We must have some way to assign intersubjectively available truth values to sentences involving logical connectives. In other words, we need some way to determine whether a given form of logical inference is `successful'. Given the principle that the world (or, a correct theoretical description of the world) does not `disobey' the correct rules of reasoning, we may hope to determine which inferences are correct by determining
how to relate propositions involving logical connectives to empirical facts. (Dickson 2001, S283)

\end{quote}

If this view were adopted, then one should not reject $\mathcal{OO}$ as logically insignificant, but rather take the non-truth-functionality of QL connectives seriously. But note that this need not entail a commitment to the claim that QL is universally true, which would in turn entail that the rule of distributivity \textit{must} be dropped. Dickson is, in fact, committed to this claim:  
	
	\begin{quote}
		
		Quantum logic is the `true' logic. It plays the role traditionally played by logic, the normative role of determining right-reasoning. Hence the distributive law is wrong. It is not wrong `for quantum systems' or `in the context of physical theories' or anything of the sort. It is just wrong, in the same way that `(p or q) implies p' is wrong. (\textit{op. cit.}, S275)
		
	\end{quote}

However, this is not the view that I have been arguing for here. In this section, I have been interested only in making the point, against Rumfitt, that distributivity \textit{can} be dropped, or in other words, that rational adjudication against CL, even though not necessary, is nonetheless possible.\footnote{To be clear, my argument does not go against Dickson's view. Just as there are no \textit{empirical} grounds that make the adoption of QL necessary, in the context of quantum mechanics, I have argued that there are no \textit{rational} grounds that make the adoption of QL impossible. Elsewhere, I argue that in the same context, but in an abductivist framework, the adoption of CL is impossible: empirical and \textit{metaphysical} grounds make the adoption of QL, indeed, necessary (see Toader 2025b).}

The second point is the following. Rumfitt's objection might appear to be a particular instance of what came to be called the adoption problem (Kripke 1974). But I think that if one looks more closely, one can see that it is not. Although this is a longer discussion, which belongs elsewhere, here is very briefly how I see this matter. Take the following formulation of the adoption problem: ``certain basic logical principles cannot be \textit{adopted} because, if a subject already infers in accordance with them, no \textit{adoption} is needed, and if the subject does \textit{not} infer in accordance with them, no \textit{adoption} is possible.” (Birman 2024: 39) Thus, QL cannot be adopted because, if a subject already infers non-distributively, no adoption of QL is needed, and if the subject does not infer non-distributively, no adoption of QL is possible. In particular, if the subject infers classically, no adoption of QL is possible. However, Rumfitt's reasons for the impossibility of adopting QL are not simply based on commitment to CL: classicality is \textit{not}, for him, a sufficient condition for rejecting the possibility of adopting QL. In fact, as we have seen above, Rumfitt adjusts the commitment to classicality, as he proposes two non-classical semantics for the evaluation of the Proof. In contrast to Kripke, Rumfitt’s reasons for the impossibility of adopting QL are based on commitment to Dummett’s stability condition: for Rumfitt, stability \textit{is} a necessary condition for the possibility of adopting QL.

\subsection{Quantum logic and meaning}

Hellman proved that the disjunction and conjunction of an orthomodular QL are non-truth-functional, and he took this to imply that their meaning is different than that of their CL counterparts (Hellman 1980). As already mentioned, this argument was directed against Putnam's revisionist claim that the \textit{classical} distributive law must be dropped in QM. Arthur Fine had given an intuitive response to this claim, maintaining that ``the sense of the distributive law in which it is said to fail is not the sense in which, as the distributive law, it is supposed to hold.'' (Fine 1972, 19) As a consequence, the transition from classical physics to QM does not require a revision of CL. Hellman's argument attempted to rigorously justify this intuitive response. 

This section revisits his argument. The main claim that I will defend is that if truth-functionality is used as a semantic constraint on logical revisionism, then if Hellman's argument is to succeed, it has to be significantly improved. To show how it should be improved, an abstract presentation of QL will first be given, one that does not interpret sentences as experimental propositions, but allows a clearer analysis of the semantic attributes of its connectives. This will be followed by a discussion of bivalent semantics for QL, which is shown to be perfectly tenable provided that one is willing to give up the truth-functionality of some of its connectives. Afterwards, an analysis of Hellman's argument will be offered and an improved argument formulated, which reinforces and makes more precise the conclusion that some QL connectives are semantically inequivalent to their CL counterparts. This new argument still depends on the meaning-variance principle originally suggested by Hellman, but fully justifies Fine's intuitive response that classical and quantum distributivity mean different things. What is thereby achieved is a refutation of Putnam’s revisionism on semantic, rather than empirical or metaphysical, grounds. 

\subsubsection{Abstract formulation of quantum logic}

In this section, QL will be presented more abstractly than in the previous sections, by means of a formalism introduced by Michael Dunn and Gary Hardegree in their classic study of algebraic methods in logic (Dunn and Hardegree 2001). The advantage of this presentation -- a rational reconstruction of orthomodular QL -- is that the interpretation of sentences is not completely fixed, like in the more standard presentation of QL given so far, where sentences are taken to express experimental propositions about magnitudes associated to quantum systems. The abstract presentation is intended to erase all such ``intensional vestiges'' in the language of QL, thereby allowing a more precise analysis of its semantic attributes, and in particular, of the truth-functionality of its connectives.

Let $L$ be a sentential language containing variables for atomic sentences and symbols for logical connectives, $\lnot$, $\land$, $\vee$, and let $S_L$ be the set of sentences generated inductively in the usual way. Then $\mathcal{S}\equiv (S_L,O_{\lnot}, O_{\land},O_{\vee})$ is the algebra of sentences whose operations on the carrier set $S_L$ are defined, for any $p,q \in S_L$, as follows:
\begin{equation*}
    \begin{split}
        &O_{\lnot}(p)=\lnot p\\
        &O_{\land}(p,q)=p \land q\\
        &O_{\vee}(p,q)=p \vee q.
    \end{split}
\end{equation*}

A logical atlas $\mathbf{A}$ is a pair $(\mathcal{P},\langle D_j \rangle)$, where $\mathcal{P}$ is a non-empty algebra and $\langle D_j \rangle$ is a family of proper subsets of the carrier set of $\mathcal{P}$. Sets $D_j$ are designated sets. Logical atlases can be used to provide a semantics for $L$. Indeed, an interpretation of $L$ in $\mathbf{A}$ is any homomorphism $h : \mathcal{S} \rightarrow \mathcal{P}$. The elements of the carrier set of $\mathcal{P}$ can be thought of as propositions, and each designated set $D_j$ can be thought of as a possible set of true propositions. An interpretation of a language into an atlas thus homomorphically assigns propositions to sentences, in accord with the principle of compositionality.  Since an atlas can contain multiple designated sets, an interpretation does not need to fully determine which propositions are true and which are not. A triple $(L, \mathbf{A}, I)$, where $I$ is the set of all interpretations of $L$ in $\mathbf{A}$, is called an \textit{interpretationally constrained language}. 

Let a truth valuation on $L$ be any function $v : S_L \rightarrow \{0,1\}$. This is assumed to be bivalent by definition: each sentence in the language is assigned either 0 (`false') or 1 (`true'), thereby excluding third values or value gaps and gluts. But there are no compositional restrictions on truth value assignments to compound sentences, i.e. truth-functionality is not assumed. In other words, $v$ is not assumed to be a homomorphism. (I will come back to this later.) The identification of designated sets with sets of true propositions suggests a natural way of defining a class of valuations induced by an atlas. For each interpretation $i \in I$ and index $j$, one can define a valuation $v^{(i)}_j$ as:

\begin{equation*}
    v^{(i)}_j(p) = 1 \quad \text{if and only if} \quad i(p) \in D_j.
\end{equation*}

 Then, a class of valuations, $C^*_{\mathbf{A}}$, induced by $\mathbf{A}$, is the set of all such valuations: 
\begin{equation*}
    C^*_{\mathrm{A}}=\left\{v^{(i)}_j, \quad \forall i \in I, \forall j\right\}.
\end{equation*}

 To introduce a logical consequence relation, also induced by $\mathbf{A}$, we say that a set of premises $\Gamma \subseteq S$ $\mathbf{A}$-implies a conclusion $p \in S_L$, or $\Gamma \vDash_{\mathbf{A}} p$, if the following holds:
\begin{equation*}
    \forall v \in C^*_{\mathbf{A}}: \quad \text{if} \quad \left(\forall \gamma \in \Gamma, v(\gamma)=1 \right) \quad \text{then} \quad v(p)=1.
\end{equation*}

 Thus, a sentence is $\mathbf{A}$-valid, or $\vDash_{\mathbf{A}} a$, if $v(a)=1$ for all valuations $v$ in $C^*_{\mathbf{A}}$. 

Having defined the relevant formal notions, let us turn to QL. Let $S_\mathcal{H}$ be the set of closed subspaces of a Hilbert space $\mathcal{H}$, and the algebra associated to it be $\mathcal{A_H} \equiv (S_\mathcal{H},^{\perp},\cap,\cup)$, where $^{\perp}$, $\cap$, and $\cup$ are orthocomplementation, set-theoretic intersection, and linear span, respectively. Let $\mathbf{A}_{\mathcal{H}}=(\mathcal{A_H},\langle D_H \rangle)$ be a quantum atlas, where the family of designated subsets $\langle D_H \rangle$, whose index ranges over all 1-dimensional subspaces $K \in S_\mathcal{H}$, is defined as
\begin{equation*}
    \forall Q \in S_\mathcal{H}: \quad Q \in D_K \quad \text{if and only if} \quad K \subseteq Q.
\end{equation*}

 A quantum interpretationally constrained language is any triple $(L, \mathbf{A}_{\mathcal{H}}, I)$, where $L$ is a sentential language, $\mathbf{A}_{\mathcal{H}}$ is a quantum atlas, and $I$ is the set of all interpretations of the language into $\mathbf{A}_{\mathcal{H}}$. In particular, since they are homomorphisms, the interpretations $i \in I$ satisfy
\begin{equation*}
    \begin{split}
         &i(\lnot p)=i(p)^{\perp}\\
         &i(p \land q)=i(p)\cap i(q)\\
        &i(p \vee q)=i(p) \cup i(q),
    \end{split}
\end{equation*}  
for any sentences $p,q \in S_L$. Further, we can associate to any atlas $\mathbf{A}_{\mathcal{H}}$ an induced class of valuations $C^*_{\mathbf{A}_{\mathcal{H}}}$ and an induced logical consequence relation $\vDash_{\mathbf{A}_{\mathcal{H}}}$, which will help us define the relation of QL consequence.\footnote{To simplify notation, $C^*_{\mathcal{H}}$ and $\vDash_{\mathcal{H}}$ will henceforth replace $C^*_{\mathbf{A}_{\mathcal{H}}}$ and $\vDash_{\mathbf{A}_{\mathcal{H}}}$.}

Let $\vDash$ be a consequence relation on the set of sentences $S_L$. We say that $\vDash$ is a QL consequence relation if there exists a quantum atlas $\mathbf{A}_\mathcal{H}$, such that $\vDash$ and $\vDash_\mathcal{H}$ coincide. QL consequence relations can also be given the following characterization, in terms of the partial order on the algebra $\mathcal{A_H}$ associated to $\mathcal{H}$, reminiscent of the one originally introduced by Birkhoff and von Neumann: for a Hilbert space $\mathcal{H}$, a set of premises $\Gamma$ $\mathcal{H}$-implies conclusion $p \in S_\mathcal{H}$ if for all homomorphisms $i$ from the algebra of sentences into $\mathcal{A_H}$, the following holds
\begin{equation*}
    \bigcap_{\gamma \in \Gamma} i(\gamma)\subseteq i(p).
\end{equation*}

Since the operations $\cap$ and $\cup$ on $S_\mathcal{H}$ are non-distributive, it follows that QL consequence relations violate distributivity.\footnote{See above section 3.3.1 for Birkhoff and von Neumann's semantic proof of this violation, and see section 4.1.1 for a formal version of the proof.} The failure of distributivity in QL is due to the fact that experimental propositions in QM form a non-distributive lattice, arguably as a consequence of the fact that $\mathcal{A_H}$ is non-commutative.

Note that we have defined the QL consequence relations $\vDash_\mathcal{H}$ relative to a Hilbert space $\mathcal{H}$: there may, in principle, be as many QL consequence relations as there are (non-isomorphic) Hilbert spaces. As noted above, already Birkhoff and von Neumann argued that modularity holds only in finite-dimensional Hilbert spaces (1936, 832). There is thus more than one QL consequence relation depending on the Hilbert space on which the semantics is built, which justifies keeping the index `$\mathcal{H}$' in `$\vDash_\mathcal{H}$'. However, since infinite-dimensional Hilbert spaces are all mutually isomorphic, then there is only one QL consequence relation for QM$_\infty$. This violates distributivity because experimental propositions in QM$_\infty$ form a non-distributive, orthomodular lattice.

Since I will be presently concerned with the question about the truth-functionality of QL connectives, let us now introduce an $\mathcal{H}$-class $C$ as a class of valuations on $L$ that obeys, for all $\Gamma \cup \left\{p\right\} \subseteq S_L$, the following:
\begin{equation*}
    \Gamma \vDash_\mathcal{H} p \quad \quad \text{iff} \quad \quad \left[\forall v \in C: \quad \text{if} \quad \left( \forall \gamma \in \Gamma: v(\gamma)= 1 \right) \quad \text{then} \quad v(p)= 1 \right],
\end{equation*}

where $\vDash_\mathcal{H}$ is induced by some quantum atlas $\mathbf{A}_\mathcal{H}$. Any $\mathcal{H}$-class thus defines the same consequence relation as the class $C^*_\mathcal{H}$ induced by $\mathbf{A}_\mathcal{H}$: in fact, $C^*_\mathcal{H}$ is just one particular $\mathcal{H}$-class. This raises the following question: are all $\mathcal{H}$-classes isomorphic, i.e. are they all equivalent to $C^*_\mathcal{H}$, for a given atlas $\mathbf{A}_\mathcal{H}$? In other words, does the collection of all $\mathcal{H}$-classes have more than one member (up to isomorphism)? The answer is that there is more than one class of valuations compatible with one and the same QL consequence relation: for any quantum atlas $\mathbf{A}_\mathcal{H}$, there are at least two $\mathcal{H}$-classes. For consider a valuation $\tilde{v}_\mathcal{H}$ given by:
\begin{equation*}
    \tilde{v}_\mathcal{H}(p)=1 \quad \text{if and only if $p$ is an $\mathcal{H}$-tautology (i.e. $\vDash_\mathcal{H} p $)}.
\end{equation*}

The two different classes of valuations, that is $C^*_\mathcal{H}$ and $C^*_\mathcal{H}\cup \{\tilde{v}_\mathcal{H}\}$, both determine the same relation $\vDash_\mathcal{H}$, and are thus both $\mathcal{H}$-classes.\footnote{See below, section 5.3.1, for an analogous situation in the case of CL, which was presented already in Carnap 1943.} But does any of these $\mathcal{H}$-classes make all connectives truth-functional (TF)?

Let us say, quite generally, that $C$ makes conjunction TF if there exists a function $f_{\land}$, such that for all $v \in C$, and any $p,q \in S_L$:
\begin{equation*}
    v(p \land q)=f_{\land}(v(p),v(q)).
\end{equation*}
Analogously, for the truth-functionality of disjunction and negation. Let us also say that $C$ is a TF class of valuations if it makes all three connectives in $\left\{\lnot,\land,\vee\right\}$ TF. We will now see that, for any quantum atlas $\mathbf{A}_\mathcal{H}$, the induced class $C^*_\mathcal{H}$ makes conjunction TF, but negation and disjunction non-TF. 

The truth-functionality of $\land$ follows immediately from the fact that, in any Hilbert space, a 1-dimensional subspace is contained in two other subspaces if and only if it is contained in their intersection. Since the conjunction of two sentences is interpreted as the intersection of the subspaces associated (under the same interpretation) to those sentences, this implies that a conjunction is true if and only if both its conjuncts are true, in accord with the normal truth table for conjunction. 

In order to illustrate the non-truth-functionality of disjunction and negation, take the following example. Consider three 1-dimensional subspaces $K_1,K_2,K \in S_\mathcal{H}$, for $\mathcal{H}$ an arbitrary Hilbert space. Suppose that $K_1,K_2,K$ respectively contain vectors $\psi_1$, $\psi_2$ and $\frac{1}{\sqrt{2}}\left(\psi_1+\psi_2\right)$, where $\psi_1$ and $\psi_2$ are orthogonal. Take an arbitrary sentence $p \in S_L$ and an interpretation $i \in I$, such that $i(p)=K_1$ and $i(\lnot p)=K_2$. It is easy to see that for the induced valuations $v^{(i)}_K$, we have $v^{(i)}_K(p)=v^{(i)}_K(\lnot p)=0$ and $v^{(i)}_K(p \vee \lnot p)=1$. On the other hand, if $K'$ is a 1-dimensional subspace containing a vector which does not lie in the span of $\psi_1$ and $\psi_2$, then we have $v^{(i)}_{K'}(p)=v^{(i)}_{K'}(\lnot p)=0$ and $v^{(i)}_{K'}(p \vee \lnot p)=0$. This shows that disjunction is not TF, because some false disjuncts combine into a true disjunction, whereas others combine into a false disjunction. Similarly, some false sentences, when negated, become true, whereas others stay false. Therefore, the $\mathcal{H}$-classes induced by quantum atlases are not TF, for they make only conjunction TF.

Can there be any TF $\mathcal{H}$-classes? As proved by David Malament (2002), not only is the answer negative, but any $\mathcal{H}$-class must make at least two connectives non-TF: there is no $\mathcal{H}$-class that makes more than one connective in $\left\{\lnot,\land,\vee\right\}$ truth-functional. Thus, there is no truth-functional $\mathcal{H}$-class of valuations on $L$. Notice that the classes $C^*_\mathcal{H}$ induced by atlases $\mathbf{A}_\mathcal{H}$ accordingly make only conjunction TF. Are there any $\mathcal{H}$-classes that make disjunction or negation TF? 

Perhaps surprisingly, the answer is negative. To see this, it is enough to notice that for any Hilbert space $\mathcal{H}$, the induced QL consequence relation $\vDash_\mathcal{H}$ validates the following arguments, which are the semantic versions of the $\land$-introduction and $\land$-elimination rules: for any $p,q \in S_L$, 

\begin{center}
    
$p \land q \vDash_\mathcal{H} p$, $p \land q \vDash_\mathcal{H} q$, and $\left\{p,q\right\}\vDash_\mathcal{H} p \land q$.

\end{center}

It is easy to see that the validity of these three arguments implies that for any $\mathcal{H}$-class, $p \land q$ is true if and only if both $p$ and $q$ are true, thus forcing conjunction to behave classically, in accordance with its normal truth table. Since, as just noted, at most one connective can be TF, and since in QL only conjunction \textit{can} be TF, since it \textit{must} be TF, there is no $\mathcal{H}$-class that makes either disjunction or negation TF. Thus, every $\mathcal{H}$-class makes conjunction TF, but negation and disjunction non-TF.


The abstract presentation of QL, in this section, can be easily compared with the more standard, less abstract one in section 3.3.1 above. Whereas the homomorphic map $h : S_\mathcal{QL} \rightarrow S_\mathcal{LA}$ assigns, as we have seen there, a fixed meaning to each sentence in $S_\mathcal{QL}$ thereby enabling an interpretation as an experimental proposition about a quantum system, a quantum interpretationally constrained language $(L, \mathbf{A}_\mathcal{H}, I)$ contains all homomorphic interpretations $i \in I$ of the sentential language $L$ in the quantum atlas $\mathbf{A}_\mathcal{H}\equiv(\mathcal{A_H},\langle D_K \rangle)$. Each designated set $D_K$ in turn contains all those propositions that are true under the valuation $v_{\psi}$, for $\psi \in K$. The main difference between the two presentations of QL is thus that the abstract one constrains only partially the association between sentences and propositions, rather than completely fixing it. But clearly, if one takes the closed subspaces of $\mathcal{H}$ to represent ``experimental propositions'' about a quantum system, as Birkhoff and von Neumann did, and the elements of $\mathcal{H}$ (or its 1-dimensional subspaces) to represent its possible quantum states, then $I$ contains all ways of compositionally assigning experimental propositions to the sentences of $L$, and the induced class of valuations $C^*_\mathcal{H}$ contains ``$\psi$-relative'' valuations that make true exactly those propositions that can be verified with certainty in an appropriate measurement of the system prepared in state $\psi$.

\subsubsection{Bivalence and metaphysical realism}

We are now in a position to analyze Hellman’s meaning-variance argument, and show exactly how it should be improved, to defeat Putnam's logical revisionism. But before embarking in this analysis, it is worth revisiting briefly the often misunderstood issue of bivalence and realism in QM. 

It is well known that a realist understanding of QM -- that is, roughly, one that considers it as a true, or at least approximately true, theory about physical reality -- is still an open issue in contemporary philosophy of science. As noted in section 4.1.1 above, an important attempt at settling this issue was famously put forward by Putnam, who deemed the adoption of QL as a necessary ingredient for a tenable realist view of the quantum world, devoid of unpalatable metaphysical hypotheses supposedly implied by theories that concord with CL (such as non-local hidden-variable theories). Putnam's initial proposal, that all interpretational conundra of QM would disappear upon acceptance of QL, and in particular that QL would allow one to think of all observables in QM as having definite values at all times, underwent extensive and serious scrutiny. In particular, concerns have been raised about the possibility of providing the language of QL with a bivalent semantics.

Indeed, bivalence can be considered as a prerequisite for metaphysical realism, at least in the present context, since it maintains that any fact expressed by a sentence either obtains or does not obtain in the physical reality: for instance, it is either the case or not the case that exactly 1971.7 $\times$ 10$^{21}$ hydrogen atoms are currently in my room. However, as pointed out by Michael Friedman and Clark Glymour (1972), Putnam's proposal is faced with immediate difficulties due to the Kochen-Specker (KS) theorem, which states that, for Hilbert spaces of dimension $d>2$, there is no valuation that assigns a definite value to all magnitudes, while respecting the functional relationships among the latter (Kochen and Specker 1967). This entails that there is no homomorphism from any Hilbert lattice -- the lattice of all closed subspaces on a Hilbert space -- to the two-element Boolean algebra $\{0,1\}$, when $d>2$. Nevertheless, Friedman and Glymour made it clear that, while the KS-theorem does present problems for realism, at least as Putnam had conceived it, it certainly does not entail that the language of QM fails to admit a bivalent semantics. This was also expressed even more clearly by William Demopolous: 

\begin{quote}
    
There are two different accounts of indeterminism which are historically important. The first, which apparently goes back to Aristotle, rejects bivalence: A theory is indeterministic if it assumes that there are propositions whose truth value is indeterminate. The second, represented by the quantum theory, retains bivalence while rejecting semi-simplicity [i.e., a property equivalent to truth-functionality]. ... This [latter] form of indeterminism implies that there is no Boolean representation of the properties obtaining at a given time; yet for any property \textit{P} it is completely determinate whether or not \textit{P} holds. (Demopoulos 1976, 76sq) 

\end{quote}

 Thus, for example, it is one thing to say that it is true that ``This photon will decay tomorrow or this photon will not decay tomorrow'', while each of the disjuncts is neither true nor false, and another thing to say that it is true that ``This photon passed through the upper slit or this photon passed through the lower slit'', when each of the disjuncts is false. But the two different accounts of indeterminism, noted by Demopoulos, and illustrated by this example, have often been conflated.\footnote{See, e.g., Bell and Hallett 1982, 368. Others are led to make the same conflation by their reading of the Jauch-Piron theorem: ``Jauch and Piron show that any so-called orthomodular lattice (in particular any Hilbert lattice) admits total homomorphisms onto $\{0, 1\}$ iff it is distributive. Note that this means that any form of quantum logic \textit{must give up bivalence}" (Bacciagaluppi 2009, 56).} 

That a bivalent semantics for QL is actually unproblematic also follows from the previous section, where a bivalent semantics for QL has been explicitly constructed. While any such semantics must, with respect to its support to realism about QM, be restricted by the KS-theorem, some bivalent semantics may actually fare better than others in this respect. In fact, Friedman and Glymour, after presenting potential candidates for a realist bivalent semantics, immediately dismissed the ones that make negation non-classical, for ``dereliction of duty'' (1972, 20). They deemed it unacceptable for a realist semantics to ascribe the same truth value both to a sentence and to its negation: on their view, a bivalent semantics is compatible with realism only if negation obeys the classical truth table, i.e. for any valuation $v$ and any sentence \textit{p} in the language, $v(p)\neq v(\lnot p)$. 

However, one can raise doubts about this requirement, especially in the present context, in which the truth-functionality of connectives is precisely what is at stake. It is not clear why one should endorse this view about negation without also endorsing parallel ones that would require any realist semantics to make conjunction and disjunction classical as well, thereby conflating realist and classical semantics.\footnote{Some ideas along these lines can be found in Dummett 1976.} Why is it unpalatable for a realist to hold that both a sentence and its negation have the same truth value, while at the same time being at ease with false disjuncts making true disjunctions or true conjuncts making false conjunctions? What makes negation special? What makes its classicality an essential aspect of realism? 

Following up on our discussion in the previous section, note that since any $\mathcal{H}$-class of valuations makes only conjunction TF, it necessarily makes negation non-TF and, thus, non-classical. Therefore, if one endorsed Friedman and Glymour's view that negation must be classical, as Hellman did in his argument to be presently discussed, then no $\mathcal{H}$-class could provide a semantics compatible with realism. In other words, a bivalent semantics that could support realism about QM would require different classes of valuations than the $\mathcal{H}$-classes introduced above.

Taking stock, since QL can be considered as the logic of experimental propositions in QM, and bivalence is widely understood as a prerequisite for realism, a bivalent semantics for QL is necessary for the possibility of realism about QM. Even though the KS-theorem presents difficulties for a ``naive'' realist understanding of QM, one can preserve bivalence, and thereby advocate a weaker form of realism, by giving up the truth-functionality of some QL connectives. And even if one assumes that the truth-functionality of negation is a necessary condition for a semantics compatible with realism, realism could still be maintained if one introduced classes of valuations other than the $\mathcal{H}$-classes. Thus, although realism about QM and bivalence in QL may not look like an ideal couple, their marriage can nevertheless be saved, although many would consider it rather costly from a semantic point of view.

\subsubsection{Hellman's meaning-variance argument}

The main point of Hellman's argument is that QL cannot be considered a revision of CL, in the sense of asserting different truths about the same connectives, since the two logics are semantically inequivalent, i.e., their corresponding connectives do not mean the same thing: 

\begin{quote}
    
[T]he opponent against whom Putnam argued was a rather dogmatic conventionalist who was rather prone to put more weight on the notion of `meaning' than scientific scrutiny should allow. What I want to do here is focus on a more precise `meaning-change' argument, one which makes absolutely minimal reliance on the problematic word, `meaning', and which, as far as I can see, a proponent of Putnam's view can neither defeat nor bypass. (Hellman 1980, 494) 

\end{quote}

In the remainder of this section, I will first give a careful reconstruction of Hellman's argument, followed by the observation that his argument actually falls short of a refutation of Putnam's revisionism. Afterwards, in light of the discussion in the previous sections, I will explain how the argument should be improved, and why, in a way that finally clarifies the semantic grounds for a successful refutation of revisionism. 

Note that Hellman stipulated a condition for meaning invariance, which would presumably be acceptable to a conventionalist without dogmas: 

\begin{quote}
If $\alpha$ and $\beta$ are synonymous sentential connectives, then (a) if one is a truth-functional connective, then so is the other, and (b) if $\alpha$ and $\beta$ are truth-functional, they have the same truth tables. (\textit{op. cit.}, 495)
\end{quote}

Hellman went on to prove that \textit{if} QL negation is TF, \textit{then} QL disjunction and QL conjunction are non-TF, which, due to clause (a) in the condition just stated implies that at least \textit{some} QL connectives and their classical counterparts are not synonymous, i.e., not semantically equivalent.\footnote{Helmann actually gave a semantic proof only for the claim that if QL negation is TF, then QL disjunction is non-TF, and stated that the claim that if QL negation is TF, then QL conjunction is non-TF can be proved analogously.} 


Hellman's argument can then be reconstructed as follows:

\smallskip

1. Two connectives have the same meaning only if they are either both TF or both non-TF.

2. If QL negation is TF, then QL disjunction and QL conjunction are non-TF.

3. Thus, QL negation and CL negation differ in meaning, or CL disjunction (conjunction) and QL disjunction (conjunction) differ in meaning.

4. Thus, some QL connectives differ in meaning from their CL counterparts.

\smallskip

Importantly, besides having the benefit of relying minimally on the problematically vague notion of `meaning', Hellman maintained that his argument is independent of the issues concerning realism, in that it is based on a purely formal semantic result that exhibits the difference in the TF status of the classical and quantum connectives: ``the non-truth-functionality argument is entirely distinct from [those that] argued that QL could not satisfy the demands of realism.'' (\textit{op.cit.}, 496) 

Note that while the argument establishes that at least some QL connectives differ in meaning from their CL counterparts, it does not specify which ones do so unconditionally. Moreover, conclusion 3 is compatible with QL negation being non-TF and thus differing in meaning from CL negation, while leaving QL conjunction and QL disjunction TF and, thus, possibly semantically equivalent with their CL counterparts. Consequently, since the distributive law concerns conjunction and disjunction, but not negation, Hellman's argument does not refute Putnam's revisionist claim that classical distributivity fails in QM. In other words, the argument does not justify Fine's claim quoted in the introduction to section 4.2. But the results from section 4.2.1 can help us improve on the above argument and finally properly justify Fine's claim.  

First, note that if one assumes QL negation to be TF, then premise 2 is an immediate consequence of the theorem discussed above, due to Malament, that any viable class of valuations can make at most one QL connective TF. Now, while the antecedent of premise 2, stating that negation is TF, can arguably be justified for semantics that are intended to support realism -- as we have also seen already in section 4.2.2 -- this assumption is not warranted in the present context. For here, CL connectives and their QL counterparts are contrasted on purely semantic, non-metaphysical grounds. Moreover, the antecedent of premise 2 cannot be maintained without invalidating the quantum-logically valid $\land$-introduction argument, $\left\{a,b\right\} \vDash a \land b$. Indeed, recall that any viable $\mathcal{H}$-class of valuations makes conjunction TF, but disjunction and negation non-TF. Therefore, the antecedent of premise 2, whose truth would be needed if Hellman's argument were to back Fine's claim against Putnam's revisionism, is not only unwarranted, but also false.\footnote{Note that this point is distinct from Rumfitt's rule-circularity objection (discussed in section 4.1) that the quantum logician should not assume CL in the metalanguage of the Proof (i.e., the proof that distributivity fails in QM). For in this case, Hellman's proof for the non-truth-functionality of QL disjunction (and conjunction) would not go through.}

Secondly, there is another qualification that needs to be made here: namely, it is not entirely clear what is meant in the above argument by saying that a sentential connective `c' is TF. Does it mean that (i) \textit{there exists} a viable class of valuations that makes `c' TF, or that (ii) \textit{any} viable class of valuations makes `c' TF? Since the argument tacitly assumes that classical connectives are TF, option (ii) is automatically excluded, since the existence of Carnap's non-standard class $C^*_\mathcal{CL} \cup \left\{\tilde{v}_\mathcal{CL} \right\}$ would make even CL disjunction and CL negation non-TF.\footnote{See section 5.3.1 for a presentation of this non-standard class of valuations for $\mathcal{CL}$.} Therefore, the only chance for a Hellman-type argument to convincingly establish meaning-variance on the basis of a difference in the TF status of QL connectives and their CL counterparts is to stick to option (i). 

Having noted all this, one can formulate an improved argument, which nevertheless assumes the same condition for meaning invariance stipulated by Hellman, but is better than his argument since it provides a definite verdict concerning Putnam's and Fine's antithetical claims. Let us say that a sentential connective `c' is TF if there is at least one viable class of valuations that makes `c' TF. Here is the argument:

\smallskip

1'. Two connectives have the same meaning only if they are either both TF or both non-TF.

2'. QL disjunction and QL negation are non-TF.

3'. Thus, CL disjunction (negation) and QL disjunction (negation) differ in meaning.

4'. Thus, some QL connectives differ in meaning from their CL counterparts.

\smallskip

 This argument can be understood as a completion of Hellman's, in that, while it maintains the same general conclusion (i.e. the one expressed in 4'), it also specifies which QL connectives are not semantically equivalent with their CL counterparts: in particular, since disjunction -- which figures in the distributive law -- is among these connectives, the argument fully justifies Fine's claim.

Note, however, that this assumes that the only viable classes of valuations for QL are $\mathcal{H}$-classes. Dropping this assumption would raise a problem both for Hellman's argument and for the argument just given. Alternative classes of valuations might make it the case that there is no difference anymore in the TF status of CL and QL connectives, at least at the level of individual connectives. Putnam's revisionism would remain standing, for then the sense of the distributive law in which it is said to fail might well be the very sense in which it is supposed to hold. Thus, the soundness of a Hellman-type argument that implies the semantic inequivalence of CL and QL connectives depends on whether one accepts only $\mathcal{H}$-classes as viable classes of valuations for QL. If one does, then a sound argument that establishes meaning-variance can be given: in particular, an argument that establishes a difference in meaning between CL and QL negation, and between CL and QL disjunction.\footnote{See Horvat and Toader 2023, for some preliminary thoughts on an alternative class of valuations for QL, and its possible consequences for a Hellman-type meaning-variance argument against Putnam's revisionism.}

Nevertheless, this new argument, as presented in this section, fares clearly better than Hellman's own. Granted all the required assumptions, Putnam's revisionism is finally rebutted for semantic reasons: the sense of the distributive law in which it is said to fail is not the sense in which it is supposed to hold. Fine was right about it. Now we know the reason why he was right. 

Furthermore, we now can easily see how this deflates Rumfitt's own objection against Putnam's revisionism -- the objection according to which the Proof that the distributive law fails in QM is rule-circular. Granted all the required assumptions, including Hellman's condition for meaning invariance, the Proof turns out to be not rule-circular, and this is precisely because the semantic attributes of QL disjunction are different than those of CL disjunction. But then, of course, the Proof fails to establish what Putnam took it to establish, in the first place. In any case, as I have already argued in section 4.1.3, Rumfitt's objection can be successfully addressed even if the Proof were taken to establish what Putnam thought it did.

\subsection{Applied classical mathematics and quantum logic}

A more general attempt to defeat logical revisionism has been made, more recently, by Timothy Williamson, who has argued that a wide class of deviant logics faces undesirable consequences due to the applicability of classical mathematics in the natural and social sciences (Williamson 2018). This applicability would provide abductive support
for classical logic over its non-classical competitors. 

The class that is specifically targeted by Williamson's argument (henceforth, WA) includes logics motivated by reasons \textit{external} to mathematics, as exemplified by QL and many-valued logics, which indeed have their origins in empirical considerations. The logics that reject some classically-valid argument or law due to issues that are \textit{internal} to mathematics -- like the intuitionist's rejection of the law of excluded middle, motivated by constructivist views about mathematics -- are not within the targeted class. So called \textit{pure} logics, that is,
unapplied mathematical objects that carry ‘logic’ in their name only because they resemble the
mathematical structures of classical logic, are not in that class, either.

The goal of this section is to show that WA is unsound and explain why this is so. WA picks up on an apparent tension between applying \textit{classical} mathematics to a certain domain (e.g. an empirical domain) and simultaneously endorsing a \textit{non-classical} logic in reasoning about that same domain. It then purports to elevate this tension to an actual inconsistency, which is supposed to corner the deviant logician into a freezing dilemma: either she is to rebuild mathematics in her own logic, or she is forced to forgo the applicability of mathematics to empirical domains, thereby losing (at least some of) the explanatory power characteristic of the natural and social sciences. As this outline already suggests, the hard work in WA lies in elevating that tension to an inconsistency. But it is precisely at this point that WA fails: the alleged tension, and thus the alleged inconsistency, does not exist. 

After reconstructing WA and clarifying what exactly would be the undesirable consequences threatening deviant logics, and then Williamson's reasons for believing that these threats are significant, I will focus on QL as a particular logic targeted by WA. In this case, a crucial premise of WA will turn out to be false. The same premise arguably fails also in the case of a logic of vagueness, so the criticism could be generalized, but this generalization is not attempted here.

\subsubsection{Williamson's attack on deviant logics}

A typical deviant logician's response to the fact that mathematical practice seems to involve a constant appeal to CL principles is well expressed in the following passage, quoted by Williamson as well: ``[M]athematical practice is consistent with these reasoning steps [i.e. the ones present in mathematical reasoning] being instances of \textit{mathematical} principles of reasoning, not generalizable to all other discourses. \textit{A fortiori}, they may very well be principles of reasoning that are permissible for mathematics, but not for theorizing about truth.'' (Hjortland 2017, 652–3) This response suggests that just because certain reasoning principles are validated in mathematical discourse, the deviant logician may contend that the same principles do not need to extend to other regions of discourse, e.g. those concerned with truth or, for that matter, mountains or quantum-mechanical phenomena. While this is tenable as long as the mathematical and the non-mathematical are kept mutually isolated, Williamson believes that the applicability of mathematics to non-mathematical domains, as exemplified in the natural and social sciences, raises the following problem. 

Let $\mathcal{D}$ be a set of terms denoting the non-mathematical objects that the deviant logician takes to require a non-classical treatment, e.g. heaps, electrons, or metalinguistic terms. Also, suppose the deviant logician wants to keep mathematics classical. Williamson argues that this is problematic for several reasons. First of all, the deviant logician cannot consistently substitute elements of $\mathcal{D}$ for variables in the statements of some classical mathematical theorems. For example, the many-valued logician cannot apply the theorem $\forall x,y (x=y \vee x \neq y)$ to, say, mountains without contradicting her thesis that the identity of mountains is possibly undetermined. Secondly, the same reason prevents the deviant logician from appealing to an isomorphism holding between the objects denoted by $\mathcal{D}$, on the one hand, and some classical purely mathematical objects, on the other hand: e.g. an isomorphism between a collection of mountains and a pure set (as defined by ZFC set theory). 

Furthermore, not only does the deviant logician have trouble with the application of classical mathematical \textit{theorems}, but she also faces problems with the use of mathematical \textit{reasoning} in science, for she cannot substitute elements of $\mathcal{D}$ for variables used in classical derivations from scientific hypotheses. Since Williamson takes such limitations to imply the impossibility of applying classical mathematics to the non-mathematical objects denoted by the terms in $\mathcal{D}$, and since he also takes it to be uncontroversial that classical mathematics can be applied - and is routinely applied - to the world outside pure mathematics, he concludes that ``in a non-classical world, pure mathematics is no safe haven for classical logic.'' (Williamson 2018, 21) 

As a consequence, the deviant logician would be forced to rebuild mathematics within her own logic. For example, a quantum logician would have to recover classical mathematics in a quantum-logical framework in order to make sense of the applicability of classical mathematics to quantum phenomena. However, the task of rebuilding mathematics within a new logic is notoriously difficult, and Williamson thinks it would rather unavoidably lead to \textit{ad hoc} premises.\footnote{Examples of such \textit{ad-hoc} premises appear in Hartry Field's recapture of the least number principle in a logic without the law of excluded middle (Field 2008). Williamson finds Field's recovery strategy unacceptable for a couple of reasons: unlike the least number principle, which is routinely derived from mathematical induction, Field's recovered version of that principle is a postulate; also, that version is underivable from a suitable schema for mathematical induction. Such reasons for blocking classical recapture are inconsequential for my analysis in this section.} This would, in turn, make the scientific explanations that involve applications of classical mathematics -- and it is hard to find many that do not -- more costly, since they must include extra assumptions, e.g., the \textit{ad-hoc} premises required for classical recapture (\textit{op. cit.}, 19). 

Therefore, Williamson concludes, even though the path of rebuilding mathematics within a non-classical logic is the one that a staunch deviant logician ought to take, it nevertheless leads to an abductively implausible view -- implausible, because the usual path trodden by the classical logician offers an elegant and less explanatorily costly alternative.

WA can be reconstructed as follows. Again, $\mathcal{D}$ is a set of terms denoting non-mathematical objects that are tentatively taken to require a non-classical logic $\mathcal{L_D}$. Suppose mathematics can be (or even has been) successfully applied to the objects referred to by $\mathcal{D}$. Then WA takes the form of the following dilemma:

\smallskip

1. $\mathcal{L_D}$ can be taken either \textit{to extend} to mathematics or \textit{not to extend} to mathematics.


2. If $\mathcal{L_D}$ is taken \textit{not to extend} to mathematics, then it is inconsistent to hold that mathematics can be applied to the objects referred to by $\mathcal{D}$. 




3. If $\mathcal{L_D}$ is taken \textit{to extend} to mathematics, then most explanations in science become more costly.




4. Therefore, maintaining that reasoning about the objects referred to by $\mathcal{D}$ can be adequately captured by $\mathcal{L_D}$ is either inconsistent or abductively implausible.

\smallskip

However, this argument is not sound, because Premise 2 turns out to be false, at least when WA is restricted to the case of QL.\footnote{Neil Tennant has recently defended a similar position regarding WA, by arguing that premise 3 fails in the case of his preferred substructural logic (Tennant 2022). Let us also note that WA may as well be countered by maintaining, as Quine suggested, that changing the logic changes the meaning of the logical constants, a view that I discussed in the previous section. On this understanding of deviant logics and of how they relate to CL, it is clear that no tension with classical mathematics can arise in the first place. However, Williamson is not concerned with such a view (\textit{op. cit.}, 12).} 

QL satisfies the presuppositions of WA, and is thus allegedly threatened by the argument. Williamson, himself, suggests this much: ``[S]ince quantum mechanics applies mathematics ubiquitously, those who propose non-distributive quantum logic as a serious alternative to classical logic are not excused from the need to reconstruct mathematics on the basis of their quantum logic.'' (\textit{op. cit.}, 20) Obviously, he is here pushing the second horn of WA against the quantum logician, while at the same time noting, in all fairness, the best attempt so far, by Michael Dunn, to classically recapture mathematics within non-distributive QL (Dunn 1980). Presumably, Williamson assumes that the quantum logician could stand no chance against the first horn of WA. That is, he appears to assume as established already that it is inconsistent to maintain, on the one hand, that reasoning about quantum phenomena can be adequately captured by QL, \textit{and} to believe at the same time, on the other hand, that classical mathematics can be applied to quantum phenomena. 

But as we will presently see, this alleged inconsistency is clearly dissolved once a closer look is taken at both of these hands.

\subsubsection{Quantum models and idealized tests}

It is a matter of scientific practice that, in their modelling of quantum phenomena, physicists regularly apply mathematical tools from linear algebra, functional analysis, group theory, probability theory, and other classical mathematical theories. Before looking at how these are applied in practice, recall the characterization of quantum models that I have given in chapter 2, as structures of classical mathematical objects referred to in the standard axioms:

\smallskip

1. The state space of a physical system corresponds to an infinite-dimensional complex Hilbert space $\mathcal{H}$, such that the quantum state of a physical system is a mathematical function $\psi : T \rightarrow \mathcal{H}$ (or an equivalence class of unit-norm vectors of $\mathcal{H}$), defined at each time instant $t \in T$, where $T \subseteq \mathbb{R}$ denotes a time interval. 

2. The set $\mathcal{A}$ of dynamical quantities of a physical system (e.g., position, momentum, spin, polarization, etc.) corresponds to the set of self-adjoint operators acting on $\mathcal{H}$, such that the possible values of any variable are contained in the spectrum of its corresponding operator. 

3. The unitary time evolution of a physical system is described by the Schrödinger equation $i \partial_t \ket{\psi}_t=H\ket{\psi}_t$, for some operator (the Hamiltonian) $H \in \mathcal{A}$. 

4. For each unit-norm vector $\ket{\phi} \in \mathcal{H}$, $\mu_{\phi}$ is a measure on the set of subspaces of $\mathcal{H}$ such that for an arbitrary subspace $S \subseteq \mathcal{H}$: $\mu_{\phi}(S)=\bra{\phi}P_S\ket{\phi}$, where $P_S$ is the projector on $S$.

\smallskip

How, more exactly, are these mathematical objects used in the practice of QM? When a physicist is confronted with the task of constructing a local quantum model for a certain experiment of duration $T$, she will start by assigning a Hilbert space $\mathcal{H}$, a set of self-adjoint operators $\mathcal{A}$ and an ``initial'' quantum state $\ket{\psi_{t_0}} \in \mathcal{H}$ to the system involved in the experiment. These assignments are of course not arbitrary, but highly constrained by the conditions set up in the given experiment, conditions that are taken to justify the ``introduction'' of a quantum model. For example, if at the beginning of the experiment, a device is activated that is capable of emitting a single electron of a certain energy, this warrants an assignment of particular objects $\mathcal{H}, \mathcal{A}$ and $\ket{\psi_{t_0}}$ to the electron. 

Furthermore, in order to characterize the dynamical evolution of the target system, the modeler assigns a particular Hamiltonian operator $H \in \mathcal{A}$ that generates the function $\psi$ according to the Schrödinger equation. Up to this point, the physicist has been merely assigning mathematical objects to her experiment, according to the rules and constraints of standard QM. The important step that brings the model closer to empirical reality is provided by the Born rule, briefly explained as follows. 

Suppose that at a certain ``final'' time $t_f$, the physicist performs a measurement of a collection of dynamical variables associated to the system of interest (e.g. the position and the spin of an electron). Let $A\equiv \left\{A_1,...,A_n \right\} \subset \mathcal{A}$, for some $n \in \mathbb{N}$, be the set of self-adjoint operators that correspond to these variables. But note, importantly, that according to standard QM, there is no possible experiment that could implement a simultaneous measurement of so-called ``incompatible'' variables, i.e. variables associated to mutually non-commuting operators (e.g., a particle's position and momentum). As a consequence of this impossibility, the operators in $A$ obey canonical commutation relations $[A_i,A_j]=0$, for all $i,j=1...,n$. 

Next, for each $i$, let $\mathcal{M}^{A_i}_{\Delta^{(i)}} \subseteq \mathcal{H}$ be the eigenspace corresponding to a subset $\Delta^{(i)} \subseteq \text{Spec}(A_i)$ of $A_i$'s spectrum. The Born rule then predicts (or prescribes) the probability $\text{Pr}^{A}_{t_f}(\Delta^{(1)},...,\Delta^{(n)})$ for the outcomes of the measurement of the $n$ dynamical variables performed at time $t_f$ to be contained respectively in subsets $\Delta^{(1)},...,\Delta^{(n)}$:
\begin{equation*}
    \text{Pr}^{A}_{t_f}(\Delta^{(1)},...,\Delta^{(n)})=\mu_{\psi_{t_f}}(\mathcal{M}^{A_1}_{\Delta^{(1)}}...\mathcal{M}^{A_n}_{\Delta^{(n)}})=\bra{\psi_{t_f}}P^{A_1}_{\Delta^{(1)}}...P^{A_n}_{\Delta^{(n)}}\ket{\psi_{t_f}},
\end{equation*}
where $P^{A_i}_{\Delta^{(i)}}$ is the projector on subspace $\mathcal{M}^{A_i}_{\Delta^{(i)}}$, for each $i$. The Born rule thus enables the physicist to use the quantum model -- and more precisely, its Hilbert space measures $\mu_{\phi}$ -- to generate (classical) probability distributions $\text{Pr}^{A}_{t_f}$ that can then be compared to statistical data models extracted from actual experiments. In other words, the application of the Born rule justifies the ``elimination'' of the quantum model.

Despite the empirical success of QM, i.e., the fact that quantum-mechanical modeling of the sort just described yields unprecedented predictive and explanatory power with respect to the microscopic world, various problems (such as the measurement problem, Bell's nonlocality, contextuality, etc.) leave the interpretational problems of QM still far from being resolved. Some realist interpretations urge us, for instance, that the lesson we ought to take from the empirical success of QM is that physical interaction can be non-local, or that our ``world'' is just one among the many worlds that partake in the Everettian ``multiverse''.\footnote{Cf., again, Lewis 2016 for a useful review of QM interpretations and their metaphysical problems.} As I noted in section 4.1, others have argued that what QM is primarily teaching us is that the ``one true logic'', or at least the logic we ought to adopt in our reasoning about the microscopic world, is Birkhoff and von Neumann's non-distributive QL (or some of its variants).\footnote{Cf. Birkhoff and von Neumann 1936. For more recent presentations, see e.g. Dalla Chiara \textit{et al.} 2004. Some physicists continue to assert that ``real quantum mechanics is not so much about particles and waves as it is about the nonclassical logical principles that govern their behavior.'' (Susskind and Friedman 2014, 236)} 

Indeed, QL looks like a perfect target for WA, since it is typically taken to violate the classical distributive law due to the peculiarities of standard QM. There are, admittedly, many formulations and presentations of QL, and albeit most of them agree in their core rejection of the distributive law, some of them yield different relations of logical consequence, which makes it more accurate to speak of QLs, in the plural (cf. Rédei 1998). For simplicity, the focus here will be on one QL, and so on a particular relation of logical consequence. In contrast to the abstract presentation given in section 4.2.1, here the QL-consequence relation will be constructed on the basis of an ``operationalist'' semantics (Putnam 1968, 192-197, Bacciagaluppi 2009, 54-55). This should be sufficient to convince us that there is no inconsistency between adopting QL in the quantum domain and applying classical mathematics therein as well, thereby refuting premise 2 of WA.

Consider once again an experiment involving a quantum system, say a particle s, and its corresponding quantum model $\Theta$. Since in what follows the dynamical aspects of the model, i.e. those aspects that have to do with time dependence, will be irrelevant, let us turn our attention to its ``kinematical'' part, $\tilde{\Theta}\equiv\langle \mathcal{H}, \mathcal{A}, \mu_{\phi} \rangle$. As stated already, each element $A$ of $\mathcal{A}$ corresponds to a dynamical variable (e.g., position or momentum) associated to the system under consideration, and each of these variables can take values in its operator's spectrum Spec($A$). Now, roughly speaking, the quantum logician urges us that there is something peculiar about the logical relations that obtain between certain sentences that refer to the dynamical variables associated to our system. The canonical set of sentences that the quantum logician asks us to focus on is $S_\mathcal{QL}$, i.e., the set that can be generated by conjoining, disjoining and negating so-called elementary sentences, each of which takes the form $\text{s}^{A}_{\Delta}$ and specifies the range of values taken by a variable associated to our quantum system: ``The value of the variable corresponding to operator $A$ lies in interval $\Delta$'', with $A \in \mathcal{A}$ and $\Delta \subseteq \text{Spec}(A)$. As we known already from our discussions above, the quantum logical relations that obtain between these sentences disobey the rule of distributivity of conjunctions over disjunctions. Let's see how this informal semantics can be articulated in operationalist terms.

An operationalist semantics for QL can be explicitly defined in terms of idealized operational procedures that may in principle be carried out in a laboratory. First, let $\mathcal{T}$ be a possibly uncountable set, whose elements are taken to denote \textit{idealized tests}, i.e. idealized operational procedures that may hypothetically be applied to physical systems. We assume that the application of any test to any physical system may result in only one of the two mutually exclusive outcomes: either the system passes the test or it does not. Further, let a function $\tau: S_\mathcal{QL} \rightarrow \mathcal{T}$ associate one idealized test to each sentence in the canonical set. In particular, each elementary sentence of the form $\text{s}^{A}_{\Delta}$ is mapped to the test that consists in an idealized measurement of the dynamical variable associated to $A$, such that if the test is applied, it is passed by a system if and only if the outcome of the measurement is contained in $\Delta$. Furthermore, all tests associated to compound sentences are determined via the following algorithm, with $p,q \in S_\mathcal{QL}$ henceforth being arbitrary sentences:
\begin{itemize}
    \item $\tau(\lnot p)$ is passed by a system with certainty (i.e. unit probability) if and only if the same system would certainly fail to pass $\tau(p)$;
    \item $\tau(p \land q)$ is passed by a system with certainty if and only if the same system would certainly pass both $\tau(p)$ and $\tau(q)$;
    \item $\tau(p \vee q)$ is passed by a system with certainty if only if the same system would certainly pass either $\tau(p)$ or $\tau(q)$.
\end{itemize}

Now, let us define the relation of logical consequence on the sentences in $S_\mathcal{QL}$ relative to their associated tests: informally, we will say that a set of premises, $\Gamma$, $\mathcal{QL}$-implies conclusion $p$, if it is the case that any quantum system that passes each test $\tau(\gamma)$ with certainty, for all $\gamma \in \Gamma$, also passes $\tau(p)$ with certainty. In order to formalize this, let us introduce a set of valuations $\mathcal{V}\equiv \left\{v_{\phi}| \forall \phi \in \mathcal{H}, \text{s.t.} \braket{\phi|\phi}=1\right\}$, where $\mathcal{H}$ is the Hilbert space assigned to the quantum system under consideration. Each $v_{\phi}: S_\mathcal{QL}\rightarrow \left\{0,1\right\}$ is a bivalent truth valuation that assigns one of the two truth values to any experimental sentence $p$, as follows: $v_{\phi}(p)=1$ if and only if a quantum system prepared in quantum state $\ket{\phi}$ would pass test $\tau(p)$ with certainty. Whether a system passes a certain test with certainty is in turn determined by the outcome of hypothetical applications of quantum models to idealized tests. This means that the value of $v_{\phi}(p)$ is determined by the probability distribution generated from an appropriate quantum model of the test $\tau(p)$ performed on a system in state $\ket{\phi}$. 

Without getting into more details about the relation between tests, valuations and quantum models, once the definition of valuations is in place, the logical consequence relation follows canonically. For any $\Gamma \cup \left\{p\right\} \subset S_\mathcal{QL}$, $\Gamma \vDash_{QL} p$ if and only if:
\begin{equation*}\label{log cons op}
    \forall v \in \mathcal{V}: \quad \text{if} \quad \left(\forall \gamma \in \Gamma,  v(\gamma)=1 \right) \quad \text{then} \quad v(p)=1.
\end{equation*}
Having revisited some basic details about the application of classical mathematics in constructing quantum models, and the operationalist semantics for QL, we are ready to get back to WA, and in particular ready to challenge its second premise.

\subsubsection{Withstanding the attack on quantum logic}

When restricted to the case of QL, premise 2 of WA can be paraphrased as follows: ``It is inconsistent to simultaneously hold that (i) QL adequately captures reasoning about quantum phenomena, and that (ii) classical mathematics can be applied to quantum phenomena.'' But I think that this premise is false, and in this section I will explain why I take this to be the case.

As discussed above, the sentences in the canonical set $S_\mathcal{QL}$ are obtained by logically compounding elementary sentences, such as $\text{S}^{(X)}_{[1,2]}$, which reads ``The value of an electron's position lies in the interval $[1,2]$.'' The quantum logician claims, recall, that the logical relations between such sentences require a non-distributive logic. Can this claim lead to an inconsistency with the application of classical mathematics in the construction of quantum models, and if so, how? 

Note that, on the basis of the informal semantics of QL, no clear answer can be given to this question, since this semantics does not precisely determine the meaning of a sentence like $\text{S}^{(X)}_{[1,2]}$. Does this sentence, for example, specify the electron's location, just like the statement ``We are currently in Vienna'' specifies our location? Or is it rather just shorthand for the conditional statement ``If a measurement were performed, its output would indicate the electron's position somewhere in the interval $[1,2]$''? Needless to say, these two possible readings do not have the same empirical significance: the former may be understood as a metaphysical claim about instantiated properties, without any implications on what we may observe in a performed experiment, whereas the latter, as a conditional statement, has a bearing on possible observations in experiments that may be carried out in a laboratory. But the operationalist semantics fares better in this respect, since it equates the meaning of experimental statements with tests that can, at least in principle, be performed in a laboratory. Hence, we can now clearly consider the possibility that the quantum logician may be exposed to an inconsistency with the applicability of classical mathematics in QM. For it may now appear that the non-distributive logical relations between sentences in $S_\mathcal{QL}$ are in tension with the classicality of the mathematics applied in quantum modeling. 

First, we need to determine what exactly the applicability of classical mathematics in QM implies with respect to the logical relations between the sentences in $S_\mathcal{QL}$. Consider a generic class of quantum models $\Theta_{\alpha}=\langle \mathcal{H}, \mathcal{A}, T, \psi, \mu_{\phi} \rangle$, where the index $\alpha$ indicates that the ``initial'' quantum state of its corresponding model is $\ket{\alpha} \in \mathcal{H}$, i.e. that $\ket{\psi_{t_0}}=\ket{\alpha}$. With the help of the Born rule, each of these models can be used to describe a variety of experiments, as already explained above. Further, consider a particular experiment $T^{A}_{\alpha}$, which consists of the measurement of a set of dynamical variables associated to operators $A\equiv \left\{A_1,...,A_n\right\}\subset \mathcal{A}$ on a quantum system prepared in state $\ket{\alpha}$. Recall that according to QM, any set of operators corresponding to simultaneously measurable variables is necessarily commutative, which implies that $[A_i,A_j]=0$, for all $i,j=1,...,n$. Finally, let $\text{Pr}^{A}_{\alpha}$ be the classical probability distribution generated from a quantum model by an application to $T^{A}_{\alpha}$. 

Here is why the \textit{classicality} of $\text{Pr}^{A}_{\alpha}$ implies that the logical relations among \textit{a certain subset} of sentences in $S_\mathcal{QL}$ obey CL. Let $s^{(i)}_{\Delta}$ be the elementary sentence ``The value of the variable associated to $A_i$ lies in $\Delta$'', where $\Delta$ is an arbitrary subset of $\text{Spec}(A_i)$. Furthermore, let $S_\mathcal{QL}^A\subset S_\mathcal{QL}$ be the subset of sentences generated by logically compounding the elementary sentences $s^{(i)}_{\Delta}$. The experiment $T^{A}_{\alpha}$ can now be used to furnish idealized tests $\tau(s^{(i)}_{\Delta})$ for the elementary sentences $s^{(i)}_{\Delta}$. Indeed, saying that a system prepared in state $\ket{\alpha} \in \mathcal{H}$ passes test $\tau(s^{(i)}_{\Delta})$ with certainty is equivalent to asserting that
\begin{equation*}
   \text{Pr}^{A}_{\alpha}(G^{(i)}_{\Delta})=1,
\end{equation*}
where $G^{(i)}_{\Delta}$ is shorthand for $(\Omega_1,...\Omega_{i-1},\Delta,\Omega_{i+1},...,\Omega_n)$, with $\Omega_j\equiv$ Spec($A_j$). 

Furthermore, the operationalist semantics for disjunction and conjunction implies the following relations between the idealized tests associated to these connectives and the probability distribution $\text{Pr}^{A}_{\alpha}$: a quantum system prepared in state $\ket{\alpha}$ 
\begin{center}
    passes test $\tau(s^{(i)}_{\Delta_i} \land s^{(j)}_{\Delta_j})$ if and only if $\text{Pr}^{A}_{\alpha}(G^{(i)}_{\Delta_i}\cap G^{(j)}_{\Delta_j})=1$, \\
    passes test $\tau(s^{(i)}_{\Delta_i} \vee s^{(j)}_{\Delta_j})$ if and only if $\text{Pr}^{A}_{\alpha}(G^{(i)}_{\Delta_i}\cup G^{(j)}_{\Delta_j})=1$,
\end{center}

 for arbitrary $i,j=1,...,n$, and for $\Delta_i \subseteq \text{Spec}(A_i)$ and $\Delta_j \subseteq \text{Spec}(A_j)$. Note that, as a classical probability distribution, $\text{Pr}^{A}_{\alpha}$ obeys the  distributive identity

\begin{equation*}
   \text{Pr}^{A}_{\alpha}(G^{(i)}_{\Delta_i}\cap(G^{(j)}_{\Delta_j}\cup G^{(k)}_{\Delta_k}))=\text{Pr}^{A}_{\alpha}( (G^{(i)}_{\Delta_i}\cap G^{(j)}_{\Delta_j})\cup (G^{(i)}_{\Delta_i} \cap G^{(k)}_{\Delta_k})), 
\end{equation*}

 which immediately implies 

\begin{equation*}
\tau(s^{(i)}_{\Delta_i} \land (s^{(j)}_{\Delta_j} \vee s^{(k)}_{\Delta_k})) = \tau((s^{(i)}_{\Delta_i} \land s^{(j)}_{\Delta_j}) \vee (s^{(i)}_{\Delta_i} \land s^{(k)}_{\Delta_k}))
\end{equation*}

 which in turn, given the operationalist definition of the QL-consequence relation, implies the validity of the following argument

\begin{equation*}
    s^{(i)}_{\Delta_i}\land(s^{(j)}_{\Delta_j}\vee s^{(k)}_{\Delta_k}) \vDash_{QL} (s^{(i)}_{\Delta_i}\land s^{(j)}_{\Delta_j})\vee (s^{(i)}_{\Delta_i} \land s^{(k)}_{\Delta_k}).
\end{equation*}

 Thus, by assuming the operationalist semantics, the quantum logician is forced to assert the validity of the classical law of distributivity. This follows as a consequence of the classicality of the probability distributions used in quantum modeling. However, lest there be any misunderstanding, note that the quantum logician is forced to assert the validity of the classical law of distributivity \textit{only} for the subset of sentences $S_\mathcal{QL}^A$ generated by logically compounding the elementary sentences  $s^{(i)}_{\Delta}$, which refer exclusively to dynamical variables associated to elements in the set of \textit{commutative} operators $A\subset \mathcal{A}$. Does this raise an inconsistency problem for the quantum logician? 

Definitely not! It is perfectly consistent to maintain that the set $S_\mathcal{QL}$ of all experimental statements requires a non-classical logic, while admitting that there are subsets like $S_\mathcal{QL}^A$, which nonetheless validate classical logic. Importantly, it is no accident that the argument started by considering a set of commutative operators: had we considered instead a non-commutative set, such as $\left\{X,P\right\}\subset \mathcal{A}$, where $X$ and $P$ are operators corresponding to, say, a particle's position and momentum, the argument would have been blocked from the get-go. This is because there is no (idealized) experiment, adequately modeled by a joint classical probability distribution, that could simultaneously measure variables whose corresponding operators are mutually non-commuting. Briefly put, since any application of a quantum model can only be used to generate probability distributions over values of quantum-mechanically compatible variables, it follows that the classicality of such distributions can force only some subsets of sentences $S_\mathcal{QL}^A \subset S_\mathcal{QL}$, for any set of commutative operators $\mathbf{A}$, to obey CL. This is the case, indeed, \textit{according to QL}. Therefore, the alleged inconsistency between QL and the applicability of classical mathematics in QM does not exist.

But the fact that this inconsistency does not exist is not surprising at all.\footnote{Indeed, that QL does not conflict with the classicality of QM models has already been suggested by Putnam (1968), and more recently by Dickson (2001, S283), Bacciagaluppi (2009), and others.} To insist that it does, as Williamson has suggested in his WA, is actually a curiosity of sorts. For recall that QL-consequence has been defined relative to the operationalist semantics, which has itself been constructed on the basis of QM and its mathematical formalism. The logical relations between sentences in the language of the theory concern the properties of tests performable in principle on quantum systems. But these tests and their properties are not freely posited by a speculative logician: they are instead explicitly defined in terms of quantum models! This is why it is rather odd that an argument such as WA has been leveled against QL, in the first place, for this logic has been explicitly motivated by, and formally constructed on the basis of, the standard formalism of QM. That WA is unsound in the case of QL is, thus, not an accident. A further question, which will however not be discussed here, is whether there is \textit{any} non-classical logic for which WA is sound.\footnote{It can be argued that WA fails, for the same reason, i.e., because premise 2 is false, also in the case of a logic of vagueness. See Horvat and Toader 2024 for more details.} 

\newpage 

\section{Quantum inferentialism}

Quantum inferentialism, as a metasemantics of standard QM, holds that the meaning of expressions in the language of the theory is determined by their inferential properties. This rejects the widespread view that meaning is determined by representational properties. After revisiting some conceptual background, two objections are formulated, from information-theoretic reconstructions and from extensions of Wigner's friend scenario. Then the chapter focuses on the categoricity problem for quantum inferentialism. It also motivates the project of a non-inferentialist metasemantics for QBism, according to which meaning is determined by decision-theoretic properties.\footnote{This chapter is partly based on Toader 2025c.}

\subsection{Inferentialist approaches to physics}

The most detailed version of quantum inferentialism, to date, has been advocated by Richard Healey in his book, \textit{The Quantum Revolution in Philosophy} (Healey 2017). This denies that the semantic attributes of quantum expressions are determined by their representational properties, articulated by correspondence or semantic rules. It also denies that quantum expressions have any representational properties. As a pragmatist, Healey is committed to the ontological claim that QM has no beables of its own, so there are no physical entities that could be represented by QM's formalism. This entails that superposition, entanglement, and nonlocality, i.e., some of the most characteristic features of QM, are not physical, but mathematical phenomena. One might think that this commitment is the very reason why he believes that QM is a revolutionary theory. But although truly attractive since it does not conflate semantics and ontology, Healey's view is no quantum fictionalism. As will become clear in this chapter, his view is that what makes QM philosophically revolutionary -- a radical change from classical physics -- is neither its ontology (or lack thereof), nor its logic, but its metasemantics. In a nutshell, Healey's quantum inferentialism is a non-representationalist metasemantics for standard QM, conjoined with a representationalist semantics for all non-quantum expressions, including expressions in the language of classical physics. 

The early articulation of inferentialism by Wilfrid Sellars, as a critical reaction to some of Carnap's views expressed in \textit{Logical Syntax of Language} (1934), and Robert Brandom's later reflections and developments constitute the historical and conceptual background that will help us to understand Healey's pragmatist view, so that is where I will start below (section 5.1.1). I will then compare (in section 5.1.2) his quantum inferentialism to Huw Price's global expressivism, which embraces inferentialism for all scientific theories (Price 2013), and I will do so especially with regard to the following dilemma: either the representational success of singular terms in a language can be accounted for inferentially, or representationalist semantics must be allowed for at least some expressions in that language. But my main interest in this chapter is to formulate some new objections to Healey's quantum inferentialism: two of them are from information-theoretic reconstructions of QM (in section 5.2.1), and from recent extensions of Wigner's friend scenario (in section 5.2.2). After considering some possible ways in which the inferentialist might respond to these objections, I will formulate (in section 5.3.2) what I think is a more serious problem with quantum inferentialism, a categoricity problem similar to that faced by its metasemantic cousin -- logical inferentialism (section 5.3.1). But I shall also make some suggestions, on behalf of the quantum inferentialist, as to how this problem might be solved in ways that are fully compatible with inferentialism. The chapter ends with some admittedly programmatic remarks (in section 5.3.3) on how a non-inferentialist metasemantics for QBism might be developed, and why.  

\subsubsection{Global inferentialism}

This is, of course, not the place for a comprehensive critical account of inferentialism as a metasemantics of natural language. But a couple of notions are central to this view and must be clarified before we embark on an analysis of quantum inferentialism. Such is Sellars' notion of material inferences, but also the notion of inferential conservativity, which is applicable to the rules of these inferences. In this section, I will first present Sellars' argument, which he directed against Carnap, to the effect that material inferences are irreducible to logical inferences. Then, I will explain the significance of inferential conservativity, and in particular Brandom's argument \textit{against} the inferential conservativity of the material rules for the introduction and elimination of non-logical, scientific terms. Both of these central notions, and these two arguments, will help us later to better understand certain aspects of Healey's quantum inferentialism.

Sellars' paper ``Inference and Meaning" (1953) is the only one of his works cited in Healey's book, and we do well to start there, also because Carnap’s \textit{Syntax} is not only the inspiration, but also sole target of that paper, so Sellars' criticism is particularly poignant. In his preface to the \textit{Syntax}, Carnap wrote: ``Let any postulates and any rules of inference be chosen arbitrarily; then this choice, whatever it may be, will determine what meaning is to be assigned to the fundamental logical symbols.'' (Carnap 1934, xv, emphasis removed) Sellars proposed that not only logical symbols, but non-logical expressions as well have their meaning determined only by their linguistic use, and more specifically, by the rules of inference. As Brandom would later describe this view, ``Sellars takes it that `grasp of a concept is mastery of the use of a word.' He then understands the metalinguistic features in question in terms of rules of \textit{inference}, whose paradigms are Carnap's L-rules and P-rules. ... Sellars identifies his `material rules of inference' with Carnap's `P-rules'. ... Carnap’s views ... made the scales fall from Sellars’s eyes.'' (Brandom 2015, 43sq) Granting this identification and the falling of scales, what are material inferences and what exactly are their rules? What is the argument that supports their metasemantic indispensability?

Against Carnap's view that material rules, or what he called P-rules, are just thought-economical devices, ``a matter of convention and hence, at most, a question of expedience'' (Carnap 1934, 180), Sellars argued that material inferences are irreducible to logical inferences. The argument will be important for understanding, when we turn to QM, the metasemantic work that the rules of material inferences are claimed to do there. In order to explain it, consider these examples: 

\begin{quote}
    
\textit{If an object is red, then it's colored. This object is red. Thus, it's colored.} 

\textit{If something is gray, then it's a slithy tove. Findus is gray. Thus, it's a slithy tove.} 

\end{quote}

What is characteristic of such inferences is, of course, that one can assent to their conclusions once one has assented to their premises, even though one may not know what the terms mean. In the case of logical inferences, it's their logical form that matters. However, Sellars pointed out that, in the case of material inferences, it's the matter that matters, not the logical form. He maintained that unless one assents to their conclusions once one has assented to their premises, one does not know what the terms involved mean. Consider these examples: 

\begin{quote}

\textit{This object is red. Thus, it's colored.} 

\textit{Findus is on the mat. Thus, Findus is not on the roof.} 

\end{quote}

An entire network of such inferences -- what Brandom calls ``the social game of giving and taking reasons'' (Brandom 2000, 159) -- determines, according to Sellars, the meaning of all of our linguistic expressions. To defend this view, he first noted that material inferences are expressed by subjunctive conditionals, such as ``If this object were red, then it would be colored.'' and ``If Findus were on the mat, Findus would not be on the roof.'' Such subjunctive conditionals, as Brandom would later emphasize, are implicit modal statements.\footnote{For this reason, Brandom characterized inferentialism as a ``modal expressivism'', i.e., the view that modal concepts make explicit what is implicit in the use of concepts in material inferences. I will return to this idea in my discussion of quantum inferentialism, in section 5.1.2.} In particular, our two inferences are expressed by the following modal statements: ``Necessarily, if this object is red, then it is colored.'' and ``Necessarily, if Findus is on the mat, then Findus is not on the roof.'' Essentially due to this implicit modality, Sellars concluded that material inferences are irreducible to logical inferences. This is because one cannot detach the consequent of a subjunctive conditional \textit{only} by affirming its antecedent:

\begin{quote}
    
\textit{Necessarily, if this object is red, then it is colored. This object is red. Thus, this object is colored.} 

\textit{Necessarily, if Findus is on the mat, then he is not on the roof. Findus is on the mat. Thus, he is not on the roof.} 

\end{quote}

As stated, these two inferences are logically invalid.\footnote{Of course, one might object that, as stated, these inferences are mere enthymemes, which could be completed by adding an appropriate premise of the form ``If necessarily \textit{p}, then \textit{p}.'' However, I am here merely interested in reconstructing, rather than criticizing Sellars' argument. But one might wonder if this objection could also be raised against Healey's quantum inferentialism.} Moreover, some material inferences are clearly nonmonotonic, since they admit of defeasors. For example, Findus' mat actually being on the roof is such a defeasor, since in this specific case of domestic negligence, the corresponding subjunctive conditional, ``Necessarily, if Findus is on the mat, Findus is not on the roof.'' is false. But other material inferences are counterfactually robust, as it is hard to find defeasors for them.

Sellars took this argument to establish the metasemantic indispensability of material inferences: such inferences are relations between meanings, rather than relations between extensions of concepts. Expressing material inferences as implicit modal
statements makes this point explicit. He also thought that this corrected the view defended by Carnap. According to Sellars, Carnap had maintained that P-rules -- the rules of material inferences -- are, in principle, reducible to logical rules: 

\begin{quote}
        
\textit{Carnap, however, makes it clear that in his opinion a language containing descriptive terms need not be governed by extra-logical transformation rules}. Indeed, he commits himself (p. 180) to the view that for every language with P-rules, a language with L-rules only can be constructed in which everything sayable in the former can be said.  (Sellars 1953, 320; original emphasis) 

\end{quote}

It is rather difficult to endorse this reading of what Carnap actually says on page 180 of his \textit{Syntax}, and one would have to say much more to justify the step from P-rules being a matter of convention and expedience, as Carnap saw them, to P-rules being metasemantically dispensable, in the sense Sellars thought Carnap saw them. On Sellars's reading, Carnap regarded material rules as admissible only on account of ``the economy in the number of premises required for inferences'' (\textit{op. cit.}, 321). So he considered Carnap's view to be that material rules are inessential to any language, and thus metasemantically inert or dispensable. But what Carnap had maintained is that, in principle, one could stipulate only logical rules or one can adopt as material rules all sentences that are not logical rules: the choice is based on pragmatic criteria like simplicity and fruitfulness. He never said that adding or dropping P-rules would leave the expressive power of a language unchanged. 

In any case, Sellars doubted that the view that material rules are
metasemantically dispensable is correct, and thought that empiricism, or at least Carnap's empiricism, could be disconnected from it: 

\begin{quote}
    
But might it not be possible for an empiricist to hold that material rules of inference are as essential to meaning as [the logical] rules? ... P-rules are essential to any language which contains non-logical or descriptive terms.  (\textit{op. cit.}, 336)

\end{quote}

Sellars believed that his argument from the implicit modal character of material inferences successfully refutes what he took to be Carnap’s view. But Carnap's reflections on P-rules actually raise an important point for inferentialism, regarding the inferential conservativity of material rules. He noted the following: 

\begin{quote}

If P-rules are stated, we may frequently be placed in the position of having to alter the language; and if we go so far as to adopt all acknowledged sentences as valid, then we must be continuously expanding it. But there are no fundamental [as opposed to practical] objections to this. (Carnap 1934, 180)

\end{quote}

Continuous expansion of a language and continuous modification of its
semantics, although impractical, may be admissible. But on Sellars’ inferentialism, such a
modification would be unavoidable. The introduction of new vocabulary, and as a special case the
introduction of new concepts in science, always allows us to make novel material inferences. Given
the metasemantic work done by such inferences, on Sellars’ view, they will not only determine the
meaning of the new vocabulary, but will change the meaning of at least some of the old expressions
as well. This updating of semantics presupposes, in any case, that the rules are such that novel
material inferences can be made when new concepts are introduced in a language. In other words, it presupposes that the rules are inferentially non-conservative. 

Moreover, as Carnap further noted, rules of inference can also be altered: 

\begin{quote}

No rule of the physical language is definitive; all rules are laid down with the reservation that they may be altered as soon as it seems expedient to do so. This applies not only to the P-rules but also to the L-rules, including those of mathematics. In this respect, there are only differences in degree; certain rules are more difficult to renounce than others. (\textit{op. cit.}, 318)

\end{quote}

Thus, for Carnap, updating the semantics of a language as a result of changing the rules of inference would be admissible as well, while for Sellars this updating would be, once again, unavoidable. Since material inferences determine the
meaning of empirical terms, every change of rules implies a modification of the meaning of at least some of these terms in the language. But again, this presupposes that the rules are inferentially non-conservative, i.e., they allow new material inferences.

The point that I emphasize here, building on Carnap's reflections, that the semantics of a descriptive language changes whenever new material inferences or new material rules of inference are formulated for that language, has been acknowledged by Brandom. He implied, however, that inferential conservativity is nevertheless a good thing in logic: 

\begin{quote}
    
Unless the introduction and elimination rules are inferentially conservative, the introduction of the new vocabulary licenses new material inferences, and so alters the contents associated with the old vocabulary. ... Outside of logic, this is no bad thing. Conceptual progress in science often consists in introducing just such novel contents. (Brandom 2000, 68-71) 

\end{quote}

Following logical inferentialism, where logical rules of inference are rules for the introduction and the elimination of logical connectives, Brandom takes material rules of inference similarly as rules for the introduction and the elimination of descriptive terms. In a bit more detail, according to logical inferentialism, introduction rules state inferentially sufficient conditions for the employment of a logical connective, and elimination rules state inferentially necessary consequences of the employment of that connective. Likewise, the  inferentially sufficient conditions for the employment of a descriptive term are taken as its introduction rules, and the inferentially necessary consequences of the employment of that term are taken as its elimination rules. In order to block the introduction of spurious connectives in logic, e.g. tonk-like ones, as well as to avoid changing the meaning of connectives, the rules of logical inference should be inferentially conservative.\footnote{Cf. Prior 1960. Note that this condition is only contextually justified: for instance, one should allow that the language of positive logic is inferentially non-conservative, lest one blocks the introduction of new connectives in this language, e.g., negation (for a brief discussion of positive logic including a proof of its non-categoricity, see Brîncuș and Toader 2019). Note also that tonk is an extreme example of a spurious connective, since it allows one to infer everything from anything.}

In contrast, according to Brandom, the material rules of inference in a descriptive language should be inferentially non-conservative, to allow for conceptual progress, or at least for the kind of scientific progress based on the introduction of new concepts.\footnote{Brandom's point here goes against a view proposed by Michael Dummett, who upheld inferential conservativity not only for logical languages, but for descriptive languages, so that the introduction of new but spurious descriptive terms can be blocked as well. Of course, part of the issue here (as in the case of logic) revolves around the nature of spuriousness.} This point is endorsed by Healey: 
  
\begin{quote}
    
Acceptance of quantum theory affects the content of all magnitude claims, though for most such claims (especially concerning ordinary objects above a microscopic scale) the effect may be safely neglected. (Healey 2017, 210) 

\end{quote}

Healey readily admits that the introduction of QM has licensed new material inferences in physics, which on the inferentialist metasemantics that he favors not only determines the meaning of quantum expressions, but they also change the meaning of non-quantum expressions like ``Particle \textit{s} is located in region \textit{R}'', ``Findus is alive'', etc. The changes in meaning of such expressions are, he contends, negligible for most such expressions, and especially so for those that concern macroscopic objects. As we will see, however, Healey's endorsement of the inferential non-conservativity of material rules of inference in standard QM is problematic, for it blocks one possible response to the categoricity problem that I will pose further below to quantum inferentialism. 

One final point before I turn to present Healey's view in more detail. This concerns Sellars's claim that the metasemantic use of correspondence or semantic rules is actually question-begging: 

\begin{quote}
    
There is at first sight some plausibility in saying that the rules to which the expressions of a language owe their meaning are of two kinds, (a) syntactical rules, relating symbols to other symbols, and (b) semantical rules, whereby basic descriptive terms acquire extra-linguistic meaning. It takes but a moment, however, to show that this widespread manner of speaking is radically mistaken. (Sellars 1953, 335sq)

\end{quote}

He rejected the idea that semantic rules can do any metasemantic work as radically mistaken in the following way: 

\begin{quote}
    
Obeying a rule entails recognizing that a circumstance is one to which the rule applies. If there were such a thing as a `semantical rule' by the adoption of which a descriptive term acquires meaning, it would presumably be of the form `red objects are to be responded to by the noise \textit{red}'. But to recognize the circumstances to which this rule applies, one would already have to have the concept of red, that is, a symbol of which it can correctly be said that it `means red'. (\textit{loc. cit.})

\end{quote}

Sellars' rejection of the idea that semantic rules can do any metasemantic work will be important in the context of my first objection to quantum inferentialism, based on information-theoretic reconstructions of QM. That objection will specifically point out that unless one allows semantic or correspondence rules to do some metasemantic work, those reconstructions cannot account for the meaning of reconstructed QM. But, of course, that such rules do no metasemantic work is precisely what the inferentialist insists upon, as Sellars indeed did.

\subsubsection{Healey's quantum inferentialism}

The quantum inferentialist is committed to the claim that an adequate understanding of QM requires not only the ability to successfully use the theory for prediction and explanation, but it further requires that questions about semantic facts, about what determines the meaning of quantum expressions, be properly answered. Some reasons for adopting inferentialism concern what Healey regards as the failure of most viable interpretations of QM to provide adequate solutions to a host of foundational problems like the problem of ontology, measurement, completeness, nonlocality, etc.\footnote{For a clear and comprehensive account of these problems in the foundations of QM, see Norsen 2017. See chapters 4 and 6 in Healey 2017 for his own account.} Since these interpretational problems arguably assume representationalism, dropping the latter is believed to dissolve such problems, which in turn provides indirect support for inferentialism. For example, as I mentioned already, standard QM has, according to Healey, no beables of its own, and so, importantly, there exists no quantum dynamics. Also, entanglement is not a physical relation between real systems, but a mathematical relation between mathematical objects. 

As Healey emphasized, ``The most significant break marked by acceptance of quantum theory is a novel, indirect use of models to further the aims of fundamental science.'' (Healey 2017, 121) Successful prediction and explanation do not assume that quantum models provide a representation of the physical world, or of our knowledge thereof. Mathematical structures like $\langle \mathcal{H}, \mathcal{A}, T, H, \psi, \mu_{\phi} \rangle$ have no representational capacities. Instead, the inferentialist proposes that they are used as epistemic advisors, in the following sense: the Schrödinger equation tells us how quantum models are used for prescribing and updating our credences in experimental propositions, i.e., what Healey calls ``canonical non-quantum magnitude claims'' about a system. For a system $s$ and magnitude M, all such claims have the the form ``s has (M $\in \Delta$)'' which says ``The value of the magnitude M of s is in the interval $\Delta$'', or a bit more precisely, ``The value of the magnitude corresponding to operator $A$ lies in interval $\Delta$'', where $A \in \mathcal{A}$ is associated to M, and $\Delta \subseteq \text{Spec}(A)$ is an interval in A's spectrum.

Thus, Healey sees QM as $\psi$-\textit{prescriptive}. This is what makes it a non-classical theory and explains its revolutionary character: for rather than representing states of affairs, the theory prescribes credences or degrees of belief in certain non-quantum magnitude claims about physical systems. Nevertheless, as we will see presently in more detail, quantum models are to be relativized to physical situations, such that differently located agents can assign different quantum models to the same system, and thus can prescribe different credences about the same magnitude claim. And each such prescription is complete, in the sense that the model it is based on requires no additional variables. In contrast, $\psi$-ontic and $\psi$-epistemic interpretations treat QM just like classical physics, i.e., they assume that quantum expressions are, just like the language of classical physics, representational. They assume, in other words, not only a representationalist semantics, but a representationalist metasemantics as well. Healey insists that precisely because of such assumptions, these interpretations fail to account for the revolutionary character of QM. On his view, what accounts for this character is neither the lack of an ontology, nor its possibly deviant logic, but rather its semantics and metasemantics. 

But one might wonder whether, despite the proposed dissolution of quantum ontology, an inferentialist makes no other ontological commitments in QM. Recall that, as noted in the previous section, on Brandom's development of Sellars' view, inferentialism turns out to be a kind of modal expressivism. Brandom further took this to be a ``fundamental Kantian idea'': 

\begin{quote}
    
Claiming that one should be a \textit{pragmatic} modal expressivist (an expressivist about what one is \textit{doing} in applying modal vocabulary) but a \textit{semantic} modal realist (a realist about what one is \textit{saying} in applying modal vocabulary) is, I think, recognizably a development and a descendant, for this special but central case, of Kant’s claim that one should be a \textit{transcendental idealist}, but an \textit{empirical realist}. (Brandom 2015, 178)

\end{quote}

Thus, an inferentialist is an expressivist about what she is doing, but a realist about what she is saying, in applying modal vocabulary, for example in the statement ``Necessarily, if this substance turns litmus red, it is an acid.'' This is presumably because this modal statement not only expresses a material inference (\textit{This substance turns litmus red. Thus, it is an acid.}), but it also describes a modal fact. If this idea is taken seriously by the quantum inferentialist, then even though QM is intended to be prescriptive, it seems it cannot be purely prescriptive. This is because the modal statements that express its material inferences turn out to be descriptive after all, in the sense just indicated, which suggests that the quantum inferentialist must be committed to a modal ontology.

In any case, how does inferentialism actually explain the relationship between the quantum formalism and its meaning? What are the rules of material inferences that do metasemantic work in standard QM? On Healey's view, the circumstances and the consequences of the application of a quantum model determine the meaning of all quantum expressions. Circumstances of application are the inferentially sufficient conditions for the employment of a model, for the assignment of a quantum state. In other words, they are its introduction rules. Consequences of application are the inferentially necessary consequences of the employment of a model, of the assignment of a quantum state --  its elimination rules. 

In order to further clarify what these introduction and elimination rules are, note that Healey conceives of quantum models as ``informational bridges'', bridges that lead an agent from a certain set of non-quantum magnitude claims (what he calls ``backing conditions'') to a set of probability statements.  In other words, the agent's description of an experimental setup justifies the assignment of a quantum state, which in turn justifies the prescription of credences in other non-quantum magnitude claims that describe experimental outcomes. Roughly, the template of all material inferences in QM is as follows:

\begin{quote}

\textit{Backing conditions are such and such.} 

\textit{Thus, the assignment of a certain quantum state is justified.} 

\textit{Thus, the Born probabilities are such and such.} 
    
\end{quote}

Such an inference is non-logical, but nevertheless objective, i.e., independent of agents, although not independent of their physical location. One of Healey's main examples of material inferences in QM involves some cool interference experiments performed in Vienna (Juffmann \textit{et al}. 2009), an example that could be explained as above, in section 4.3.2. But here is a simpler example, featuring again my beloved cat in a Schrödinger's cat scenario:

\begin{quote}
    
\textit{Findus is purring in his box, next to a flask of poison, a radioactive atom, etc.} 

\textit{Thus,} $\Psi_{0} \longrightarrow \frac{1}{\sqrt{2}}(\psi_{notdecayed} \phi_{intact} \xi_{alive} +
\psi_{decayed} \phi_{shattered} \xi_{dead})$. 

\textit{Thus, Pr(Atom hasn't decayed, the flask of poison isn't shattered, and Findus is alive) = 0.5 and Pr(Atom has decayed, the flask of poison is shattered, and Findus is dead) = 0.5.}

\end{quote}

Inferences like this one are not logical inferences for precisely the reason already indicated by Sellars. They are expressed by subjunctive conditionals: ``If the backing conditions were such and such, then the assignment of a quantum state would be justified, and so by a legitimate application of the Born rule the prescribed credences would be such and such.'' Furthermore, such conditionals are understood as implicit modal statements: ``Necessarily, if the backing conditions are such and such, then the assignment of a quantum state is justified, and so by a legitimate application of the Born rule the prescribed credences are such and such.'' As we have discussed above, this shows that one cannot validly detach the consequent of a material inference only by affirming its antecedent. 

In standard QM, as viewed by Healey, there are two sets of rules for material inferences. The first set includes only the Born rule, which prescribes probabilities, interpreted as credences or degrees of belief in non-quantum magnitude claims, so the inferentialist understands it as an elimination rule for quantum models (but only in decoherent contexts, as Healey emphasizes, and I will return below to this qualification). The second set includes rules for the introduction of quantum models, for the assignment of quantum states. However, these rules are never explicitly stated, but they are said to be implicitly learned in the practice of QM, and more exactly, in the practice of state preparation: ``Deciding on the best available quantum model requires the expertise of the practicing physicist, which can be acquired only through extensive education and long experience.'' (Healey 2017, 135) 

Importantly, however, for any set of true backing conditions, these rules are supposed to introduce a \textit{unique} quantum model. For, as already mentioned, the assignment of a quantum state is always relative to the spatiotemporal location of the observer applying the rules of QM, and backing conditions are true in virtue of events in the backward light cone of the observer.\footnote{This is the only way in which Healey takes QM to be observer-dependent. I'll come back to this issue further below, in section 5.2.2, when I discuss his view on Wigner's friend scenarios.} Clearly then, the prescription of credences on the basis of a quantum model is taken to be counterfactually robust: ``If the backing conditions were such and such, then the assignment of a unique quantum state would be justified, and so by a legitimate application of the Born rule the prescribed credences would be such and such.'' Had there been different events in the backward light cone, a different quantum model would have been introduced, and so different credences would have been prescribed to the relevant non-quantum claims. The implicit modal statement that expresses the relevant material inference is this: ``Necessarily, if the backing conditions are such and such, then the assignment of a unique quantum state is justified, and so by a legitimate application of the Born rule the prescribed credences are such and such.'' 

Having presented this central aspect of Healey's quantum inferentialism, we should further observe that dissolving quantum ontology is neither sufficient nor necessary for this metasemantics. For on the one hand, quantum fictionalism can preserve representationalism, even though it denies the existence of quantum reality. On the other hand, one could in principle adopt the view that the meaning of quantum expressions is fully determined inferentially, despite commitment to a quantum ontology that the theory may nevertheless be taken to represent.\footnote{See section 1.2, where I mentioned the conceptual possibility of views like Bohmian inferentialism, Everretian QBism, etc., according to which semantic rules do not do any metasemantic work, but they can nevertheless specify the representational properties of quantum concepts.} Neither of these two hands is, however, Healey's own. While embracing an inferentialist metasemantics, as we have seen, he also adopts an inferentialist semantics for quantum expressions, for he denies that the standard formalism has representational properties. Nevertheless, he allows a representationalist semantics for non-quantum expressions. This looks like a retreat from Healey's professed pragmatism, so it's worth considering more closely. In particular, it turns out to be one important difference between Healey's and Price's views. 

In contrast to Price, whose global expressivism extends the scope of assertion to all vocabularies (moral, normative, modal, but also scientific vocabularies), Healey may be said to advocate a local expressivism, only for the language of QM. Expressions that do not belong to this language, such as non-quantum magnitude claims (or experimental statements about the possible values of dynamical variables), are taken to represent facts. Probabilistic statements about such claims, although they are expressions of advice concerning one's degrees of belief in non-quantum magnitude claims, are also said to be ``weakly representational''  (Healey 2017, 135), so in some sense they also represent facts.\footnote{I will return to this issue about probabilistic statements in section 5.2.2.} 

How does Healey justify the need for a representationalist semantics? The answer to this question is clear from his criticism of Price's global expressivism as a form of radical pragmatism: 

\begin{quote}
    
A radical pragmatist might deny that \textit{any} statement has the primary function of representing the world. Price... argues against what he calls Representationalism, the view that `the function of statements is to ``represent'' worldly states of affairs, and true statements succeed in doing so'. But in order to address the goals of predicting and explaining natural phenomena, science must be capable of representing those phenomena in language: in the case of quantum theory, this means assigning magnitude claims a primary representational role. (\textit{op. cit.}, 135) 

\end{quote}

At the heart of Healey's criticism of Price's view is a version of the so-called ``no-exit problem'' for global expressivism.\footnote{The no-exit problem was raised by Simon Blackburn against Price's view in the following way: ``I am much less certain about \textit{global} pragmatism, the overall rout of the representationalists apparently promised by Rorty and perhaps by Robert Brandom. The reason is obvious enough. It is what Robert Kraut, investigating similar themes, calls the no-exit problem. It points out, blandly enough, that even genealogical and anthropological stories have to start somewhere. There are things that even pragmatists need to rely upon, as they produce what they regard as their better understandings of the functions of pieces of discourse. This is obvious when we think of the most successful strategies of the pragmatist’s kind.'' (Blackburn 2013, 78) The pragmatist inferentialist, in particular, would have to allow that at least some expressions of the natural language admit of a representationalist semantics. For Price's own response to this problem, see Price 2013, 157 sq.} Taking this problem seriously, Healey acknowledged that the scientific prediction and explanation of natural phenomena require that non-quantum magnitude claims have a representationalist semantics. Magnitude claims (as well as the probabilistic statements about them) must ultimately describe facts in the world. How might a radical pragmatist avoid the adoption of representationalist semantics for at least some linguistic expressions?

In his 2008 Descartes lectures, Price distinguished two relations of representation: what he called i-representation, on the one hand, and what he called e-representation, on the other hand (Price 2013). The former relation is said to hold between assertoric statements and facts that are accessible only from within the use of language. In the sense of i-representation, as Price put it, ``something counts as a representation in virtue of its position or role in some cognitive or inferential architecture.'' (\textit{op. cit.}, 36) For example, moral, normative, and scientific statements are all i-representational: they describe facts that are accessible only from within the use of language. The totality of these facts, which can be i-represented by such statements, constitutes what Price called the i-world. The existence of the i-world is compatible with inferentialism.

Unlike moral and normative statements, scientific statements are not only i-representational; they are e-representational as well. The relation of e-representation is said to hold between scientific statements and facts that are accessible from within science. In the sense of e-representation, something counts as a representation in virtue of its covariance with an external environmental condition, or in other words, in virtue of its tracking the world. The totality of facts that are e-represented by scientific statements constitutes what Price called the e-world. The existence of the e-world is also compatible with inferentialism. This is because, according to him, 

\begin{quote}
    
the e-world is visible only from within science in precisely the same sense as the i-world is visible only from within the viewpoint of users of assertoric vocabularies in general. Indeed, the e-world simply \textit{is} the i-world of the scientific vocabulary. (Price 2013, 55)

\end{quote}

But it seems to me that Price's two-worlds view flies in the face of the problem of representational success, which was raised against Brandom's inferentialist view of the natural language.\footnote{Here is how Michael Kremer summarizes his criticism of Brandom, glossing on an important Kantian distinction: ``Brandom’s inferentialism, like the representationalism it is supposed to supplant, is, in the end, one-sided. Just as representationalism cannot provide an adequate account of representational \textit{uptake}, inferentialism cannot provide an adequate account of representational \textit{success}. Brandom’s anti-representationalist arguments are salutary insofar as they remind us that representation without inference is blind; but taken as arguments \textit{for} inferentialism, they lead us astray, causing us to forget the equally important insight that inference without representation is empty.'' (Kremer 2010, 244)}  This problem points out, essentially, that a network of material inferences -- an inferential architecture -- although perhaps sufficient \textit{metasemantically}, i.e., sufficient for determining the meaning of all descriptive expressions in a language, cannot be sufficient \textit{semantically}, i.e., it cannot be enough for fixing the reference of singular terms in that language. In the case of Price's global expressivism, even though the existence of the e-world may be compatible with inferentialism, there is no guarantee that it simply is the i-world of the scientific vocabulary. For there is no guarantee that singular terms in the language of science are successfully tracking the world  only in virtue of their position and role in the inferential architecture of science.\footnote{One might contend that tracking the world and explaining the meaning of vocabularies are distinct and unrelated endeavors. Thus, scientific terms are tracking the world, but this is independent of their position and role in the inferential architecture of science. As mentioned before, in section 1.3, this gap between semantics and metasemantics afflicts all views conjoining a representationalist semantics and a non-representationalist metasemantics for the same language (including, of course, Healey's view of non-quantum expressions in the language of physics). }

In contrast to Price, Healey avoids the no-exit problem by acknowledging that a representationalist semantics for part of the scientific language is indispensable for prediction and explanation, for tracking the world, as it were. The problem of representational success for non-quantum magnitude claims and the probabilistic statements about these claims does not arise. For even though he takes their meaning to be determined in virtue of their inferential roles, this is not what Healey believes can fix the reference of their terms. They track the world independently of their inferential roles. The problem of representational success for quantum expressions does not arise, either, for he does not think they have any representational capacities. Healey denies that there exists anything like an e-world or an i-world that QM could e-represent or i-represent. On Healey's inferentialist view, as I understand it, QM is neither i-representational, nor e-representational: there are simply no quantum facts visible from within QM.\footnote{Nevertheless, see above my suggestion that if, closely following Brandom, one takes inferentialism as a modal expressivism, then Healey might have to admit that quantum statements describe some modal facts after all. Of course, this does not block his characterization of standard QM as a prescriptive physical theory.} 


In the next section, I will present two preliminary objections to Healey's quantum inferentialism. Afterwards, in section 5.3, I will argue that quantum inferentialism must face a third, more serious problem, which had already been encountered by logical inferentialism: the problem of categoricity. This is an issue for quantum inferentialism as well because it indicates that the rules for the inferential use of quantum expressions cannot in fact precisely determine their semantic attributes, if they -- the rules -- are not categorical. This will draw attention to Healey's point that assigning different quantum models to one and the same physical system is possible only relative to different spatiotemporal locations of observers.

\subsection{Two preliminary objections to quantum inferentialism}

The reconstructionist approach to QM, which I shall refer to as reconstructionism, has been thriving in the last couple of decades. Since Lucien Hardy proposed his first reconstruction in the framework of a general probability theory (GPT), a reconstruction based on what he called ``reasonable'' axioms (Hardy 2001), others have followed up and developed variations in the GPT framework (e.g., Masanes and Müller 2011). Yet others have proposed reconstructions from different principles in different theoretical frameworks, e.g., the general C$^{*}$-algebra framework of the CBH reconstruction (Clifton, Bub, Halvorson 2003).\footnote{The CBH reconstruction has been given up in the meantime. For philosophical discussions of quantum reconstructions, see e.g., Dickson 2015, Grinbaum 2007.} 

The main epistemic benefit of reconstructionism, envisaged already by John Wheeler, has been recently emphasized by Markus Müller: ``it allows us to understand which features are uniquely quantum and which others are just general properties of probabilistic theories.'' (Müller 2021) But consider also the epistemic benefit that Healey has recently argued comes within reach if one adopts inferentialism, rather than representationalism: ``We can understand quantum theory better if we stop asking how it represents the world and ask instead how we are able to use it so successfully in predicting, explaining and controlling the world.'' (Healey 2023) Taken together, these comments suggest that there should be double epistemic benefit in adopting inferentialism as a metasemantics of \textit{reconstructed} QM: not only can we understand QM better philosophically, but we can also understand what exactly and perhaps uniquely distinguishes it from other probabilistic theories. 

However, I think that this is too good to be true, so I want to raise some doubts about this conceivable joint project (in section 5.2.1), although my goal is not to show that the project is bound to fail. Quite the opposite, in fact, I think it might be a project worth pursuing. But it just requires more work. Afterwards, I will move on to discuss (in section 5.2.2) Healey's inferentialist approach to Wigner's friend scenario (and recent extensions of the scenario), an approach intended to save a notion of objectivity in QM that avoids truth relativism -- arguably the worst case of semantic indeterminacy.

\subsubsection{Reconstructions of quantum mechanics} 

Consider the following questions: In virtue of what do expressions in the language of reconstructed QM have meaning? More specifically, is there any reason in favor of a representationalist metasemantics, rather than a non-representationalist one? Reconstructionists could, in principle, adopt representationalism by just taking up the semantic rules of a $\psi$-ontic interpretation and allow them to do metasemantic work. Of course, this requires that some such interpretation be acceptable to the reconstructivist in the first place. However, this is not the case: viable $\psi$-ontic interpretations of QM are generally rejected by reconstructionists as rationally unacceptable. 

Here is an argument to this effect developed by Jeffrey Bub (Bub 2004). Assume that the CBH information-theoretic constraints (i.e., no signalling, no broadcasting, and no bit commitment) are true, and that all theories that satisfy these constraints are quantum theories. Further, assume that there is an equilibrium distribution over particle positions. If there is an equilibrium distribution over particle positions, then any hidden variables interpretation of QM satisfies the CBH constraints. Thus, any hidden variables interpretation of QM is (equivalent to) QM. It follows that all hidden variables interpretations are equivalent to one another. However, if QM does not break this equivalence, then no hidden variables interpretation is rationally acceptable. But QM does not break this equivalence, i.e., it provides no evidence in favor of any such interpretation. Therefore, no hidden variables interpretation is rationally acceptable.

Bub suggested that this argument goes through for the many worlds interpretation as well. Nevertheless, one might be inclined to reject it: indeed, why should the rational acceptability of any quantum interpretation require that it is QM that singles it out among all the equivalent interpretations? Why should one ignore other criteria, including theoretical and pragmatic ones, beside those that depend on QM, in an attempt to break the equivalence? 

A similar argument to the same conclusion, that no hidden variables interpretation of QM is rationally acceptable, has been offered by Hardy. This starts from his five axioms: probabilities, subspaces, composite systems, simplicity, and continuity (cf. Hardy 2004a for a concise account) and goes as follows: 

\begin{quote}
    
If we really believe these axioms to be reasonable then they would also apply to hidden variables and it would follow that the hidden variable substructure must look like quantum theory. We could not then use hidden variables to solve the measurement problem (since this relies on being able to give the hidden variables a classical probability interpretation). (Hardy 2001)

\end{quote}

Assume that Hardy's axioms, formulated in a GPT language, are reasonable. This is taken to imply that they apply to any hidden variables interpretation of QM. If one grants this implication, then it follows, according to Hardy, that any such interpretation must be equivalent, in some sense, to QM. But if any hidden variables interpretation were equivalent, in that sense, to QM, then it could not solve the measurement problem. Thus, no hidden variables interpretation would be acceptable (unless other reasons compel us to accept it).

One might be inclined to reject this argument as well, perhaps on account of its vagueness: what might it mean, more precisely, to ``apply'' Hardy's axioms to an interpretation of QM? Might it mean that, just as one could derive QM from them, one could also derive from them any hidden variables interpretation? Perhaps the sense of this application may be clarified, to some extent, by pointing to what Hardy called the ``ontological embedding'' of his axioms in an interpretation of QM (Hardy 2004b). But this is not clear enough, either. Might it mean that the models of the axioms can be embedded in the models of the interpretation? If so, then this kind of embedding cannot guarantee the implied equivalence.

Despite such difficulties with these arguments against $\psi$-ontic interpretations, the point I want to emphasizes is that, for better or worse, reconstructionists strongly believe that these interpretations are not rationally acceptable. And there exist, of course, more arguments to the same effect. For example, $\psi$-ontic interpretations are sometimes rejected on the basis of Einstein's famous distinction between principle and constructive theories: 

\begin{quote}
    
Reconstructions represent a challenge for existing `$\psi$-ontic' interpretations of quantum theory by highlighting a relative deficiency of those interpretations in terms of their explanatory power. ... None of Bohmian mechanics, Everettian quantum theory, or collapse theories fill the
explanatory role of a principle theory. ... Nevertheless, ... one should demand that they give us something else in replacement in order to be considered successful: some additional element of empirical or theoretical success that goes \textit{beyond} the standard formulation of quantum theory.\footnote{Cf. Koberinski and Müller 2018, 262-265. For more on Einstein's distinction, see section 6.1.}

\end{quote}

I take the argument here to be as follows: A physical theory has the explanatory power of a principle theory only if it is reconstructed in a general framework from simple principles. It has the explanatory power of a constructive theory only if it increases the success of its predecessor(s). But no $\psi$-ontic interpretation of QM has been reconstructed in a general framework from simple principles. And no $\psi$-ontic interpretation of QM has increased the success of QM. Thus, $\psi$-ontic interpretations have neither the explanatory power of a principle theory, nor the explanatory power of a constructive theory. Therefore, $\psi$-ontic interpretations should be rejected as explanatorily defective. 

But if viable $\psi$-ontic interpretations are to be rejected as explanatorily defective (or as rationally unacceptable), then adopting their semantic rules can hardly be an option for the reconstructionist looking for a representationalist (meta)semantics. Still, one might be wondering whether, in the reconstructionist arguments just presented, one is throwing the baby (i.e., representationalism) out with the water (i.e., $\psi$-ontic interpretations). Of course, the reconstructionist could adopt a $\psi$-epistemic interpretation together with its semantic rules, and allow these rules to do metasemantic work for reconstructed QM. Alternatively, the reconstructionist could forgo all semantic rules and instead adopt inferentialism. This latter option will be my focus in what follows. But what I want to argue is that this option cannot stand on its own. This is because some semantic rules are metasemantically indispensable to GPT-reconstructions of QM, or so I will argue. 

Just to be very clear, my argument will not be against reconstructionism. Rather, I present it as a preliminary objection to inferentialism considered as a metasemantics of reconstructed QM -- preliminary, because I take it as a mere step in the development of a more robust version of quantum inferentialism, one that could be adopted for the language of a reconstructed QM as well.

Recall that, according to inferentialism, the meaning of all linguistic expressions is determined in virtue of their inferential use. Hence, the meaning of terms like ``state'', ``property'', and ``probability'' is articulated by the material inferences of the theory the language of which contains these expressions. For instance, as expressions in the language of GPT, their meaning is determined in virtue of the material inferences of GPT. As expressions in the language of QM, their meaning is determined in virtue of the material inferences of QM. But, of course, QM and GPT are not identical theories. Since the standard axioms of QM are different from the GPT principles, these two theories have different rules for material inferences. Therefore, according to inferentialism, the meaning of GPT expressions is different than that of their quantum mechanical counterparts. But this implies that GPT-reconstructions are not meaning-preserving, which should throw doubt on their adequacy as reconstructions \textit{of QM}. 

But perhaps one might think differently about the meaning of reconstructed QM. Here is one suggestion: 

\begin{quote}
    
The question of meaning, previously asked with regard to the formalism, is removed and now bears, if at all, only on the selection of the principles. No room for mystery remains in what concerns the meaning of the theory’s mathematical apparatus. One now makes sense of all of the formalism solely on the basis of the first principles, and whatever mathematical element is contained in the formalism ... it now acquires a precise meaning in virtue of the first principles. (Grinbaum 2007, 389) 

\end{quote}

The suggestion here is that GPT principles, for instance, are enough to account for the metasemantics of reconstructed QM. For the meaning of all expressions in the language of reconstructed QM is determined in virtue of the GPT principles, rather than in virtue of the rules of the reconstructed theory. I think that the are two problems with this suggestion.

The first problem, for the inferentialist, is that semantic or correspondence rules appear to be metasemantically indispensable. Indeed, reconstructionists typically use such rules to fix the information-theoretic meaning of QM terms. For example: ``A state is an equivalence class of preparation procedures.'' (Müller 2021) Quite explicitly, the meaning of the term ``state'' in the language of GPT is defined in terms of a certain class of preparation procedures. Then the meaning of its quantum mechanical counterpart is fixed by a meaning-preserving translation from the language of GPT, i.e., by a correspondence rule that associates a GPT term to its counterpart QM term. Obviously, this is a different view than that advocated by the inferentialist, since it requires that correspondence rules do metasemantic work. Correspondence rules do metasemantic work by explaining the semantic attributes of quantum concepts via the information-theoretic properties of GPT concepts.

The second, and more important problem is that the suggestion under discussion makes the material inferences of QM metasemantically idle: since the meaning of QM expressions is determined in virtue of GPT principles, then QM's own inferential rules do no metasemantic work! Naturally, this cannot be acceptable to the quantum inferentialist. If this is correct, it can cut either way: against inferentialism or against reconstructionism. However, I think the interesting philosophical work lies ahead on a reconciliation route, on which the challenge is to develop inferentialism into a workable metasemantics of reconstructed QM.

\subsubsection{Extended Wigner's friend scenarios}

Another preliminary objection to quantum inferentialism is motivated by Healey's critical approach to recent relativistic extensions of the Wigner's friend scenario, and especially to some of the philosophical consequences of such extensions (as outlined, e.g., in Brukner 2018). These extensions, as we will presently see, are sometimes taken to reinforce a philosophical view already implied by Wigner's original scenario (Wigner 1961), that is the view that there are no objective, observer-independent facts: what is a fact for some observers is not a fact for others.
This in turn is taken to imply truth-relativism: what is true for some observers is false for others. Very recently, Healey rejected such consequences (Healey 2021). What I want to do in this section is evaluate the grounds for this rejection from the perspective of quantum inferentialism. My claim will be that there is an overlooked tension between Healey's attempt to save the objectivity of facts and his inferentialism.

Consider an isolated lab: inside, there is a target system and an observer (Alice); outside the lab, there is a superobserver (Super Alice). The possibility of superobservers, i.e., observers of observers, follows from the assumption that QM is a universally applicable theory, which means applicable not only to microscopic systems, but to macroscopic ones, including human beings, as well. Suppose Alice measures a physical variable $x_{a}$ on the system and suppose she obtains the result $a$. For simplicity, the only possible values of $a$ are +1 or --1. Let the two experimental statements be $p_{a}$ and $p'_{a}$, which state ``$a$ = +1'' and ``$a$ = --1'', respectively. We then say that there is a fact about Alice's measurement if and only if $p_{a} \vee p'_{a}$ is true, where $\vee$ is an exclusive disjunction, since we also assume that all measurements have unique results. From Super Alice's perspective, however, neither $p_{a}$ nor $p_{a}'$ is true. Thus, from that perspective, in accord with QM, $p_{a} \vee p'_{a}$ is false. This already suggests that facts are observer-dependent, since what is a fact for Alice, inside the lab, is not a fact for Super Alice outside. The extensions of this scenario reinforce this claim.

Now, suppose there is a physical variable $X_{A}$ that Super Alice can measure on Alice's lab and obtain result $A$, which just as above can, for simplicity, be either +1 or --1. Let the experimental statements be $p_{A}$ and $p'_{A}$, which state ``$A$ = +1'' and ``$A$ = --1'', respectively. We say that there is a fact about Super Alice's result if and only if $p_{A} \vee p'_{A}$ is true, where $\vee$ is again an exclusive disjunction. Further, consider a second isolated lab: inside, a target system and an observer (Bob); outside this lab, a superobserver (Super Bob). Suppose Bob measures $x_{b}$ on his system and obtains result $b$ (again, either +1 or --1). Bob's experimental statements are then $p_{b}$ and $p'_{b}$, which state ``$b$ = +1'' and ``$b$ = --1'', respectively. Similarly, we say that there is a fact about Bob's result if and only if $p_{b} \vee p_{b}'$ is true. Suppose, again as above, that there is a physical variable $X_{B}$ that Super Bob can measure on Bob's lab and obtain result $B$ (either +1 or --1). Super Bob's experimental statements are $p_{B}$ and $p'_{B}$, which state ``$B$ = +1'' and ``$B$ = --1'', respectively. We say that there is a fact about Super Bob's result if and only if $p_{B} \vee p'_{B}$ is true. Finally, consider a third lab: inside, a target system and an observer (Carr); outside, a superobserver (Super Carr). Suppose Carr measures $x_{c}$ on the system and gets result $c$ (exactly, +1 or --1). Carr's experimental statements are then $p_{c}$ and $p'_{c}$, which state ``$c$ = +1'' and ``$c$ = --1'', respectively. There is a fact about Carr's result if and only if $p_{c} \vee p'_{c}$ is true. Suppose further that there is $X_{C}$ that Super Carr can measure on Carr's lab and get result $C$, and let $p_{C}$ and $p'_{C}$ state the corresponding experimental statements ``$C$ = +1'' and ``$C$ = --1''. There is a fact about Super Carr's result if and only if $p_{C} \vee p'_{C}$ is true. 

Putting everything together, we should say that there is a fact -- I will call this a \textit{superfact}, since it is supposed to be observer-independent -- about all six results $a,A,b,B,c,C$ if and only if 

\begin{center}
    
$(p_{a} \vee p_{a}') \wedge (p_{A} \vee p_{A}') \wedge (p_{b} \vee p_{b}') \wedge (p_{B} \vee p_{B}') \wedge (p_{c} \vee p_{c}') \wedge (p_{C} \vee p_{C}')$ 

\end{center}

 is true. But here a difficulty arises. If the three target systems in the three isolated, spacelike separated labs are entangled, such that the extended Wigner's friend scenario satisfies the conditions of the Greenberger-Horne-Shimony-Zeilinger (GHSZ) theorem (Greenberger \textit{et al.} 1989, 1990), then superobservers and superfacts are incompossible. In particular, if there are superfacts, then QM cannot be universally applicable.\footnote{For the application of the GHSZ theorem to Wigner's friend scenario, see Leegwater 2018, which is followed in Healey 2024.}

Here is, briefly, how the difficulty arises. Applying the GHSZ theorem, we get the following claims: If Super Alice's, Bob's, and Carr's measurements are simultaneous, then $Abc$ = +1. If Alice's, Super Bob's, and Carr's measurements are simultaneous, then $aBc$ = +1. If Alice's, Bob's, and Super Carr's measurements are simultaneous, then $abC$ = +1. The antecedents of these claims are all relativistically compatible, since each is true in some inertial frame. Multiplying on both sides of the equations in their consequents, we obtain $ABC$ = +1. However, if Super Alice's, Super Bob's, and Super Carr's measurements are simultaneous, then $ABC$ = --1. The contradiction entails that in at least one relativistic frame, QM is not applicable. 

If one wants to preserve universal applicability, the obvious option is to eliminate superfacts. To see how this can be achieved, note that, on the one hand, $ABC$ = +1 entails that $p_{A} \wedge p_{B} \wedge p_{C}$ is true, or $p_{A}' \wedge p_{B}' \wedge p_{C}$ is true, or $p_{A} \wedge p_{B}' \wedge p_{C}'$ is true, or $p_{A}' \wedge p_{B} \wedge p_{C}'$ is true. In each of these conjunctions, the number of primed sentences must be even. On the other hand, $ABC$ = --1 entails that $p_{A}' \wedge p_{B} \wedge p_{C}$ is true, or $p_{A} \wedge p_{B}' \wedge p_{C}$ is true, or $p_{A} \wedge p_{B} \wedge p_{C}'$ is true, or $p_{A}' \wedge p_{B}' \wedge p_{C}'$ is true. In each of these conjunctions, the number of primed sentences must be odd. Thus, if $ABC$ = +1 and $ABC$ = --1, then there is a truth valuation such that one of the first four conjunctions, i.e., 

\begin{center}

$p_{A} \wedge p_{B} \wedge p_{C}$, or $p_{A}' \wedge p_{B}' \wedge p_{C}$, or $p_{A} \wedge p_{B}' \wedge p_{C}'$, or $p_{A}' \wedge p_{B} \wedge p_{C}'$ 

\end{center}

 and one of the last four conjunctions, i.e.,  

\begin{center}
    
$p_{A}' \wedge p_{B} \wedge p_{C}$, or $p_{A} \wedge p_{B}' \wedge p_{C}$, or $p_{A} \wedge p_{B} \wedge p_{C}'$, or $p_{A}' \wedge p_{B}' \wedge p_{C}'$

\end{center}

 are true together. But on the same truth valuation, at least one conjunct in

\begin{center}
    
$(p_{a} \vee p_{a}') \wedge (p_{A} \vee p_{A}') \wedge (p_{b} \vee p_{b}') \wedge (p_{B} \vee p_{B}') \wedge (p_{c} \vee p_{c}') \wedge (p_{C} \vee p_{C}')$ 

\end{center}

 is false. Thus, on this truth valuation, this big conjunction is false.\footnote{The same conclusion can obviously be reached from the other possible three contradictions. For $Abc$ = +1 entails that $p_{A}' \wedge p_{b}' \wedge p_{c}$ is true, or $p_{A}' \wedge p_{b} \wedge p_{c}'$ is true, or $p_{A} \wedge p_{b}' \wedge p_{c}'$ is true, or $p_{A} \wedge p_{b} \wedge p_{c}$ is true. But $Abc$ = --1 entails that $p_{A}' \wedge p_{b} \wedge p_{c}$ is true, or $p_{A} \wedge p_{b}' \wedge p_{c}$ is true, or $p_{A} \wedge p_{b} \wedge p_{c}'$ is true, or $p_{A}' \wedge p_{b}' \wedge p_{c}'$ is true. And $aBc$ = +1 entails that $p_{a} \wedge p_{B}' \wedge p_{c}'$ is true, or $p_{a} \wedge p_{B} \wedge p_{c}$ is true, or $p_{a}' \wedge p_{B} \wedge p_{c}'$ is true, or $p_{a}' \wedge p_{B}' \wedge p_{c}$ is true. But $aBc$ = --1 entails that $p_{a}' \wedge p_{B} \wedge p_{c}$  is true, or $p_{a} \wedge p_{B}' \wedge p_{c}$ is true, or $p_{a} \wedge p_{B} \wedge p_{c}'$ is true, or $p_{a}' \wedge p_{B}' \wedge p_{c}'$ is true. Lastly, $abC$ = +1 entails that $p_{a} \wedge p_{b} \wedge p_{C}$ is true, or $p_{a}' \wedge p_{b}' \wedge p_{C}$ is true, or $p_{a}' \wedge p_{b} \wedge p_{C}'$ is true, or $p_{a} \wedge p_{b}' \wedge p_{C}'$ is true. But $abC$ = --1 entails that $p_{a}' \wedge p_{b} \wedge p_{C}$  is true, or $p_{a} \wedge p_{b}' \wedge p_{C}$ is true, or $p_{a} \wedge p_{b} \wedge p_{C}'$ is true, or $p_{a}' \wedge p_{b}' \wedge p_{C}'$ is true. Following the same reasoning steps as above, one can again show that there is a truth valuation on which $(p_{a} \vee p_{a}') \wedge (p_{A} \vee p_{A}') \wedge (p_{b} \vee p_{b}') \wedge (p_{B} \vee p_{B}') \wedge (p_{c} \vee p_{c}') \wedge (p_{C} \vee p_{C}')$ is false.} This entails that there is no superfact -- no fact for all of our six observers. 
 
There are, however, facts for some, though not all, of the six observers. By eliminating the false conjuncts in the big conjunction we obtain the following: there is a truth valuation such that $(p_{a} \vee p_{a}') \wedge (p_{b} \vee p_{b}') \wedge (p_{c} \vee p_{c}')$ is true, so there is a fact for Alice, Bob and Carr; there is a truth valuation such that  $(p_{a} \vee p_{a}') \wedge (p_{B} \vee p_{B}') \wedge (p_{C} \vee p_{C}')$ is true, so there is a fact for Alice, Super Bob, and Super Carr; there is a truth valuation such that $(p_{A} \vee p_{A}') \wedge (p_{b} \vee p_{b}') \wedge (p_{C} \vee p_{C}')$ is true, so there is a fact for Super Alice, Bob, and Super Carr; and there is a truth valuation such that $(p_{A} \vee p_{A}') \wedge (p_{B} \vee p_{B}') \wedge (p_{c} \vee p_{c}')$ is true, so there is a fact for Super Alice, Super Bob, and Carr. Hence, there are different facts for different observers. This has been taken to imply that facts are not observer-independent: ``In quantum physics the objectivity of facts is not absolute, but only relative to the observation and the observer.'' (Brukner 2022, 628) Indeed, the extended Wigner's friend scenario is sometimes taken to imply that there are no objective, i.e., observer-independent facts. This leads to truth relativism: what is true for some observers is false for others.

Unsurprisingly, such semantic indeterminacy does not sit well together with an inferentialist metasemantics of QM. Indeed, Healey has recently rejected truth relativism as a consequence of the extended Wigner's friend scenario, by arguing essentially that one can eliminate superfacts without having to accept truth-relativism. In other words, the elimination of superfacts need not entail that facts are observer-dependent. The point is quite simple. Instead of eliminating superfacts by showing that the big conjunction above is false, one could also eliminate superfacts by arguing that the big conjunction lacks a truth value. This is based on Healey's condition that the class of truth valuations should be restricted such that if a statement is assessed as true by an observer, it is true when assessed by \textit{all others who are in a position to assess it}.

To be in a position to assess an experimental statement as true or false, one needs to be properly situated in a context of assessment. Thus, for example, if there is a truth valuation such that $(p_{a} \vee p_{a}') \wedge (p_{B} \vee p_{B}') \wedge (p_{C} \vee p_{C}')$ is true for Alice, Super Bob, and Super Carr, then there should be no truth valuation such that $(p_{a} \vee p_{a}') \wedge (p_{B} \vee p_{B}') \wedge (p_{C} \vee p_{C}')$ is false for any other observer, and in particular, there should be no truth valuation such that it is false for Super Alice. Rather, one should say that $(p_{a} \vee p_{a}') \wedge (p_{B} \vee p_{B}') \wedge (p_{C} \vee p_{C}')$ lacks a truth value for this superobserver, since she is not properly situated in the relevant context of assessment regarding $(p_{a} \vee p_{a}') \wedge (p_{B} \vee p_{B}') \wedge (p_{C} \vee p_{C}')$. 

One potential problem with this attempt to save the objectivity of facts -- their observer-independence -- is that it makes logical connectives non-truth-functional. This is because the truth value of compound statements is not merely a function of the truth values of their atomic components, but also a function of spatiotemporal location -- regions that are taken as proper contexts of assessment for some, but not for other statements. As I have suggested in the previous chapter, however, truth-functionality should not be taken as an indispensable feature of logicality. So I don't think there is a real problem here.

However, and this I think is a real problem for the quantum inferentialist, the proposed restriction on the class of admissible truth valuations is not justifiable on purely inferentialist grounds. According to inferentialism, the meaning of all non-logical terms in experimental statements is to be fully determined by material inferences. But restricting truth valuations in the way suggested by Healey essentially reintroduces semantic constraints that are independent of the rules of QM. This creates a certain tension with his own inferentialist metasemantics. 

Nevertheless, I do not think that this tension is unavoidable. Just as in the case of the first preliminary objection in the previous section, I think there is interesting philosophical work to be done here, in the attempt to develop a more robust inferentialist account of objectivity in QM. Part of this work should also be concerned with the categoricity problem to which I now turn.

\subsection{The categoricity problem for inferentialism}

The categoricity problem for Healey's quantum inferentialism brings to the fore the regularity assumption behind his claim that, for a physical system and experimental setup, there is a unique quantum state that can be assigned to that system relative to its spatiotemporal location. This is the same assumption that we have discussed, in chapter 2, in my reconstruction of Einstein's argument for the incompleteness of QM. The assumption raises a difficulty for Healey's view because, as far as I can see, there is no inferentialist justification available for assigning a regular, rather than a non-regular, state to a physical system. Without a justification of regularity, multiple quantum state assignments are possible to one and the same system relative to the same spatiotemporal location. 

Once this problem is spelled out, and its metasemantic consequences, as well as the epistemological consequences regarding the objectivity of probabilistic statements, are explained (in section 5.3.2), I will consider what a quantum inferentialist might say in response, insisting on a reformulation of Schrödinger's equation as a rule of inference. I will also surmise (in section 5.3.3) what a QBist might want to say in response to the categoricity problem, and why this response could motivate the development of an alternative non-representationalist metasemantics for QBism. But I want to start by revisiting (in section 5.3.1) a categoricity problem for logical inferentialism that Carnap identified already in 1943. This will help us better appreciate both the nature and the force of the categoricity problem for quantum inferentialism.

\subsubsection{Carnap's categoricity problem}

In \textit{Formalization of Logic}, Carnap embarked on an analysis of the semantics of classical logic (CL), both propositional and quantificational (Carnap 1943). More specifically, he was interested in determining whether, in addition to its standard truth-valuational semantics, which obeys the normal truth tables, the logical calculus admits of a non-standard semantics, which deviates from the normal truth tables. Likewise, he was interested in whether the quantifiers admit of a non-standard semantics, in addition to the standard one. In particular, for the propositional calculus, Carnap constructed two different non-standard valuations: in one of them, all sentences in the language (including their negations) are true; in the other, all sentences (except classical tautologies) are false, and if a sentence and its negation are both false, then their disjunction is still true, in order to validate the law of excluded middle. The existence of such valuations proved, according to Carnap, that the calculus is incomplete, i.e., non-categorical.\footnote{Similar problems for classical logic had been discussed by others, e.g., Bernstein 1932. See also section 3.3.2 above, for Weyl's argument, in his 1940 paper, that the quantum logical calculus introduced by Birkhoff and von Neumann in 1936 is incomplete, i.e., non-categorical.} 

More exactly, in the formalism introduced in section 4.2.1, let the CL-consequence relation $\vDash_\mathcal{CL}$ be defined via the standard class of valuations $C^*_\mathcal{CL}$ determined by the normal truth tables. The class $C^*_\mathcal{CL}$ is therefore defined by the following properties, for all $v \in C^*_\mathcal{CL}$: 

\begin{center}
    
$v(\lnot p)=1$ if and only if $v(p)=1$, \\$v(p \land q)=1$ if and only if $v(p)=v(q)=1$, \\ $v(p \vee q)=0$ if and only if $v(p)=v(q)=0$. 

\end{center}

Carnap's question was whether the same consequence relation could be characterized by a different class of valuations. Is the class $C^*_\mathcal{CL}$ unique (up to truth table isomorphism)? In his answer, he showed that there are indeed non-standard classes of valuations, which characterize the same consequence relation $\vDash_\mathcal{CL}$. One such class is given by $C^*_\mathcal{CL} \cup \left\{\tilde{v}_\mathcal{CL}\right\}$, where the non-normal valuation $\tilde{v}_\mathcal{CL}$ obeys

\begin{equation*}
    \tilde{v}_\mathcal{CL}(p)=1 \quad \text{if and only if `p' is a classical tautology (i.e. $\vDash_\mathcal{CL}p$)}.
\end{equation*}

Despite being non-isomorphic, both $C^*_\mathcal{CL}$ and $C^*_\mathcal{CL} \cup \left\{\tilde{v}_\mathcal{CL}\right\}$ define the same relation of logical consequence. Carnap saw this as a problem because it implies that CL cannot have what he considered a ``full formalization'', i.e., one that precisely determines the intended class of valuations (the ones that obey the normal truth tables). This is because it turns out that $C^*_\mathcal{CL} \cup \left\{\tilde{v}_\mathcal{CL}\right\}$ makes both CL negation and CL disjunction non-truth-functional.

Carnap attempted to solve the problem, arguing that one way to do this requires a multiple-conclusions formalization of CL. In other words, he noted that one can impose the normal truth table on disjunction if one formalizes valid arguments as relations between sets of premises and sets of conclusions, and if one modifies the definition of classes of valuations accordingly. The $\vee$-elimination and $\vee$-introduction rules, 

\begin{center}
    
$a \vDash_{CL} a \vee b$, $b \vDash_{CL} a \vee b$, and $a \vee b \vDash_{CL} \left\{a,b\right\}$, 

\end{center}

then imply that a disjunction is false if and only if each of its disjuncts is false.\footnote{For discussions of multiple-conclusions calculi, see Shoesmith and Smiley 1978 and Restall 2005.} 

Barring multiple-conclusions calculi, Carnap's problem presents a challenge to the logical inferentialist thesis that the rules of inference fix the meaning of logical terms.\footnote{As we have seen in section 5.1.1, this thesis was expressed by Carnap himself (Carnap 1934, xv). See also Quine 1992, 7. Logical inferentialism has been recently defended in, e.g., Peregrin 2014 and Warren 2020. For an overview of inferentialism, and some of its problems, see Murzi and Steinberger 2017.} As a consequence, an inferentialist justification for singling out the normal, truth-valuational semantics for CL connectives is not something that comes for free, but must be earned. Proposals for doing so include semantic solutions, which in general are not acceptable from an inferentialist point of view, but also syntactic ones, which adjust or qualify the application of the inferential rules of the logical calculus.\footnote{See Brîncuș 2024 for a brief overview of recent solutions to Carnap's categoricity problem.} 

As we will see in the next section, one of these solutions -- based on the requirement that the rules of inference, and in particular Schrödinger's equation expressed as such a rule, should be open-ended, in a sense to be explained below -- could help provide a solution to a similar categoricity problem that I shall now formulate against quantum inferentialism.

\subsubsection{A categoricity problem for quantum inferentialism}
 
Logical inferentialism holds, as we have just seen, that the semantic attributes of the connectives in a language are determined by their inferential properties, i.e., by the rules of inference that govern their employment. This is, of course, a local version of inferentialism, concerned with the meaning of a particular class of terms -- the logical ones. Global versions of inferentialism, as already emphasized above, when I reviewed Sellars' view, hold that the semantic attributes of non-logical, descriptive terms are also determined by their inferential properties, i.e., by the rules for their employment in material inferences. Turning to quantum inferentialism, I  noted that Healey adopted inferentialism as a metasemantics for all vocabularies, including all non-quantum expressions (for which, however, he allowed a representationalist semantics). The semantic attributes of all linguistic expressions whatsoever are taken to be determined in virtue of the inferences in which the expressions are used. More specifically, it is the rules of introduction and elimination of each expression that is supposed to explain its meaning. And we have seen that the template of material inferences in QM is as follows: 

\begin{quote}

\textit{Backing conditions are such and such.} 

\textit{Thus, the assignment of a unique quantum state is justified.} 

\textit{Thus, the Born probabilities are such and such.} 
    
\end{quote}

Such inferences are not logical inferences, as they can be expressed by subjunctive conditionals: ``If the backing conditions were such and such, then the assignment of a unique quantum state would be justified, and so the Born probabilities would be such and such.'' They are counterfactually robust and should be understood as explicit modal claims: ``Necessarily, if the backing conditions are such and such, then the assignment of a unique quantum state is justified, and so the Born probabilities are such and such.'' According to quantum inferentialism, an agent who does not endorse such modal claims does not understand what quantum mechanical expressions mean. For, recall, it is material inferences that determine the meaning of all non-logical terms, in the ``social game of giving and asking for reasons''. In QM, legitimate applications of the introduction and eliminations rules for quantum expressions articulate the meaning of these expressions. 

Importantly, on Healey's view, the introduction rules are supposed to justify, for any set of backing conditions, relative to a physical situation, the assignment of a \textit{unique} a quantum state -- a unique local model -- to a physical system. But what is it that really justifies this uniqueness claim? I think that what might be taken to justify this claim, in standard QM, is the Stone-von Neumann theorem, already discussed in chapter 2. This establishes that the Hilbert space representations of the Weyl algebra (generated by the CCRs) of a system with a finite number of degrees of freedom are unitarily equivalent to its Schrödinger representation. Unitary equivalence is just uniqueness up to an isometric isomorphism, and implies that for any set of backing conditions, relative to a physical situation, the quantum state assigned to the physical system would be unique up to this isomorphism. But the point that I want to make now is that the inferentialist cannot properly justify unitary equivalence. 

As noted already in section 2.2.2., a crucial assumption of the Stone-von Neumann theorem is that representations must be regular. However, on Healey's approach to QM, no inferentialist justification of regularity is readily available. Recall that for a Weyl algebra $\mathfrak{A}$, i.e., the C$^{*}$-algebra generated by the Weyl CCRs, a faithful representation $(\mathcal{H}, \pi)$ is a $^{*}$-isomorphism $\pi : \mathfrak{A} \rightarrow B(\mathcal{H})$, which is irreducible if no (nontrivial) subspace of $\mathcal{H}$ is invariant under the operators in $\pi(\mathfrak{A})$, and it is regular if the operators in $\pi(\mathfrak{A})$ are weakly continuous. Any two faithful, irreducible, and regular representations $(\mathcal{H}_{1}, \pi_{1})$ and $(\mathcal{H}_{2}, \pi_{2})$ of $\mathfrak{A}$ are unitarily equivalent iff there is a unitary intertwiner $\mathfrak{A} \ni U:\mathcal{H}_{1}\rightarrow\mathcal{H}_{2}$ which means that $UU^{*}=U^{*}U=1$ and $\pi_{1}(A)$ = $U\pi_{2}(A)U^{*}$ for all elements $A\in\mathfrak{A}$. But as we have seen, there is no such intertwiner for representations like the position representation (with position eigenstates, but no operator for momentum) or the momentum representation (with momentum eigenstates, but no operator for position). Such non-regular representations are unitarily inequivalent to the Schrödinger representation, and also to one another. Thus, the uniqueness (up to isomorphism) of any local model, assigned to a system relative to a physical situation, fails in QM. This poses a categoricity problem for a representationalist metasemantics.

But furthermore, non-regularity poses a categoricity problem for an inferentialist metasemantics as well. This is because, if there is no inferentialist way of eliminating non-regular states, then the modal claims that make explicit what is implicit in the material inferences of QM would turn out to be all false! If there is no inferentialist way of eliminating non-regular states, no set of backing conditions would be sufficient to justify assigning to a system, relative to a physical situation, a unique state. For it would not be sufficient to justify assigning a regular state in the Schrödinger representation, rather than, say, a position state (or a momentum state) in a non-regular representation. But in this case, taking once again Hilbert space representations as local models, it follows that the meaning of quantum expressions cannot be determined precisely by the rules of material inferences because these rules allow the introduction of mutually non-isomorphic models. 

One might think that this could be fixed by making modal claims more explicit than before: ``Necessarily, if the backing conditions are such and such, then the assignment of either a regular or a non-regular quantum state
is justified, and so the Born probabilities are such and such.'' But this reformulation only emphasizes the problem more clearly, for the latter modal claim actually expresses distinct material inferences -- divergent informational bridges -- starting from the same set of backing conditions, relative to the same system and physical situation, but leading to different Born probabilities. The Born probabilities will be different depending on whether the assigned state is regular or non-regular, as in the example I will presently give. So which one of these material inferences determines the meaning of quantum expressions? It seems reasonable to believe that, according to quantum inferentialism, different material inferences determine different meanings for the same expressions. But then we have a categoricity problem for quantum inferentialism, for the same QM rules of material inference allow different meanings for the same QM expressions. The problem shows that semantic attributes cannot be determined precisely in the way the inferentialist claims they are determined: \textit{inferentially}. 

If this categoricity problem can admit of no inferentialist solution, then quantum expressions turn out to be semantically indeterminate. But furthermore, this same problem also has undesirable epistemological consequences for Healey's view of QM, regarding the objectivity of facts described by probabilistic statements and, thus, for his prescriptivist interpretation of Born probabilities. Before I suggest what a quantum inferentialist might do in order to address the categoricity problem, let me spell out the metasemantic and epistemological consequences of the categoricity problem by returning to Alice in Wigner's friend scenario.\footnote{Just in case what I have said so far may not be worrisome enough, recall the fact mentioned in chapter 2, that the Stone-von Neumman theorem fails in QFT, which allows an infinity of unitarily inequivalent Hilbert space representations. Also consider related worries expressed by Abhay Ashtekar: ``How can there be an \textit{unique} diffeomorphism state? Surely, quantum gravity admits an infinite number of diffeomorphism invariant states!'' (Ashtekar 2009, 1929) If it is reasonable to believe that the non-uniqueness of quantum states allows a plurality of material inferences starting from the same set of backing conditions and relative to the same physical situation, then inferentialism has a categoricity problem in these quantum theories as well.}

On an inferentialist approach, Alice would follow the rules for the introduction of quantum states, and given the experimental setup in her lab, she would assign a quantum state to her target system. Then, following a legitimate application of elimination rules, this state would yield probabilities for her experimental statements $p_a$ and $p'_a$ via the following material inference: 

\begin{quote}

\textit{Backing conditions are such and such.} 

\textit{Thus, the assignment of a unique quantum state is justified.}

\textit{Thus, the Born probabilities are ``$Pr(p_a)$ = 0.5'' and ``$Pr(p'_a$) = 0.5''.} 
    
\end{quote}

These two probabilistic statements are, on Healey's view, prescriptions for Alice's degrees of belief. She is therefore justified to believe that $a$ = +1 and $a$ = --1 are equally likely to obtain. But suppose that Alice realizes that, given the assumptions of the Stone-von Neumann theorem, there is a variety of quantum states that can be assigned to her system, including non-regular ones. Accordingly, she is considering extending the class of her models by introducing nonregular $\ket{\psi}^{nr}$ besides the regular ones $\ket{\psi}^r$. As she well knows, there is no unitary transformation between such states. This allows her to formulate distinct material inferences. 

One material inference might look like this:

\begin{quote}

\textit{Backing conditions are such and such.}

\textit{Thus, the assignment of $\ket{\psi}^{r}$ is justified.}

\textit{Thus, the Born probabilities are ``$Pr(p_a)$ = 0.5'' and ``$Pr(p'_a$) = 0.5''.} 

\end{quote}

But the other material inference might look as follows:

\begin{quote}

\textit{Backing conditions are such and such.}

\textit{Thus, the assignment of $\ket{\psi}^{nr}$ is justified.}

\textit{Thus, the Born probabilities are ``$Pr(p_a)$ = 0.49'' and ``$Pr(p'_a$) = 0.51''.} 
    
\end{quote}

If it is reasonable to believe that these distinct material inferences determine different meanings for the expressions used in them, then this is definitely a problem for an inferentialist metasemantics of QM.

There are two easy ways out, which we need to acknowledge and reject before we consider more serious ones. First, the quantum inferentialist might be inclined to ignore the categoricity problem and insist on the uniqueness of a quantum state assignment, by assuming regularity of states from the get-go. That is, she could take the Stone-von Neumann theorem as a fundamental result, but discount its problematic assumptions, since it seems safe to believe that in practice they are insignificant in any case. Secondly, the quantum inferentialist might want to bite the bullet of semantic indeterminacy, on the ground that this type of indeterminacy is unlikely to have any consequences for the practice of QM. There are even logical inferentialists who are ready to do so in the face of Carnap's categoricity problem:

\begin{quote}
    
After all, \textit{inferential rules} are the only thing that matter to an inferentialist. If criticized for \textit{ignoring} something vital –- namely that inferentialism is not able to render the \textit{truly} logical constants –- the inferentialist can reply that such criticism is on a par with criticizing an atheist for ignoring the secrets of the Holy Trinity. ... True, inferentialism and classical logic do not form an ideal couple, but this does not undermine inferentialism as such. (Peregrin 2014, 179)

\end{quote}

Paraphrasing, one could similarly say that only the rules of material inference matter to a quantum inferentialist. If criticized for being unable to single out the genuinely quantum states, $\ket{\psi}^{r}$ or $\ket{\psi}^{nr}$, the inferentialist can reply that such criticism is on a par with criticizing an atheist for ignoring the secrets of the Holy Trinity, and then admit that inferentialism and QM do not form an ideal couple. 

Needless to say, however, the first easy way out is begging the very question at stake here, while the second turns out to recommend a very big bite: indeed, I think that Peregrin's suggestion is tantamount to giving up inferentialism as metasemantics, i.e., as an explanation of semantic facts. I will come back to this kind of response to the categoricity problem, in the next section, where I look ahead at a non-representationalist, but also non-inferentialist metasemantics for QBism.

Let us now turn to the epistemological consequences of the categoricity problem for Healey's view. Since on his account Born probabilities give prescriptions for Alice's degrees of belief, if she can materially infer different probabilistic statements, and they are all justified on the basis of her application of QM, then she has an epistemological problem: what is Alice now really justified to believe? What set of credences is she supposed to adopt? Furthermore, recall that on Healey's account, non-quantum magnitude claims or experimental statements, like Alice's $p_{a}$ and $p_{a}'$, have a primary representational role. This is, recall, Healey's solution to the no-exit problem, discussed above. What it means is that, even if they do not do any metasemantic work, we can specify their truth conditions: 

\begin{center}

``$a$ = +1'' is true iff $a$ = +1, \\ ``$a$ = --1'' is true iff $a$ = --1, 

\end{center}

 in the same way in which ``The atom has decayed, the flask has shattered, and Findus is alive'' is true if and only if the atom has decayed, the flask has shattered, and Findus is alive. Similarly, and even if their primary role is to provide prescriptions, probabilistic claims about non-quantum magnitude claims are also representational, though only ``weakly representational'' (Healey 2017, 135), so we can specify their truth conditions as well:

\begin{center}

``$Pr(p_a) = 0.5$'' is true iff $Pr(p_a) = 0.5$, \\ ``$Pr(p_a) = 0.49$'' is true iff $Pr(p_a) = 0.49$. 

\end{center}

But despite Healey's emphasis on the objectivity of such probabilistic claims, which of these truth conditions obtain appears to be a fact that depends on Alice's choice between $\ket{\psi}^{r}$ and $\ket{\psi}^{nr}$. More generally, if an observer can extend her class of models as I suggested, then her choice of a model for a system yields different probabilistic claims about the same experimental statement. This implies, against Healey's view, that a significant category of facts is observer-dependent after all. 

What else might a quantum inferentialist say in response to the categoricity problem? As far as I can see, there is a variety of options, some better than others. So let me take them in turn, considering each with respect to whether they are compatible with inferentialism. 

First, the inferentialist can insist \textit{on physical grounds}, that for any given experimental setup, relative to a physical situation, there is always a uniquely correct state -- a regular state -- which can be assigned to a target system. Indeed, one might wonder, are non-regular states even admissible in QM? As mentioned above in chapter 2, they are arguably physically significant and may be taken to give a precise form to Bohr's complementarity. However, as we have noted there, their elimination might be justified on the ground that the unitary time evolution described by the Schrödinger equation does not sit well on non-regular representations, which are dynamically mutually inaccessible (Feintzeig \textit{et al.} 2019, 2022). 

Nevertheless, Healey cannot justify regularity in this way, by just endorsing the elimination of non-regular states for dynamical reasons. This is not only because his dissolution of quantum ontology indeed blocks the inferentialist's access to this justification. For even if one allowed QM to have its own beables, this justification for the elimination of non-regular states would not be acceptable to the inferentialist: for the elimination of non-regular states should be justified \textit{inferentially}, that is decided only on the basis of QM's material rules of inference, not by appealing to features of quantum models, but exclusively to their introduction and elimination rules. Furthermore, if non-regular states turn out to be useful in QM, then \textit{any} reason for eliminating them as spurious concepts would conflict with Brandom's inferential non-conservativity argument, endorsed by Healey. 

Secondly, the inferentialist might want to restrict, \textit{on pragmatist grounds}, the probability valuations of non-quantum magnitude claims such that an agent cannot assign different probability values to the same claim. Here, as should be obvious, pragmatism would again conflict with inferentialism. As we have seen above, in the context of a relativistic extension of Wigner's friend scenario, the inferentialist illicitly restricts the truth valuations of experimental statements to solve the objectivity problem raised by the elimination of superfacts, and thereby avoid truth-relativism. But such restrictions do not help the inferentialist against truth-relativism, since they are semantic constraints, rather than constraints that follow from the inferential rules. For the same reason, pragmatist restrictions on probability valuations should not be appealed to in the present context, either. 

My own suggestion of a potential solution to the categoricity problem, on behalf of the quantum inferentialist, takes seriously the idea that the elimination of non-regular states should be justified \textit{inferentially}, that is decided only on the basis of QM's material rules of inference, and not by appealing to features of quantum models and dynamics or to restrictions on probability valuations of magnitude claims. The suggestion is simply that the inferentialist should explicitly reformulate the Schrödinger equation as a rule of material inference, that is more precisely as a rule for the introduction of quantum states. Moreover, the inferentialist should take the Schrödinger equation as a permanent or open-ended rule, i.e., she should justify its preservation for all mathematically possible extensions of the language of standard QM. This, by itself, should block any assignment of non-regular states to physical systems. It seems to me that this is the most promising way of solving the problem of categoricity for quantum inferentialism and, thereby, averting the semantic indeterminacy caused by this problem.\footnote{This inferentialist strategy, which takes rules to be open-ended in the sense explained in the text, has been applied already in the metasemantics of classical logic and mathematics, and it is arguably a successful strategy. More exactly, inferentialists about CL have taken introduction and elimination rules for the logical connectives to be open-ended, i.e., valid for all possible extensions of the language of CL, and inferentialists about PA have taken first-order mathematical induction to be an open-ended rule for the introduction of arithmetical concepts, i.e., a rule valid for all possible extensions of the language of PA. For more on open-endedness in philosophy of logic and mathematics, see e.g. Warren 2020, Murzi and Topey 2021. An alternative strategy, for the case of QM, is to change the standard formalism by replacing the Weyl algebra with a different mathematical structure, as proposed in Feintzeig and Weatherall 2019, Feintzeig 2022. But such alternatives will have to be discussed elsewhere.} 

But this is not enough. The quantum inferentialist should adopt a similar strategy not only with respect to introduction rules, but with respect to elimination rules as well. The only quantum rule that Healey indicates must be understood inferentially, as an elimination rule for quantum states, is the Born rule. But I think that the inferentialist should attempt to reformulate decoherence explicitly as an elimination rule as well. This is because it is not compatible with inferentialism to apply decoherence, as Healey does, as a semantic constraint on the application of the Born rule, i.e., as a constraint that helps eliminate the magnitude claims that, in a certain physical situation, are meaningless. Healey argues that the Born rule should only be applied to stably decohered states, for instance, to position states that decohere at detection, i.e, after a suitable interaction with another system. This allows the selection of magnitude claims, e.g., about position, that have a determinate truth value, and are thus meaningful as a guide for an agent’s expectations. However, as already mentioned, this should be unacceptable to the inferentialist: just like the Schrödinger equation, decoherence should be formulated as an inferential rule, rather than employed as a semantic constraint. 

Healey, himself, suggests what might be considered as a normative approach to the categoricity problem: 

\begin{quote}
    
There are at least three respects in which ... probabilistic statements ... are objective: There is widely shared agreement on them within the scientific community. A norm is operative within that community requiring resolution of any residual disagreements. This norm is not arbitrary but derives directly from the scientific aims of prediction, control, and explanation of natural phenomena. ... Flouting this norm would leave one unable to account for the patterns of statistical correlation among events described by true magnitude claims to which quantum theory is ultimately responsible. (Healey 2017, 135)

\end{quote}

The demand that I think the quantum inferentialist should satisfy is that this norm, whatever it is, be compatible with an inferentialist metasemantics, i.e., with the view that the semantic attributes of all linguistic expressions in QM are determined by quantum-mechanical rules of material inference. This is the ultimate criterion for the admissibility of any solution to the categoricity problem for quantum inferentialism. 

Finally, the inferentialist might, of course, also want to insist that the very question about the categoricity of QM is well posed only for a rational reconstruction of QM, i.e., a reformulation of a theory in a formal (ideally, first-order) language, rather than for its standard Hilbert space formalism. In other words, the inferentialist could demand that the metasemantic analysis should start by first verifying that QM can be rationally reconstructed as a formal system in which QM rules are expressed formally. Only after a rational reconstruction has been given could one verify whether its semantics is unique up to the relevant isomorphism. After all, Carnap's metasemantic analysis of CL also started with formal axiomatization, before its outcome would be articulated as a categoricity problem for logical inferentialism. Since Healey rejects, as we will see in section 6.2, the Carnapian task of a rational reconstruction of any scientific theory, it seems plausible to believe that this should be enough to reject the categoricity problem as a problem for standard QM. 

But I think that this response -- insisting on rational reconstruction -- would not be justified. The notion of categoricity that I have articulated in section 2.2.2, perfectly suitable for standard QM, is enough to express the problem. Moreover, as I have argued there, if standard QM has a categoricity problem, then a rationally reconstructed QM would also have a categoricity problem. The formalization of a scientific theory cannot really give the impression of what Tarski called ``a closed and organic unity'' in case the theory itself lacks that kind of unity. But if it does, then this tells us something about formalization rather than about the scientific theory. However, I do not think that the project of a rational reconstruction of QM should be blocked. Quite the contrary, I will argue further below, in section 6.2, when I will revisit the problem of the viability of a rational reconstruction of QM, that Healey's own challenge to the Carnapian task can be met on inferentialist grounds.

\subsubsection{The meaning of QBism}

I noted above that a quantum inferentialist facing the categoricity problem might be inclined to consider an easy way out, simply ignoring it and biting the bullet of semantic indeterminacy. One reason for doing so could be that any semantic indeterminacy of the standard formalism is considered inconsequential for the practice of QM. But I have also maintained that this response would be detrimental to quantum inferentialism \textit{as a metasemantics}, in the same way in which it is detrimental to logical inferentialism \textit{as a metasemantics} to contend that since an inferentialist is exclusively concerned with inferential rules, any semantic indeterminacy of logical connectives can be ignored. In both cases, the philosophical endeavor to explain semantic facts in a satisfactory manner would be cut short.

Note, however, that although detrimental to quantum inferentialism, this response might not be damaging to other neo-Bohrian approaches to QM, like QBism. Even if a metasemantics could be developed in this case, a QBist might nevertheless insist that scientific practice does not actually require any explanation of semantic facts. Of course, the categoricity problem would still loom behind like the shadow of an unfinished philosophical business, but a QBist might feel free to ignore its consequences, i.e., free to discount the apparently insignificant detail that the meaning of QM cannot be precisely determined by its rules. And unlike quantum inferentialism, QBism might be thought immune to the objection that it would remain ``unable to nail down a claim's content sufficiently to explain how this may be unambiguously communicated in public discourse''. Unambiguous public communication, a QBist might be inclined to say, is practically unachievable anyway.

Be that as it may, the question I will take up in this section is whether it is possible to develop a metasemantics for QBism at all; and if so, how. I will first recall what is perhaps the most characteristic feature of QBism as an interpretation of standard QM, a feature that has attracted a lot of philosophical criticism (including, more recently, a certain phenomenological buzz around it). Then I will note that although, just like quantum inferentialism, QBism can be seen to distinguish between the semantics and ontology of QM and to reject any representational capacities of quantum models, it regards the rules of QM, and especially the Born rule, in a general normative sense, rather than specifically as inferential rules. Moreover, Chris Fuchs has intriguingly claimed that Healey's inferentialism ``is so very far from QBism in what it takes to be a norm ... that it essentially shares nothing in common with QBism except the word ‘normative'.'' (Fuchs 2023, 132, f16) 

I will focus my exploratory analysis on this normativity issue, with the avowedly subversive intention to redirect some philosophical attention from the controversial epistemological aspects of QBism to its unsettled metasemantic commitments. For I think that although it is clear that it favors a non-representationalist semantics for QM, and even if one might think it set to ignore the categoricity problem and the ensuing semantic indeterminacy of quantum expressions, QBism should not avoid providing an explanation of the semantic facts of QM. Not everything in the philosophy of science can be dictated by the achievements or limitations of scientific practice.

Arguably the most contentious tenet of QBism is that quantum states are personal judgments, which implies that QM is a ``single user theory'', in the sense that any agent assigns her personal quantum state to a physical system.\footnote{For presentations of QBism, see e.g. Fuchs, Mermin, and Schack 2014, Fuchs and Schack 2015, Fuchs 2018, Fuchs and Stacey 2019. My discussion in this section refers primarily to the most recent presentation, in Fuchs 2023.} This dismisses quite bluntly the uniqueness of a quantum state assignment that we have discussed above in the case of Healey's inferentialism. It also immediately suggests that an inferentialist metasemantics would not be acceptable to a QBist: for if QM is a single user theory, how could its meaning ever be determined in a multiple user inferential architecture, in a ``social game of giving and asking for reasons''? This QBist tenet has been typically perceived as an attack on scientific objectivity, and has been the focus of intense philosophical criticism.\footnote{For a recent discussion, see e.g. Glick 2021. But recall that, as I have argued in the previous section, inferentialism might also have a rather difficult time saving objectivity in QM.} Nevertheless, QBism regards the rules of QM, and especially the Born rule – ``the primary empirical statement of the theory'', along with probability theory as a whole, in an \textit{objectively normative} sense, even though not specifically as inferential rules. This strong emphasis on the objective normativity of the quantum formalism is another tenet that is currently taken to characterize QBism. But as it stands, it is rather unclear what this might mean. More explanation should be in order.

To be sure, Fuchs emphasized that what is normative in QM, as interpreted by QBism, is the relation expressed by the Born rule, rather than its individual relata: ``The normative thrust of quantum theory is that it is a kind of glue for probability assignments over and above the requirements of raw probability theory. From the QBist perspective, it is this glue which indicates the physical content of quantum theory.'' (Fuchs 2023, 90) Perhaps more exactly, what this may be taken to say is that the meaning of QM is purportedly determined in virtue of the normative properties of the relation between probability assignments specified by the Born rule. 

Note that the emphasis on normative properties clearly distinguishes QBism from interpretations that typically attribute descriptive properties to quantum expressions: ``Particularly in QBism, the quantum formalism plays a \textit{normative} role for its users; it does not play a descriptive role concerning exactly how the world is. Its focus is on how a user \textit{should} gamble.'' (\textit{op. cit}, 82) But the normative properties that the QBist attributes to QM also distinguish QBism from inferentialism, since the formalism is said to provide gambling rules, rather than inferential rules. 

One immediate problem that QBism had to deal with is that the standard formalism does not appear to have the normative character attributed by QBism. This is why it has to be reformulated in a way that may bring this character to light: ``This [i.e., the normative character] is most easily seen by rewriting the quantum formalism in a form that is purely probabilistic in nature, without operators on complex vector spaces, etc.'' (\textit{op. cit.}, 116) Once this has been properly done, one would then clearly see what the QBist already saw in the standard formalism: rules for an agent's gambling. One indispensable constraint on such rules is, of course, that their use must be consistent. Consistency is here understood in the sense of Bayesian coherence -- a condition that ensures that a Dutch book, i.e., a bet that an agent is guaranteed to lose, is avoided. 

Now, the QBist reformulation of QM raises several important questions: What exactly are the relevant normative properties of quantum rules, as understood by QBism? What kind of language is the purely probabilistic reformulation of these rules? How can this achieve what it is designed to achieve – bringing to light the relevant normative properties? Furthermore, could an inferentialist metasemantics be extended, at least in principle, to account for the meaning of expressions in this language? Or should a metasemantics for QBism be developed independently, based on the purely probabilistic structure of the QBist reformulation of QM? To answer these questions, an extended analysis of the language of this reformulation must be given. I do not attempt to do this here, but I want at least to motivate such an analysis, and indicate a place where this could start. 

One place where the analysis could start is the crucial observation that, when reformulated as a relation between probabilities, without any reference to a quantum state, the Born rule can be considered a decision-theoretic device. This is why QBism is properly understood as revealing the decision-theoretic structure of QM. Indeed, QBism sees QM as an extension of decision theory that helps an agent make better decisions by providing guidance as to how she ought to adjust her personal judgments. More specifically, as a decision-theoretic device, the Born rule provides guidance as to how an agent should gamble on the expected consequences of her actions on a physical system, always under a Dutch-book coherent application. This supports Fuchs' claim that the normative properties attributed to QM by QBism are different from those attributed to it by Healey's inferentialism, and suggests that the QBist understands the meaning of QM as determined in virtue of the decision-theoretic properties, rather than the inferential properties, of quantum expressions. 

Although this aspect has been so far overlooked, as far as I can tell, it seems to me significant that a QBist reformulation of the standard formalism incorporates not only the language of decision theory, but includes deontic modals, i.e.,
terms like “ought” and “should”. This is the case, for example, in expressions like “I ought to
believe that the atom hasn't decayed, the flask of poison isn't shattered, and the cat is alive with
a probability of 50\%” and “I should expect the particle's spin to be up with a probability of 60\%”,
etc. This strongly suggests, I think, that the project of developing a metasemantics for QBism, i.e., an account of its reformulation of the quantum rules that clarifies the meaning of its linguistic expressions and explains why the latter have that meaning,  could be based on, or at least could begin with, an explanation of the meaning of its deontic modals. 

Despite the differences noted above, would it be possible to take up inferentialism as a metasemantics for QBism?  Can inferentialism account for the meaning of linguistic expressions that
include deontic modals? The inferentialist understands deontic modals as metaconceptual devices that make explicit one's commitments to practical inferences, that is
“language-exit” transitions from linguistic expressions to expressions of intentions or actions.\footnote{Cf. Brandom 2000. On this view, deontic modals are on a par with conceptual modals, which
make explicit one's commitments to theoretical inferences, that is transitions from linguistic
expressions to other linguistic expressions (as we have seen above, in the discussion of Healey's inferentialism).} However, this view has been criticized as unable to explain the meaning of a large class of deontic modals.\footnote{For a criticism of Brandom's conception of deontic modals, see e.g., Chrisman 2016a, 2016b.} For example, in an expression that includes deontic necessity, like “I ought to win the lottery”, it is not clear that the meaning of “ought” can be explained by its role as a metaconceptual device that makes explicit my commitment to a practical inference: for despite my genuine intention to win the lottery, I cannot infer to an expression of action, i.e., I cannot infer that since I ought to win the lottery, I will win the lottery. All I can infer is that it would be highly preferable -- ideal -- that I win the lottery. 

Would such limitations of inferentialism further block the attempt to inferentially explain the meaning of QBist deontic modals? Might this very attempt encounter additional obstacles pertaining to
the QBist reformulation of the standard formalism? Can the meaning of “should” in an expression like “I
should expect the particle's spin to be up with a probability of 60\%'' be explained by its role as a
metaconceptual device that makes explicit my commitment to a “language-exit” inference? Can
I infer that since I should expect the particle's spin being up with a probability of 60\%, I will
gamble on the particle's spin to be up? Or I am only justified to infer that it would be highly preferable -- ideal -- that I gamble on the particle's spin being up? If an inferentialist metasemantics for QBism turns out to be unworkable, what might be the alternative? Assuming that decision-theoretic properties cannot be reduced to inferential ones, could one explain non-inferentially the meaning of QBist deontic modals on the basis of the explicitly decision-theoretic structure of the QBist reformulation of QM? 

One strategy for developing a non-inferentialist metasemantics for QBism could be to start from the observation that deontic modals are information-sensitive, i.e., relative to states of information and, in the case of QBism, relative to a quantum state understood as a personal probability distribution. For example, suppose that relative to an agent's initial personal judgments and her personal probability distribution, the expectation of, say, particle
spin up may be greater than the expectation of particle spin down. This is decision-theoreticly
equivalent to saying that she ought to gamble on particle spin being up rather than on particle spin being down, which indicates that the agent's deontic modals are sensitive to the decision-theoretic structure of her expectations. The meaning of deontic modals can then perhaps be explained in virtue of their role in making explicit an agent's commitment to a highly preferable -- ideal -- betting behavior, in the transition from the particular structure of her personal probability distribution to her exact expression of this commitment.

Along the same line, note that the Born rule, as reformulated in QBism, is not merely a decision-theoretic device; rather, it may be considered a choice function articulated by the quantum-theoretic extension of decision theory, that is a function from decision problems into sets of preferable actions that are admissible according to QM. In other words, the Born rule behaves like a \textit{deontic choice function}.\footnote{See Kolodny and MacFarlane 2010, for a general analysis of deontic choice functions} A deontic choice function maps an initial information state to another information state that is as deontically ideal as possible, given the particular structure of the agent's probability distribution. As such, the Born rule bridges the gap between an agent's personal quantum state and her deontically
ideal betting behavior, i.e., how the agent should gamble in the context of all available betting
options. 

This strategy could further clarify the exact role of the Born rule in determining the meaning of quantum
expressions and, if pursued, it might help develop a full-fledged, non-representationalist, but also non-inferentialist, metasemantics for QM, as interpreted by QBism. Once again, according to this view, it would be the legitimate application of the Born rule as a deontic choice function, rather than as an inferential rule, that articulates the meaning of quantum expressions. This could be an important \textit{philosophical} development of QBism, which might in the end also help us better understand not only the controversial epistemological aspects of QBism itself, but perhaps pave the way for a better philosophical appreciation of neo-Bohrian approaches to QM, more generally.

\newpage 

\section{The logical perfection of physical theories}

Whether we adopt a representationalist or an inferentialist metasemantics
for standard quantum mechanics (QM), categoricity turns out to be a problem.
One might be inclined to think that the ensuing semantic indeterminacy of QM is
inevitable. This chapter aims to oppose this inclination: first, categoricity, as a property apt to characterize physical theories, should be pursued, at least as a metasemantic ideal; and secondly, despite almost unanimous rejection, the rational reconstruction of QM, as a Carnapian language, should not really be thought impossible as a basis for further metasemantic investigations.\footnote{This chapter is partly based on Toader 2025c and Horvat and Toader 2025.}

\subsection{Categoricity as a metasemantic ideal}

In section 1.1 of the book, I mentioned Hilbert's distinction between two development strategies based on the axiomatic method: a progressive strategy, leading from the axioms of a theory to what can be derived from them by following its own rules, and a regressive strategy, leading to deeper axioms, uncovered by a critical examination of that theory. In the philosophy of physics, this distinction is usually attributed to Einstein, who in his article for \textit{The London Times} (Einstein 1919), similarly distinguished between constructive and principle theories, or rather, between progressive and regressive strategies for developing theories. Of course, the British audience of the \textit{Times} would have been already familiar with that very distinction, since Russell had advocated and extensively discussed it in writing, as we will presently see.\footnote{For Russell's influence on Hilbert and his school, see Mancosu 2010, chapter 4. As for Einstein, he declared: ``I owe innumerable happy hours to the reading of Russell's works, something which I cannot say of any other contemporary scientific writer, with the exception of Thorstein Veblen.'' (Einstein 1944, 18sq) For an account of Einstein's distinction, see Giovanelli 2021.} But the British scientists would have been familiar with the two strategies at least since Peacock's 1833 \textit{Report} (see above, section 3.1.1), where he noted

\begin{quote}
    
a marked distinction between those sciences which, like algebra and geometry, are founded upon assumed principles and definitions, and the physical sciences: in one case we consider those principles and definitions as \textit{ultimate} facts, from which our investigations proceed in one direction only ... whilst in the physical sciences there are no such \textit{ultimate} facts which can be considered as the natural or the assumable limits of our investigations. It is true, indeed, that ... we assume certain facts or principles ... making them the foundation of a system of propositions ... such assumed first principles, however vast may be the superstructure which is raised upon them, form only one or more links in the great chain of propositions, the termination and foundation of which must be for ever veiled in the mystery of the first cause. ... The first principles, therefore, which form the foundation of our mathematical reasonings in the physical sciences ... can never cease to be more or less the subject of examination and inquiry at any point of our researches. (Peacock 1833, 186) 
\end{quote}

 On Peacock's view, the continuous examination of the principles of a physical theory is demanded by their very nature, since they cannot be considered as ultimate facts. But the result of examination may be expected to take the examiner as close as possible to principles that can be assumed as such facts. Reconstructions of theories attempted in this vein are, thus, metaphysically and epistemologically motivated. What I want to suggest in this chapter is that starting with Russell's examination of the principles of mathematics, reconstructions came to be motivated metasemantically as well (or even primarily so). Before we turn to Russell, let us look again at Einstein's famous distinction, since it is his notion of the logical perfection of a theory that I take to be at the center of this kind of motivation. 
 
Here is the most widely quoted passage from the \textit{Times} article mentioned above: 

\begin{quote}
    
We can distinguish various kinds of theories in physics. Most of them are constructive. They attempt to build up a picture of the more complex phenomena out of the materials of a relatively simple formal scheme from which they start out. ... When we say that we have succeeded in understanding a group of natural processes, we invariably mean that a constructive theory has been found which covers the processes in question. Along with this most important class of theories there exists a second, which I will call `principle-theories.' These employ the analytic, not the synthetic, method. ... The advantages of ... the principle theory are logical perfection and security of the foundations. (Einstein 1919, 228)

\end{quote}

On Einstein's view, both of these strategies (or methods) start from (or apply to) the same formal scheme -- the same set of axioms -- but they are not exclusive or mutually inconsistent, only driven by different ideals, which lead them accordingly in different directions. Einstein always emphasized logical perfection as an ideal of physics: ``our notions of physical reality can never be final. We must always be ready to change these notions -- that is to say, the axiomatic basis of physics -- in order to do justice to perceived facts in the most perfect way logically.'' (Einstein 1931, 266) Such passages have always raised a nagging question: what \textit{is} logical perfection? Unfortunately, Einstein never seems to have paused to define this notion more explicitly (Dongen 2010, 59).

My suggestion is that we can get an idea of what Einstein might have had in mind if we consider Russell's own reconstruction of arithmetic from logical principles. This reconstruction, we will presently see, was primarily driven by the need to provide a unique semantics for arithmetic, i.e., to give arithmetical expressions a precisely determined meaning, rather than allowing an infinity of models of the arithmetical axioms. This view transpires most explicitly in Russell's \textit{Introduction to Mathematical Philosophy}, where he wrote:

\begin{quote}

Mathematics is a study which, when we start from its most familiar portions, may be pursued in either of two opposite directions. The more familiar direction is constructive, towards gradually increasing complexity: from integers to fractions, real numbers, complex numbers; from addition and multiplication to differentiation and integration, and on to higher mathematics. The other direction, which is less familiar, proceeds, by analysing, to greater and
greater abstractness and logical simplicity; instead of asking what can be defined and deduced from what is assumed to begin with, we ask instead what more general ideas and principles can be found, in terms of which what was our starting-point can be defined or deduced. It is the fact of pursuing this opposite direction that characterises mathematical philosophy as opposed to ordinary mathematics. (Russell 1919, 1)\footnote{Of course, Russell had been saying similar things for at least a decade: ``In every science, we start with a body of propositions of which we feel fairly sure. These are our empirical premises, commonly called the facts, which are generally got by observation. We may then ask either: What follows from these facts? or, what do these facts follow from?'' (Russell 1907, 573) Note also that, like Russell in 1919, Einstein sometimes referred to logical simplicity as, presumably, a form of logical perfection (e.g., in Einstein 1949).}
\end{quote}

Russell compared these two pursuits with the use of different tools, ``one to take us forward to the higher mathematics, the other to take us backward to the logical foundations'' (\textit{op. cit.}, 2). After discussing Peano's achievements with respect to the latter pursuit, Russell famously noted ``some of the reasons why Peano's treatment is less final than it appears to be'' (\textit{op. cit.}, 7). First and foremost, ``Peano's three primitive ideas --  namely, `0,' `number,' and `successor' -- are capable of an infinite number of different interpretations, all of which will satisfy the five primitive propositions.'' (\textit{loc. cit.}) In more detail:

\begin{quote}
    
Every progression verifies Peano's five axioms. It can be proved, conversely, that every series which verifies Peano's five axioms is a progression. Hence these five axioms may be used to define the class of progressions: `progressions' are `those series which verify these five axioms'. ...  The progression need not be composed of numbers: it may be composed of points in space, or moments of time, or any other terms of which there is an infinite supply. Each different progression will give rise to a different interpretation of all the propositions of traditional
pure mathematics; all these possible interpretations will be equally true. In Peano’s system there is nothing to enable us to distinguish between these different interpretations
of his primitive ideas.'' (\textit{op. cit.}, 8-9)\footnote{In his \textit{Principles of Mathematics}, Russell had already criticized Dedekind on similar grounds (Russell 1903, §242), which as is well known, attracted Cassirer's criticism (Cassirer 1953, 39sq).}

\end{quote}

 Following Frege, Russell believed that one could do better than this, that one should be able to derive the Peano axioms from logical principles, which then Russell hoped would make it possible to single out the intended interpretation of arithmetic: ``our numbers should have a definite meaning, not merely ... certain formal properties.'' (\textit{op. cit.}, 10) Mathematical philosophy, as he conceived of it, is a project that aims at determining the ``definite'', i.e., intended, meaning of arithmetical terms, as opposed to the ``indefinite'' meaning they would have if interpretations could be determined up to isomorphism only. Summarizing Frege's and his own achievements, Russell concluded: 

\begin{quote}
    
We have thus reduced Peano’s three primitive ideas to ideas of logic: we have given definitions of them which make them definite, no longer capable of an infinity of different meanings, as they were when they were only determinate to the extent of obeying Peano’s five axioms. (\textit{op. cit.}, 24)\footnote{Interestingly, Russell also suggested a more general project in metasemantics: ``What are the possible meanings of a law expressed in terms of which we do not know the substantive meaning, but only the grammar and syntax?'' (\textit{op. cit.}, 55)}

\end{quote}

Russell's hope to single out the intended interpretation of the arithmetical axioms, by pursuing their reconstruction from logical principles, turned out to be illusory: one can \textit{only} single out the intended interpretation of the arithmetical axioms up to isomorphism \textit{at best} (for more on this, see section 2.2.1 above). But the significant point here is that Russell's primary motivation for pursuing the reconstruction of arithmetic was clearly metasemantic. 

Taking this seriously, one may indeed consider categoricity as a metatheoretical ideal capable of reinforcing a metasemantic motivation for the reconstructionist strategy that Einstein envisaged. Categoricity can, and I think should, be pursued in physics, not only in logic and mathematics, as a form of logical perfection. More recently, Boris Zilber and his collaborators have attempted to rekindle philosophical interest in the logical perfection of scientific theories, based on the notion of categoricity:

\begin{quote}
    
Although the expression [``logical perfection''] is quite often used (informally) in mathematical practice and even sometimes in more formal discussion around mathematics, we construe it here for the first time as an independent philosophical notion. Informal use of the expression often happens in the form of an (implicit) aesthetic criterion; it is arguably one of the strongest drivers of mathematical activity, as one of the main tests for its relevance. Since the advent of mathematical logic as an independent discipline, it
has become possible to investigate the formal notion of categoricity by mathematical means. We use this notion as the main base of our notion of logical perfection. (Morales \textit{et al.} 2021)
\end{quote}

The logical perfection of theories that Zilber investigates is formally defined in terms of the modern model-theoretic notion of \textit{k}-categoricity, which applies to a theory just in case, for any \textit{k}, all its models of cardinality \textit{k} belong to the same isomorphism class. This is different than the usual notion of categoricity, which is independent of cardinality constraints, but it seems more suitable for theories like QM, given the necessity to consider local, rather than global, models of quantum systems (as discussed in section 2.2.2). 

Importantly, the notion of \textit{k}-categoricity is apt to characterize first-order theories. Zilber's programmatic claim about physics is that one should not give up the attempt to reconstruct or reformulate quantum theories in a first-order language: ``a formal theorem like Löwenheim-Skolem could not undermine the quest for a univocal [i.e., categorical] theory of physics.''\footnote{Lecture in Bogotá, May 13, 2020, quoted in Kennedy 2022, 82. See also Zilber 2016, 2024.} Nevertheless, one may still harbor doubts as to whether the standard Hilbert space formalism can be given a first-order formalization. Such doubts may, however, be diminished if one considers minimal extensions of first-order logic as the framework of such formalizations.

Take, for instance, continuous first-order logic, which is an extension of first-order logic whose vocabulary consists of constant symbols, \textit{n}-ary function symbols and \textit{n}-ary relation symbols, and which replaces the set of possible truth values from \{0, 1\} to the bounded interval [0, 1], and Boolean functions as connectives by continuous functions from [0, 1]$^{n}$ to [0, 1] (Ben  Yaacov \textit{et al.} 2008). Negation, disjunction, conjunction, and the conditional are defined as follows: 

\begin{center}
    
$\neg a := 1-a $

$a \vee b := min(a,b)$

$a \wedge b := max(a,b)$

$a \rightarrow b := min(1-a,b)$. 

\end{center}

First-order quantifiers $\forall x$ and $\exists x$ are replaced by the operations $sup_{x}$ and $inf_{x}$, respectively, which are defined on the complete linear ordering of elements in the unit interval. Continuous first-order logic allows, for example, a formalization of C$^{*}$-algebras (Farah \textit{et al}. 2021). Similarly, one can search for a formalization of Hilbert space representations (or Rieffel's G-modules) as continuous first-order models. Then an appropriately formulated version of the Stone-von Neumann theorem might be rigorously interpreted as an appropriately conceived categoricity result for QM. 

But such attempts to formalize or rationally reconstruct QM must surely give us, philosophers, pause. Have we not learned already that the 20th century project of formalizing scientific theories has been an utter failure? Are philosophers of science not in agreement that the Carnapian project of rationally reconstructing physical theories like QM is generally uninformative and perhaps incoherent as well? 

In the next section, I want to suggest that this need not be the case. More specifically, I will argue that at least two of the most widely endorsed criticisms of the Carnapian project could actually be met, provided that one is willing to follow Sellars in giving up representationalism altogether.

\subsection{Quantum mechanics as a Carnapian language}

Let me start by considering some of Carnap's remarks on QM, as given in his ``Indeterminism in Quantum Physics” -- the last chapter of \textit{Philosophical Foundations of Physics: An Introduction to Philosophy of Science}, where he considered whether ``a change in the form of logic used in physics'' (Carnap 1966, 288) can be motivated by QM. Unlike the early Putnam, however, Carnap was never ready to take lessons in logic from QM. The reasons he gave for this are as follows: ``Physicists seldom present their theories in a form that logicians would like to see. ... the postulates of the entire field of physics stated in a systematic form that would include formal logic.'' (\textit{op. cit.}, 291) 

How did he think this might ever be achieved? A systematic presentation of the entire field of physics should be achieved, Carnap continued, through ``the application of modern logic and set theory, and the adoption of the axiomatic method in its modern form, which presupposes a formalized language system.'' (\textit{loc. cit.}) The suggestion here is, of course, that questions about the logic of QM cannot be properly addressed until a rational reconstruction of the whole of physics becomes available, which seems easy to reject as preposterous. However, a watered-down suggestion, which only requires a rational reconstruction of QM, might be more difficult to reject. But first, what did Carnap actually mean by rational reconstruction?  

The notion of rational reconstruction first appears in his \textit{Aufbau}: ``A theory is \textit{axiomatized} when all statements of the theory are arranged in the form of a deductive system whose basis is formed by the axioms, and when all concepts of the theory are arranged in the form of \textit{a constructional system} whose basis is formed by the fundamental concepts.'' (Carnap 1928, \S 2)\footnote{For the development of the notion of rational reconstruction, see Beaney 2013. For a discussion focused on Carnap on rational reconstruction, see Demopoulos 2007. For my own reading of Carnap's reconstructivist project in the \textit{Aufbau}, see Toader 2015.} Thus, a theory is rationally reconstructed if its Hilbert-style axiomatization -- a framework of concepts -- is turned into a constructional system. This requires the application of what Susan Stebbing would call ``directional analysis'' to uncover the fundamental concepts of the system.\footnote{Cf. Stebbing 1935, where she characterized the system of Russell and Whitehead's \textit{Principia Mathematica} as a directional postulate system, rather than an axiomatic system. In his ``Axiomatic Thinking'', Hilbert famously wrote: ``When we assemble the facts of a definite, more-or-less comprehensive field of knowledge, we soon notice that these facts are capable of being ordered. This ordering always comes about with the help of a certain \textit{framework of concepts}... [which] is nothing other than the \textit{theory} of the field of knowledge.'' (Hilbert 1918, \S 2)} 

After 1934, a rational reconstruction is supposed to use the tools developed in the \textit{Logical Syntax}: one can then say that a theory is rationally reconstructed if it is constructed as a Carnapian language: a consequence relation is first defined, logical and descriptive terms must then be distinguished, etc. As is well known, Carnap's view on the rational reconstruction of scientific theories evolved from the early partial interpretation approach, to an approach based on the formulation of Ramsey and Carnap sentences, culminating with his mature view that deploys Hilbert’s epsilon operator.\footnote{This evolution has been the subject of extended analysis in contemporary philosophy of science. For a recent account, see Demopoulos 2022.}

As I noted in section 5.3.1, Carnap argued in 1943 that submitting a rational reconstruction of CL -- Hilbert and Bernays formalization of classical logical reasoning -- to metasemantic evaluation shows that this is not categorical, since it lacks a unique semantics up to a isomorphism. This entails, as we noted, that the meaning of logical connectives is not fixed by its formal rules, since some of them turn out to be not truth-functional under valuations that Carnap constructed. This categoricity problem led him to propose a revision of the rational reconstruction that formalizes valid arguments by allowing multiple conclusions. 

Taking the hint, one may surmise that a rational reconstruction of QM should also be required as a basis for a metasemantic analysis of QM: investigating whether this rational reconstruction is categorical should tell us something about the meaning of quantum expressions. More exactly, if it turned out that no rational reconstruction of QM is categorical, then this would imply that its rules do not fix the semantic attributes of quantum expressions. Quantum inferentialism would, thus, be false. 

The inferentialist could reply that since there cannot be any viable rational reconstruction of QM, the objection based on the categoricity problem against quantum inferentialism misfires. As I noted already (in section 2.2.2), the categoricity problem does not actually depend on the existence of viable rational reconstructions of QM, since a notion of categoricity apt to characterize standard QM can be specified. Nevertheless, why would the inferentialist, or anyone for that matter, think that there cannot be any viable rational reconstruction of QM? 

Clearly enough, standard QM, i.e., the axiomatization given by von Neumann in 1932, can hardly be considered a Carnapian constructional system, although it certainly looks like a Hilbertian framework of concepts. This is because its basic concepts are not fundamental in the sense required by Carnap -- they are not the result of a directional analysis -- and there is no systematic construction of concepts, either. In contrast, recent re-axiomatizations of QM do seem to require a directional analysis, since they derive standard QM from information-theoretic principles, considered to be more fundamental than the von Neumann axioms (see section 5.2.1). But these information-theoretic reconstructions are not formulated as Carnapian languages, and do not consider the definition of a consequence relation as a central task. 

Perhaps this should not be surprising after all. Maybe the inferentialist is right to think that there cannot be any viable rational reconstruction of QM. However, consider Healey's reasons for rejecting the Carnapian project:

\begin{quote}
    
For Carnap, the task of the philosopher is to seek clarification of what a scientific theory says by means of a logical reconstruction of the theory within a precisely defined language. ... there is now a consensus among philosophers of science that the labor involved in re-expressing a significant theory in a formal language and then giving its semantics by means of correspondence rules would make this neither a practicable nor a useful technique for revealing its structure and function. (Healey 2017, 123)

\end{quote}

Rational reconstruction is said to be neither practicable, nor useful, but note that Healey does not take it to be impossible. More importantly, note that a rational reconstruction is understood to require representationalism: the semantics of its language is supposed to be given by correspondence rules, which  presumably are also supposed to do metasemantic work. But why must the Carnapian project be wedded to representationalism? An alternative that may be immediately suggested is that a rational reconstruction is possible, and might be even practicable and useful, if it would not be taken to require representationalism, but would, instead, adopt inferentialism. Healey, for one, might definitely embrace this suggestion. 

But others reject Carnap's project precisely because they understand it to require non-representationalism, at least for theoretical terms (which is what Carnap had already suggested in his 1927 paper, ``Proper and Improper Concepts''):

\begin{quote}

The non-representationalist strategy ... is not new, nor is it specific to quantum theory. It is, rather, the central idea in the logical-positivist and logical-empiricist pictures of science. ... It is almost universally accepted today that these approaches are not viable. But the predominant reason, historically, that they fell from grace was ... the increasingly clear realization -- notably ... in the recognized failure of Carnap’s project in the \textit{Aufbau} (1928) -- that observation is theory-laden.'' (Wallace 2020, 92)

\end{quote}

Rational reconstruction is considered not viable because it is taken to assume a sharp division between observational terms (having a representationalist semantics) and theoretical terms (having a non-representationalist semantics). The problem, of course, is that no such division is defensible. But, again, why must the Carnapian project be wedded to this division? The alternative that may be immediately suggested is, again, that a viable rational reconstruction should simply not require non-representationalism only for theoretical terms: if inferentialism were globally adopted, for \textit{all} terms, then no problematic division between observational and theoretical terms would be required.


One further obstacle for the very possibility of a rational reconstruction of QM, other than the problems that would follow from the adoption of global inferentialism, should also be considered. Inspired by a more general and seemingly devastating criticism raised by Michael Potter (2000, 277) against the rational reconstruction of mathematical theories, this obstacle can be presented in the form of the following argument:

\smallskip

1. It is an experimental fact that QM makes adequate empirical claims, i.e., there is inductive evidence for this fact.  

2. Thus, we can know that QM makes adequate empirical claims.

3. Suppose a rational reconstruction of QM can be given.

4. We cannot know that QM makes adequate empirical claims unless one proves the consistency of the rational reconstruction of QM.

5. Due to Gödel's second incompleteness theorem, the consistency of the rational reconstruction of QM cannot be proved (in an adequate, informative way). 

6. Thus, we cannot know that QM makes adequate empirical claims.

7. Therefore, QM cannot be rationally reconstructed.

\smallskip

However, I think that one might be able to overcome this obstacle by making a simple distinction between two types of knowledge: on the one hand, fallible knowledge, which is based on inductive evidence (premise 2), and on the other hand, infallible knowledge, which is based on mathematical proof (premises 4 and 6). If we make this distinction, the conclusion of the argument can be resisted. 

Then it might be open to us after all to say that we could learn something about QM from a rational reconstruction of QM. Submitting the latter to metasemantic analysis by investigating its categoricity would perhaps allow us, for example, to further examine the validity of quantum inferentialism (or any other non-representationalist explanation of semantic facts). If this examination failed, it would not fail because a rational reconstruction of QM cannot be given.

In closing the book, I want to emphasize that a fuller metasemantic analysis of QM, and more generally, of any mathematical and scientific theory, should actually take us all the way back to Hilbert’s own considerations, in the late 1920s, concerning the principle of permanence and the metatheoretical ideal of completeness as categoricity.

\subsection{Back to Hilbert!}

In a letter to Weyl, dated 6 November 1929, from which I extracted the motto for my book, Schrödinger thanked him for sending the first edition, from 1928, of \textit{The Theory of Groups and Quantum Mechanics} (Weyl 1930), but complained about the introduction of a new algebraic formalism in physics: 

\begin{quote}
    
Dear mathematicians ... Building new conceptual structures is fun, it is your very own sphere, but for us the physics is still far too deep in the darkness to hope that we could work successfully with such complicated, unfamiliar instruments. ... You have to understand the physiology and biology and eventually the phylogeny of a mathematical apparatus, then you have something.\footnote{The German passage is this: ``Liebe Mathematiker... Neue Begriffsgebäude aufbauen macht Spass, es ist Eure allerureigenste Sphäre, aber für uns liegt das Physikalische noch viel zu Tief im Dunkel, als dass wir hoffen könnten, in dieser Finsternis mit solch komplizierten, ungewohnten Instrumenten erfolgreich arbeiten zu können. .... Man muss die Physiologie und Biologie und womöglich die Phylogenie eines mathematischen Apparatus verstehen, dann hat man etwas davon.'' (Manuscript Hs 91:730, located in the Archive at the ETH Library in Zurich)}

\end{quote}

These conditions for the genuine understanding of a mathematical formalism -- what Schrödinger called its physiology, biology, and phylogeny -- suggest that he believed an answer should be given to the question about the relationship between the algebraic formalism just introduced by Weyl in QM and its content -- the ``blood of empirical reality'', as Carnap famously put it (Carnap 1927). This is essentially the question that, as I noted at the very outset, in section 1.1, Hilbert had asked in his 1917 lecture in Zurich. Schrödinger expressed, I take it, a concern that many, including Weyl and Carnap, among others, thought justified.

I have interpreted Hilbert's question as pointing to the metasemantics of classical logic and mathematics, but also that of quantum physics. And I have maintained that an investigation of the metasemantics of standard QM and non-distributive QL would help provide a better philosophical understanding of these theories. The investigation has focused on two problems, which I called, for both systematic and historical reasons, the categoricity problem and the permanence problem. 

This led me to propose, among other things, a new reconstruction of Einstein's argument for the incompleteness of QM, which suggests that his debate with Bohr could be understood metasemantically, as a divergence about the indispensability of categoricity as a condition on the relationship between the mathematical apparatus of QM and its physical meaning. It also led me to uncover the 19th century philosophical and mathematical background to Bohr's correspondence principle, which helps to understand the latter as grounded in the principle of permanence, and to further illuminate some of his obscure claims about the relationship between the mathematical apparatus of QM and its physical meaning. From Einstein and Bohr, the investigation led me to consider the categoricity problem and the permanence problem, as well as their metasemantic implications for QL, in the works of von Neumann and Weyl. And from this, further, to current debates in contemporary philosophy of logic and physics.

Hilbert himself was, of course, fully aware of both of these problems. What he called the ``higher-order'' principle of permanence was at the center of his foundational debate about mathematics with Brouwer's intuitionism. As Bernays put it: 

\begin{quote}
    The system of arithmetic is built on a foundation of conceptions which have decisive significance for scientific systematization in general: namely the principle of conservation (or `permanence') of laws, which appears in this connection as the postulate of the unlimited applicability of the usual logical forms of judgment and inference, and the demand for a purely objective conception of the theory through which the latter is freed from every reference to our knowledge. (Bernays 1930/1931) 
\end{quote}

But an account of the role of permanence as conservation in Hilbert's views on the foundations of mathematics, and especially their potential influence on Bohr's approach to QM, and particularly on his understanding of the correspondence principle (discussed in section 3.2), still needs to be given.  

The metatheoretical ideal of categoricity, which Hilbert had formulated already in his axiomatizations of geometry and analysis, was expressed also in connection with QM: ``The goal is to formulate the physical requirements so completely that the analytical apparatus is uniquely determined.'' (Hilbert \textit{et al.} 1928) Hilbert's view, I take it, is that if the axioms or rules of QM are complete, then the quantum model for a physical system is uniquely determined (up to isomorphism). If this is a correct reading of his view, then the completeness of QM that Hilbert was looking for, in 1928, comes close to the notion of completeness that Einstein was looking for, in 1935 (discussed in section 2.3). Of course, due to the limitations of first-order logic, especially within the context of his mathematical finitism, Hilbert very soon gave up categoricity, in an important lecture in Bologna (Hilbert 1929). But I strongly suspect he would not have allowed such limitations to undermine the quest for a precise determination of the meaning of the formalism of a scientific theory. 

Another book on metasemantics must be written, one that includes a detailed discussion of Hilbert's own views on the matter, including his views on permanence and categoricity. One would then be totally justified, I believe, to attribute to him the very claim that metasemantics, as an investigation of the foundations of semantics, contributes to a better philosophical understanding of scientific theories.

\newpage

\section{References}

Acuña, P. (2021) von Neumann’s Theorem Revisited. \textit{Foundations of Physics}, 51, 1--29


Arens, R. and V. S. Varadarajan (2000) On the concept of Einstein-Podolsky-Rosen states and their structure. \textit{Journal of Mathematical Physics}, 41, 638--651

Ashtekar, A. (2009) Some surprising implications of background independence in canonical quantum gravity. \textit{General Relativity and Gravitation}, 41, 1927--1943

Bacciagaluppi, G. (2009) Is Logic Empirical? In D. Gabbay, D. Lehmann, and K. Engesser (eds.) \textit{Handbook of Quantum Logic}, Elsevier, 49--78 

Bacciagaluppi, G. and E. Crull (2024) \textit{The Einstein Paradox. The Debate on Nonlocality and Incompleteness in 1935}, Cambridge University Press



Barrett. T. W. and H. Halvorson (2016) Morita equivalence. \textit{The Review of Symbolic Logic}, 9, 556--582

Beaney, M. (2013) Analytic Philosophy and History of Philosophy: The Development of the Idea of Rational Reconstruction. In E. H. Reck (ed.) \textit{The Historical Turn in Analytic Philosophy}, Palgrave-Macmillan, 231--260


Bell, J. and M. Hallett (1982) Logic, quantum logic and empiricism. \textit{Philosophy of Science}, 49, 355--379

Bellomo, A. (2025) Peacock’s Principle of Permanence and Hankel’s Reception. In HOPOS: The Journal of the International Society for the History of Philosophy of Science 15, doi:10.1086/734645

Beltrametti, E. G. and G. Cassinelli (1981) \textit{The Logic of Quantum Mechanics}, Addison-Wesley

Ben Yaacov, I., A. Berenstein, C. W. Henson, and A. Usvyatsov (2008) Model theory for metric structures. In Z. Chatzidakis, D. Macpherson, A. Pillay, and A. Wilkie (eds.) \textit{Model theory with applications to algebra and analysis}, vol. 2, Cambridge University Press, 315--427

Bernays, P. (1930/1931) ``Die Philosophie der Mathematik und die Hilbertsche Beweistheorie'', \textit{Bl\"atter f\"ur deutsche Philosophie}, 4, 326--67.

Bernays, P. (1966) Remarks About Formalization and Models. In E. Nagel, P. Suppes and
A. Tarski (eds.) \textit{Logic, Methodology and Philosophy of Science}, Proceeding of the 1960 International Congress, 176--180

Bernstein, B. A. (1932) Relation of Whitehead and Russell's theory of deduction to the Boolean logic of propositions. Bulletin of the American Mathematical Society 38, 589--593 

Birkhoff, G. and J. von Neumann (1936) The Logic of Quantum Mechanics. \textit{Annals of Mathematics} 37, 823--843

Birman, R. (2024) The Adoption Problem and the Epistemology of Logic. \textit{Mind} 133: 37--60

Blackburn, S. (2013) Pragmatism: all or some? In H. Price, \textit{Expressivism, Pragmatism and Representationalism}, Cambridge University Press, 67--84

Bohm, D. (1985) On Bohr's views concerning the quantum theory. In A. French and P. Kennedy (eds.) \textit{Niels Bohr: A Centenary Volume}, Harvard University Press, 153--159

Bohr, N. (1920) On the series spectra of the elements. Lecture before the German Physical Society in Berlin, 27 April 1920, in J. R. Nielsen (ed.) \textit{Niels Bohr Collected Works, Vol. 3: The Correspondence Principle (1918–1923)}, North-Holland Publishing, 241--282

Bohr, N. (1935) Can Quantum-Mechanical Description of Physical Reality Be Considered Complete? \textit{Physical Review}, 48, 696--702

Bohr, N. (1939) The causality problem in atomic physics (Report drafted and submitted by N. Bohr). \textit{New theories in physics: Conference organized in collaboration with the International Union of Physics and the Polish Intellectual Co-operation Committee, Warsaw, May 30th-June 3rd, 1938}, 11--45

Bokulich, A. (2008) \textit{Reexamining the Quantum-Classical Relation. Beyond Reductionism and Puralism}, Cambridge University Press

Bokulich, A. and P. Bokulich (2020) Bohr's Correspondence Principle. \textit{The Stanford Encyclopedia of Philosophy} (Fall 2020 Edition), Edward N. Zalta (ed.), URL = <https://plato.stanford.edu/archives/fall2020/entries/bohr-correspondence/>.

Bokulich, P. and A. Bokulich (2005) Niels Bohr’s generalization of classical mechanics. \textit{Foundations of Physics}, 35, 347--371


Brandom, R. B. (2000) \textit{Articulating Reasons. An Introduction to Inferentialism}, Harvard University Press

Brandom, R. B. (2015) \textit{From Empiricism to Expressivism}, Harvard University Press

 
Brîncuș, C. C. (2024) Inferential Quantification and the $\omega$-Rule. In A. P.  d'Aragona (ed.) \textit{Perspectives on Deduction}, Springer, 345--372

Brîncuș, C. C. and I. D. Toader (2019) Categoricity and Negation. A Note on Kripke’s Affirmativism. In I. Sedlar and M. Blicha (eds.) \textit{The Logica 2018 Yearbook}, College London Publications, 57--66 

Brukner, Č. (2018) A No-Go Theorem for Observer-Independent Facts. \textit{Entropy}, 20(5), 350

Brukner, Č. (2022) Wigner’s friend and relational objectivity. \textit{Nature Reviews Physics}, 4, 628--630

Bub, J. (2004) Why the quantum?
\textit{Studies in History and Philosophy of Modern Physics}, 35, 241--266

Bunge, M. (1973) \textit{Philosophy of Physics}, D. Reidel

Burgess, A. and B. Sherman (2014) \textit{Metasemantics: New Essays on the Foundations of Meaning}, Oxford University Press

Button, T. and S. Walsh (2018) \textit{Philosophy and Model Theory}, Oxford University Press

Carnap, R. (1927) Eigentliche und uneigentliche Begriffe. \textit{Symposion: Philosophische Zeitschrift für Forschung und Aussprache}, 1, 355--374

Carnap, R. (1928) \textit{Der logische Aufbau der Welt}, Eng. tr. as \textit{Logical Structure of the World}, University of California Press, 1967

Carnap, R. (1934) \textit{Logische Syntax der Sprache}, Eng tr. as \textit{Logical Syntax of Language}, K. Paul, Trench, Trubner $\&$ Co. Ltd., 1937

Carnap, R. (1943) \textit{Formalization of Logic}, Harvard University Press

Carnap, R. (1966) \textit{Philosophical Foundations of Physics: An Introduction to Philosophy of Science}, Basic Books

Cassirer, E. (1953) \textit{The philosophy of symbolic forms}, vol. 1, Yale University Press


Chrisman, M. (2016a) Metanormative Theory and the Meaning of Deontic Modals. In N. Charlow and M. Chrisman (eds.) \textit{Deontic Modality}, Oxford University Press, 395--424

Chrisman, M. (2016b) \textit{The Meaning of `Ought'}, Oxford University Press

Clifton, R. and H. Halvorson (2001) Are Rindler Quanta Real? Inequivalent Particle Concepts in Quantum Field Theory. \textit{The British Journal for the Philosophy of Science}, 52, 417--470

Clifton, R., J. Bub, and H. Halvorson (2003) Characterizing Quantum Theory in Terms of Information-Theoretic Constraints. \textit{Foundations of Physics}, 33, 1561--1591

Dalla Chiara, M., R. Giuntini, and R. Greechie (2004) \textit{Reasoning in Quantum Theory. Sharp and Unsharp Quantum Logics}, Kluwer

Darrigol, O. (1986) The Origin of Quantized Matter Waves. \textit{Historical Studies in the Physical and Biological Sciences}, 16, 197--253

Darrigol, O. (1997) Classical Concepts in Bohr’s Atomic Theory (1913–1925). \textit{Physis: Rivista Internazionale di Storia della Scienza}, 34, 545--567




Demopoulos, W. (1976) The Possibility Structures of Physical Systems. In W. L. Harper and C. A. Hooker (eds.) \textit{Foundations of Probability Theory, Statistical Inference, and Statistical Theories of Science}, D. Reidel, 55--80

Demopoulos, W. (2007) Carnap on the Rational Reconstruction of Scientific Theories. In M. Friedman and R. Creath (eds.) \textit{The Cambridge Companion to Carnap}, Cambridge University Press, 248--272

Demopoulos, W. (2022) \textit{On Theories. Logical Empiricism and the Methodology of Modern Physics}, ed. by M. Friedman, Harvard University Press

Detlefsen, M. (2005) Formalism. In S. Shapiro (ed.) \textit{The Oxford Handbook of Philosophy of Mathematics and Logic}, Oxford University Press, 236--317

Detlefsen, M. (2014) Completeness and the ends of axiomatization. In J. Kennedy (ed.) \textit{Interpreting Gödel. Critical Essays}, Cambridge University Press, 59--77

Dickson, M (1998) \textit{Quantum chance and non-locality}, Cambridge University Press

Dickson, M. (2001) Quantum Logic Is Alive $\wedge$ (It Is True $\vee $ It Is False). \textit{Philosophy of Science}, 68, S274--S287

Dickson, M. (2015) Reconstruction and Reinvention in Quantum Theory. \textit{Foundations of Physics}, 45, 1330--1340

Dirac, P.A.M. (1925) The Fundamental Equations of Quantum Mechanics. \textit{Proceedings of the Royal Society of London. Series A}, 109, 642--653

Dirac, P.A.M. (1945) On the Analogy Between Classical and Quantum Mechanics. \textit{Reviews of Modern Physics}, 17, 195--199

Dongen, J. v. (2010) \textit{Einstein's Unification}, Cambridge University Press

Dummett, M. (1976) Is Logic Empirical? \textit{Truth and Other Enigmas}, Duckworth, 269--289

Dummett, M. (1987) Reply to John McDowell. In B. M. Taylor (ed.) \textit{Michael Dummett: Contributions to Philosophy}, Nijhoff International Philosophy Series, vol. 25, Springer, 253--268

Dunn, J. M. (1980) Quantum Mathematics. \textit{Proceedings of the the Philosophy of Science Association}, 512--531

Dunn, J. M. and Hardegree, G. (2001) \textit{Algebraic methods in philosophical logic}, Oxford University Press.

Einstein, A. (1919) What is the theory of relativity? \textit{The Times}, 28 November 1919, repr. in Einstein 1954, 227--232

Einstein, A. (1931) Maxwell's Influence on the Evolution of the Idea of Physical Reality. In \textit{James Clerk Maxwell: A Commemoration Volume}, Cambridge University Press, repr. in Einstein 1954, 266--270

Einstein, A. (1936) Physics and Reality. \textit{The Journal of the Franklin Institute}, 221, repr. in Einstein 1954, 290--323

Einstein, A. (1944) Remarks on Bertrand Russell's Theory of Knowledge. In P. A. Schilpp (ed.) \textit{The Philosophy of Bertrand Russell}, repr. in Einstein 1954, 18--24 

Einstein, A. (1949) Remarks Concerning the Essays Brought together in this Co-operative Volume. In P. A. Schilpp (ed.) \textit{Albert Einstein: Philosopher-Scientist}, repr. in Einstein 1954, 665--688

Einstein, A. (1954) \textit{Ideas and opinions}, Bonanza Books

Einstein, A., B. Podolski, and N. Rosen (1935) Can Quantum-Mechanical Description of Physical Reality Be Considered Complete? \textit{Physical Review}, 47, 777--780

Eisenthal, J. (2021) Hertz's \textit{Mechanics} and a unitary notion of force. \textit{Studies in History and Philosophy of Science}, 90, 226–-234

Farah, I., B. Hart, M. Lupini, L. Robert, A. Tikuisis, A. Vignati, and W. Winter (2021) Model theory of C$^{*}$-algebras. \textit{Memoirs of the American Mathematical Society}, 271

Faye, J. (2017) Complementarity and Human Nature. In J. Faye and H. J. Folse (eds.) \textit{Niels Bohr and the Philosophy of Physics: Twenty-First-Century Perspectives}, Bloomsbury, 115--131

Feintzeig, B. (2022) Reductive Explanation and the Construction of Quantum Theories. \textit{British Journal for the Philosophy of Science}, 73, 457--486

Feintzeig, B., J. B. Le Manchak, S. Rosenstock, and J.O. Weatherall (2019) Why Be regular?, part I. \textit{Studies in History and Philosophy of Modern Physics}, 65, 122--132

Feintzeig, B. \& J.O. Weatherall (2019) Why Be regular?, part II. \textit{Studies in History and Philosophy of Modern Physics}, 65, 133--144

Feyerabend, P. (1962) Problems of Microphysics. In S. Gattei and J. Agassi (eds.) \textit{Paul K. Feyerabend: Physics and Philosophy. Philosophical Papers}, vol. 4, Cambridge University Press, 2016, 99-187

Feyerabend, P. (1995) \textit{Killing Time}, The University of Chicago Press.

Field, H. (2008) \textit{Saving truth from paradox}, Oxford University Press

Fine, A. (1971) Probability in quantum mechanics and in other statistical theories.
In M. Bunge (ed.), \textit{Problems in the Foundations of Physics}, Springer, 79--92

Fine, A. (1972) Some Conceptual Problems with Quantum Theory. In Robert G. Colodny (ed.) \textit{Paradigms and Paradoxes: The Philosophical Challenge of the Quantum Domain}, University of Pittsburgh Press, 3--31

Fine, A. (1981) Einstein's Critique of Quantum Theory: The Roots and Significance of EPR. In \textit{The Shaky Game}, 1986, 26--39

Freire Jr., O., G. Bacciagaluppi, O. Darrigol, Th. Hartz, Ch. Joas, A. Kojevnikov, and O. Pessoa Jr. (2022) \textit{The Oxford Handbook of the History of Quantum Interpretations}, Oxford University Press

French, Steven (2016) \textit{The Structure of the World. Metaphysics and Representation}, Oxford University Press 

Friedman, M and C. Glymour (1972) If quanta had logic. \textit{Journal of Philosophical Logic}, 1, 16--28 

Fuchs, C. A. (2018) Notwithstanding Bohr, the Reasons for QBism. \textit{Mind and Matter}, 15, 245–300

Fuchs, C. A. (2023) QBism, Where Next? In Ph. Berghofer and H. A. Wiltsche (eds.) \textit{Phenomenology and QBism
New Approaches to Quantum Mechanics}, Routledge, 78--143

Fuchs, C. A. and A. Perez (2000) Quantum Theory Needs No `Interpretation'. \textit{Physics Today} 53, 70--71

Fuchs, C. A., N. D. Mermin, and R. Schack (2014) An Introduction to QBism with an Application to the Locality of Quantum Mechanics. \textit{American Journal of Physics}, 82, 749–754 

Fuchs, C. A. and R. Schack (2015) QBism and the Greeks: Why a Quantum State Does
Not Represent an Element of Physical Reality. \textit{Physica Scripta}, 90, 015104 

Fuchs C. A. and B. C. Stacey (2019) Are Non-Boolean Event Structures the Precedence or
Consequence of Quantum Probability? arXiv:1912.10880

García-Carpintero, M. (2012a) Foundational Semantics I: Descriptive Accounts. \textit{Philosophy Compass}, 7, 397--409 

García-Carpintero, M. (2012b) Foundational Semantics II: Normative Accounts. \textit{Philosophy Compass}, 7, 410--421


Giovanelli, M. (2021) `Like thermodynamics before Boltzmann.' On the emergence of Einstein's distinction between constructive and principle theories. \textit{Studies in History and Philosophy of Modern Physics}, 71, 118--157

Gleason, A. M. (1957) Measures on the Closed Subspaces of a Hilbert Space. \textit{Journal of Mathematics and Mechanics}, 6, 885--893

Glick, D. (2021) QBism and the limits of scientific realism. \textit{European Journal of Philosophy of Science}, 11, 53

Glymour, C. (1970) Theoretical realism and theoretical equivalence. In R. Buck and R. Cohen (eds.) \textit{Boston Studies in Philosophy of Science}, 7, 1971, D. Reidel, 275--288

Gömöri, M. and G. Hofer-Szabó (2021) On the meaning of EPR’s Reality Criterion. \textit{Synthese} 199, 13441--13469

Greenberger, D. M., M. A. Horne, and A. Zeilinger (1989) Going beyond Bell's Theorem In Kafatos, M. (ed.) \textit{Bell's Theorem, Quantum Theory, and Conceptions of the Universe}, Kluwer, 69--72

Greenberger, D. M., M. A. Horne, A. Shimony, and A. Zeilinger (1990) Bell's theorem without inequalities. \textit{American Journal of Physics}, 58, 1131--1143

Gregory, D. (1840) On the real nature of symbolic algebra. \textit{Transactions of the Royal Society of Edinburgh}, 14, 208--216. In W. Ewald (ed.) \textit{From Kant to Hilbert: A Source Book in the Foundations of Mathematics}, vol. 1, Oxford University Press, 1996, 323--330

Grinbaum, A. (2007) Reconstruction of Quantum Theory. \textit{British Journal of Philosophy of Science}, 58, 387--408


Halvorson, H. (2000) The Einstein-Podolsky-Rosen State Maximally. \textit{Letters in Mathematical Physics}, 53, 321--329

Halvorson, H. (2001) On the Nature of Continuous Physical Quantities in Classical and Quantum Mechanics. \textit{Journal of Philosophical Logic}, 30, 27--50

Halvorson, H. (2004) Complementarity of representations in quantum mechanics, \textit{Studies in History and Philosophy of Modern Physics}, 35, 45--56


Hankel, H. (1867) \textit{Theorie der complexen Zahlensysteme}, Leopold Voss

Hardy, L. (2001) Quantum theory from five reasonable axioms.  arXiv:quant-ph/0101012

Hardy, L. (2004a) Why is nature described by quantum theory? In J. D. Barrow, P. C. W. Davies and C. L. Harper, Jr. (eds.) \textit{Science and Ultimate Reality}, Cambridge University Press, 45--71

Hardy, L. (2004b) Quantum ontological excess baggage. \textit{Studies in History and Philosophy of Modern Physics}, 35, 267--276

Healey, R. (2017) \textit{The quantum revolution in philosophy}, Oxford University Press

Healey, R. (2024) Scientific Objectivity and its Limits. \textit{British Journal for Philosophy of Science}, 75 (3), 639-662

Healey, R. (2023) Pragmatism, Relativism and Quantum Theory. Talk at IQOQI, Vienna, 17.03.2023

Hellman, G. (1980) Quantum logic and meaning. \textit{Proceedings of the Philosophy of Science Association}, 493--511

Hertz, H. (1899) \textit{The principles of mechanics presented in a new form}, Macmillan and Co. 

Hilbert, D. (1918) Axiomatisches Denken. \textit{Mathematische Annalen}, 78, 405--415. Eng. tr. in W. B. Ewald (ed.) \textit{From Kant to Hilbert}, vol. 2, Oxford University Press, 1996, 1105--1115

Hilbert, D. (1929) Probleme der Grundlegung der Mathematik. \textit{Mathematische Annalen}, 102, 1--9. Eng. tr. in P. Mancosu (ed.) \textit{From Brouwer to Hilbert. The Debate on the Foundations of Mathematics in the 1920s}, Oxford University Press, 1998, 227--233

Hilbert, D., L. Nordheim, J. v. Neumann (1928) Über die Grundlagen der Quantenmechanik, \textit{Mathematische Annalen}, 98, 1--30

Hjortland, O. T. (2017) Anti-exceptionalism about logic. \textit{Philosophical Studies}, 174, 631--658

Hodges, W. (2016) A Strongly Differing Opinion on Proof-Theoretic Semantics?
In T. Piecha and P. Schroeder-Heister (eds.) \textit{Advances in Proof-Theoretic Semantics}, Trends in Logic 43, Springer Nature, 173--188

Hollings, C., U. Martin, and A. Rice (2017) The Lovelace–De Morgan mathematical correspondence: A critical re-appraisal. \textit{Historia Mathematica}, 44, 202--231

Horvat, S. and I. D. Toader (2023) Quantum logic and meaning. \\ 
arXiv:2304.08450

Horvat, S. and I. D. Toader (2024) An Alleged Tension between Non-classical Logics and Applied Classical Mathematics, \textit{The Philosophical Quarterly}, pqad125, https://doi.org/10.1093/pq/pqad125

Horvat, S. and I. D. Toader (2025) Carnap on Quantum Mechanics. In \textit{Carnap Handbuch}, ed. by Christian Damböck and Georg Schiemer, J. B. Metzler

Howard, D. (1985) Einstein on Locality and Separability, \textit{Studies in History and Philosophy of Science}, 16, 171--201 

Howard, D. (1990) `Nicht Sein Kann Was Nicht Sein Darf', Or the Prehistory of the EPR, 1909-1935: Einstein's Early Worries About the Quantum Mechanics of Composite Systems. In A. I. Miller (ed.) \textit{Sixty-Two Years of Uncertainty: Historical, Philosophical, and Physical Inquiries into the Foundations of Quantum Mechanics}, Plenum Press, 61--111

Howard, D. (1992) Einstein and \textit{Eindeutigkeit}: A Neglected Theme in the Philosophical Background to General Relativity. In J. Eisenstaedt and A. J. Kox (eds.) \textit{Studies in the History of General Relativity}, Birkhäuser, 154--243 

Howard, D. (1994) What makes a classical concept classical? Towards a reconstruction  of Niels Bohr's philosophy of physics. In J. Faye and H. J. Folse (eds.) \textit{Niels Bohr and Contemporary Philosophy}, Kluwer, 201--230

Howard, D. (2007) Revisiting the Einstein-Bohr Dialogue. \textit{Iyyun: The Jerusalem Philosophical Quarterly}, 56, 57--90

Howard, D. (2011) The physics and metaphysics of identity and individuality. \textit{Metascience}, 20, 225--231

Howard, D. (2012) The Trouble with Metaphysics (unpublished manuscript)

Howard, D. (2021) Complementarity and decoherence. In G. Jaeger, D. Simon, A. V. Sergienko, D. Greenberger, and A. Zeilinger (eds.) \textit{Quantum arrangements: Contributions in honor of Michael Horne}, Springer, 151--175

Hughes, R. I. G. (1989) \textit{The Structure and Interpretation of Quantum Mechanics}, Harvard University Press

Hughes, R. I. G. (2010) \textit{The Theoretical Practices of Physics. Philosophical Essays}, Oxford University Press

Hüttemann. A. (2009) Pluralism and the Hypothetical in Heinrich Hertz’s
Philosophy of Science. In  M. Heidelberger and G. Schiemann (eds.) \textit{The Significance of the Hypothetical in the Natural Sciences}, de Gruyter, 145--168 

J\"ahnert, M. (2019) \textit{Practicing the Correspondence Principle in the Old Quantum Theory}, Springer

Jammer, M. (1966) \textit{The Conceptual Development of Quantum Mechanics}, McGraw-Hill

Juffmann, Th., S. Truppe, Ph. Geyer, A. G. Major, S. Deachapunya, H. Ulbricht, and M. Arndt (2009) Wave and Particle in Molecular Interference Lithography. \textit{Physical
Review Letters}, 103, 263601

Kaplan, D. (1989) Afterthoughts. In J. Almog, J. Perry, and H. Wettstein (eds.) \textit{Themes from Kaplan}, Oxford
University Press, 565--614

Kennedy, J. (2022) \textit{Gödel, Tarski and the Lure of Natural Language}, Cambridge University Press 

Koberinski, A. and M. P. Müller (2018) Quantum Theory as a Principle Theory: Insights from an Information-Theoretic Reconstruction. In M. Cuffaro and S. Fletcher (eds.) \textit{Physical Perspectives on Computation, Computational Perspectives on Physics}, Cambridge University Press, 257-280

Kochen, S. and E. P. Specker (1967) The Problem of Hidden Variables in Quantum Mechanics. In C. A. Hooker (ed.) \textit{The Logico-Algebraic Approach to Quantum Mechanics}, I, D. Reidel, 1975, 293--328

Kolodny, N. and J. MacFarlane (2010) Ifs and Oughts. \textit{Journal of Philosophy}, 107, 115--143

Kremer, M. (2010) Representation or Inference: Must We Choose? Should We? In B.Weiss and J. Wanderer (eds.) \textit{Reading Brandom: On Making It Explicit}, Routledge, 227--246

Kremer, M. (2012) Russell's Merit.
In J. L. Zalabardo (ed.) \textit{Wittgenstein's Early Philosophy}, Oxford University Press, 195--240

Kripke, S. (1974) The Question of Logic. \textit{Mind} 133 (2024): 1--36

Kuby, D. (2021) Feyerabend's Reevaluation of Scientific Practice: Quantum Mechanics, Realism and Niels Bohr. In K. Bschir and J. Shaw (eds.) \textit{Interpreting Feyerabend: Critical Essays}, Cambridge University Press, 132--156

Lalo\"e, F. (2019) \textit{Do We Really Understand Quantum Mechanics?}, Cambridge University Press

Lambert, K. (2013) A Natural History of Mathematics: George Peacock and the Making of English Algebra. \textit{Isis}, 104, 278--302


Landsman, N. P. (2006) When champions meet: Rethinking the Bohr–Einstein debate. \textit{Studies in History and Philosophy of Physics}, 37, 212--242

Landsman, K. (2017) \textit{Foundations of Quantum Theory. From Classical Concepts to Operator Algebras}, Springer

Lawrence, R. (2021) Frege, Hankel, and Formalism in the \textit{Foundations}. \textit{Journal for the History of Analytical Philosophy}, 9, 5-27

Leegwater, G. (2018) When Greenberger, Horne and Zeilinger meet Wigner's Friend. arXiv:1811.02442
 
Lehner, Ch. (2014) Einstein's Realism and His Critique of Quantum Mechanics. In M. Janssen and Ch. Lehner (eds.) \textit{The Cambridge Companion to Einstein}, Cambridge University Press, 306--353

Lewis, P. J. (2016) \textit{Quantum ontology: A guide to the metaphysics of quantum mechanics}, Oxford University Press



Malament, D. B. (2002) Notes on Quantum Logic (unpublished manuscript)

Mancosu, P. (2010) \textit{The adventure of reason: interplay between philosophy of mathematics and mathematical logic, 1900-1940}, Oxford University Press

Masanes, L. and M. P. Müller (2011) A derivation of quantum theory from physical requirements. \textit{New Journal of Physics}, 13, 063001

Maudlin, T. (2010) The Tale of Quantum Logic. In Y. Ben-Menahem (ed.) \textit{Hilary Putnam}, Cambridge Universoty Press, 156--187

Maudlin, T. (2016) The Metaphysics of Quantum Theory. \textit{Belgrade Philosophical Annual}, 29, 5--13 

Maudlin, T. (2019) \textit{Philosophy of Physics: Quantum Theory}, Princeton University Press. 

Maudlin, T. (2022) The Labyrinth of Quantum Logic. In J. Conant and S. Chakraborty (eds.) \textit{Engaging Putnam}, de Gruyter, 183--205

McGrath, J. H. (1978) A formal statement of the Einstein-Podolsky-Rosen argument. \textit{International Journal of Theoretical Physics}, 17, 557--571

Mitsch, C. (2022) Hilbert-style axiomatic completion: On von Neumann and hidden variables in quantum mechanics. \textit{Studies in History and Philosophy of Science}, 95, 84--95

Morales, J. A. C., A. Villaveces and B. Zilber (2021) Around Logical Perfection. \textit{Theoria}, 87, 971--985

Mostowski, A. (1965) Thirty years of foundational studies. In his \textit{Foundational Studies. Selected Works}, vol. 1, ed. by K. Kuratowski \textit{et al.}, Elsevier

Müller, M. P. (2021) Probabilistic Theories and Reconstructions of Quantum Theory. arXiv:2011.01286v4

Murgueitio Ram\'irez, S. (2020) Separating Einstein's separability. \textit{Studies in History and Philosophy of Modern Physics}, 72, 138--149

Murzi, J. and F. Steinberger (2017) Inferentialism. In B. Hale, C. Wright, and A. Miller (eds.), \textit{A companion to the philosophy of language}, Blackwell, 197--224 

Murzi, J. and B. Topey (2021) Categoricity by Convention. \textit{Philosophical Studies}, 178, 3391--3420

Norsen, T. (2017) \textit{Foundations of Quantum Mechanics}, Springer

Peacock, G. (1833) Report on the Recent Progress and Present State of Certain Branches of Analysis. \textit{Report of the British Association}, 185--352

Peirce, Ch. S. (1868) Some Consequences of Four Incapacities. \textit{The Journal of Speculative Philosophy}, 2, 140--157. Repr. in C. Hartshorne and P. Weiss (eds.) \textit{The Collected Papers of Charles Sanders Peirce}, vol. 5, Harvard University Press 

Peirce, Ch. S. (1877) The Fixation of Belief. \textit{Popular Science Monthly}, 12, 1--15. Repr. in C. Hartshorne and P. Weiss (eds.) \textit{The Collected Papers of Charles Sanders Peirce}, vol. 5, Harvard University Press 

Peregrin, J. (2014) \textit{Inferentialism: Why Rules Matter}, Springer

Perovic, S. (2008) Why were Matrix Mechanics and Wave Mechanics considered equivalent? \textit{Studies in History and Philosophy of Modern Physics}, 39, 444--461

Perovic, S. (2021) \textit{From Data to Quanta: Niels Bohr's Vision of Physics}, The University of Chicago Press

Potter, M. (2000) \textit{Reason's Nearest Kin: Philosophies of arithmetic from Kant to Carnap}, Cambridge University Press

Price, H. (2013) \textit{Expressivism, Pragmatism and Representationalism}, Cambridge University Press


Prior, A. N. (1960) The Runabout Inference-Ticket. \textit{Analysis}, 21, 38--39

Putnam, H. (1968) Is Logic Empirical? In R. S. Cohen and M. W. Wartofsky (eds.) \textit{Boston Studies in the Philosophy of Science}, 5, D. Reidel, 216--241, repr. as ``The logic of quantum mechanics'' in his \textit{Mathematics, Matter and Method. Philosophical Papers}, vol. 1, Cambridge University Press, 174--197


Putnam, H. (1991) Il principio di indeterminazione e il progresso scientifico. \textit{Iride}, 7, 9--27

Putnam, H. (1994) Michael Redhead on quantum logic. In P. Clark and B. Hale (eds.) \textit{Reading Putnam}, Blackwell, 265--280

Putnam, H. (2012). The Curious Story of Quantum Logic. In M. De Caro and D. Macarthur (eds.) \textit{Philosophy in the Age of Science: Physics, Mathematics, and Skepticism}, Harvard University Press, 162--177.


Pycior, H. (1981) George Peacock and the British Origins of Symbolic Algebra. \textit{Historia Mathematica}, 8, 23--45

Quine, W. V. O. (1992) Structure and Nature. \textit{The Journal of Philosophy}, 89, 5--9


Radder, H. (1991) Heuristics and the Generalized Correspondence Principle. \textit{The British Journal for the Philosophy of Science}, 42, 195--226

Read, S. (1997) Completeness and Categoricity: Frege, G\"odel and Model Theory. \textit{History and Philosophy of Logic}, 18, 79--93

Rédei, M. (1998) \textit{Quantum Logic in Algebraic Approach}, Kluwer

Rédei, M. (2014) Hilbert's 6th Problem and Axiomatic Quantum Field Theory. \textit{Perspectives on Science}, 22, 80--97 

Resnik, M. (1996) Structural Relativity. \textit{Philosophia Mathematica}, 3, 83--99

Restall, G. (2005) Multiple Conclusions.
In P. Hájek, L. Valdés-Villanueva and D. Westerståhl (eds.) \textit{Logic, Methodology and Philosophy of Science}, College Publications, 189--205

Rieffel, M. A. (1972) On the Uniqueness of the Heisenberg Commutation Relations. \textit{Duke Mathematical Journal}, 39, 745--752

Roe, J. (2003) \textit{Lectures on coarse geometry}, University Lecture Series, American Mathematical Society, vol. 31 

Rosenberg, J. (2004) A selective history of the Stone–von Neumann Theorem. In R. S. Doran and R. V. Kadison (eds.) \textit{Operator Algebras, Quantization, and Noncommutative Geometry: A Centennial Celebration Honoring John von Neumann and Marshall H. Stone}, American Mathematical Society, 331--354

Ruetsche, L. (2011) \textit{Interpreting Quantum Theories. The Art of the Possible}, Oxford University Press

Rumfitt, I. (2015) \textit{The Boundary Stones of Thought}, Oxford University Press


Russell, B. (1903) \textit{Principles of Mathematics}, W. W. Norton

Russell, B. (1907) The Regressive Method of Discovering the Premises of Mathematics. In G. H. Moore (ed.) \textit{The Collected Papers of Bertrand Russell: Toward ``Principia Mathematica'' 1905-08}, Routledge, 2014, 571--580

Russell, B. (1919) \textit{Introduction to Mathematical Philosophy}, George Allen \& Unwin

Sauer, T. (2007) An Einstein manuscript on the EPR paradox for spin observables, \textit{Studies in History and Philosophy of Physics}, 38, 879--887

Sellars, W. (1953) Inference and Meaning. \textit{Mind}, LXII, 313--338

Shoesmith, D. J. and T. Smiley (1978) \textit{Multiple Conclusion Logic}, Cambridge University Press

Skolem, Th. (1955) Peano's Axioms and Models of Arithmetic. \textit{Studies in Logic and the Foundations of Mathematics}, 16, 1--14


Stalnaker, R. (1997) Reference and Necessity. In B. Hale, C. Wright, and A Miller (eds.) \textit{A Companion To Philosophy of Language}, Blackwell, 2017, 902--919 

Stebbing, L. S. (1935) The Method of Analysis in Metaphysics. \textit{Proceedings of the Aristotelian Society}, 33, 65--94

Stein, H. (1970) Is there a Problem of Interpreting Quantum Mechanics? \textit{Noûs}, 4, 93--103

St\"oltzner, M. (2002) Bell, Bohm, and Von Neumann: Some Philosophical Inequalities Concerning No-Go Theorems and the Axiomatic Method. In T. Placek and J. Butterfield (eds.) \textit{Non-locality and Modality}, Springer, 37--58

Stone, M. H. (1930) Linear Transformations in Hilbert Space. \textit{Proceedings of the National Academy of Sciences of the USA}, 16, 172--175


Suppes, P. (1993) \textit{Models and Methods in the Philosophy of Science: Selected Essays}, Springer

Susskind, L. and A. Friedman (2014) \textit{Quantum Mechanics. The Theoretical Minimum}, Basic Books

Tanona, S. (2004) ``Idealization and Formalism in Bohr’s Approach to Quantum Theory'', in \textit{Philosophy of Science}, 71, 683--695.

Tappenden, J. (2019) Infinitesimals, Magnitudes, and Definition in Frege. In P. A. Ebert and M. Rossberg (eds.) \textit{Essays on Frege's Basic Laws of Arithmetic}, Oxford University Press, 235--263

Tarski, A. (1934) Some methodological investigations on the definability of concepts. In \textit{Logic, Semantics, Metamathematics}, Hackett, 298--319

Teller, P. (1979) Quantum Mechanics and the Nature of Continuous Physical Quantities. \textit{The Journal of Philosophy}, 76, 345--361

Tennant, N. (2022) On the Adequacy of a Substructural Logic for Mathematics and Science. \textit{The Philosophical Quarterly}, 72, 4, 1002-1018 

Toader, I. D. (2015) Objectivity and understanding: a new reading of Carnap’s \textit{Aufbau}. \textit{Synthese}, 192, 1543--1557

Toader, I. D. (2018) `Above the Slough of Despond': Weylean Invariantism and Quantum Physics. \textit{Studies in History and Philosophy of Modern Physics}, 61, 18--24

Toader, I. D. (2021a) On the Categoricity of Quantum Mechanics. \textit{European Journal for Philosophy of Science}, 11, 17

Toader, I. D. (2021b) Permanence as a Principle of Practice. \textit{Historia Mathematica}, 54, 77--94

Toader, I. D. (2023) Einstein Completeness as Categoricity. \textit{Foundations of Physics}, 53, 39

Toader, I. D. (2024) Is Bohr's Correspondence Principle just Hankel's Principle of Permanence? \textit{Studies in History and Philosophy of Science}, 103, 137--145

Toader, I. D. (2025a) Distribution can be Dropped: Reply to Rumfitt, \textit{Analysis}, doi:10.1093/analys/anae093

Toader, I. D. (2025b) Distribution must be Dropped: Reply to Williamson, talk to the ASL Logic Colloquium, Technical University of Vienna, 7-11.07.2025

Toader, I. D. (2025c) Quantum Mechanics as a Carnapian Language. In G. Schiemer (ed.) \textit{The Legacy of the Vienna Circle}, Springer, doi:10.1007/978-3-031-80568-4

Uffink, J. (2020) Schrödinger’s reaction to the EPR paper. In M. Hemmo and O. Shenker 9eds.) \textit{Quantum, Probability, Logic. The Work and Influence of Itamar Pitowsky}, Springer, 545--566

van Fraassen, B. C. (1987) The Semantic Approach to Scientific Theories. In N. J. Nersessian (ed.) \textit{The Process of Science. Contemporary Philosophical Approaches to Understanding Scientific Practice}, Springer, 105--124

van Fraassen, B. C. (1991) \textit{Quantum Mechanics: An Empiricist View}, Clarendon Press

van Fraassen, B. C. (2008) \textit{Scientific Representation. Paradoxes of Perspective}, Oxford University Press.

Vaught, R. L. (1974) Model theory before 1945. In L. Henkin \textit{et al.} (eds.) \textit{Proceedings of the Tarski Symposium}, American Math. Society, 153--172

von Neumann, J. (1925) An axiomatization of set theory. In J. van Heijenoort (ed.) \textit{From Frege to G\"{o}del: A Source Book in Mathematical Logic, 1879-1931}, Harvard University Press, 1967, 393--413

von Neumann, J. (1931) Die Eindeutigkeit der Schr\"{o}dingerschen Operatoren. \textit{Mathematische Annalen}, 104, 570--578

von Neumann, J. (1932) \textit{Mathematische Grundlagen der Quantenmechanik}, Springer. Eng. tr. \textit{Mathematical Foundations of Quantum Mechanics} Princeton University Press, 2018

von Neumann, J. and O. Morgenstern (1944) \textit{Theory of Games and Economic Behavior}, Princeton University Press

Wallace, D. (2013) A prolegomenon to the ontology of the Everett interpretation. In A. Ney and D. Z. Albert (eds.) \textit{The Wave Function: Essays on the Metaphysics of Quantum Mechanics}, Oxford Unversity Press, 203--222

Wallace, D. (2020) On the Plurality of Quantum Theories: Quantum Theory as a Framework, and its Implications for the Quantum Measurement Problem. In S. French and J. Saatsi (eds.) \textit{Scientific Realism and the Quantum}, Oxford University Press, 78-102

Warren, J. (2020) \textit{Shadows of Syntax. Revitalizing Logical and Mathematical Conventionalism}, Oxford University Press

Weatherall, J. O. (2019) Part 1: Theoretical equivalence in physics. \textit{Philosophy Compass}, 14, e12592

Werner, R. F. (1999) EPR states for von Neumann algebras. arXiv:quant-ph/9910077

Weyl, H. (1930) \textit{Gruppentheorie und Quantenmechanik}, 2nd ed., Hirzel. Eng. tr. as \textit{The Theory of Groups and Quantum Mechanics}, Dover, 1931

Weyl, H. (1940) The Ghost of Modality. In M. Farber (ed.) \textit{Philosophical Essays in Memory of Edmund Husserl}, Cambridge University Press, 278--303

Weyl, H. (1949) \textit{Philosophy of Mathematics and Natural Science}, Princeton University Press, 2009

Weyl, H. (1953) Scientific Method (unpublished manuscript, Hs 91a:30 in the Archive at the ETH Library in Zurich)

Weyl, H. (1954) Erkenntnis und Besinnung (Ein Lebensr\"{u}ckblick). \textit{Studia Philosophica}, repr. in K. Chandrasekharan (ed.) \textit{Gesammelte Abhandlungen}, vol. IV, Springer, 631--649

Whewell, W. (1840) \textit{The Philosophy of the Inductive Sciences, Founded Upon Their History}, John W. Parker

Wigner, E. (1961) Remarks on the Mind-Body Question. In J. Mehra (ed.) \textit{Philosophical Reflections and Syntheses}, The Collected Works of Eugene Paul Wigner (2876, vol. B/6), 247--260


Williamson, T. (2018) Alternative logics and applied mathematics. \textit{Philosophical Issues}, a supplement to \textit{Noûs}, 28, 399--424

Zermelo, E. (1930) \"{U}ber Grenzzahlen und Mengenbereiche. Neue Untersuchungen \"{u}ber die Grundlagen der Mengenlehre. \textit{Fundamenta Mathematicae}, 16, 29-47 

Zilber, B. (2016) The semantics of the canonical commutation relations. \\ arXiv:1604.07745

Zilber, B. (2024) On the logical structure of physics. arXiv:2410.01846

\end{document}